\pdfoutput=1
\documentclass{aa}  

\newcommand{\micro}{$\mu$m}
\newcommand{\spc}{ }
\newcommand{\co}{\element[][13]{CO}}
\newcommand{\MH}{H$_{2}$}
\newcommand{\msun}{M$_{\odot}$}
\newcommand{\DEG}{$^{\circ}$}
\newcommand{\B}{$\beta$}
\newcommand{\Reff}{$R_{\rm eff}$}

\usepackage{lscape}             % For rotated long tables
\usepackage{rotating} % For rotating tables
\usepackage{longtable} %For tables which span several pages

\usepackage[varg]{txfonts}
\usepackage{natbib,twoopt}
\usepackage[breaklinks=true,hidelinks]{hyperref} %% to avoid \citeads line fills

\bibpunct{(}{)}{;}{a}{}{,}             %% natbib format for A&A and ApJ
\makeatletter
  \newcommandtwoopt{\citeads}[3][][]{\href{http://adsabs.harvard.edu/abs/#3}%
    {\def\hyper@linkstart##1##2{}%
     \let\hyper@linkend\@empty\citealp[#1][#2]{#3}}}
  \newcommandtwoopt{\citepads}[3][][]{\href{http://adsabs.harvard.edu/abs/#3}%
    {\def\hyper@linkstart##1##2{}%
     \let\hyper@linkend\@empty\citep[#1][#2]{#3}}}
  \newcommandtwoopt{\citetads}[3][][]{\href{http://adsabs.harvard.edu/abs/#3}%
    {\def\hyper@linkstart##1##2{}%
     \let\hyper@linkend\@empty\citet[#1][#2]{#3}}}
  \newcommandtwoopt{\citeyearads}[3][][]%
    {\href{http://adsabs.harvard.edu/abs/#3}
    {\def\hyper@linkstart##1##2{}%
     \let\hyper@linkend\@empty\citeyear[#1][#2]{#3}}}
\makeatother

\graphicspath{{./}{/home/emma/gradu/A_A_paper/A_A_paper_Oct_2020/pictures/}{/home/emma/gradu/data/All_data/PICTURES/contours_with_YSO/with_confirmed_YSOs/}{/home/emma/gradu/data/All_data/PICTURES/surface_brightness_maps/no_edge/}{/home/emma/gradu/data/All_data/PICTURES/NH2_SCUBA_contours/}{/home/emma/gradu/A_A_paper/NRO/NRO_new/pics2/}{/home/emma/gradu/A_A_paper/NRO/pics/full_fields/}}

\usepackage{graphicx}
\begin{document}

   \title{Characterization of dense \textit{Planck}\thanks{\textit{Planck} (\url{http://www.esa.int/Planck}) is a project of   the European Space Agency -- ESA -- with instruments provided by two scientific consortia funded by ESA member states (in particular the lead countries: France and Italy) with contributions from NASA (USA), and telescope reflectors provided in a collaboration between ESA and a scientific Consortium led and funded by Denmark.} clumps observed with \textit{Herschel}\thanks{{\it Herschel} is an ESA space observatory with science instruments provided by European-led Principal Investigator consortia and with important participation from NASA.} and SCUBA-2 }

   \author{E. Mannfors
          \inst{1}
          \and
          M. Juvela\inst{1}
          \and
          L. Bronfman\inst{2}
          \and
          D.J. Eden\inst{3}
          \and
          Jinhua He\inst{4,5,2}
          \and
          Gwanjeon Kim\inst{6}
          \and
          Kee-Tae Kim\inst{7,8}
          \and
          H. Kirppu\inst{1}
          \and
          T. Liu\inst{9}
          \and
          J. Montillaud\inst{10}
          \and
          H. Parsons\inst{11}
          \and
          Patricio Sanhueza\inst{12}
          \and
          Hsien Shang\inst{14}
          \and
          A. Soam\inst{15}
          \and
          K. Tatematsu\inst{6,13}
          \and
          A. Traficante\inst{16}
          \and
          M. S. V{\"a}is{\"a}l{\"a}\inst{14}
          \and
          Chang Won Lee\inst{7,8}
          }
   \institute{Department of Physics, P.O. box 64, FI- 00014, University of Helsinki, Finland\\
              \email{emma.mannfors@helsinki.fi}
              \and
              Astronomy Department, Universidad de Chile, Camino el Observatorio 1515, Las Condes, Santiago de Chile, Chile % Leonardo Bronfman, Jinhua He #3
              \and
              Astrophysics Research Institute, Liverpool John Moores University, IC2, Liverpool Science Park, 146 Brownlow Hill, Liverpool L3 5RF, UK % David Eden      
              \and
              Yunnan Observatories, Chinese Academy of Sciences, Phoenix Mountain, East Suburb of Kunming, 650216, Yunnan, PR China % Jinhua He #1
              \and
              Chinese Academy of Sciences South America Center for Astronomy, National Astronomical Observatories, CAS, Beijing 100101, China % Jinhua He #2 
              \and
              Nobeyama Radio Observatory, National Astronomical Observatory of Japan, National Institutes of Natural Sciences, Nobeyama, Minamimaki, Minamisaku, Nagano 384-1305, Japan % K. Gwanjeong, Tatematsu      
              \and
              Korea Astronomy and Space Science Institute, 776 Daedeokdae-ro, Yuseong-gu, Daejon 34055, Republic of Korea % Chang Won Lee 1, Kee-Tae Kim 1
              \and
              University of Science \& Technology, 176 Gajeong-dong,Yuseong-gu, Daejeon, Republic of Korea % Chang Won Lee 2, Kee-Tae Kim 2
              \and
              Shanghai Astronomical Observatory, Chinese Academy of Sciences, 80 Nandan Road, Shanghai, 200030, P.R. China % Tie Liu      
              \and
              Institut UTINAM -- UMR 6213 -- CNRS -- Univ. Bourgogne Franche Comté, OSU THETA, 41bis avenue de l’Observatoire, 25000 Besançon, France % Montillaud                    
              \and
              East Asian Observatory, 660 N. A'oh\={o}k\={u} Place, University Park, Hilo, HI 96720, USA % Harriet Parsons
              \and
              National Astronomical Observatory of Japan, National Institutes of Natural Sciences, 2-21-1 Osawa, Mitaka, Tokyo 181-8588, Japan % Patricio Sanhueza,
              \and
              Department of Astronomical Science, The Graduate University for Advanced Studies, SOKENDAI, 2-21-1 Osawa, Mitaka, Tokyo 181-8588, Japan %  Tatematsu
              \and
              Institute of Astronomy and Astrophysics, Academia Sinica. 11F of Astronomy-Mathematics Building, AS/NTU No.1, Section 4, Roosevelt Rd,Taipei 10617, Taiwan % Hsien Shang, Miikka Väisälä   
              \and
              SOFIA Science Center, Universities Space Research Association, NASA Ames Research Center, Moffett Field, California 94035, USA   % Archana Soam 
              \and
              IAPS-INAF, via Fosso del Cavaliere 100, I-00133, Rome, Italy % Alessio Traficante 
             }

   \date{Received Month Day, 1996; accepted Mo dd, 1997}

  \abstract
   {Although the basic processes of star formation (SF) are known, more research is needed on SF across multiple scales and environments. The Planck all-sky survey provided a large catalog of Galactic cold clouds and clumps that have been the target of several follow-up surveys.}
   {We aim to characterize a diverse selection of dense, potentially star-forming cores, clumps, and clouds within the Milky Way in terms of their dust emission and SF activity.}
   {We studied 53 fields that have been observed in the JCMT SCUBA-2 continuum survey SCOPE and have been mapped with \textit{Herschel}. We estimated dust properties by fitting \textit{Herschel} observations  with modified blackbody functions, studied the relationship between dust temperature and dust opacity spectral index $\beta$, and estimated column densities. We extracted clumps from the SCUBA-2 850\,\micro\spc maps with the FellWalker algorithm and examined their masses and sizes. Clumps are associated with young stellar objects found in several catalogs. We estimated the gravitational stability of the clumps with virial analysis. The clumps are categorized as unbound starless, prestellar, or protostellar.}
   {We find 529 dense clumps, typically with high column densities from (0.3--4.8)$\times 10^{22}$\,cm$^{-2}$, with a mean of (1.5$\pm$0.04)$\times10^{22}$\,cm$^{-2}$, low temperatures ($T\sim $10 --20\,K), and estimated submillimeter \B =1.7$\pm$0.1. We detect a slight increase in opacity spectral index toward millimeter wavelengths. Masses of the sources range from 0.04\,\msun\ to 4259\,\msun. Mass, linear size, and temperature are correlated with distance. Furthermore, the estimated gravitational stability is dependent on distance, and more distant clumps appear more virially bound. Finally, we present a catalog of properties of the clumps.}
   {Our sources present a large array of SF regions, from high-latitude, nearby diffuse clouds to large SF complexes near the Galactic center. Analysis of these regions will continue with the addition of molecular line data, which will allow us to study the densest regions of the clumps in more detail.}


   \maketitle

\section{Introduction}

Star formation (SF) processes involve all scales within the interstellar medium (ISM). Large filaments in molecular clouds (MCs) fragment into parsec-scale clumps and subparsec cores, which can eventually form young stellar objects \citep[YSOs;][]{origin_universality_of_stellar_IMF}. SF processes are affected by the combined effects of turbulence, magnetism, thermal pressure, and gravity. There is evidence that the relative importance of these forces changes in structures from filaments to cores \citep{structure_Nessie, 2018ApJ...859..151L,gravity_B_turbulence_G34, 2019ApJ...883...95S,2020MNRAS.491.4310T}. In addition, there is evidence that larger mass reservoirs are required for SF in energetic environments such as the Galactic center \citep{2019MNRAS.488.3972T}. Regions with second-generation SF, such as the edges of bubbles, are also likely to have higher column and volume density than the general ISM \citep{2012MNRAS.421..408T,2012ApJ...751...68L,2016ApJ...818...95L,2019MNRAS.487.1517L}, leading to an increased occurrence of SF in these regions \citep{2012ApJ...755...71K,2017A&A...605A..35P}. 
Changing environments make it necessary to study the interplay of these forces under different conditions. Regions of SF within the Milky Way provide an opportunity to study SF from small high-latitude clouds to the dense MCs within the Galaxy's spiral arms. We seek to extend the range of SF research by studying potentially star-forming clouds up to 4.5\,kpc in distance, a sample that  includes high-mass SF regions.

Filamentary structures are believed to be ubiquitous within the ISM \citep[e.g.][]{filaments_Aquila,GBS_initial_highlights,filaments_IC, filaments_swan,filaments_synthetic, filaments_TMC, filaments_Taurus_B211, filaments_to_cores,2015MNRAS.450.4043W}. Fragmentation creates clumpy condensations of radii under 2\,pc as well as smaller cores of radii under 0.5\,pc \citep{origin_universality_of_stellar_IMF, 2014MNRAS.439.3275W, 2015ApJ...804..141Z}. Whether a core will form stars in the future depends on the relative strength of gravity, thermal pressure, turbulence, and magnetic energy.

Nonthermal motions inside large MCs are supersonic \citep{physics_of_SF}, and kinetic and magnetic energy dominate at large scales \citep{magnetic_field,2017ApJ...846..122P}. The magnetic field dominates in the difuse ISM \citep{2005ApJ...624..773H,2013MNRAS.433.1675B}, while the effect of gravity and external pressure is significant on smaller scales \citep{theory_of_SF,2010ApJ...714..680M, GBS_Ophiuchus_MC}. Magnetic energy can be crucial to the formation of filaments, with filaments being either contained by magnetic flux or contracting along magnetic field lines \citep{doi:10.1146/annurev-astro-081811-125514, 2018A&A...620A..26J,2019MNRAS.487.3631G,gravity_B_turbulence_G34, 2019ApJ...883...95S}. Fragmentation at core scales is likely rapid enough that it does not depend on the larger environment \citep{2013A&A...557A.120K,2013A&A...554A..55H, Orion_MC_fragment, structure_Nessie}. In clouds that have already begun SF, feedback from protostars is effective at diminishing future SF, and at lower densities, destroying the parent cloud \citep{2015ApJ...798...32N,2015ApJ...801...33T,2018A&A...615A..94F}. The relative strengths of these forces depend on the scale at which they are observed and on the surrounding environment. To understand these various forces, it is necessary to incorporate multiple scales of SF into our research.

Cores that show signs of YSOs already are classified as protostellar (PS). Of the starless cores, those that are gravitationally bound and will likely form stars in the future are classified as prestellar (PRE), and those that are too diffuse for gravitational collapse as unbound starless (SL). In cases in which we refer to all starless cores, we use the acronym SL PRE. In time, unbound starless cores disperse unless they lose the internal support against collapse or more mass is accreted onto them.  We seek to answer if the star-forming status of dense clumps can be predicted from continuum data, by observing the properties available through continuum observations: temperatures, column densities, and masses.

Objects within the cold ISM are studied by observing dust and molecular lines; the latter method is important to the study of cloud stability and to that of general  kinematics \citep{1998A&A...336..150M,2000prpl.conf...59A,2007ApJ...666..982E,cepheus_GBS}. The spectral energy distribution (SED) of an accreting protostar peaks at under 100\,\micro\spc \citep{catalog_in_Lupus_cluster}, and thus emission in far-infrared (IR) wavelengths can be used to determine the presence of YSOs. At wavelengths of over 100\,\micro, dust emission from the cold, dense ISM is detectable, which is useful for estimating the temperature and density of the clouds. This emission can be approximated by modified blackbody (MBB) column emission \citep{JCMT_gould_belt_survey,Planck_all_sky_model}, which gives the opportunity to estimate the temperature of the emitting grains. According to theory, at far-IR and submillimeter wavelengths opacity spectral index $\beta$ should be mostly independent of temperature and restricted to the values of 1--2 \citep{doi:10.1002/9783527618156}. However, a dependence of spectral index on temperature was first detected by \citet{beta_pronaos}, and spectral index has been found to have values above 2 in cold regions \citep{GCC_VI}. Recent cm observations of Orion have shown significant mm excess compared to shorter wavelengths, lowering the estimated value of the opacity spectral index \citep{2019arXiv190505221M}. Uncertainty on the grain sizes and optical properties of the cold dust still exists. Understanding the behavior of this parameter is important because the value of the spectral index affects the shape of the MBB function and thus the derived dust properties.

The European Space Agency (ESA) \textit{Herschel} space observatory has been crucial to observing the cold ISM. Key surveys using \textit{Herschel} include Hi-GAL (the \textit{Herschel} infrared Galactic Plane Survey) (PI: S. Molinari), which mapped the inner Galactic plane \citep{Hi_gal} and the Gould Belt Survey (PI: P. Andr\'{e}), which imaged major SF regions within 500\,pc of the Sun in clouds within the Gould Belt \citep{2007PASP..119..855W,GBS_initial_highlights}. The Gould Belt Survey has resulted in multiple studies about nearby low-mass SF regions \citep[e.g.][]{JCMT_gould_belt_survey, dense_cores_in_Aquila, GBS_Ophiuchus_MC,cepheus_GBS, dense_cores_corona_australis,catalog_in_Lupus_cluster}. Cold, dense Galactic regions have been observed as part of the Galactic cold cores (GCC) survey \citep[PI: M. Juvela;][]{planck_early_XXIII, planck_early_XXII,PGCC_catalog} (see Sect. \ref{sec:observations}), using fields selected from the \textit{Planck} Galactic Cold Clumps (PGCC) catalog \citep{planck_early_XXIII,planck_early_XXII, PGCC_catalog}.

Many PGCC sources have been imaged with ground-based observatories, as in the SCOPE (SCUBA-2 Continuum Observations of Pre-protostellar Evolution) survey (PI T. Liu), which used the SCUBA-2 (Submillimetre Common-User Bolometer Array 2) instrument \citep{SCUBA_instrumentation} on the James Clerk Maxwell Telescope (JCMT; see Sect. \ref{sec:observations}). Preliminary analysis of SCUBA-2 sources combined with \textit{Herschel} observations has been performed \citep{sample_PGCC,SCOPE_survey} and a catalog of detected clumps has recently been published \citep{SCOPE_catalogue}.

In this paper, we analyze 53 high-column density fields from the SCOPE survey that have also been mapped with \textit{Herschel}. With this sample we seek to understand SF by analyzing star-forming regions in a wide range of Galactic environments. We combine \textit{Herschel} data with JCMT SCUBA-2 data to detect clumps and to study the dust properties of these dense environments and their SF status. This study of the properties of clumps, their dust properties, and SF activity also serves as preparation for future investigations where the chemical properties of the clumps will be studied.

This paper is structured as follows: Sect. \ref{sec:observations} describes the observation and reduction of these data and Sect. \ref{sec:methods} the methodology used in analysis. Section \ref{sec:results} presents the results of analysis of the dust properties and of the dense clumps, which are further discussed in Sect. \ref{sec:discussion}.

%__________________________________________________________________

\section{Observations\label{sec:observations}}

In this paper, we use \textit{Herschel} and JCMT dust continuum observations to study high-column-density clumps in 53 fields. Around half of these fields also have Taeduk Radio Astronomy Observatory (TRAO)\footnote{\url{https://radio.kasi.re.kr/trao/main_trao.php}} CO line data. These regions were originally selected from the PGCC catalog and are also the target of future chemical studies. We present instrumental characteristics for all bands used in Appendix \ref{sec:appendix_instrument_specs}.

\subsection{Target selection \label{sec:observations_target_selection}}

\begin{figure}
        \resizebox{\hsize}{!}{\includegraphics{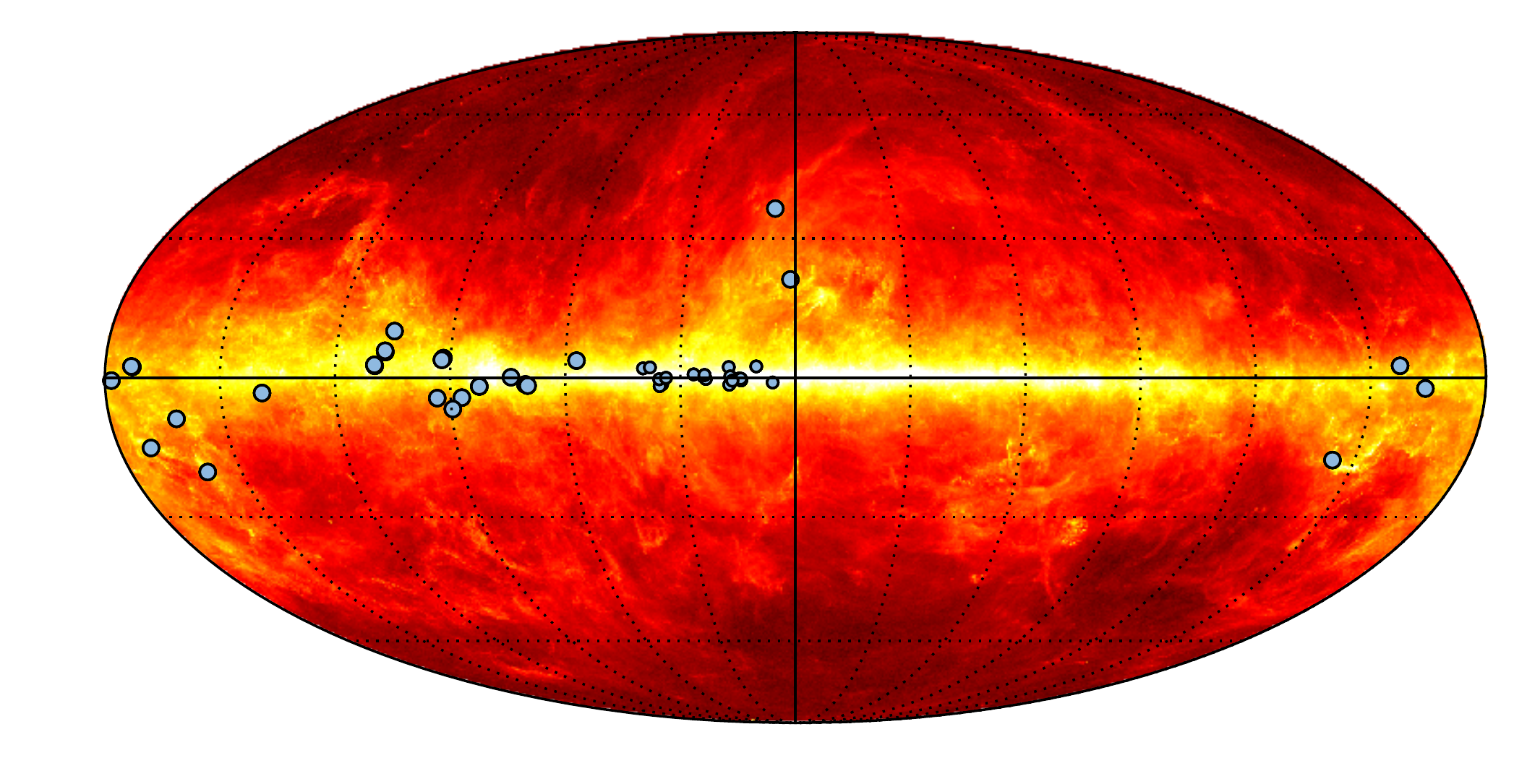}}
        \caption{Fifty-three fields denoted on the 857 GHz \textit{Planck} surface brightness map of the Galaxy. Some SCUBA-2 fields are located on the same \textit{Herschel} field, and the markers overlap at this scale. \label{fig:MW}}
\end{figure}

The PGCC has led to several follow-up projects, including the TOP (TRAO Observations of \textit{Planck} cold clumps; PI: T. Liu) survey, which aims to observe 2000 sources that have high column densities with TRAO in  $^{12}$CO and $^{13}$CO J = 1--0 transitions. Preference was given to sources that had also been imaged with \textit{Herschel}, but the objects are located across all Galactic latitudes and longitudes. From the TOP sources, 1235 PGCCs with high column densities, generally over $10^{21}$\,cm$^{-2}$ in \textit{Planck} measurements, were imaged with the JCMT on Mauna Kea in Hawai'i in 850\,\micro\spc continuum emission for the SCOPE survey \citep{SCOPE_survey,SCOPE_catalogue}. For a complete sample, some PGCCs with lower column densities at high Galactic latitude were also imaged. The purpose of the TOP-SCOPE joint survey is to study the formation and evolution of dense clumps and filaments in a variety of environments \citep{SCOPE_survey}. The regions are located across the whole sky, although mostly concentrated around the Galactic plane (see Fig. \ref{fig:MW}).

These fields are also being observed with the Nobeyama 45\,m radio telescope (PI: Ken'ichi Tatematsu) in DNC, HN$^{13}$C, N$_{2}$D$^{+}$, N$_{2}$H$^{+}$, CCS, HC$_{3}$N, and cyclic-C$_{3}$H$_{2}$. The purpose of the Nobeyama survey is to characterize the chemical evolution of the cores and find the earliest signs of SF, as in \citet{tatematsu_astrochem_properties}. The densest clumps in these 53 fields have also been observed in NH$_{\rm 3}$ lines with the Effelsberg 100 m radio telescope (PI: Viktor T\'{o}th). These molecular line observations are discussed in \citet{Tatematsu2020} and \citet{Kim2020}.

\subsection{SCUBA-2 observations \label{sec:observations_SCUBA_2}}

The SCUBA-2 850 \micro\spc observations were made as part of the SCOPE program \citep[][; project ID: M16AL003]{SCOPE_survey}. As the average angular size of PGCC sources is 8\arcmin\spc \citep{SCOPE_survey}, the fields were observed with the constant velocity (CV) daisy scan pattern, which is suitable for sources up to 12\arcmin\spc in size \citep{2014SPIE.9153E..03B}\footnote{\url{https://www.eaobservatory.org/jcmt/instrumentation/continuum/scuba-2/observing-modes/}}. The average observing time was 16 minutes per field \citep{SCOPE_survey}. Data reduction was performed using an iterative mapmaking technique described in \citet[][]{SCUBA_mapmaking}. This method compares detector beams and removes any low-frequency signals that are correlated between the beams, which removes contamination from the atmosphere but also any large-scale structure above 200\arcsec\, \citep{SCOPE_survey}. In the pipeline, a mean flux conversion factor of 554\,Jy\,pW$^{-1}$\,beam$^{-1}$ was used to convert data from pW to Jy\,beam$^{-1}$ by \citet{SCOPE_survey}. SCUBA-2 data show compact sources with high column densities and can thus be used to find dense cores in the fields. Owing to its resolution of $\sim$ 14\arcsec, SCUBA-2 can detect dense cores with size of $\sim$ 0.1\,pc to within 2\,kpc. As noise ranged from 6--10\,mJy\,beam$^{-1}$ in the central 3\arcmin\spc to 10--30\,mJy\,beam$^{-1}$ out to 12\arcsec\spc \citep{SCOPE_survey}, we assume an uncertainty of 10\,mJy\,beam$^{-1}$ for the SCUBA-2 data, as in \cite{sample_PGCC}.

\subsection{\textit{Herschel} observations \label{sec:observations_Herschel}}

We used data observed with two instruments on ESA's \textit{Herschel} space observatory \citep{ESA_Herschel}: SPIRE \citep[Spectral and Photometric Imaging Receiver;][]{SPIRE} and PACS \citep[Photodetector Array Camera and Spectrometer;][]{PACS}. The 250, 350, and 500 \micro\spc data were imaged simultaneously using the SPIRE photometer and the 70, 100, and 160 \micro\spc data using PACS. The PACS imaged either 70 and 160 \micro\spc or 100 and 160 \micro\spc maps simultaneously. 

Our sample consists of 53 TOP-SCOPE fields, 45 of which have both \textit{Herschel} PACS 160 \micro\spc and SPIRE data, 3 only SPIRE data, and 1 only PACS 160 \micro\spc data. In addition, 29 fields have PACS 70 \micro\spc data and 25 PACS 100 \micro\spc data. The observations are summarized in Table \ref{tbl:data_coverage} in Appendix \ref{sec:appendix_data_coverage}. Angular resolution of the \textit{Herschel} instruments ranges from 5.6\arcsec\spc at 70\,\micro\spc to 35.4\arcsec\spc at 500\,\micro. Unlike the SPIRE fields used, the PACS fields do not have a constant FWHM size due to different scanning velocities; a faster scan results in elongation of the effective beam.

\textit{Herschel} data were collected from the \textit{Herschel} Science Archive\footnote{\url{http://archives.esac.esa.int/hsa/whsa/}}. The data were reduced by the official pipeline to level 2.5 or 3, and the SPIRE maps include intensity zero-point corrections, which are obtained from comparison to \textit{Planck} measurements. The PACS images are not zero point corrected by the pipeline. All data were converted to units of MJy\,sr$^{-1}$.

\begin{figure}
        \resizebox{\hsize}{!}{\includegraphics{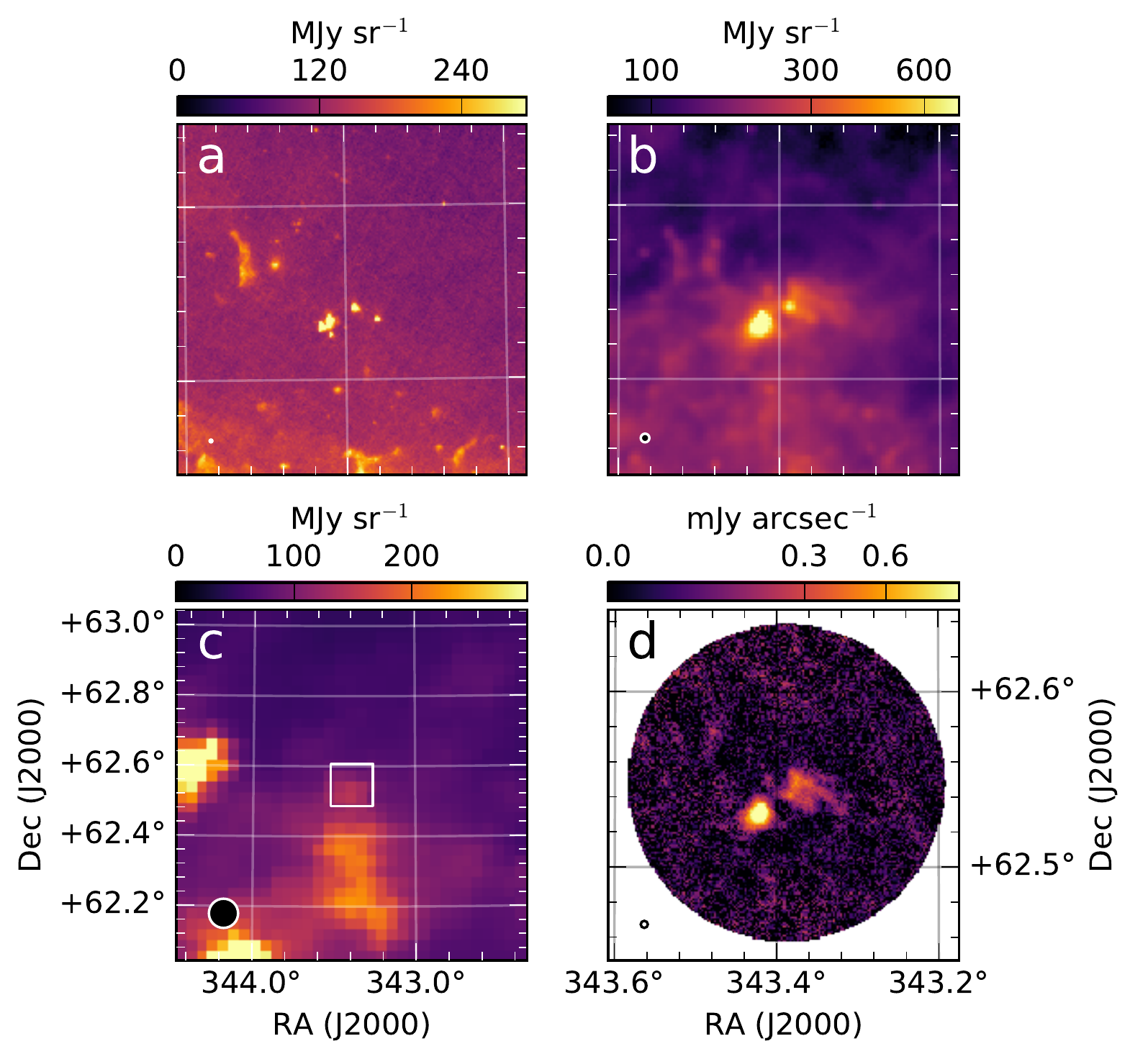}}
        \caption{Surface brightness maps of field G109.81+02.72, part of the Cepheus MC, in PACS 70\,\micro\spc (a), SPIRE 250\,\micro\spc (b), \textit{Planck} 857\,GHz ($\sim$350\,\micro) (c), and SCUBA-2 850\,\micro\spc (d). SCUBA-2 data are represented in units of mJy\,arcsec$^{-2}$, all others are in MJy\,sr$^{-1}$. The \textit{Planck} image is 1$\times$1\,degrees in size. All other images cover a smaller area of 0.2$\times$0.2\,degrees, as indicated by the rectangle drawn on the \textit{Planck} image. The beam is shown in the lower left corner of each frame.  \label{fig:70_850_um}}
\end{figure}

Extended emission is visible in the SPIRE observations (Fig. \ref{fig:70_850_um}b), which are useful for determining cloud characteristics such as temperature and column density. Hotter protostellar sources are visible at shorter wavelengths, and 70 or 100 \micro\spc detection are a strong indicator of ongoing SF. Figure \ref{fig:70_850_um}a shows an example of several protostellar sources at 70\,\micro,\spc which are forming inside the clumps visible in 850\,\micro\spc (Fig. \ref{fig:70_850_um}d). 

\subsection{TRAO observations}

The TRAO data were observed as part of the TOP key science program (PI: T. Liu), which observed $\sim$2000 PGCCs in \element[][12]{CO}(J=1-0) and \co(J=1-0), as well as $\sim$10\% of these fields in \element[18][]{C}\element[][]{O}(J=1-0). The \element[][12]{CO} and \co\spc lines were mapped simultaneously using on-the-fly mapping, with maps being $\leq $30\arcmin$\times$30\arcmin\spc in size. These data were regridded to a pixel size of 24\arcsec\spc and smoothed to 0.33\,km\,s$^{-1}$, and the baseline was removed with a first- or third-order polynomial. The front end of the TRAO telescope is the SEcond QUabbin Optical Image Array (SEQUOIA-TRAO), a $4\times 4$ array receiver with spatial separation of 89\arcsec, and the back end a fast Fourier transform spectrometer with $4096\times 2$ channels at a resolution of 15\,kHz \citep{2019ApJ...877..114C}. Full TRAO data reduction steps are available  in \citet{SCOPE_survey}, but we have listed instrumental specifics in Table \ref{tbl:TRAO_instrument_specs} in Appendix \ref{sec:appendix_instrument_specs} for \element[][12]{CO} and \co\spc observations.

\section{Methods\label{sec:methods}}

\subsection{Distance estimates}
\label{sec:methods_distance_estimate}

In addition to the distances from literature, we calculated distances to 27 fields based on $^{12}$CO TRAO line data, using a kinematic distance code written by Chris Beaumont\footnote{\url{http://www.ifa.hawaii.edu/users/beaumont/code/kdist-code.html}} and the rotation curve of \citet{1993A&A...275...67B}. This method gives two solutions for heliocentric distance \textit{d}; $d_{\rm min} - \Delta d, d_{\rm min} + \Delta d$, where $$d_{\rm min} = R_{\rm 0}\cos(l)$$ $$\Delta d =\sqrt{R^{2}-(R_{0} \sin(l))^{2}}, $$ 
using $R_{\rm 0}$ = 8.8\,kpc. 
One realistic positive solution exists for objects outside the solar circle and two solutions for objects inside of the solar circle \citep{2011A&A...526A.151R}. For sources inside the solar circle, we choose the lower solution for a field's distance. The derived galacto- and heliocentric distances for these fields are listed in Table \ref{tbl:kinematic_distance} in Appendix \ref{sec:appendix_distance}. For fields that have distances estimated in the literature using other methods other than than kinematic estimates, we use those distances instead owing to the uncertain nature of kinematic distances.

\subsection{Dust temperature and column density maps \label{sec:methods_general_ISM}}
To estimate temperature, column density, optical depth, and intensity of the fields, the observed data were fitted with a MBB function. The MBB takes the form 
\begin{equation}
\label{eq:MBB}
F_{\nu} = F_{\nu_{0}}\dfrac{B_{\nu}(T)}{B_{\nu_{0}}(T)}\Bigg(\frac{\nu}{\nu_{0}} \Bigg)^{\beta},
\end{equation}
where $F_{\nu}$ is the flux density at a frequency $\nu$ and $F_{\nu_{0}}$ the flux density at a reference frequency $\nu_{0}$  = 250\,\micro. When using a fixed value for the opacity spectral index, we fixed it to $\beta=1.8$, which should be appropriate for dense regions \citep{GCC_VI}.

Unless otherwise stated, analysis of dense clumps used only SPIRE bands. Before fitting the MBB, the maps of the three wavelengths were convolved to a 35.4\arcsec\spc resolution using a Gaussian beam with beam sizes listed in  Appendix \ref{sec:appendix_instrument_specs} and reprojected onto the same pixels. We did not use SCUBA-2 data in the MBB fitting of maps because, as a result of filtering, these do not contain extended emission. The SCUBA-2 measurements were used only in clump analysis, where we compared the fluxes to MBB fits made to the \textit{Herschel} data. Shorter wavelength PACS data are not used for analysis of clumps because PACS data exist only for 45 out of the 53 fields. The shorter wavelengths are sensitive to temperature variations and therefore their inclusion could also increase bias in the derived physical quantities (i.e., lead to higher dust temperature and lower column density estimates) \citep{beta_error_1,2012A&A...541A..33J}. We compare the results of SED fits including PACS and SCUBA-2 data with those that only use SPIRE data in Appendix \ref{sec:appendix_SED_fits_using_PACS_SCUBA}.

A least-squares fit of the MBB function was performed on each pixel. Based on the fits, we made maps of the dust color temperature \textit{T}, intensity $I_{250}$, and the optical depth $\tau_{250}$ at 250\,\micro. This intensity map is referred to as the $I_{250}$ map, whereas the original observed 250 \micro\spc map is referred to as the SPIRE 250 \micro\spc map. The $I_{250}$ map is the best fit of the MBB at each pixel. In the future we refer to dust color temperature simply as temperature. Optical depth at 250\,\micro\spc was calculated by
\begin{equation}
\tau_{250} = I_{250} / B_{\rm 250}(T),
\label{eq:tau}
\end{equation}
where $B_{\rm 250}$(\textit{T}) is the value of the Planck function at 250\,\micro. 
Column density $N$(\MH) was estimated using 
\begin{equation}
N(\rm H_{2}) = \tau_{250} / \kappa_{250} / \mu_{\rm H_2},
\label{eq:NH2}
\end{equation}
where $\mu_{\rm H_2} = 2.8 m_{\rm H}$ is the total mass relative to the H$_{2}$ molecule. We assume dust opacity $\kappa_{\nu}$ of \citet{kappa_beckwith}, which takes into account the gas-to-dust mass ratio
\begin{equation}
\label{eq:kappa}
\kappa_{\nu} = 0.1(\nu/1000 \text{ GHz})^{\beta} \rm cm^{2}g^{-1},
\end{equation}
using $\beta$ = 1.8. Optical depth and column density are quantities averaged over the effective beam, 35.4\arcsec.

\subsection{Clump detection \label{sec:methods_FellWalker} }

The FellWalker algorithm from the Starlink: FindClumps\footnote{\url{http://starlink.eao.hawaii.edu/docs/sun255.htx/sun255ss5.html}} package was used to extract clumps from SCUBA-2 850 \micro\spc maps. FellWalker is a watershed algorithm, which uses uphill paths rather than contours to define the clumps \citep{FellWalker}. FellWalker has been used in other SCUBA-2 surveys \cite[e.g.][]{GBS_Serpens,2017MNRAS.469.2163E,GBS_IC_5146,sample_PGCC,SCOPE_catalogue}, as well as in, for example,  \cite{california_MC},
using \textit{Herschel} column density maps. The input values used in the algorithm are presented in Appendix \ref{sec:appendix_FellWalker_input}. The RMS noise values are calculated from the reference regions (Table \ref{tbl:SCUBA_centers_1}, Appendix \ref{sec:appendix_scuba_centers}). FellWalker, along with ClumpFind and several other algorithms, are included in Starlink's CUPID--package. We compare the performance of some of these algorithms in Appendix \ref{sec:appendix_clumpFind}. 

\subsection{Clump properties \label{sec:methods_clump_chars}}

Clump properties were extracted using Starlink's EXTRACTCLUMPS algorithm from the CUPID package using 850 \,\micro\spc data as well. 
Up to 15 clump properties were derived, either directly with EXTRACTCLUMPS or extracted from \textit{Herschel} maps using those parameters. The characteristics are listed in Table \ref{tbl:characteristic_list}. The 850\,\micro\spc clump maps are at the resolution of the SCUBA-2 data, 14.6\arcsec, and the temperature and column density maps at the resolution of the SPIRE 500\,\micro\spc data, 35.4\arcsec. 
\begin{table}
        \caption{Characteristics calculated for each clump.  \label{tbl:characteristic_list}}
        \begin{tabular}{ll}
                \hline
                \hline
                Coordinates & Position angle \\
                %\hline
                Angular size &  Effective radius $R_{\rm eff}$ \\
                Flux density $F_{\rm \nu}$ & Temperature \textit{T} \\
                Intensity at 250\,\micro\spc $I_{\rm 250}$ & Mean column density \textit{N}(H$_{2}$) \\
                Mass \textit{M} & Peak column density \textit{N}(H$_{2}$)$_{850}$ \\
                Virial mass $M_{\rm vir}$ & Virial alpha $\alpha_{\rm vir}$\\
                Number of YSOs $N_{\rm YSO}$ & Number of confirmed YSOs $N_{\rm YSO,\Delta C}$ \\
                Clump type \\
                \hline
        \end{tabular}
        \tablefoot{Mean column density is calculated from \textit{Herschel} SPIRE data and peak column density from SCUBA-2 data. See Sect. \ref{sec:methods_YSO_ass_intensity_YSO_determination} for explanation on the difference between $N_{\rm YSO}$ and $N_{\rm YSO,\Delta C}$. }
\end{table}

Central coordinates, position angle, and angular sizes of the major and minor axes of the clump are given by the EXTRACTCLUMPS routine. Average values of dust temperature $\langle T \rangle$, surface brightness $\langle I_{250} \rangle$, and column density $\langle N(\rm H_{2}) \rangle$ over the clump area were calculated using the \textit{Herschel} SPIRE maps. The source flux density $F_{\rm \nu}$ is calculated by multiplying the average 850\,\micro\spc surface brightness with the clump solid angle, $F_{\rm 850}$ = $\langle I_{\rm 850} \rangle\Omega$. Volume density is estimated by dividing mass by the clump volume, mean molecular weight $\mu$, and atomic hydrogen mass $m_{\rm H_{2}}$ \citep{SCOPE_survey} as follows:
\begin{equation}
\label{eq:volume_density}
n = \dfrac{M}{\frac{4}{3}\pi R_{\rm eff}^{3}\mu m_{\rm H_{2}}}.
\end{equation}

Radius and mass are calculated as in \citet{SCOPE_survey}. 
Effective radius is written as
\begin{equation}
\label{eq:R_eff}
R_{\rm eff} = \sqrt{ab},
\end{equation}  
where \textit{a} and \textit{b} are the deconvolved spatial major and minor axis FWHMs, which are given in arcsec by EXTRACTCLUMPS. Masses of clumps were calculated from Eq. (\ref{eq:MBB}) as follows: 
\begin{equation}
\label{eq:mass}
M = \dfrac{F_{\rm \nu}d^{2}}{\kappa_{\rm \nu}B_{\rm \nu}(T)},
\end{equation}
where $F_{\rm \nu}$ is the flux density of the clump, \textit{d} the distance to the field, $\kappa_{\rm \nu}$ the dust opacity, and $B_{\rm \nu}(T)$ is the Planck function at the clump average dust temperature, setting the temperature of any clumps without temperature estimates or with \textit{T}$<$ 10\,K to the mean overall clump temperature.

\subsection{Gravitational stability \label{sec:methods_grav_stability}}

We estimate gravitational boundedness for all clumps with mass estimates, assuming that they are isolated systems supported against gravity by thermal pressure and turbulence. The level of turbulence within the clouds is estimated using \co\spc line width measurements from \cite{ke_wang_sampling_data_release}. Velocity dispersion $\sigma_{\rm tot}$ is calculated from the full width at half maximum (FWHM) line width $\Delta V$ of the \co\spc molecule and is written as
\begin{equation}
\label{eq:virial_sigma}
\sigma_{\rm tot} = \sqrt{\dfrac{k_{\rm B}T_{\rm kin}}{\overline{m}} + \Bigg( \dfrac{\Delta V^{2}}{8\rm ln(2)} - \dfrac{k_{B}T_{\rm kin}}{m}\Bigg)},
\end{equation}
where $k_{\rm B}$ is the Boltzmann constant, $\overline{m}$ = 2.33\,u the mean molecular mass (total mass divided by the number of particles), and \textit{m} = 29 u the mass of the \co\spc molecule. We assume kinetic temperature is equal to dust temperature, a good assumption at the density of star-forming cores \citep{goldsmith_dust_temp}. Virial mass is then calculated as 
\begin{equation}
\label{eq:virial_mass}
M_{vir} = \dfrac{k R_{\rm eff}\sigma^{2}}{G},
\end{equation}
where \textit{k} is a constant that depends on the density distribution. Assuming the density profile of the core scales with distance as $r^{-n}$ \citep{virial_mass_k}, \textit{k} can be calculated as 
\begin{equation}
k = \dfrac{5-2n}{3-n}.
\end{equation}
As in \citet{CO_line_analysis}, we assume \textit{n} = 1.5 and thus use a value of k = 1.333. Virial mass ratio $\alpha_{\rm vir}$ is calculated, with a critically bound core having observed mass equal to its virial mass. We define $\alpha_{\rm vir} = M_{\rm vir}/M_{\rm obs}$, and require prestellar cores to have $\alpha_{\rm vir} \leq 1$, as in \citet{dense_cores_corona_australis}.

\subsection{YSO association with clumps\label{sec:methods_YSO_ass}}

\subsubsection{Catalogs used \label{sec:methods_YSO_ass_catalogs}}

We searched for potential YSOs from several catalogs. We used AKARI and Herschel point source catalogs, the YSO candidate catalog of \cite{Vizier_Marton_catalog}, and the Spitzer catalog of \citet{2015AJ....149...64G}. As coverage between catalogs varies, the fields covered by each catalog are listed in Table \ref{tbl:data_coverage}.

The Akari/IRC point-source catalog (PSC)\footnote{\url{https://skyview.gsfc.nasa.gov/current/cgi/query.pl}} is based on the 9\,\micro\spc and 18\,\micro\spc observations of the AKARI satellite \citep{AKARI_IRC_documentation} and the AKARI/FIS bright-source catalog (BSC) data from 65--160\,\micro\spc observations \citep{AKARI_FIS_documentation}. The \textit{Herschel} PACS point source catalogs (hereafter PACS catalogs) include point sources at 70, 100, and 160\,\micro, as well as extended source and rejected source catalogs \citep{PACS_point_source_catalogue}. For YSO detection, we used the 70 and 100\,\micro\spc catalogs. The YSO candidates in the \citet{Vizier_Marton_catalog} catalog are based on source classification using the WISE (Wide-field IR Survey Explorer) satellite \citep{2010AJ....140.1868W} data and the 2MASS (Two Micron All-Sky Survey) catalog \citep{2006AJ....131.1163S}. The Marton catalog is divided into two sections: Class I and II YSO candidates, and Class III and more evolved YSO candidates. We used both of these. The Gutermuth catalog was created using the MIPSGAL 24\,\micro\spc Galactic Plane Survey matched with archival data from the 2MASS, GLIMPSE, and WISE surveys \citep{2015AJ....149...64G}. We searched the catalog for those sources with spectra matching those of Class I and II YSOs. Sources from different catalogs were merged if their separations were smaller than 2\arcsec.

The detection limit of the \textit{Herschel} catalogs depends on the wider environment of the field, but also on the mass of the protostar. For YSOs with mass 1\,\msun, a circumstellar mass of $\geq$ 0.5\,\msun\spc is required for detection; YSOs of 10\,\msun\spc require circumstellar masses of only $\geq $ 0.0025\,\msun\spc \citep{2013A&A...549A..67G}. The AKARI catalogs are supposed to have uniform detection limits over the whole sky, of $\sim$0.7\,Jy for PSC, and $\sim$1.8\,Jy for the BSC \citep{AKARI_IRC_documentation,AKARI_FIS_documentation}. In the different WISE bands a S/N of 20 was achieved at $\sim$12\,mag \citep{2012wise.rept....1C}. In the 2MASS source catalogs, a 10$\sigma$ detection level of $\geq$ 15 was achieved over the entire sky \citep{2006AJ....131.1163S}. The sensitivity of the MIPSGAL point sources is $\sim$2 and 75\,mJy at 24 and 70\,\micro, respectively \citep{2009PASP..121...76C}.

\subsubsection{Associating YSOs with clumps \label{sec:methods_YSO_ass_ass}}

Not all YSOs in a field are physically associated  with the detected clumps. The method for choosing clump association was based on spatial matching. If any of the clump's pixels are within the FWHM beam size of SCUBA-2 (14.6\arcsec) of a YSO, then the YSO is considered potentially associated with that clump and the clump is characterized tentatively as protostellar. 
If a YSO is near the boundary of two or more clumps, it is associated with the clump with most pixels within an 14.6\arcsec\spc distance of it. Otherwise, the YSO is considered a field YSO.

\subsubsection{Confirming YSO candidates\label{sec:methods_YSO_ass_intensity_YSO_determination} }

To seek further corroboration of the physical connection between YSO candidates and clumps, we compare the mean intensity of a circle within 12\arcsec\spc of the YSO in PACS 160\,\micro\spc with that in SPIRE 250\,\micro. Before this analysis the PACS images are converted to MJy\,sr$^{-1}$ and both fields are convolved to 20\arcsec\spc resolution. The average signal in a reference region chosen from a low-intensity region of each map is used to subtract the background. Furthermore, other YSO candidates in the field are masked to prevent contamination. The increased temperature at the YSO location results in increased short-wavelength emission, which should be visible as higher intensity at 160\,\micro. This is compared to a reference annulus between 36\arcsec\spc and 84\arcsec\spc of the YSO. 
We define the change in \textit{Herschel} intensity ratios $\Delta C$ as the difference between the ON and OFF areas. To be significant, $\Delta C$ is required to be significantly positive compared to the fluctuations, that is, 
\begin{equation}
\label{eq:dmean}
\Delta C = \Bigg(\dfrac{I_{\rm 160}}{I_{\rm 250}}\Bigg)_{\rm ON} - \Bigg(\dfrac{I_{\rm 160}}{I_{\rm 250}}\Bigg)_{\rm OFF} > \xi\times \sigma\Bigg( \dfrac{I_{\rm 160}}{I_{\rm 250}}\Bigg),
\end{equation}
where $\xi$ is a threshold value that takes into account the convolution of the original data and $I_{\rm 160}$ and $I_{\rm 250}$ are the 160\,\micro\spc and 250\,\micro\spc intensities, respectively. The quantity $\sigma$ refers to the standard deviation of the ratios. We estimate $\xi$ by simulating maps of white noise and taking the 95\% probability value of the distribution of $\Delta C$. We find a value $\xi$ = 12.98.

In figures and tables, we refer to clumps that have additional confirmation from a significantly positive $\Delta C$ as PS-C, and those that do not as PS. We note that \textit{Herschel} intensity ratio does not provide absolute confirmation of the YSO, only a stronger probability of association with the clump in question.

\section{Results\label{sec:results}}

In this section we present the results of our analysis of these data, which are discussed in more detail in Sect. \ref{sec:discussion}. The fields show characteristics we would expect from dense regions. In this paper, all sources are called clumps even though they span a large range of linear sizes, from nearby cores to entire distant clouds. Owing to the resolution of the SCUBA-2 instrument, individual cores are not resolved in distant fields.

\subsection{Distances of the examined fields \label{sec:results_distances}}

Distances to 23 fields were estimated in \citet{GCC_IV_cold_submm}, where the authors use kinematic distance and extinction modeling. Distances to ten fields have been estimated in the PGCC \citep{PGCC_catalog}; four of these also in \cite{GCC_IV_cold_submm}. The distance estimates in the two catalogs are consistent within their uncertainties. 

Twenty-seven new kinematic distances were derived from line-of-sight velocities of \element[][12]{CO} spectra from TRAO (See Sect. \ref{sec:methods_distance_estimate}). The derived galacto- and heliocentric kinematic distances are listed in Table \ref{tbl:kinematic_distance}, with all distances used in this paper in Table \ref{tbl:distances_GCCIV_2} in Appendix \ref{sec:appendix_distance}. Assuming uncertainties of $\sim$5\,km\,s$^{-1}$ in velocity dispersion \citep{1984ApJ...281..624S,1985ApJ...295..422C,2017AJ....154..140W}, we derive a final uncertainty in $v_{\rm LSR}$ of 50\%. Adopting uncertainties of 0.2\,kpc for  $R_{\rm 0}$, 8\,km\,s$^{-1}$ for $v_{\rm 0}$ \citep{2014ApJ...783..130R}, and 5\arcmin\spc for \textit{l} and \textit{b} (the resolution of Planck), we derive an uncertainty of $\sim$51\% for galacto- and heliocentric distances. As we are unable to solve the near-far kinematic distance ambiguity, we increase the uncertainty in heliocentric distance to $^{+100}_{-51}$\%. \citet{GCC_IV_cold_submm} provide uncertainty estimates for their derived distances, which are on average 53\%, and we adopt an uncertainty of 30\% for the distances from the PGCC that do not have previous uncertainty estimates. The fields in this sample are located at distances of 0.1-4.5\,kpc, with a mean at around 1\,kpc. Near $\ell$ $\sim$ 0\DEG, many fields are likely to be associated with the Sagittarius arm and near $\ell \sim$ 180\DEG\spc with the Perseus arm \citep[see Fig. 3 in][]{galactic_structure}. We also note that while we analyzed the projected two-dimensional distance between clumps, it is possible that these clumps are more distant along the line of sight. To solve this question, molecular line data of all of the clumps would be required.

\subsection{The general ISM \label{sec:results_general_ISM}}

\begin{figure}
        \resizebox{\hsize}{!}{\includegraphics{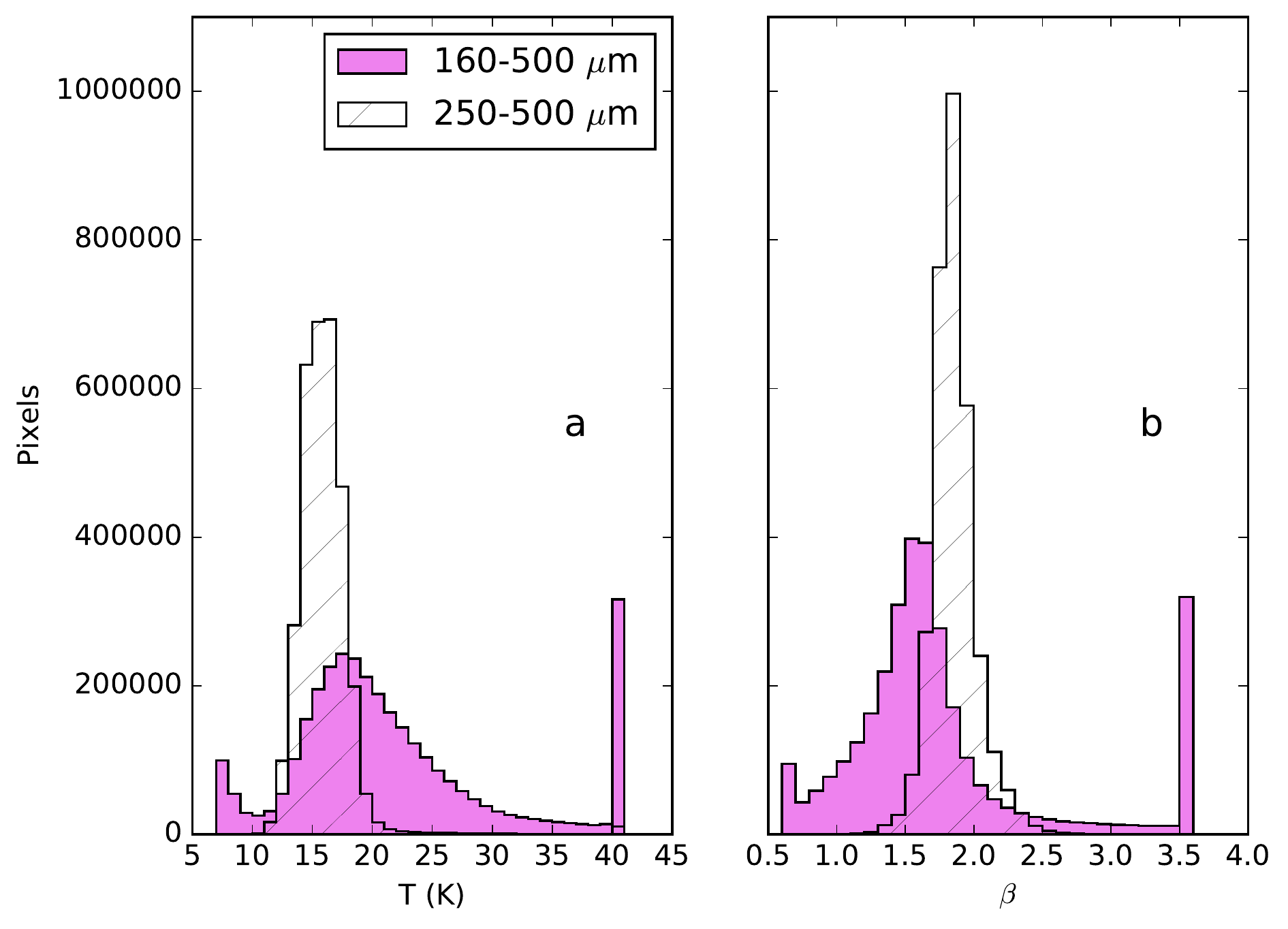}}
        \caption{Temperature (a) and spectral index (b) histograms for the entire fields, computed with SED fits of the 250-500\,$\mu$m (hatched histogram) and 160-500\,$\mu$m (purple histograms) data. Only areas covered by both PACS and SPIRE observations are used. As the MBB script fits temperatures between 5-40\,K, and spectral index between 0.6-3.5, thus the first and last bins include all pixels with T or $\beta$ outside of these ranges.  \label{fig:fullfield_T_beta}   }
\end{figure}

We calculated the intensity, dust temperature, column density, and optical depth for the 45 \textit{Herschel} fields with full SPIRE and PACS 160\,\micro\spc data. Analysis was performed with constant $\beta = 1.8$ as well as varying $\beta$. The PACS and SPIRE fields were background subtracted. The coordinates of the reference regions used for zero-point correction are listed in Appendix \ref{sec:appendix_scuba_centers}. The same analysis was performed using only the SPIRE 250--500\,\micro\spc fields relying on the zero points set in the archival maps. Derived temperature, column density, and spectral index maps are in Appendix \ref{sec:appendix_extended_emission} and the average values are listed in Table \ref{tbl:full_field_values}. We plot histograms of temperature and opacity spectral index for all fields with SPIRE and PACS data in Fig. \ref{fig:fullfield_T_beta}. As PACS and SPIRE coverage varies within individual fields, this figure includes only pixels that have data at all wavelengths.

\begin{figure}
        \resizebox{\hsize}{!}{\includegraphics{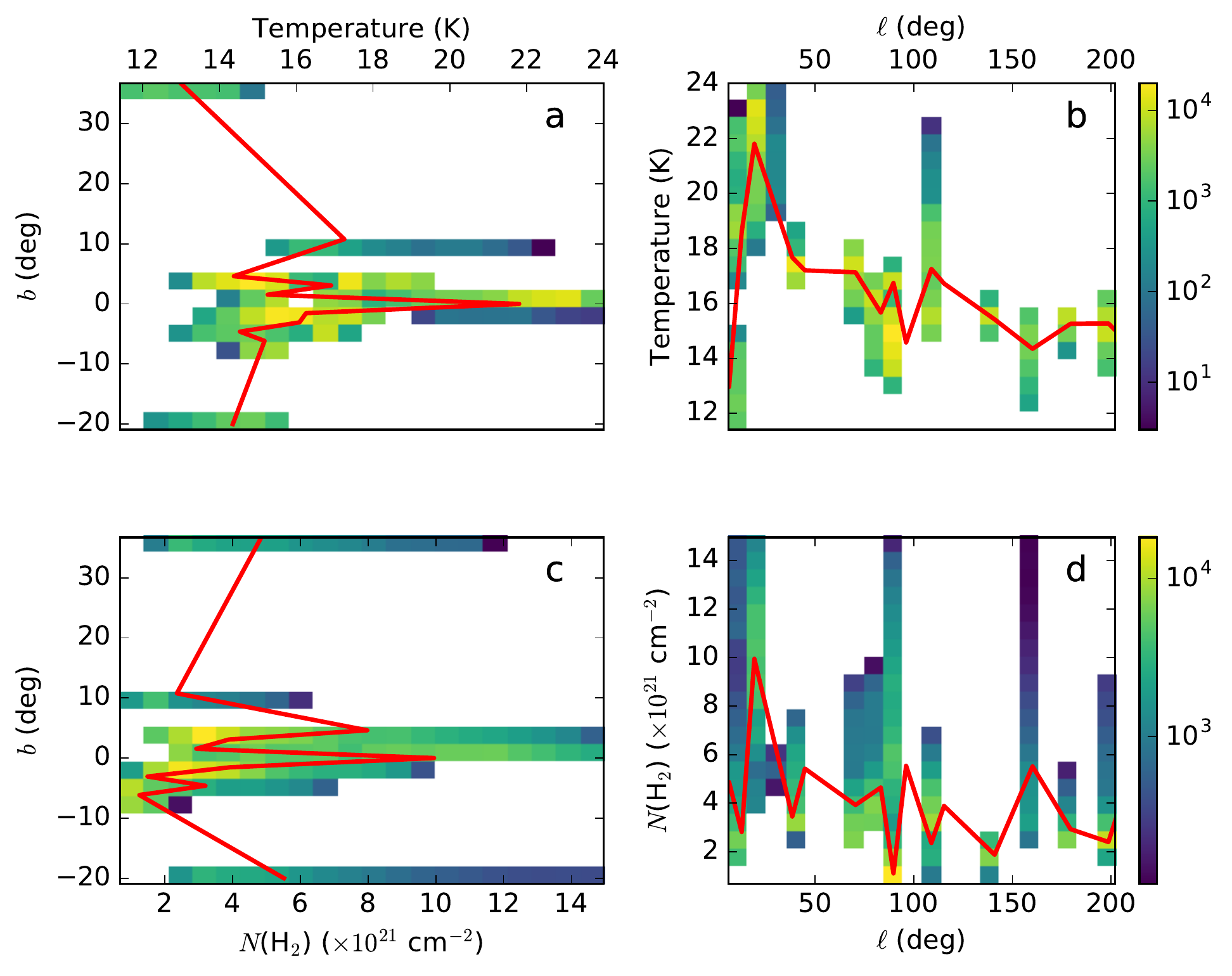}}
        \caption{Average properties of the 53 fields. The frames show 2D histograms of (a) temperature plotted against Galactic latitude \textit{b,} (b) temperature plotted against Galactic longitude $\ell$, (c) column density plotted against Galactic latitude \textit{b}, and (d) column density plotted against Galactic longitude $\ell$. The red line plots the averages for each bin. The plots are based on fits to SPIRE data only.  \label{fig:galactic_coordinates_vs_temp} }
\end{figure}

The average color temperature of the fields is 18.8\,K and the average column density $N({\rm H}_2)=$ 0.4$\times 10^{21}$\,cm$^{-2}$. Using only 250--500\,\micro\spc data results in lower temperature, higher optical depth, and higher column density estimates. This is expected as 160\,\micro\spc emission is more sensitive to warm dust along the line of sight. The temperature derived from 160--500\,\micro\spc data ranges from 7\,K to 40\,K, and column densities from 2.7$\times10^{19}$ to 1.5$\times10^{22}$\,cm$^{-2}$. We plot temperature and column density as a function of Galactic latitude \textit{b} and Galactic longitude $\ell$ in Fig. \ref{fig:galactic_coordinates_vs_temp}. Both mean temperature and column density are highest at low Galactic latitudes, although the high-latitude bins ($|b|> 10$\degr) often contain only one or two nearby fields.

The mean spectral index of the fields is around 1.7--1.9; the standard deviation under 0.5. \B\spc ranges from 0.6 -- 3.5. Variations of opacity spectral index do not, in general, have a large effect on the properties derived for the regions. Only the changes in optical depth and column density (probabilities \textit{p} = 0.09 and \textit{p} = 0.001 for SPIRE and SPIRE+PACS, respectively for the null hypothesis of no change) are significant (see Table \ref{tbl:beta_t_test}). Thus we use maps with constant spectral index for analysis of dense clumps.

\begin{table*}
        \centering
        \caption{ Parameters of the full \textit{Herschel} fields \label{tbl:full_field_values}}
        \begin{tabular}{l|lllll}
                \hline
                \hline
                Wavelength & $\langle T \rangle$ & $\langle I \rangle$  & $\langle  N(\rm H_{2})\rangle$ & $\langle  \tau\rangle$ & $\langle  \beta\rangle$\\
                used  &  (K) &  (MJy\,sr$^{-1}$) &  (10$^{21}$\,cm$^{-2}$) & (10$^{-3}$)&\\
                \hline
                250-500\,\micro & 17.1 $\pm$ 3.1 & 261.2 $\pm$ 480.9 & 2.6 $\pm$ 2.9 & 1.7 $\pm$ 1.9 & 1.8 \\ % & B1.8\_ ML
                
                160-500\,\micro  & 20.3 $\pm$ 6.7 & 142.5 $\pm$ 338.8 & 1.3 $\pm$ 2.2 & 0.9 $\pm$ 1.5 & 1.8\\ % & B1.8\_ ML
                \hline
                250-500\,\micro  & 16.5 $\pm$ 3.5 & 261.0 $\pm$ 481.1 & 3.1 $\pm$ 3.2 & 2.1 $\pm$ 2.1 & 1.9 $\pm$ 0.3\\ %  & ML
                
                160-500\,\micro  & 22.7 $\pm$ 8.5 & 142.2 $\pm$ 338.2 & 1.2 $\pm$ 2.2 &  0.8 $\pm$ 1.5 & 1.7 $\pm$ 0.7 \\% & bgsub\_ ML
                \hline
                
                Average:  & 17.7 $\pm$ 4.8  &  239.7 $\pm$ 460.9 & 2.6 $\pm$ 3.0 & 1.7 $\pm$ 2.0 &  1.9 $\pm$ 0.4\\
                \hline
                
        \end{tabular}   
        \tablefoot{Mean temperature \textit{T}, intensity \textit{I}$_{250}$, column density \textit{N}(H$_{2}$), optical depth $\tau_{250}$, and opacity spectral index $\beta$ calculated for the full map, for pixels with value over 0, using the various methods. The first two maps have constant spectral index. }
\end{table*}

\begin{table}
        \caption{Effect of constant vs. variable spectral index on derived field properties. }
        \label{tbl:beta_t_test}
        \begin{tabular}{l|ll||l|ll}
                \hline
                \hline
                \multicolumn{3}{l||}{SPIRE maps only} & \multicolumn{3}{l}{SPIRE and PACS maps} \\
                Variable & \textit{t} & \textit{p} & Variable & \textit{t} & \textit{p} \\
                \hline
                \textit{T} & 0.69  & 0.497  & T & -0.13 & 0.895 \\
                \textit{I} & $<10^{-2}$ & 0.998  & I & $<10^{-2}$ & 0.999 \\
                \textit{N}(\MH) &  -1.73  & 0.091   & N(\MH) & -3.57 & 0.001 \\
                $\tau$ & -1.73 & 0.091  & $\tau$ & -3.57 & 0.001 \\

                \hline
        \end{tabular}
        \tablefoot{Student-t test comparing the mean values of maps with constant and variable spectral index, for SPIRE maps only (columns 1-3) and SPIRE+PACS 160\,\micro\spc maps (columns 4-6). }
\end{table}

\subsection{The temperature-$\beta$ relation \label{sec:results_general_ISM_T_beta}}

The temperature dependence of opacity spectral index \B\spc is a much-studied feature, which provides information on the chemical composition, structure, and size distribution of interstellar dust grains \citep{1994A&A...291..943O,1994A&A...288..929K, 1998ApJ...496.1058M,2005ApJ...633..272B,2007A&A...468..171M,2011A&A...525A.103C,2011A&A...527A.111J}. However, as instrumental noise causes false anticorrelation between temperature and dust opacity spectral index \citep{beta_error_1,beta_error_2,2012A&A...541A..33J,2013A&A...556A..63J}, how much of this anticorrelation is physical is uncertain.

\begin{figure}
        \resizebox{\hsize}{!}{\includegraphics{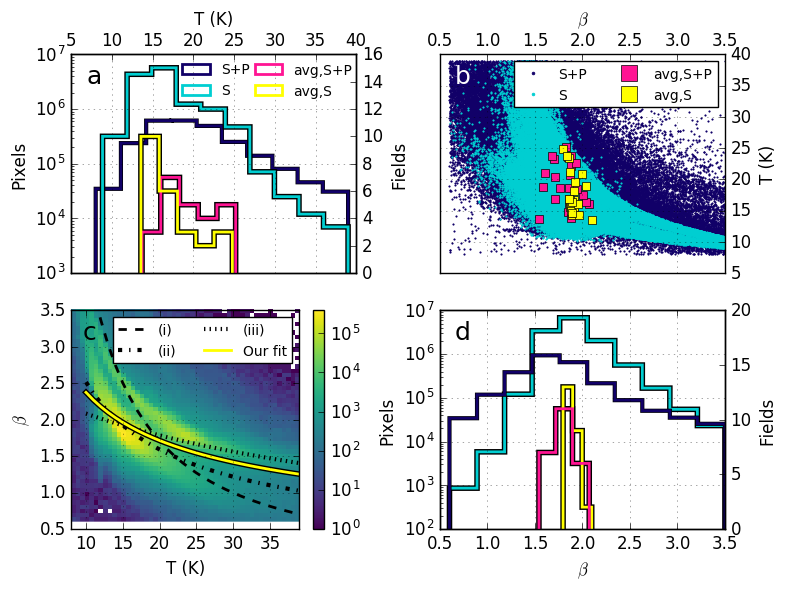}}
        \caption{(a) Histogram of field temperature. (b) Spectral index plotted against temperature. (c) 2D histogram of temperature plotted against spectral index. (d) Histogram of field spectral index. The values are calculated from background-subtracted data in the ranges 250--500\,\micro\spc (light blue, S) and 160--500\,\micro\spc (dark blue, S+P). In frames (a) and (d), these histograms correspond to the left-hand y-axis. Every tenth value of those pixels with $9 < T < 39$\,K and $0.6 < \beta < 3.4$ is included in frame (b) and every pixel in the other frames. Yellow and pink stand for averages of each field (\textit{n} = 24) with the four-map (avg, S+P) and three-map (avg, S) combinations, respectively, and in frames (a) and (d) correspond to the right-hand y-axis. $\beta$-$T$ relations from the literature are also plotted in (c), where (i) $\beta=1.71\times(T/20 K)^{-1.33}$  \citep{beta_HI_Gal}, dashed line, (ii) $\beta=11.5\times T^{-0.66}$ \citep{beta_archeops}, dash-dotted line, and (iii) $\beta=1/(0.4+0.008\times T)$ \citep{beta_pronaos}, dotted line. The relation derived for our data is plotted on frame (c) in yellow. \label{fig:beta_T_plot}}
\end{figure}

\begin{figure}
        \resizebox{\hsize}{!}{\includegraphics{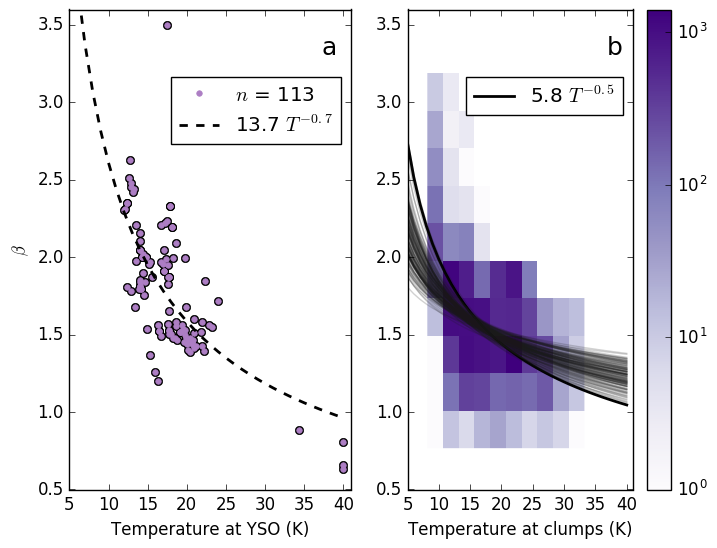}}
        \caption{The temperature-$\beta$ relation, taking into consideration only those pixels where (a) a YSO is located (\textit{n} = 113 pixels) and (b) at the location of clumps (\textit{n} = 15\,920 pixels). The color bar corresponds to the number of pixels in the case of b. We plot the calculated relations for 113 random points from the clumps sample in gray in frame b. \label{fig:beta_T_clumps_YSO}}
\end{figure}

We examined the (T, \B) values from the 160-500\,\micro\spc and 250-500\,\micro\spc fits with the function
\begin{equation}
\label{eq:beta_T}
\beta(T;A, \alpha) = A\times T^{-\alpha}\end{equation}
\citep{beta_archeops}. Fitting data in the range \textit{T}=8-39\,K and $\beta$=0.6-3.5 resulted in $\beta=7.01\times T^{-0.47}$ (Table \ref{tbl:beta_T}, Fig.\ref{fig:beta_T_plot}). We also examined the relation separately at the YSO and clump locations using 160-500\,\micro\spc data (Fig. \ref{fig:beta_T_clumps_YSO}). The relation appears to be steeper at the location of YSOs compared to at dense clumps or over the entire fields. However at the location of YSOs we have only 113 pixels, compared to 15\,920 at clumps. To test whether the differing numbers of points had an effect on the derived slope of the relation, we randomly selected 113 pixels from the clump locations and calculated the temperature-$\beta$ relation, repeating this 137 (= 15\,920/133) times. We find a mean slope of $\langle\alpha \rangle = 0.3 \pm 0.1 $, showing that the temperature-$\beta$ relation is steeper at the location of YSOs. To estimate the accuracy of all these derived relations, we compare our results with simulations where the true spectral index values are independent of temperature.

\begin{figure}
        \resizebox{\hsize}{!}{\includegraphics{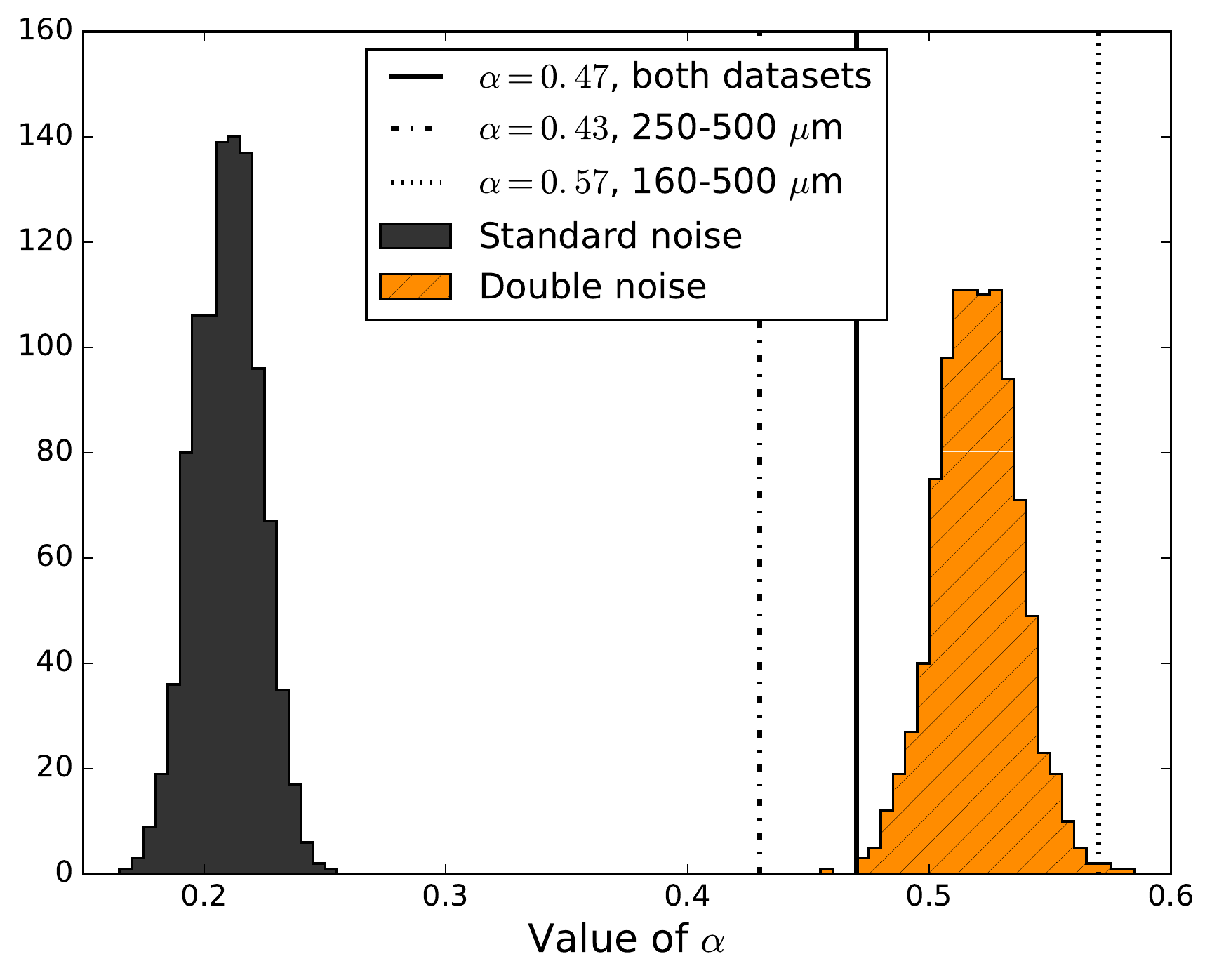}}
        \caption{ Histogram of derived values of $\alpha$ using 1000 sets of 1000 SEDs each, with assumed observational noise (black) and double assumed noise (orange). Our derived relations are plotted with a black dotted line.   \label{fig:alpha_hist}}
\end{figure}

To estimate the effects of this contamination by noise, we performed Monte Carlo simulations using 1000 data points with the mean and standard deviation after addition of noise matching that of our data. We fixed the value of $\beta$ to 1.8 and calculate intensities from Eq. (\ref{eq:MBB}). To this intensity we added Gaussian noise at 2\% of intensity for SPIRE wavelengths and 5\% at 160\,\micro, the predicted uncertainties for the two instruments \citep{spire_handbook}. We calculated the new temperature and spectral index for this noisy data using the methodology in Sect. \ref{sec:methods_general_ISM}, and we calculated the \textit{A} and $\alpha$ parameters, which are compared to those of our derived relation. We repeated this 1000 times. The histogram of derived $\alpha$ is plotted in gray in Fig. \ref{fig:alpha_hist}. We find 100\% of derived values of $\alpha$ to be below 0.47, which suggests that some of the T--$\beta$ anticorrelation is not due to noise. We also perform an additional test, by doubling noise to 4\% of intensity at 250--500\,\micro\spc and 10\% at 160\,\micro, and plot this in orange in Fig. \ref{fig:alpha_hist}. With this, we find 99.9\% of values above 4.7.

\begin{table}
        \caption{Parameters of the fitted $\beta = A\times T^{-\alpha}$ relations }
        \label{tbl:beta_T}
        \begin{tabular}{l|ll}
                \hline
                \hline
                Dataset & A & $\alpha$\\
                \hline
                250--500\,\micro &  6.32 & 0.43 \\ % 3.3 minutes
                160--500\,\micro & 9.30 & 0.57 \\
                Both datasets & 7.01 & 0.47 \\
                At YSOs & 13.7 & 0.72 \\
                At clumps & 5.76 & 0.46\\
                \hline
        \end{tabular}
        \tablefoot{Coefficients derived for Eq. (\ref{eq:beta_T}) using $\chi^{2}$ method. The last two rows use 160--500-\,\micro\spc data, taking into account only those pixels at YSOs or clumps, and for which both temperature and spectral index are above zero.  }
\end{table}

\subsection{ Herschel spectra and 850\,$\mu$m measurements \label{sec:results_Herschel_spectra_and_850um} }

Because of previous reports of spectral index variations as a function of frequency \citep{1995ApJ...451..188R,2012A&A...537A.113P,GCC_VI,2019arXiv190505221M}, we wished to extend the SED analysis to SCUBA-2 data at 850\,$\mu$m. In the comparison with satellite data, a major concern is the spatial filtering that the sky-noise reduction introduces into the SCUBA-2 maps. This could be taken into account by filtering the Herschel observations so that the loss of low spatial frequencies would be similar to the filtering in the ground-based observations \citep{JCMT_gould_belt_survey, sample_PGCC}. However, we concentrate here only on the cores at small spatial scales where the effects of filtering are smaller. \citet{JCMT_gould_belt_survey} simulated flux loss due to SCUBA-2 filtering, finding that on scales below 2.5\,\arcmin\, objects are fully recovered. We also examined flux loss using \textit{Herschel} 250\,\micro\spc maps, comparing the clump fluxes extracted from the original maps with maps run through the SCUBA-2 reduction pipeline. The approach is therefore the same as in \citet{JCMT_gould_belt_survey}, but in this work we use the actual structures in the maps and  similar values for the aperture and reference-annulus size as in the SCUBA-2 analysis. In the sample of 15 clumps, the filtering caused a flux loss of (1.3$\pm$6.2)\%, where the flux loss is under 5\% in 80\% of cases. The filtering thus introduces some scatter but only a small bias, which is statistically compatible with zero.

We convolved the 250--850\,$\mu$m maps to a common 40$\arcsec$ resolution and color corrected them assuming a MBB spectrum $B_{\nu}\rm (T=17\,K)\nu^{1.8}$. We constructed SEDs from clump peak values from which the local background was subtracted. The background estimates correspond to the median over an ellipse that starts at a distance of 0.7 times the clump FWHM (using the major and minor axis FWHM values) and is 30$\arcsec$ wide. The SCUBA-2 maps should recover spatial scales up to $\sim 200\arcsec$ and because the median FWHM size of the clumps is less than 1$\arcmin$, the systematic effects from the spatial filtering should remain small compared to other sources of uncertainty. We do not apply any aperture corrections because we are only interested in the SED shape and all input data are already at the same resolution and even use the same pixelization. 

We selected the most reliable set of cores by setting a threshold for the minimum signal-to-noise ratio (S/N) in each band. The 250-500\,$\mu$m data were fitted with MBBs, estimating the uncertainties with a Markov chain Monte Carlo procedure. Figure~\ref{fig:SED} in Appendix \ref{sec:appendix_herschel_spectra} shows the fits for nine random clumps with S/N $>3$ in all bands. The green boxes indicate the 5\%-95\% confidence interval for the extrapolated 850\,$\mu$m value. In most cases the observed 850\,$\mu$m point falls inside the box.

\begin{figure*}
        \sidecaption
        \includegraphics[width=12cm]{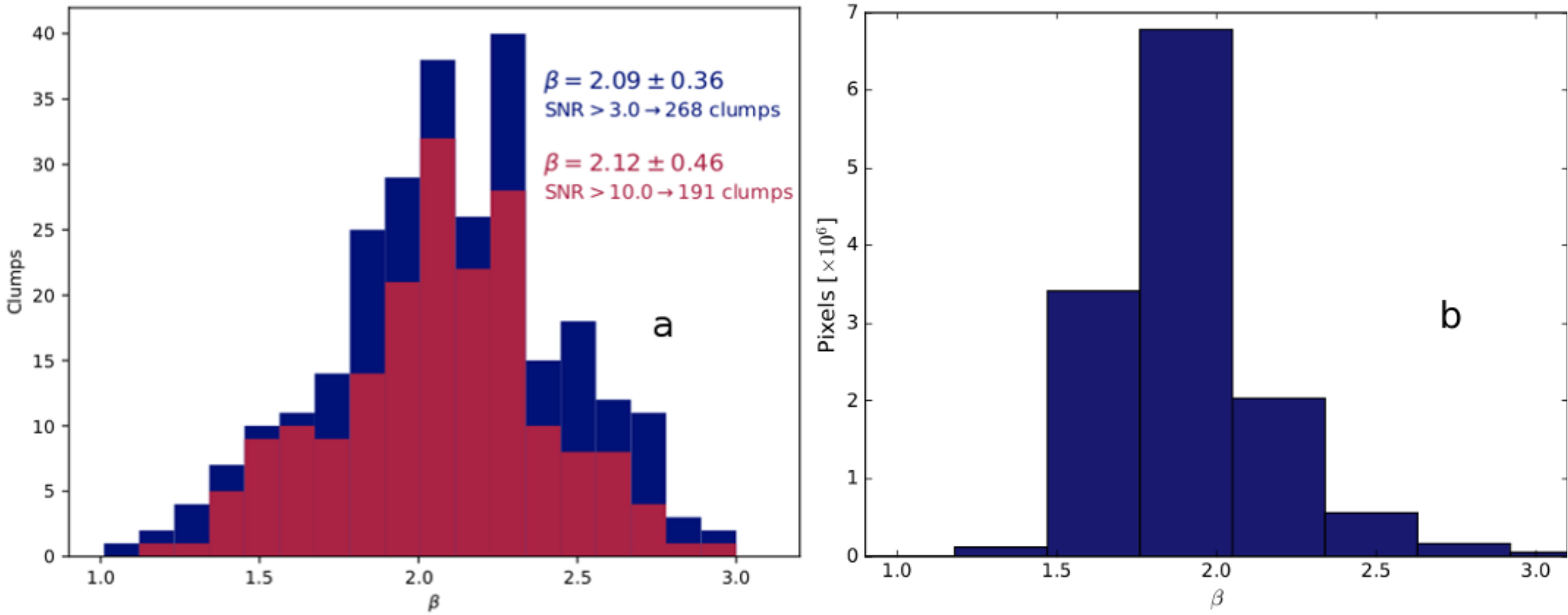}
        \caption{(a) Values of opacity spectral index for clumps with S/N $>3$ (blue) and S/N $>6$ (red). (b) Values of derived opacity spectral index over all pixels over the fields in in our sample.  \label{fig:SPIRE_beta} }
\end{figure*}

\begin{figure}
        \resizebox{\hsize}{!}{\includegraphics{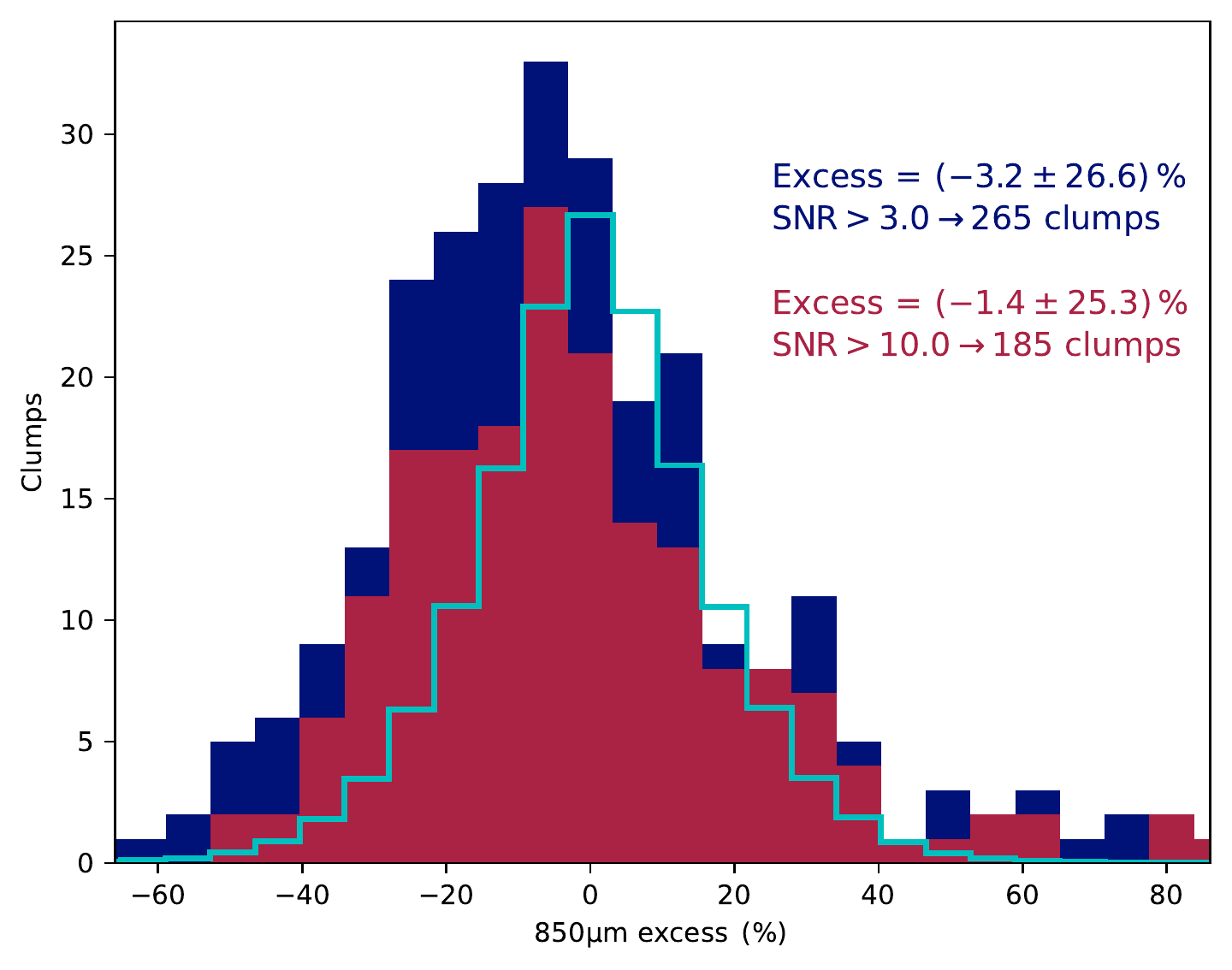}}
        \caption{SCUBA-2 850\,\micro\spc excess fit for clumps with S/N $>3$ (blue) and S/N $>6$ (red). The turquoise line is the predicted distribution for the S/N $>6$ cores that is calculated assuming a zero mean.  \label{fig:excess} }
\end{figure}

The distributions of spectral index values (Fig.~\ref{fig:SPIRE_beta}a) are consistent with those derived in Sect.~\ref{sec:results_general_ISM} (Fig.~\ref{fig:SPIRE_beta}b), although here the values represent the background-subtracted emission of the clump centers only. Figure~\ref{fig:excess} shows the distributions of the residuals when the SED prediction is subtracted from the 850\,$\mu$m observation. These peak close to zero. Moreover, the plot also includes the predicted distribution for the S/N $>6$ cores, which is calculated assuming a zero mean and for individual clumps the error distributions predicted by the uncertainties of the 850\,$\mu$m observed and SED-predicted values. The sample shows neither indications of a systematic spectral index variation beyond 500\,$\mu$m nor any source-to-source variations beyond the observational uncertainties. We detect no correlation between column density and excess. Our detected 850\,\micro\, excess is consistent with zero, although a marginal excess may exist when SCUBA-2 flux loss is taken into account. In Appendix \ref{sec:appendix_SED_fits_using_PACS_SCUBA} we further compare the temperatures, spectral index values, and masses derived for the clumps using different combinations of the 160--850\,$\mu$m bands. We find that the inclusion of the 160\,$\mu$m point systematically decreases the temperature estimates, while the inclusion of the 850\,$\mu$m point slightly increases the spectral index estimates.

\subsection{Detected clumps \label{sec:results_detected_clumps}}
FellWalker analysis found a total of 529 clumps in the 53 fields. Each field contains between 1 and 35 clumps, with an average of 10 clumps per field, although in general fields closer to the plane of the Galaxy also contain more clumps. The clumps are generally found on filamentary structures. The clumps found in each field are plotted on top of SCUBA-2 intensity maps in Appendix \ref{sec:appendix_YSO_clump}.

Clumps are cold and dense, with temperatures around 10--25\,K and average column density on the order of $10^{22}$\,cm$^{-2}$. Clump masses range from $10^{-2}$ to $10^{3}$\,\msun, with an average mass on the order of $10^{2}$\,\msun. The flux density is approximately 2\,Jy. Properties are estimated as described in Sect. \ref{sec:methods_clump_chars}, and are discussed further in Sect. \ref{sec:results_clump_properties}. We present a catalog of derived clump properties in Appendix \ref{sec:appendix_clump_properties}.

\subsection{YSO detections\label{sec:results_clump_association}}

The YSOs that are not associated with any clump are referred to as field YSOs. Clumps that are spatially associated with at least one YSO are tentatively classified as protostellar. The possibilities of misclassifying clumps as protostellar are discussed further in Sect. \ref{sec:results_virial_anal}. We also plot clump-associated YSOs on top of the clumps in Appendix \ref{sec:appendix_YSO_clump}. The mean distance to field YSOs is slightly higher: 2\,kpc compared to 1.8\,kpc for clump-associated YSOs. This is to be expected, as nearby clumps are generally high-latitude clumps with fewer field YSOs. Over half of the YSO candidates located in the fields are not associated with any clump.

\subsection{Virial analysis and clump classification \label{sec:results_virial_anal}}

We use virial analysis to predict whether the self-gravity of the clump is strong enough to prevent dissipation due to internal turbulence (Sect. \ref{sec:methods_grav_stability}). We use \co\spc line data from \citet{ke_wang_sampling_data_release}, which covers 13 clumps in three fields: G105.44+09.88, G159.23-20.09, and G202.31+02.53. This sample contains 10 potentially protostellar clumps, of which one is confirmed. The other three clumps are unbound starless or prestellar. The line width was estimated from the spectra plotted in Appendix \ref{sec:appendix_CO_line_graphs}. For the 13 clumps we calculate $\langle\Delta V\rangle $ = 0.9 $\pm$ 0.03\,m\,s$^{-1}$ and $\sigma_{\rm avg}$ = 0.43 $\pm$ 0.01\,km\,s$^{-1}$. As the line widths for the 13 clumps are similar enough, we use $\sigma_{\rm avg}$ to estimate the critical virial mass for the rest of the clumps. This is very similar to the nonthermal velocity dispersion for dense clumps of 0.49\,km\,s$^{-1}$ derived by \citet{2018A&A...614A..83J}. However, we note that this is a tentative classification because the local environment strongly affects the kinematics of a clump. For this reason, we also include values derived for all starless clumps in Table \ref{tbl:starless_vs_prestellar_values}. 

We classify starless clumps as prestellar if they are gravitationally bound, with $\alpha_{\rm vir} \leq$ 1.0. Through propagation of error we find that uncertainty in virial mass is approximately 100\%. Out of 234 starless clumps with distance estimates, we estimate 210 (90\%) to be virially bound. Extrapolated into the whole sample, virial analysis predicts $\sim$37 unbound and $\sim$331 prestellar clumps, although of course magnetic forces, for example, play a role in core evolution. 

By requiring that YSOs have 160\,\micro\spc intensity significantly higher than 250\,\micro\spc intensity (Sect. \ref{sec:methods_YSO_ass_intensity_YSO_determination}), we find that 50 out of 108 (46\%) YSOs that are located on \textit{Herschel} PACS 160\,\micro\spc and SPIRE 250\,\micro\spc fields can be confirmed as YSOs. A further 106 (51\%) out of 214 total YSOs do not have \textit{Herschel} 160\,\micro\spc and 250\,\micro\spc data. Only 242 (46\%) of all clumps have both SPIRE 250\,\micro\spc and PACS 160\,\micro\spc data, as well as mass estimates. Out of these, we find 28 (12\%) protostellar and 167 (69\%) starless clumps, out of which 145 (60\%) are classified prestellar. A further 47 (19\%) clumps are tentatively protostellar but without further confirmation from \textit{Herschel} intensity ratio $\Delta C$.

Uncertainty in number of YSOs associated with a clump depends on a range of factors, including the completeness of the surveys used, the distance, the extinction, and the YSO class. We assume an uncertainty in the position across all catalogs of 6.3\arcsec\spc (22\% uncertainty), a chance of misclassifying other sources as YSOs of 10\%, and of missing YSOs of 10\%, resulting in final uncertainty of 26\% in number of YSOs.

\begin{sidewaystable*}
        \centering
        \caption{ Estimated mean parameters for different clump categories \label{tbl:starless_vs_prestellar_values}}
        \begin{tabular}{llllllllllll}
                \hline
                \hline
                (1) & (2) & (3) & (4) & (5) & (6) & (7) & (8) & (9) & (10) & (11) & (12)\\
                Clump & $\langle T\rangle$ &$\langle F_{\nu}(850\,\mu{\rm m}) \rangle$ & $\langle N$(H$_{2})\rangle$ & $\langle N$(H$_{2})_{\rm 850}\rangle$ & Ratio & $\langle \beta\rangle$ & $\langle M\rangle$  & $\langle M_{\rm vir}\rangle$ & $\langle d\rangle$ & \multicolumn{2}{c}{$N_{\rm clumps}$}   \\
                 type & (K) & (Jy) & ($10^{21}$cm$^{-2}$) & ($10^{21}$cm$^{-2}$) & (peak/mean) &  & (\msun)  & (\msun)& (kpc) & &  \\
                \hline

                                \multicolumn{12}{l}{ All fields with distance, \textit{n} =  395  clumps} \\ \hline
                All  &  17.8  $\pm$  3.6  &  2.4  $\pm$  3.9  &  14.5  $\pm$  6.7  &  30.0  $\pm$  39.1  &  2.2  $\pm$  4.5  &  1.7  $\pm$  0.1  &  135.7  $\pm$  316.0  &  9.8  $\pm$  7.4  &  2.4  $\pm$  1.3  &  264  &  395 \\ 
                PS-C  &  16.5  $\pm$  3.3  &  2.8  $\pm$  2.7  &  17.8  $\pm$  8.1  &  52.9  $\pm$  52.2  &  2.1  $\pm$  5.7  &  1.7  $\pm$  0.1  &  176.8  $\pm$  300.6  &  9.4  $\pm$  7.1  &  2.3  $\pm$  1.6  &  36  &  36 \\ 
                PS  &  17.7  $\pm$  3.3  &  4.2  $\pm$  5.8  &  15.7  $\pm$  7.8  &  43.1  $\pm$  56.5  &  2.5  $\pm$  2.4  &  1.8  $\pm$  0.1  &  233.4  $\pm$  552.5  &  11.0  $\pm$  8.1  &  2.2  $\pm$  1.2  &  52  &  90 \\ 
                PRE  &  18.2  $\pm$  3.6  &  2.0  $\pm$  3.0  &  14.1  $\pm$  5.7  &  23.7  $\pm$  26.0  &  2.8  $\pm$  2.1  &  1.7  $\pm$  0.1  &  107.0  $\pm$  171.1  &  10.2  $\pm$  7.0  &  2.6  $\pm$  1.2  &  154  &  245 \\ 
                SL  &  17.3  $\pm$  4.4  &  0.4  $\pm$  0.2  &  9.9  $\pm$  3.7  &  11.0  $\pm$  5.5  &  1.2  $\pm$  0.6  &  1.8  $\pm$  0.1  &  1.5  $\pm$  1.0  &  2.3  $\pm$  1.2  &  0.7  $\pm$  0.2  &  22  &  24 \\ 
                SL PRE  &  18.1  $\pm$  3.7  &  1.8  $\pm$  2.9  &  13.6  $\pm$  5.6  &  22.5  $\pm$  25.1  &  2.2  $\pm$  1.9  &  1.8  $\pm$  0.1  &  97.6  $\pm$  166.0  &  9.5  $\pm$  7.1  &  2.4  $\pm$  1.2  &  176  &  269 \\ 
                PS PS-C  &  17.2  $\pm$  3.4  &  3.8  $\pm$  5.1  &  16.5  $\pm$  8.0  &  45.9  $\pm$  55.5  &  2.2  $\pm$  5.0  &  1.7  $\pm$  0.1  &  217.2  $\pm$  494.5  &  10.6  $\pm$  7.9  &  2.2  $\pm$  1.3  &  88  &  126 \\ 
                \hline

        \multicolumn{12}{l}{ $d < 0.5$ kpc , \textit{n} =  13  clumps} \\                 \hline
                All  &  11.9  $\pm$  1.0  &  1.1  $\pm$  1.0  &  14.1  $\pm$  5.8  &  20.9  $\pm$  11.2  &  2.0  $\pm$  1.0  &  1.8  $\pm$  0.1  &  1.7  $\pm$  1.8  &  1.4  $\pm$  0.7  &  0.4  $\pm$  0.1  &  8  &  13 \\ 
                PS  &  --  &  1.0  $\pm$  0.6  &  --  &  15.3  $\pm$  5.5  &  --  &  --  &  1.4  $\pm$  0.8  &  1.6  $\pm$  0.6  &  0.4  $\pm$  0.0  &  0  &  3 \\ 
                PRE  &  12.1  $\pm$  1.2  &  2.1  $\pm$  1.0  &  15.2  $\pm$  5.2  &  34.9  $\pm$  9.1  &  2.5 $\pm$ 0.9  &  1.8  $\pm$  0.1  &  3.5  $\pm$  2.1  &  1.9  $\pm$  0.8  &  0.4  $\pm$  0.1  &  4  &  4 \\ 
                SL  &  11.7  $\pm$  0.7  &  0.5  $\pm$  0.1  &  13.1  $\pm$  6.1  &  14.3  $\pm$  3.5  &  1.5  $\pm$  0.9  &  1.8  $\pm$  0.1  &  0.7  $\pm$  0.5  &  1.0  $\pm$  0.5  &  0.3  $\pm$  0.2  &  4  &  6 \\ 
                SL PRE  &  11.9  $\pm$  1.0  &  1.1  $\pm$  1.0  &  14.1  $\pm$  5.8  &  22.6  $\pm$  11.9  &  1.5  $\pm$  0.9  &  1.8  $\pm$  0.1  &  1.8  $\pm$  2.0  &  1.4  $\pm$  0.8  &  0.3  $\pm$  0.2  &  8  &  10 \\  
                \hline

                \multicolumn{12}{l}{ $0.5 \leq d < 1.5$ kpc , \textit{n} =  99  clumps} \\               \hline
                All  &  15.8  $\pm$  3.5  &  1.9  $\pm$  2.7  &  13.8  $\pm$  6.6  &  34.6  $\pm$  45.9  &  2.3  $\pm$  2.1  &  1.7  $\pm$  0.1  &  13.7  $\pm$  18.8  &  3.4  $\pm$  1.5  &  0.8  $\pm$  0.1  &  99  &  99 \\ 
                PS-C  &  14.0  $\pm$  1.4  &  2.7  $\pm$  2.7  &  16.6  $\pm$  8.1  &  66.0  $\pm$  62.4  &  2.2  $\pm$  2.4  &  1.7  $\pm$  0.1  &  23.0  $\pm$  24.4  &  3.6  $\pm$  1.5  &  0.8  $\pm$  0.1  &  19  &  19 \\ 
                PS  &  17.0  $\pm$  3.8  &  3.5  $\pm$  3.8  &  15.0  $\pm$  8.6  &  40.2  $\pm$  51.5  &  2.4  $\pm$  1.5  &  1.8  $\pm$  0.1  &  20.7  $\pm$  24.7  &  3.9  $\pm$  1.9  &  0.8  $\pm$  0.1  &  21  &  21 \\ 
                PRE  &  14.7  $\pm$  2.7  &  1.4  $\pm$  1.9  &  13.9  $\pm$  4.4  &  28.2  $\pm$  33.6  &  3.6  $\pm$  2.3  &  1.7  $\pm$  0.1  &  11.0  $\pm$  11.5  &  3.5  $\pm$  1.4  &  0.9  $\pm$  0.1  &  41  &  41 \\ 
                SL  &  18.6  $\pm$  3.8  &  0.4  $\pm$  0.3  &  9.2  $\pm$  2.4  &  9.9  $\pm$  5.6  &  1.1  $\pm$  0.5  &  1.8  $\pm$  0.1  &  1.8  $\pm$  0.9  &  2.7  $\pm$  1.1  &  0.8  $\pm$  0.1  &  18  &  18 \\ 
                SL PRE  &  15.9  $\pm$  3.6  &  1.1  $\pm$  1.7  &  12.5  $\pm$  4.5  &  22.6  $\pm$  29.4  &  2.4  $\pm$  2.1  &  1.7  $\pm$  0.1  &  8.2  $\pm$  10.5  &  3.2  $\pm$  1.4  &  0.9  $\pm$  0.1  &  59  &  59 \\ 
                PS PS-C  &  15.6  $\pm$  3.3  &  3.1  $\pm$  3.4  &  15.8  $\pm$  8.4  &  52.5  $\pm$  58.4  &  2.3  $\pm$  2.1  &  1.7  $\pm$  0.1  &  21.8  $\pm$  24.6  &  3.8  $\pm$  1.7  &  0.8  $\pm$  0.1  &  40  &  40 \\ 
                \hline

                \multicolumn{12}{l}{ $1.5 \leq d < 3.0$ kpc , \textit{n} =  168  clumps} \\              \hline
                All  &  18.5  $\pm$  2.5  &  2.9  $\pm$  4.8  &  14.8  $\pm$  6.4  &  30.5  $\pm$  39.3  &  2.7  $\pm$  8.7  &  1.7  $\pm$  0.1  &  114.3  $\pm$  203.1  &  9.5  $\pm$  4.5  &  2.3  $\pm$  0.2  &  63  &  168 \\ 
                PS-C  &  15.7  $\pm$  0.5  &  1.2  $\pm$  0.6  &  8.9  $\pm$  1.3  &  23.1  $\pm$  12.0  &  2.7  $\pm$  10.1  &  1.6  $\pm$  0.2  &  63.3  $\pm$  42.7  &  9.5  $\pm$  1.5  &  2.2  $\pm$  0.2  &  3  &  3 \\ 
                PS  &  18.2  $\pm$  3.1  &  4.4  $\pm$  6.4  &  13.9  $\pm$  6.7  &  43.8  $\pm$  55.4  &  2.6  $\pm$  3.3  &  1.7  $\pm$  0.1  &  171.8  $\pm$  274.5  &  11.0  $\pm$  4.7  &  2.2  $\pm$  0.2  &  15  &  47 \\ 
                PRE  &  18.8  $\pm$  2.2  &  2.3  $\pm$  3.8  &  15.5  $\pm$  6.3  &  25.4  $\pm$  29.5  &  2.8  $\pm$  1.7  &  1.7  $\pm$  0.1  &  92.6  $\pm$  163.7  &  8.9  $\pm$  4.4  &  2.3  $\pm$  0.2  &  45  &  118 \\ 
                PS PS-C  &  17.8  $\pm$  3.0  &  4.3  $\pm$  6.3  &  13.1  $\pm$  6.4  &  42.6  $\pm$  54.1  &  2.7  $\pm$  8.9  &  1.7  $\pm$  0.1  &  165.3  $\pm$  267.5  &  10.9  $\pm$  4.6  &  2.2  $\pm$  0.2  &  18  &  50 \\ 
                \hline

                \multicolumn{12}{l}{ $d \geq 3.0$ kpc , \textit{n} =  115  clumps} \\             \hline
                All  &  19.9  $\pm$  2.8  &  2.4  $\pm$  3.3  &  15.2  $\pm$  6.9  &  26.2  $\pm$  33.7  &  1.7  $\pm$  1.5  &  1.8  $\pm$  0.2  &  287.3  $\pm$  493.9  &  16.8  $\pm$  7.9  &  4.0  $\pm$  0.4  &  94  &  115 \\ 
                PS-C  &  19.9  $\pm$  2.1  &  3.2  $\pm$  2.9  &  21.2  $\pm$  7.0  &  41.6  $\pm$  33.8  &  1.5  $\pm$  1.1  &  1.7  $\pm$  0.1  &  409.7  $\pm$  376.8  &  17.4  $\pm$  4.0  &  4.2  $\pm$  0.3  &  14  &  14 \\ 
                PS  &  18.3  $\pm$  2.6  &  4.7  $\pm$  6.2  &  18.2  $\pm$  6.9  &  48.8  $\pm$  66.5  &  2.6  $\pm$  2.5  &  1.8  $\pm$  0.2  &  657.7  $\pm$  1005.4  &  20.4  $\pm$  9.8  &  4.1  $\pm$  0.3  &  16  &  19 \\ 
                PRE  &  20.3  $\pm$  2.9  &  1.8  $\pm$  1.9  &  13.1  $\pm$  5.8  &  18.4  $\pm$  12.2  &  1.8  $\pm$  1.4  &  1.8  $\pm$  0.2  &  180.6  $\pm$  194.7  &  15.9  $\pm$  7.6  &  4.0  $\pm$  0.4  &  64  &  82 \\ 
                PS PS-C  &  19.1  $\pm$  2.5  &  4.1  $\pm$  5.1  &  19.6  $\pm$  7.1  &  45.7  $\pm$  55.2  &  1.7  $\pm$  1.5  &  1.8  $\pm$  0.1  &  552.5  $\pm$  810.7  &  19.1  $\pm$  8.0  &  4.2  $\pm$  0.3  &  30  &  33 \\ 
                \hline

        \end{tabular}
        \tablefoot{    Mean temperature $T$, flux density $F_{\nu}$, column density $\langle N$(H$_{2})\rangle$, peak column density $N$(H$_{2})_{\rm 850}$, ratio of peak to mean \textit{N}(\MH), mean $beta$, observed mass \textit{M}, virial mass $M_{\rm vir}$, and distance \textit{d} of unbound SL, PRE, and PS-C, and unconfirmed (PS) clumps. Only fields with distance estimates are considered. Average temperatures and column densities are further limited to fields with SPIRE data. The number of clumps of each type with distance estimates and \textit{Herschel} SPIRE data are listed in column 11 and the total number of clumps with distance estimates in column 12. Estimates for all starless (SL plus PRE) and all protostellar (PS plus PS-C) clumps are also included.  } 
\end{sidewaystable*}

\begin{figure}
        \resizebox{\hsize}{!}{\includegraphics{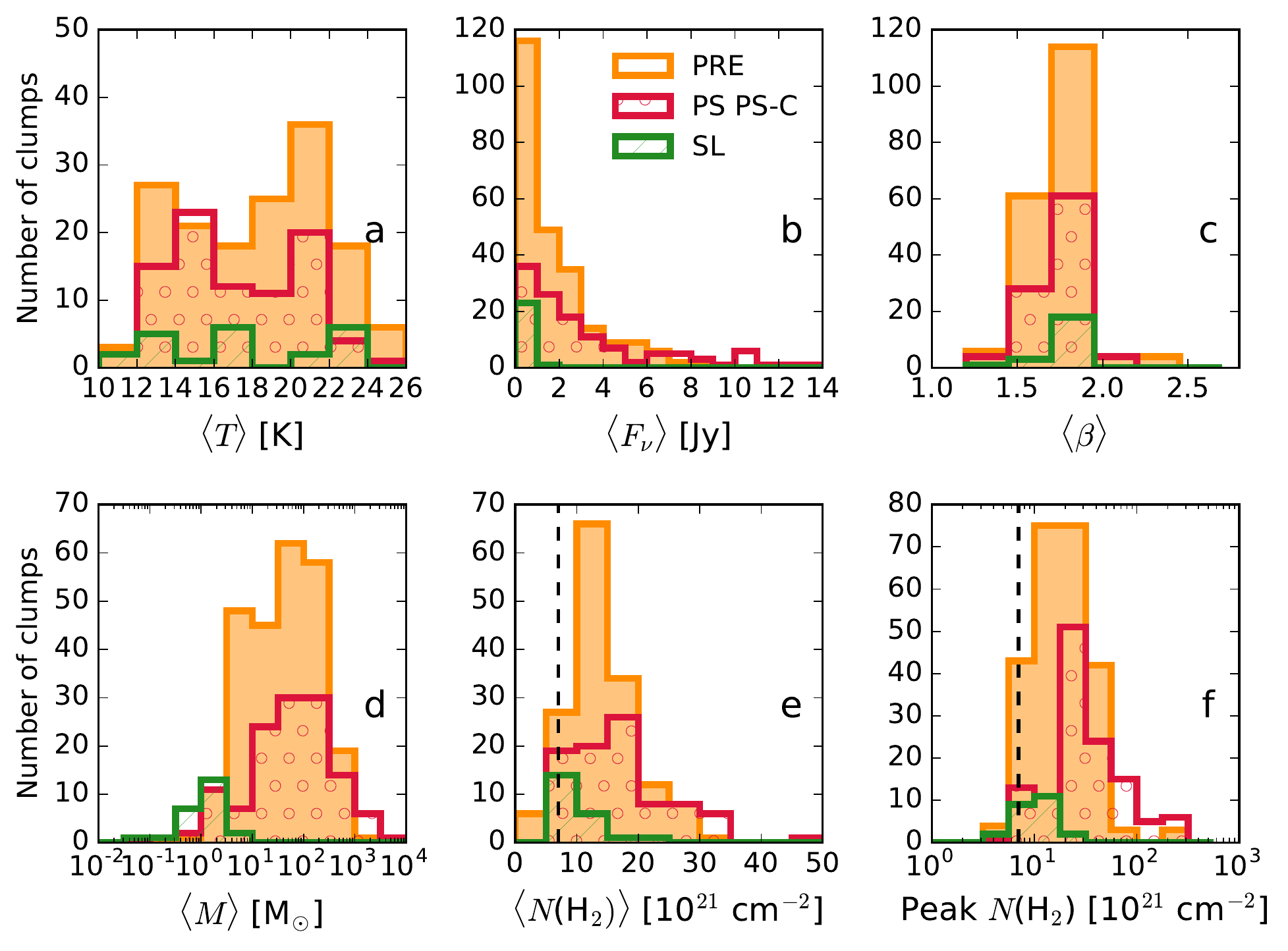}}
        \caption{Histograms for properties of unbound starless (SL), prestellar (PRE), and all protostellar (PS PS-C) clumps. (a) Mean temperature. (b) Flux density. (c) Mean clump opacity spectral index (d) Mass, plotted on a logarithmic scale.  (e) Mean clump column density. (f) Peak clump column density. The dotted line in (e) and (f) correspond to $A_{\rm v}$ = 7. 
         \label{fig:comparisons_both}  }
\end{figure}

\subsection{Clump properties\label{sec:results_clump_properties}}

We present the full catalog of derived clump properties in Appendix \ref{sec:appendix_clump_properties}. Mean temperature, flux density, opacity spectral index, mass, and mean and peak column density for all unbound starless (SL), prestellar (PRE), and protostellar (PS) clumps with distance is shown in Table \ref{tbl:starless_vs_prestellar_values} and plotted in Fig. \ref{fig:comparisons_both}. The confirmed protostellar (PS-C) clumps are included under the PS bar in Fig. \ref{fig:comparisons_both}; but these are listed separately in Table \ref{tbl:starless_vs_prestellar_values}, which also includes derived values for all starless (SL PRE) and all protostellar (PS PS-C) clumps.

We divided our clumps into four distance bins (bin I: $d < 0.5$\,kpc, bin II: $0.5 \leq d < 1.5$\,kpc, bin III: $1.5 \leq d < 3.0$\,kpc, and bin IV: $d \geq 3.0$\,kpc). We chose these distance bins because clump distances fall on four distinct peaks, shown in Fig. \ref{fig:clump_chars} in Appendix \ref{sec:appendix_clump_data}. As a consequence of the large range of distances, we do not include analysis of regions with no distance estimates, which includes 193 clumps in 18 fields.

\subsubsection{Dust opacity spectral index in clumps \label{sec:results_clump_properties_beta}}

Dust emission within clumps was examined based on the pixel-by-pixel MBB fits of \textit{Herschel} 250--500\,\micro\spc data. The mean opacity spectral index over the clumps (Table \ref{tbl:starless_vs_prestellar_values}, column 7) is lower by $\sim$0.2 compared to the full fields. The opacity spectral index value does not significantly differ between the prestellar and starless clumps or between the distance bins (Fig. \ref{fig:comparisons_beta} in Appendix \ref{sec:appendix_clump_data}). Overall, we find \B\, to be fairly consistent, with a standard deviation in all categories of $\leq$ 0.2.

\subsubsection{Clump sizes \label{sec:results_clump_properties_size}}

Clump angular effective radii range from 1.7 to 49.4\arcsec, with a mean of (15.2 $\pm$ 6.8)\arcsec, and show no dependence on distance (Fig. \ref{fig:comparisons_Reff}). Mean spatial effective radius is (0.2 $\pm$ 0.1)\,pc, with a range from 0.006--0.6\,pc. As the uncertainties of linear size are dominated by distance, 30\% uncertainties are adopted for the effective radius and the SCUBA-2 FWHM, 14.6\arcsec, for angular radii. The uncertainty in linear size  is discussed further in Sect. \ref{sec:discussion_clump_properties_size}. The clump sizes are similar above the threshold of the SCUBA-2 beam size, most likely because of selection effects: large-scale structure is filtered out and smaller-scale structure is not resolved by SCUBA-2. The sizes of the unresolved small clumps are unreliable. There is, in general, no significant difference in the angular sizes of different clump categories.

The average area of a clump is approximately 0.25\,arcmin$^{2}$. Clumps are generally round and have a mean aspect ratio of 0.75. More massive clumps are also likely to be rounder, whereas smaller cores have a large variation in aspect ratio. This may be due to deconvolution of clumps close to beam size.

\subsubsection{Clump temperatures \label{sec:results_clump_properties_T}}

Based on 250--500\,\micro\spc \textit{Herschel} data, the average temperature over a clump ranges from 10 to 25\,K and has a mean of 17.8$\pm$3.6\,K over all clumps.  The total error in \textit{Herschel} intensities $I_{250}$ is estimated to be below 10\%, resulting in temperature errors of the order of 1\,K as in \cite{GCC_IV_cold_submm}.  

The temperature is strongly correlated with distance (Fig. \ref{fig:comparisons_T}). In distance bins I and II, the mean temperature is under 16\,K, rising to almost 20\,K in more distant bins. We do not see large differences in temperature between clump categories. The largest differences are found in bin II, in which there is a difference of 4.6\,K between SL and PS-C clumps. In the largest distance bins, bins III and IV, PRE clumps have higher mean temperatures. PS clumps have generally higher temperature than PS-C clumps.

\subsubsection{Column density \label{sec:results_clump_properties_NH2}}

Based on 250--500\,\micro\spc data, mean column density over all the clumps is (1.5$\pm 0.7)\times 10^{22}$\,cm$^{-2}$, corresponding to visual extinction $\langle A_{\rm v}\rangle$ = 15.0 $\pm 7.0$ mag, using the relation \textit{N}(\MH)/$A_{\rm v}=10^{21}$ cm$^{-2}$mag$^{-1}$ \citep{1978ApJ...224..132B,Perseus_core_forming_clump}. The uncertainty of the dust opacity $\kappa$ is at least 50\%, leading to a similar uncertainty in $N$(\MH). 
The highest column density is found in PS-C clumps in bin IV, whereas SL clumps are the most diffuse. The PRE clumps have lower column density than PS-C or PS clumps, also suggesting that these clumps include matter from the diffuse ISM. There is little change in column density as a function of distance (Fig. \ref{fig:comparisons_NH2}).

\begin{figure}
        \resizebox{\hsize}{!}{\includegraphics{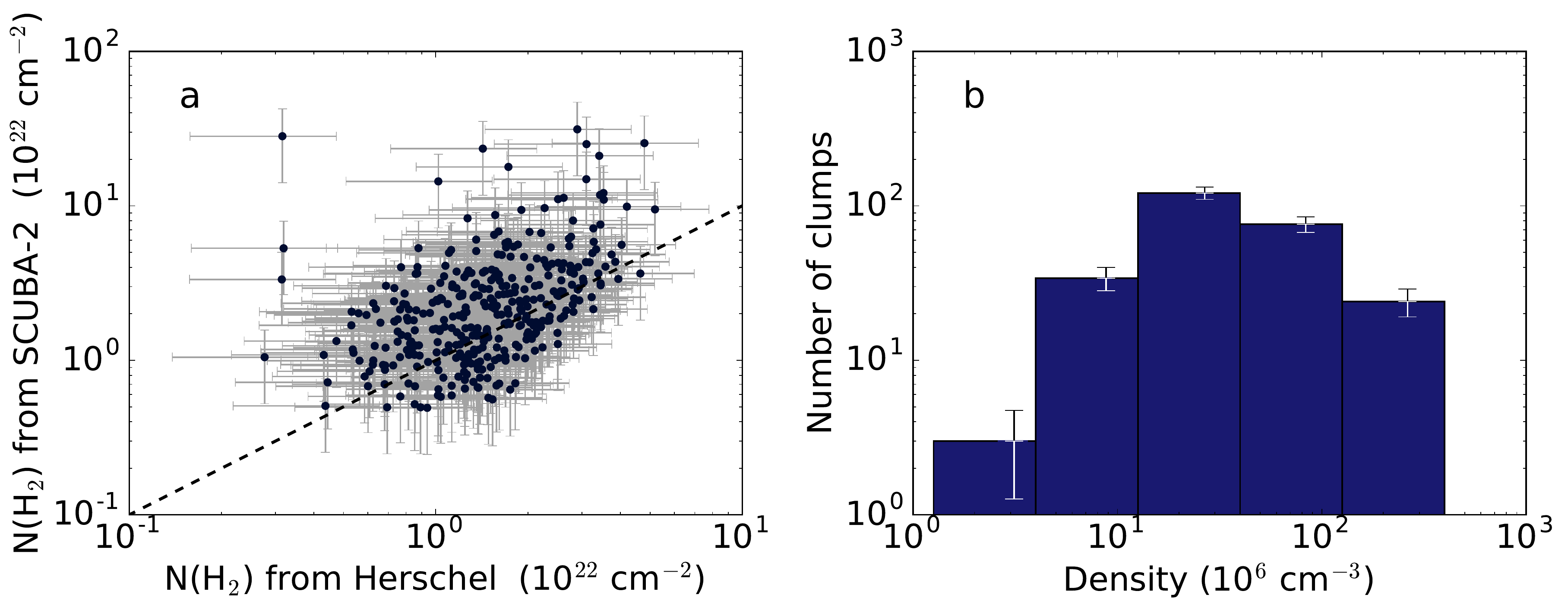}}
        \caption{(a) column density from \textit{Herschel} (\textit{x-axis}) plotted against column density from SCUBA-2 peak values (\textit{y-axis}). The black dotted line shows where both are equal. The clumps for which temperature was assumed to be unrealistic, set to 18\,K, are denoted with red squares and the rest with dark blue circles. (b) Histogram of peak volume density of each clump with uncertainty. \label{fig:NH2_scuba} } 
\end{figure}

For comparison, peak column densities were derived from the higher resolution SCUBA-2 surface brightness measurements. For clumps without 250--500\,\micro\spc temperature estimates (or with unrealistically low estimates below 10\,K), the mean clump temperature of 18\,K was used. We plot SCUBA-2 peak column density against \textit{Herschel} column density in Fig. \ref{fig:NH2_scuba}a. Column density was estimated by calculating optical depth $\tau$ from Eq. (\ref{eq:tau}) and column density using Eq. (\ref{eq:NH2}). We calculate volume density from Eq. (\ref{eq:volume_density}). The average of clump peak column densities is (3.0\,$\pm $\,0.2)$\times 10^{22}$\,cm$^{-2}$, twice as high as that estimated from \textit{Herschel} (Fig. \ref{fig:NH2_scuba}a), with a maximum of (3.1\,$\pm $\,1.6)$\times 10^{23}$\,cm$^{-2}$. These values give a better picture of the densest regions of the clouds. 

Highest peak column densities are found in protostellar clumps; PS-C clumps have the highest peak column density in bin II and PS clumps in other distance bins (Table \ref{tbl:starless_vs_prestellar_values}). The SL clumps are the most diffuse, and PRE clumps have peak column densities over double that of SL clumps. There is a not much difference in mean peak column density over the distance bins. Average volume density calculated with peak column density (Eq. (\ref{eq:volume_density})) is (6.1 $\pm $ 0.04)$\times 10^{7}$ cm$^{-3}$, and values range from (1.1 $\pm $ 0.7)$\times 10^{4}$ cm$^{-3}$ to (1.3 $\pm $ 0.8)$\times 10^{9}$ cm$^{-3}$ (Fig. \ref{fig:NH2_scuba}b).

\subsubsection{Clump masses \label{sec:results_clump_properties_M}}

Clumps have a wide range of masses, from 0.04\,\msun\spc to 4259\,\msun\spc based on \textit{Herschel} temperature and SCUBA-2 flux, with a mean mass of (136$^{+316}_{-136}$)\,\msun. Uncertainty in mass is dominated by distance, but is also affected by uncertainty in $\kappa$, $\beta$, and flux density. We assume a total uncertainty in mass of 80\%.

Mass increases significantly with distance (Fig. \ref{fig:comparisons_M}), proving that the most distant clumps are really unresolved clouds, which likely contain many smaller clumps and cores. In all distance bins, SL clumps are the least massive by as much as a factor of 10 when compared to protostellar clumps in bin II. Protostellar clumps are the most massive across all bins; PS-C clumps are generally smaller than PS clumps. The masses of PRE clumps are of the same order as those of protostellar clumps.

Virial masses are generally smaller than observed masses,  have a mean of (9.8 $\pm$ 7.4)\,\msun , and a smaller range of (0.3  $\pm$ 0.5) -- (38.5 $\pm$ 53.9)\,\msun. Few nearby cores are virially bound and no distant clouds are unbound. There is an increase of virial mass with distance by about one order of magnitude (Fig. \ref{fig:comparisons_Mvir}). As our \co\spc data comes from fields at distances of $\sim$1.7, 0.8, and 0.7\,kpc, our virial masses are likely to be more precise in bin II, where our derived virial masses most closely match our observed masses. We derive a final uncertainty of 140\% on virial mass.

\section{Discussion \label{sec:discussion}}

\subsection{Properties of the extended ISM \label{sec:discussion_ISM}}

Various relations between temperature and spectral index can be found in the literature, from the inner Galaxy \citep{beta_HI_Gal}, the outer Galaxy 
\citep{beta_archeops}, the solar neighborhood \citep{beta_pronaos}, and the entire Galaxy \citep{Planck_all_sky_model}. Even after accounting for error, anticorrelation has been found to exist at the scale of cold clouds and near the Galactic poles \citep[e.g.][]{beta_pronaos, beta_archeops, beta_HI_Gal,Planck_all_sky_model,GCC_VI}; however, a positive correlation has been found in some Bok globules and the overall Galactic plane \citep{2012ApJ...752...55K,Planck_all_sky_model}, as well as in small, dense regions near strong point sources \citep{GCC_VI}. Our simulations strongly suggest that not all the anticorrelation is due to noise.

Our derived mean temperature of 18.8\,K over the fields is slightly lower than all-sky models, and our mean column density approximately four times higher, as expected for star-forming regions, which are protected from the interstellar radiation field. In the literature, an average dust temperature of 19.7\,K has been derived for the whole sky, which is dominated by diffuse regions \citep{Planck_all_sky_model}. The mean temperatures of large-scale filaments in the Milky Way are approximately 19--21\,K, and have column densities of approximately 10$^{21}$\,cm$^{-2}$ \citep{2018ApJ...864..153Z}. In the case of dense clumps, \citet{GCC_VI} find mean temperatures of 16.9\,K for SPIRE-only data, and 18.4\,K for fits including also SPIRE 160\,\micro. 
\citet{GCC_VI} also derive slightly lower temperatures from 250-500\,\micro\, than from 160-500\,\micro\spc \textit{Herschel} data.

We derive a mean opacity spectral index of 1.9 over the fields using 250--500\,\micro\spc data and 1.7 by including also 160\,\micro\spc data. \citet{GCC_VI} derive slightly higher spectral index of 1.89 for 166 GCC fields, using \textit{Herschel} and \textit{Planck} data. The omission of shorter wavelengths increases the mean spectral index; in our sample there is a mean of $\Delta\beta \approx$ 0.06.  \citet{GCC_VI} find slightly larger difference; however, their sample includes a larger range of wavelengths.

\citet{2017MNRAS.471..100E} find no relation between temperature and Galactocentric radius, but  find a slight increase for median bolometric temperature. \citet{GCC_VI} also do not find significant latitude- or longitude-dependence on temperature, but find higher spectral index at higher latitudes. However, \citet{2017MNRAS.471.2730M} find a decreasing radial profile of dust temperature using \textit{Herschel} data, while \citet{2018MNRAS.473.1059U} find a mild decrease using SCUBA-2 observations. Many nearby spiral galaxies have mean cold dust temperatures of approximately 25\,K near the center ($\ell\sim0$), which drops $\sim$10\,K toward the edge of the galaxies \citep{2014A&A...561A..95T,2015ApJ...804...46M,2018A&A...612A..81R}. We also find lower temperatures away from the plane and center of the Galaxy, although not as pronounced; of course, the extragalactic surveys study the entire cold ISM and not the densest clumps. \citet{SCOPE_catalogue} also find high column densities around $10^{22}$\,cm$^{-2}$ near the plane of the Galaxy for their observed SCOPE cores.

\subsection{YSO association \label{sec:discussion_catalogs}}

There are several possibilities why over half of YSO candidates located within the SCUBA-2 fields are not associated with any clumps. First, they may be more evolved YSOs that have already left the densest cores; however, these would likely already be more evolved stars. Second, some cores may have lower flux and thus not have been found by the FellWalker algorithm. There are also clouds  outside the clumps, although these typically have lower column densities and are generally not star-forming. In the case of large, smooth clumps, the filtering of large-scale structure may also prevent detection. Finally, some YSO candidates in the catalogs may have been incorrectly classified sources. In addition, as the YSO association is based on the match of projected positions on the sky, YSOs may not be located at the distance of the clumps. This is especially likely for fields near the plane of the Galaxy, which are likely to have several SF regions and IR sources along the line of sight.

\subsection{Clump classification \label{sec:discussion_clump_classification}}

We find mean virial mass of 9\,\msun, and mean $\alpha_{\rm vir}$ of $\sim$0.5, which is a lower virial mass and mean $\alpha_{\rm vir}$ than found in \citet{2017MNRAS.466..340C} for clumps found within 10\,kpc. Owing to increased difficulty of detecting fainter clumps, a larger percentage of distant starless clumps are classified as prestellar \citep{catalog_in_Lupus_cluster}. 
According to the literature, mostly using \textit{Herschel} observations, only 20\% of starless cores were classified as prestellar in nearby clouds with distance under 200\,pc \citep{census_dense_cores_Taurus,catalog_in_Lupus_cluster, dense_cores_corona_australis,PGCC_properties_in_L1495}. This percentage rises to nearly 60\% at distances between 200 and 500\,pc \citep{mass_dist_in_orion_A,dense_cores_in_Aquila} and is 80--100\% at distances over 1\,kpc \citep{vela_C_SF,hi_gal_observations,thermal_Hi_gal}. This trend is noted across a wide range of distances in \citet{GCC_IV_cold_submm} and is also clearly visible in our data; Fig. \ref{fig:mass_vs_r_d}c shows that $\alpha_{\rm vir}$ is significantly lower, and thus clumps appear to be more gravitationally bound in distant fields. 
Furthermore, the smaller $\alpha_{\rm vir}$ at larger distances may be caused by uncertainties in distance, as $\alpha_{\rm vir} \propto M_{\rm obs}^{-1} \propto d^{-1}$. Finally, resulting from our lack of dense gas tracers, we are unable to observe the kinematic environments of the clumps, and it is likely that we underestimated the turbulence within some regions.

Clumps classified as protostellar should be already undergoing SF. In addition to the possibility of misclassifying starless clumps as protostellar, quiescent cores classified as starless may have hidden SF, which is only visible through interferometric observations \citep{2019arXiv190907985S,2019arXiv190908916L}. 
Many studies have used a detection in 70\,\micro\spc as a definitive sign of protostars \citep[eg.][]{vela_C_SF,dense_cores_in_Aquila,thermal_Hi_gal} or evolved clumps \citep{2019A&A...621L...7G}. However, especially low-mass SF clumps have been found to be quiet in 70\,\micro\spc \citep{70_um_problems}, and some 70\,\micro -quiet clumps have been found to have associated water masers, a sign of SF activity \citep{2017A&A...599A.139K}.

In this study, we looked for additional evidence for protostellar clumps by looking at the increase in the 160\,\micro\spc to 250\,\micro\spc ratio. With this clarification, we find that just under one half of the YSOs can be confirmed. However, \citet{2019arXiv190903781M} find that dust temperature often also drops toward the center of protostellar clumps. Furthermore, they find several "warm starless cores," finding that cores without SF can also show significant increase in temperature toward the center. Their sample was at a distance of 760\,pc and the 160\,\micro\spc to 250\,\micro\spc ratio might be a more sensitive test of the protostellar status in more nearby fields. We perform this test only on those clumps with a candidate YSO from one or more catalogs, and thus are less likely to misclassify warm starless cores as protostellar. Warm starless cores are also often more diffuse, and given the requirement of a SCUBA-2 detection and thus high column density of all our sources, it is unlikely that our sample includes warm starless cores misclassified as protostellar clumps.

\subsection{Clump properties \label{sec:discussion_clump_properties}}

Figure \ref{fig:clump_chars} in Appendix \ref{sec:appendix_clump_data} shows histograms and the correlations of clump distance, aspect ratio, mass, temperature, column density, and volume density (Fig. \ref{fig:clump_chars}). A strong correlation ($ |r|\geq 0.4 $), is found between mass and distance, temperature and distance, volume density and distance, and temperature and volume density. A weak correlation ($ 0.2 \leq |r| \leq 0.4 $) is found between mass and the following quantities: temperature, column density, volume density, and aspect ratio.

\begin{figure*}
        \resizebox{\hsize}{!}{\includegraphics{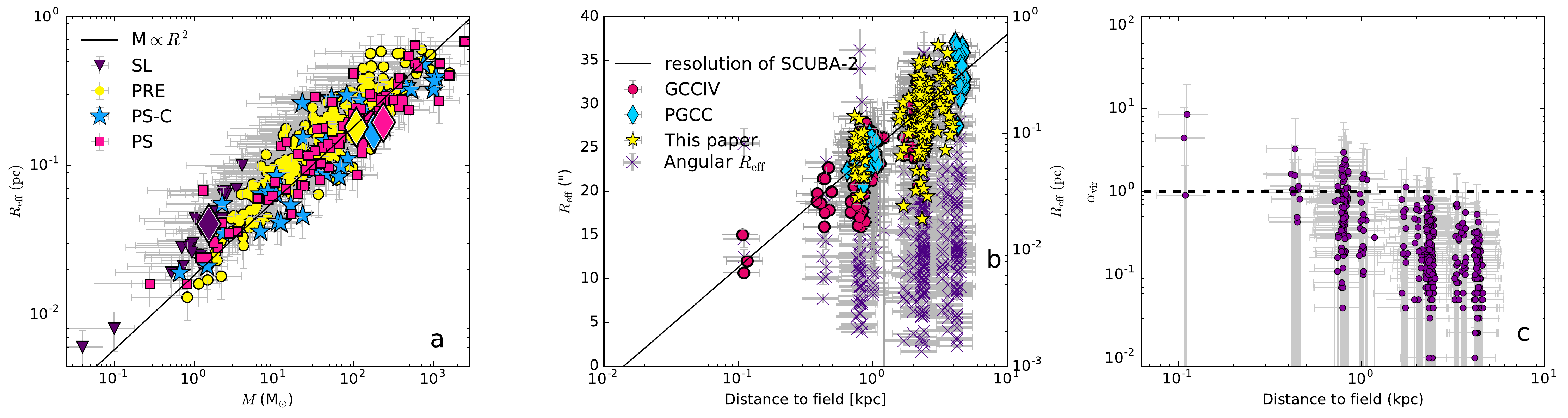}}
        \caption{(a) Effective clump radius as a function of clump mass. The symbols correspond to unconfirmed PS (pink squares), PS-C (blue stars), PRE  (yellow circles), and unbound SL (purple triangles) clumps. Mean mass and radius are plotted with diamonds of the same color. The black line shows $M\propto R^{2}$. (b) Angular size (left axis, purple crosses) and spatial size (right axis) of the clumps as a function of distance on a logarithmic scale. The distances are from GCCIV \citep{GCC_IV_cold_submm} (pink circles), the PGCC catalog \citep{PGCC_catalog} (blue diamonds), and this paper  (yellow stars). Random jitter of $\leq$ 10\% has been added to distances to improve readability. The black line corresponds to the resolution of the SCUBA-2 instrument. (c) Clump $\alpha_{\rm vir}$ (blue circles) plotted as a function of distance. Random jitter of $\pm$ 3\% of distance has been added for readability. The black dashed line is the limit of gravitational instability.   \label{fig:mass_vs_r_d} }
\end{figure*}

We plot mass against \Reff, physical and angular size against distance, and $\alpha_{\rm vir}$ against distance in Fig. \ref{fig:mass_vs_r_d}a, b, and c, respectively. The strong relation between distance and other parameters is apparent in the images.

\subsubsection{Clump spectral index \label{sec:discussion_clump_properties_beta}}

The derived opacity spectral index of the clumps in our samples are in the range 1.3--2.4, with a mean around 1.7 using 250--500\,\micro\spc data. Our sample agrees with current understanding of dense SF regions, which generally are understood to have $\beta\approx 1.8$ \citep{planck_early_XXII,planck_early_XXIII,Planck_thermal_dust, GCC_VI}. Our values are also consistent with those found in the $\lambda$ Orionis cloud of \B $\simeq$1.65, and slightly lower than those found in Orion A and Orion B \citep{2018ApJS..236...51Y}.  
In the literature, the highest values of opacity spectral index have been found in SL clumps \citep{GCC_VI}. Although all averages of opacity spectral index in our sample are similar within the uncertainties, our SL clumps also show slightly higher mean values. \citet{GCC_VI} find values up to \B = 2.2 using the full sample of GCC clumps.

We extended the SED analysis of our clumps to 850\,\micro, which showed a slight increase in opacity spectral index as a function of wavelength. This is in contrast to previous studies, in which opacity spectral index generally decreases beyond 500\,\micro\spc \citep{1995ApJ...451..188R,2012A&A...537A.113P,GCC_VI}. We also do not find any 850\,\micro\, excess, but rather a slight loss around 5\%.

\subsubsection{Clump sizes \label{sec:discussion_clump_properties_size}}

There is a correlation between distance and derived radius, both in the literature and in our sample, with a derived relation of $R_{\rm eff} \textrm{[pc]} =  7.1 \times d \textrm{[kpc]} +  1.0 $. This relation is set primarily by the resolution of the SCUBA-2 instrument on the lower end, and spatial filtering on the upper end. The clump angular effective radius is similar for all clumps regardless of distance (Fig. \ref{fig:mass_vs_r_d}b). The black line represents the resolution of the SCUBA-2 instrument; the deconvolved sizes of some clumps are below the beam size. Typical core sizes in the literature are $\sim 0.01$\,pc at a distance of 200\,pc \citep{Perseus_core_forming_clump}, but rise to  $\sim 0.1$\,pc at distances over 5\,kpc \citep{2017MNRAS.466..340C,SCOPE_survey,thermal_Hi_gal}. Our derived radii and distances match the literature. In addition, more massive clumps are also likely to be rounder, whereas smaller cores have a large variation in aspect ratio. As in our sample, the peak aspect ratio of all SCOPE sources is also close to 1 compared to all PGCC sources, which are more extended or filamentary \citep{SCOPE_catalogue}. However, many of our clumps fall on filamentary structures, as can be seen from the SCUBA-2 surface brightness maps in Appendix \ref{sec:appendix_YSO_clump}.

\subsubsection{Clump temperatures \label{sec:discussion_clump_properties_T}}

The mean clump temperature in our study of 17.8\,$\pm$\,3.6\,K is similar to the mean temperatures of 3000 MALT90 high-mass clumps at distances from $\sim$3--10\,kpc \citep{2015ApJ...815..130G,2017MNRAS.466..340C}. The temperatures of 15.5$\pm$\,3.5\,K in clumps with $d<$1.5\,kpc (distance bins I and II) are consistent with those found within nearby regions \citep[e.g. Corona Australis and Cepheus;][]{dense_cores_corona_australis,cepheus_GBS} and all PGCCs and other GCC studies \citep{sample_PGCC,SCOPE_catalogue}, but a few degrees higher than those found in the Lupus cluster and some infrared dark clouds \citep{2013ApJ...773..123S,2015MNRAS.451.3089T, 2017ApJ...841...97S,catalog_in_Lupus_cluster}. In the literature prestellar clumps have been found to be colder than protostellar clumps \citep{2015ApJ...815..130G,GCC_IV_cold_submm,2017MNRAS.466..340C,2017MNRAS.471..100E,thermal_Hi_gal}. In our sample, the mean clump temperatures are generally equal within their uncertainties. The peak temperatures of a few clumps are over 20\,K, suggesting presence of YSO formation. As our sample derives from the Planck cold clumps sample, we do not see temperatures as high as those found in, for example, the MALT90 sample \citep{2015ApJ...815..130G}.

\begin{figure*}
        \sidecaption
        \includegraphics[width=12cm]{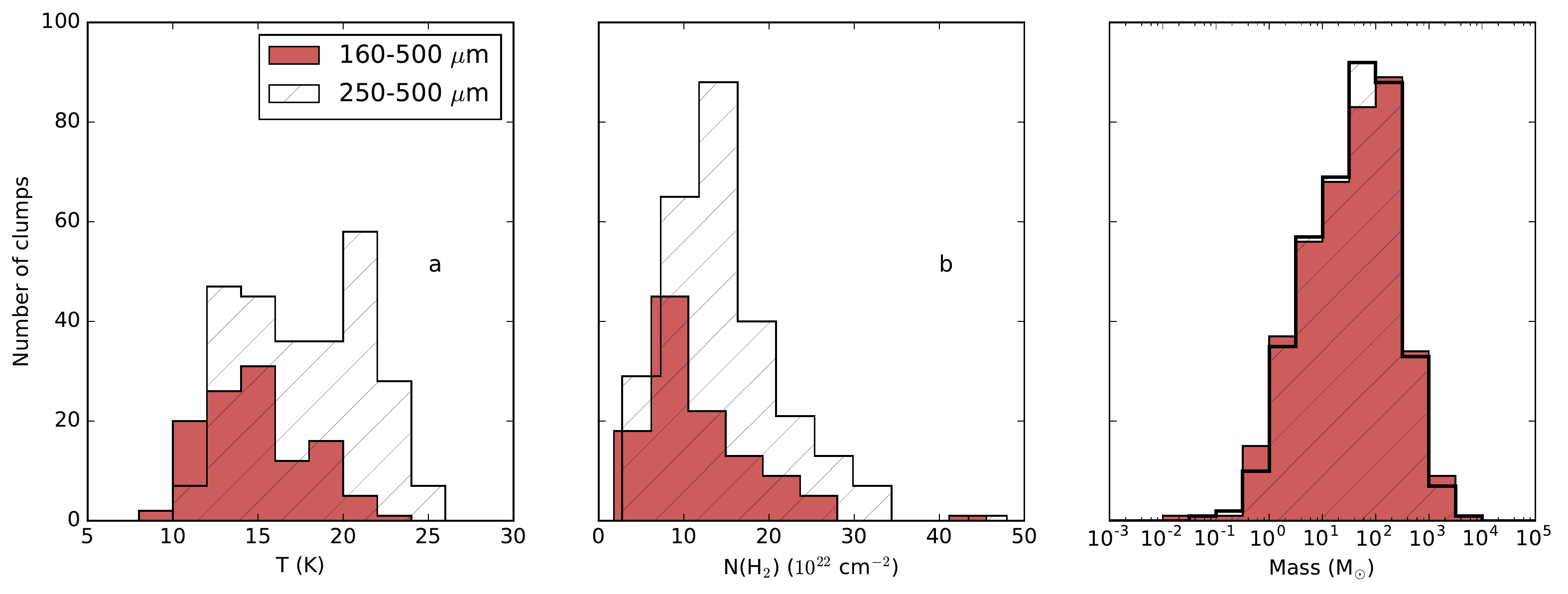}
        \caption{ Histograms of (a) mean clump temperature, (b) mean column density, and (c) mass calculated using 250--500\,\micro\, data (black dashes) and 160--500\,\micro\, data (pink) using a constant \B\, = 1.8. Clumps are not separated by type or distance for this analysis. Mass estimation assumes \textit{d}=1\,kpc and $F_{\rm 850}$ = 2\,Jy for clumps without these values. \label{fig:temp_comparison} }
\end{figure*}

A strong increase in temperature as a function of distance is visible in our sample. The mean dust temperature increases with distance as a consequence of the difficulty in observing distant colder sources, as simulated by  \citet{GCC_IV_cold_submm}. Furthermore, nearby fields are likely to be dominated by a single dense core, whereas the clumps detected in distant fields are clouds that include also more diffuse, warmer material \citep{GCC_VI}. Temperature within ATLASGAL clumps has been found to increase with increasing helio- and galactocentric distance \citep{2017MNRAS.471..100E,2018MNRAS.473.1059U}, similar to our sample. As line-of-sight temperature variations tend to increase the measured color temperatures \citep{GCC_IV_cold_submm}, it is likely that this analysis overestimates the physical mass-averaged dust temperature. This effect gets worse with the inclusion of shorter wavelengths such as PACS 160\,\micro\spc  estimates \citep{beta_error_1,beta_error_2, sample_PGCC}, thus using only SPIRE wavelengths may give a less biased estimate. We calculated temperatures, column densities, and masses for clumps using 160--500\,\micro\, data (Fig. \ref{fig:temp_comparison}). The mean temperatures are $\sim$3\,K lower and the mean column densities $\sim 3\times 10^{21}$\,cm$^{-2}$ lower than those derived using 250--500\,\micro\, data. Assuming a distance of 1\,kpc and a flux of 2\,Jy, clump masses derived using 160--500\,\micro\, data are $\sim4$\,\msun\, higher.

\subsubsection{Column density \label{sec:discussion_clump_properties_NH2}}

The mean column density over the extracted clumps, calculated from \textit{Herschel} data is approximately (1.5\,$ \pm $\,0.7)$\times 10^{22}$\,cm$^{-2}$, and is highest for protostellar clumps. The peak column densities calculated from SCUBA-2 850 \micro\spc intensities are higher by about 10\% for SL clumps and almost three times for PS-C clumps (Sect. \ref{sec:results_clump_properties_NH2}). The column densities derived for our sample are higher than those found by other studies using only \textit{Herschel} data \citep[e.g.][]{2010A&A...518L..91D,dense_cores_in_Aquila,2015ApJ...815..130G,catalog_in_Lupus_cluster} because SCUBA-2 targets are selected for high column density. We also find significantly higher volume density than in \citet{2017MNRAS.466..340C}, where our derived median volume density, derived from peak column density, is larger than their maximum \textit{n}(\MH). \citet{2017MNRAS.466..340C} also find volume density decreasing with increased dust temperature; while this trend is also visible in our results it is not significant. As these sources may be more extended along the line of sight than the distance corresponding to a single SCUBA-2 beam, our method likely overestimates the mean volume density. Especially in highly crowded regions near the plane of the Galaxy, it is possible for irregularly-shaped diffuse clouds to overlap along the line of sight, creating an illusion of a single dense clump. Even isolated clumps may face this issue because the random orientation of irregular 3D structures may affect derived clump properties. These cloud structures require observation of 3D chemistry, kinematics, and polarization to resolve \citep{2019arXiv190906004G}. However, while high-density prestellar cores exist \citep[e.g. L1544;][]{1999ApJ...518L..41O}, these high values suggest the sources have the potential to be directly involved in SF.

A density threshold for SF was first proposed by \citet{1959ApJ...129..243S} and has been observed \citep{2009ApJS..181..321E,2010ApJ...723.1019H,2010ApJ...724..687L,2012ApJ...745..190L}. Using $A_{\rm v, BG} =$ 7 mag as the threshold for SF \citep{2014MNRAS.444.2396C,dense_cores_in_Aquila}, approximately 88\% of PS-C,  85\% of PS, 100\% of PRE, and 87\% of SL clumps with estimates for column density are found above this threshold; however, these percentages apply to the mean clump column densities. In our sample, we thus do not detect significant differences between clump categories, similar to \citet{thermal_Hi_gal}. Commonly, in the literature prestellar cores are associated with high background column density. About 90\% of prestellar cores observed with \textit{Herschel} in the Aquila complex are found in regions with $A_{\rm v} \geq $ 7, and all prestellar cores in Taurus on filaments with $A_{\rm v} \geq $ 5  \citep{dense_cores_in_Aquila,census_dense_cores_Taurus}. A similar high fraction of prestellar cores with high column density has been found in Corona Australis \citep{dense_cores_corona_australis}.

In our sample, the ratio between the peak and mean column density ranges from 0.4--69, where 90\% of ratios are between 0.5--4.8; this is a wider range than that found for the 3000 MALT90 clumps of \citet{2015ApJ...815..130G}. We also find that prestellar clumps have the steepest column density profile, unlike, for example, \citet{2015ApJ...815..130G}, who find the steepest profiles in protostellar clumps.

\subsubsection{Clump masses \label{sec:discussion_clump_properties_mass}}

In our sample, we derive masses between 0.04 -- 4259\,\msun. Within all but bin II, prestellar clumps have larger masses; however, the overall distribution is dominated by the distance dependence. The mass of clumps compared to their radius is plotted for unbound starless, prestellar, and protostellar clumps in Fig. \ref{fig:mass_vs_r_d}a, and this shows a strong linear correlation caused by the finite angular resolution of the SCUBA-2 instrument. The clump masses derived by also including 850\,\micro\, data (Table \ref{tbl:SED_SCUBA}, Appendix \ref{sec:appendix_SED_fits_using_PACS_SCUBA}) are $\sim$15\% higher, but show similar patterns to masses derived using only SPIRE data.

In previous \textit{Herschel} studies, typical core masses of under 0.2\,\msun\spc have been found in the Lupus and Perseus MCs at a distance of 200\,pc \citep{catalog_in_Lupus_cluster,Perseus_core_forming_clump}, and under 10\,\msun\spc in the Aquila complex at 300\,pc \citep{dense_cores_in_Aquila}. At up to 4\,kpc distances, significantly larger masses between 27 and 2000\,\msun\spc were found \citep{SCOPE_survey}; in a sample of Hi-GAL clumps at distances under 18\,kpc, the mean mass was 850\,\msun\spc \citep{thermal_Hi_gal}, and in a sample of MALT90 clumps at distances under 10\,kpc, the mean mass was on the order of $10^{3}$\,\msun\spc \citep{2017MNRAS.466..340C}. Higher masses are found toward the center of the Galaxy, as in a sample of 48 \textit{Planck} clumps, no sources with $M > 10^{4}$\,\msun\spc were found at Galactocentric distances over 6\,kpc \citep{2016A&A...591A.105Z}; sources closer to the Galactic center have larger heliocentric distances. As in our study, Hi-GAL protostellar clumps were found to have higher mass and density than prestellar clumps \citep{thermal_Hi_gal}.

%\pagebreak
Distant starless clumps are more likely to be classified as prestellar than unbound because unbound clumps are more difficult to detect owing to lower flux \citep{catalog_in_Lupus_cluster}. Within 500\,pc of the Sun, 20--60\% of starless sources have been found to be prestellar \citep{mass_dist_in_orion_A,dense_cores_in_Aquila, census_dense_cores_Taurus,catalog_in_Lupus_cluster, dense_cores_corona_australis,PGCC_properties_in_L1495}, although mostly based on \textit{Herschel} studies. At distances of $\sim$ 1\,kpc, 80--100\% of starless sources are prestellar \citep{vela_C_SF,hi_gal_observations}. This of course depends strongly on clump detection methods and on the sensitivity of the observations.

The derived characteristics of our sources match these previous studies very well. In general, clumps in nearby fields under 0.5\,kpc have mass under 5\,\msun\spc and radius under 0.1\,pc, fitting the definition of cores. The most distant clumps are likely entire bound clouds and have radii nearer to 1\,pc and mass up to 4$\times 10^{3}$\,\msun, which themselves contain unresolved clumps and cores. We also note this strong correlation between virial $\alpha$ and distance, seen in Fig. \ref{fig:mass_vs_r_d}, where all clumps with $d\gtrsim$2\,kpc are classified as gravitationally bound.

\section{Summary and conclusions \label{sec:conclusions}}

We have characterized 529 dense clumps selected from the JCMT SCOPE survey. The clumps were selected from 53 fields mapped by the \textit{Herschel} satellite and cover a range of distances from 0.1 to 4.5\,kpc. We classify the clumps as protostellar, prestellar, or unbound starless based on their estimated virial masses, the presence of YSOs, and dust temperature (Sect. \ref{sec:results_virial_anal}). We find the following properties for these dense clumps. 
\begin{enumerate}
        \item The gravitational stability was estimated for 336 clumps with distance estimates. A total of 242 (46\%) clumps have both mass estimates and measurements of the \textit{Herschel} 160\,\micro\spc /250\,\micro\spc intensity ratio. 
        \begin{enumerate}
                \item  Using those 242 clumps, we find 167 (69\%) starless clumps, out of which 145 are classified as prestellar, and 75 (31\%) protostellar clumps (PS), 28 of which we are able to further confirm by an increase in the \textit{Herschel} 160\,\micro\spc versus 250\,\micro\spc intensity ratio.              
                \item The reliability of the classification is affected by the lack of measurements of molecular line data (for estimates of turbulent support and external pressure) and magnetic fields.  
                \item We find that many prestellar clumps are likely to be entire bound structures, which themselves contain smaller clumps and cores. This is because of their significantly higher mean distance compared to the other categories. 
        \end{enumerate}
        \item Gravitational stability is also dependent on distance; all clumps beyond 2\,kpc are classified as gravitationally bound. 
\end{enumerate}

These fields have high column density and low temperature, and there is little effect of internal heating. The characteristics of these sources show strong dependence on distance. They span a wide range of objects, from individual nearby high-latitude cores to distant star-forming clouds near the Galactic center. Calculated distances to sources can be improved in the future by utilizing new observations, such as GAIA data. 
We do not find it possible to resolve significant differences between the various categories of clumps with our data. Chemical evolution surveys are necessary to confirm the evolutionary status of these sources.

The mean temperature of dense clumps is 17.8\,K, where clumps with distance $d < 1.5$\,kpc have $\langle$\textit{T}$\rangle \simeq 15.5$\,K and more distant clumps have $\langle$\textit{T}$\rangle \simeq $19.4\,K. All starless (SL PRE) clumps have temperatures higher by 1\,K than all protostellar (PS PS-C) clumps, but also have higher mean distances by $\sim$0.4\,kpc.  We attribute the strong temperature dependence on distance to three factors: the increasing difficulty of detecting distant cold sources, the fact that these distant fields are closer to the Galactic center, and that distant sources cover a larger spatial area and thus include more diffuse matter. 

The clump masses range from 0.04 to 4259\,\msun. However, the mean virial mass only increases by one order of magnitude, leading to a higher percentage of bound sources in distant fields. The mean column density over the extracted clumps, calculated from \textit{Herschel} data, is (1.5\,$ \pm $\,0.7)$\times 10^{22}$\,cm$^{-2}$. The peak column densities calculated from SCUBA-2 850\,\micro\spc intensities are approximately two times higher (Sect. \ref{sec:results_clump_properties_NH2}), and the densest clumps have a peak column density of $\sim 10^{23}$\,cm$^{-2}$.  Prestellar clumps tend to have the steepest column density profiles and unbound starless clumps have the flattest. Clump opacity spectral index shows no dependence on distance or the SF state of the clump, with the mean \B\spc = 1.7 $\pm$ 0.1. We find that among dense clumps the assumption of a constant spectral index can be accurate.

We have characterized the dust properties of the ISM in the fields containing the SCUBA-2 clumps. Both the temperature and column density are higher in the plane of the Galaxy and drop toward $|b| = 10$\degr. We also find higher \textit{T} and \textit{N}(\MH) near the Galactic center. Monte Carlo simulations suggest that only part of the detected T-\B\, anticorrelation is caused by observational noise. We derive a temperature-$\beta$ relation of $$\beta = 7.01\times T^{-0.47},$$ for the temperature range 10-30\,K.

\begin{acknowledgements}

We thank the anonymous referee for their helpful comments. 
E.M. and M.J. acknowledge the support of the Academy of Finland Grant No. 285769. 
E.M. is funded by the University of Helsinki doctoral school in particle physics and universe sciences (PAPU). 
L.B. acknowledges support from CONICYT project Basal AFB-170002. 
J.H. thanks the National Natural Science Foundation of China under grant Nos. 11873086 and U1631237 and support by the Yunnan Province of China (No.2017HC018). This work is sponsored in part by the Chinese Academy of Sciences (CAS) through a grant to the CAS South America Center for Astronomy (CASSACA) in Santiago, Chile. 
P.S. was financially supported by a Grant-in-Aid for Scientific Research (KAKENHI Number 18H01259) of Japan Society for the Promotion of Science (JSPS).
A.S. acknowledges financial support from the National Science Foundation grant AST-1715876. 
CWL is supported by Basic Science Research Program through the National Research Foundation of Korea (NRF) funded by the Ministry of Education, Science and Technology (NRF-2019R1A2C1010851). Tie Liu acknowledges the support from the international partnership program of Chinese academy of sciences through grant No.114231KYSB20200009 and the support from National Natural Science Foundation of China (NSFC) through grant NSFC No.12073061.\\

The James Clerk Maxwell Telescope is operated by the East Asian Observatory on behalf of The National Astronomical Observatory of Japan; Academia Sinica Institute of Astronomy and Astrophysics; the Korea Astronomy and Space Science Institute; Center for Astronomical Mega-Science (as well as the National Key R\& D Program of China with No. 2017YFA0402700). Additional funding support is provided by the Science and Technology Facilities Council of the United Kingdom and participating universities in the United Kingdom and Canada. Additional funds for the construction of SCUBA-2 were provided by the Canada Foundation for Innovation. 
SCUBA-2 data were collected under program IDs M16AL003 and M15AI05.  M16AL003 corresponds to the SCOPE survey, M15AI05 to data collected during the pilot program. %\\
This research used the facilities of the Canadian Astronomy Data Centre operated by the the National Research Council of Canada with the support of the Canadian Space Agency. 
The James Clerk Maxwell Telescope is operated by the Joint Astronomy Centre on behalf of the Science and Technology Facilities Council of the United Kingdom, the Netherlands Organisation for Scientific Research, and the National Research Council of Canada.

\end{acknowledgements}

\bibliographystyle{aa} % style aa.bst
\bibliography{paper_bib} % your references Yourfile.bib

\appendix

\section{Instrument specifications \label{sec:appendix_instrument_specs}}
In the following tables we present properties of the instruments used in our study. Table \ref{tbl:instrument_specs} lists the parameters of the continuum observations and Table \ref{tbl:TRAO_instrument_specs} the parameters of the line observations.

\begin{table}
        \caption{Continuum instrumental characteristics \label{tbl:instrument_specs}}
                \begin{tabular}{llll}
                        \hline
                        \hline
                         & Central  & Beam  & Pixel\\
                        Band & Wavelength & FWHM & size\\
                         & (\micro)  & (\arcsec) & (\arcsec)\\
                        \hline
                        PACS blue\tablefootmark{a}  & 70   & 5.6 &  3.2 \\
                        PACS green\tablefootmark{a}  & 100  & 6.8 & 3.2  \\
                        PACS red\tablefootmark{a}  & 160  & 12.0 & 3.2 \\
                        SPIRE PSW\tablefootmark{b} & 250   & 17.9 & 6.0 \\
                        SPIRE PMW\tablefootmark{b} & 350 & 24.2 & 10.0 \\
                        SPIRE PLW\tablefootmark{b} & 500  & 35.4 & 14.0\\
                        SCUBA-2\tablefootmark{c}\tablefootmark{d} & 850\tablefootmark{c} & 14.6\tablefootmark{d} & 4.0\tablefootmark{c} \\
                        \hline
                                
                \end{tabular}
                \tablefoot{ Central wavelength, beam FWHM, and pixel size of the 7 bands used in this paper.  \\ \tablefoottext{a}{Sources from \citet{pacs_handbook}.} \\ \tablefoottext{b}{ Sources from \citet{spire_handbook}.} \\ \tablefoottext{c}{ Sources from \citet{SCUBA_instrumentation}.}   \\ \tablefoottext{d}{ Sources from \citet{SCUBA_calibration}.}  }
\end{table}

\begin{table*}
        \caption{TRAO molecular line instrumental characteristics \label{tbl:TRAO_instrument_specs}}
                \begin{tabular}{llllll}
                        \hline
                        \hline
                         & Main-beam  & Beam  & Pixel & System & Typical\\
                        Band & Efficiency $\eta_{\rm B}$ & FWHM & size & Temperature & Sensitivity \\
                        J=(1-0)  & (\%)  & (\arcsec) & (\arcsec) & (K) & ($T_{\rm A}^{*}$)\\
                        \hline
                        \element[][12]{CO} &  54 & 45 & 24.0 & 500 & 0.5\tablefootmark{a}\\
                        \element[][13]{CO} & 51  &  47 & 24.0 & 250 & 0.2\tablefootmark{a}\\
                        \hline
                                
                \end{tabular}
                \tablefoot{ Instrumental parameters for the \element[][12]{CO} and \element[][13]{CO} TRAO observations.  \\ \tablefoottext{a}{Typical sensitivity is calculated at a spectral resolution of 0.33\,km\,s$^{-1}$ over the mapping field for the typical 40 minute integration time \citep{SCOPE_survey}. }     } 
\end{table*}

\section{SED fits using PACS 160\,\micro\, and SCUBA-2 850\,\micro\, data \label{sec:appendix_SED_fits_using_PACS_SCUBA}  }

\begin{table*}
        \caption{SED fits with all wavelength combinations \label{tbl:SED_SCUBA} }
                \begin{tabular}{l|llll|llll}
                        \hline
                        \hline

                Band (\micro) & (160--500) & (250--500)  & (250--850) & (160--850)  &  (160--500) & (250--500)  & (250--850) & (160--850)    \\ 
                Clump & \multicolumn{4}{c|}{ $\langle T\rangle$ }    & \multicolumn{4}{c}{ $\langle N$(H$_{2})\rangle$  }  \\ 
                 type & \multicolumn{4}{c|}{(K)}   & \multicolumn{4}{c}{($10^{21}$cm$^{-2}$)}  \\ 
                \hline
                All & 15.5 $\pm$ 5.3 & 16.1 $\pm$ 6.3 & 14.3 $\pm$ 6.4 & 14.9 $\pm$ 5.3& 18.3 $\pm$ 56.8 & 15.4 $\pm$ 31.4 & 21.5 $\pm$ 30.3 & 20.5 $\pm$ 57.4 \\ 
                PS-C & 15.9 $\pm$ 1.3 & 16.2 $\pm$ 0.3 & 12.7 $\pm$ 0.2 & 14.2 $\pm$ 0.6& 6.3 $\pm$ 2.4 & 8.2 $\pm$ 5.4 & 19.4 $\pm$ 13.1 & 10.2 $\pm$ 5.1 \\ 
                PS & 14.8 $\pm$ 3.6 & 17.1 $\pm$ 7.1 & 16.0 $\pm$ 7.7 & 13.8 $\pm$ 3.9& 7.4 $\pm$ 3.0 & 6.8 $\pm$ 4.3 & 16.5 $\pm$ 24.7 & 16.2 $\pm$ 15.7 \\ 
                PRE & 16.1 $\pm$ 5.5 & 16.5 $\pm$ 6.4 & 15.1 $\pm$ 6.7 & 15.6 $\pm$ 5.5& 20.9 $\pm$ 66.0 & 17.0 $\pm$ 36.1 & 22.2 $\pm$ 33.7 & 20.7 $\pm$ 62.7 \\ 
                SL & 14.0 $\pm$ 4.8 & 14.6 $\pm$ 5.7 & 11.9 $\pm$ 4.4 & 12.8 $\pm$ 4.5& 12.6 $\pm$ 18.4 & 12.4 $\pm$ 14.4 & 20.3 $\pm$ 18.5 & 20.9 $\pm$ 42.7 \\ 
                SL PRE & 15.6 $\pm$ 5.4 & 16.0 $\pm$ 6.3 & 14.3 $\pm$ 6.3 & 14.9 $\pm$ 5.4& 18.7 $\pm$ 57.6 & 15.9 $\pm$ 32.2 & 21.8 $\pm$ 30.6 & 20.7 $\pm$ 58.3 \\ 
                PS PS-C & 15.1 $\pm$ 3.2 & 17.0 $\pm$ 6.6 & 15.6 $\pm$ 7.3 & 13.9 $\pm$ 3.3& 7.1 $\pm$ 2.9 & 7.0 $\pm$ 4.5 & 16.9 $\pm$ 23.6 & 14.5 $\pm$ 13.8 \\ 
                \hline
                Band (\micro) &  (160--500) & (250--500)  & (250--850) & (160--850)  &  (160--500) & (250--500)  & (250--850) & (160--850) \\ 
                Clump & \multicolumn{4}{c|}{ $\langle M\rangle$} & \multicolumn{4}{c}{ $\beta$   } \\ 
                type  & \multicolumn{4}{c|}{(\msun)} & \multicolumn{4}{c}{  } \\ 
                \hline
                        All & 154 $\pm$ 276 & 137 $\pm$ 258 & 149 $\pm$ 262 & 139 $\pm$ 249& 1.88 $\pm$ 0.66 & 1.86 $\pm$ 0.63 & 2.2 $\pm$ 0.66 & 2.03 $\pm$ 0.62 \\ 
                PS-C & 10 $\pm$ 2 & 10 $\pm$ 3 & 16 $\pm$ 5 & 12 $\pm$ 3& 1.44 $\pm$ 0.07 & 1.45 $\pm$ 0.09 & 2.03 $\pm$ 0.17 & 1.77 $\pm$ 0.03 \\ 
                PS & 14 $\pm$ 9 & 31 $\pm$ 40 & 34 $\pm$ 38 & 45 $\pm$ 68& 1.5 $\pm$ 0.47 & 1.74 $\pm$ 0.32 & 2.04 $\pm$ 0.85 & 1.84 $\pm$ 0.55 \\ 
                PRE & 209 $\pm$ 309 & 175 $\pm$ 287 & 191 $\pm$ 290 & 176 $\pm$ 275& 1.95 $\pm$ 0.68 & 1.92 $\pm$ 0.67 & 2.18 $\pm$ 0.68 & 2.03 $\pm$ 0.65 \\ 
                SL & 18 $\pm$ 29 & 17 $\pm$ 31 & 20 $\pm$ 31 & 18 $\pm$ 30& 1.73 $\pm$ 0.59 & 1.7 $\pm$ 0.54 & 2.29 $\pm$ 0.52 & 2.07 $\pm$ 0.54 \\ 
                SL PRE & 159 $\pm$ 279 & 143 $\pm$ 264 & 156 $\pm$ 268 & 144 $\pm$ 254& 1.89 $\pm$ 0.66 & 1.87 $\pm$ 0.64 & 2.21 $\pm$ 0.65 & 2.04 $\pm$ 0.62 \\ 
                PS PS-C & 13 $\pm$ 8 & 29 $\pm$ 38 & 31 $\pm$ 36 & 41 $\pm$ 65& 1.49 $\pm$ 0.40 & 1.7 $\pm$ 0.31 & 2.04 $\pm$ 0.79 & 1.82 $\pm$ 0.47 \\ 
        
                \hline
                \end{tabular}
                \tablefoot{ Results of the MBB fit to different wavelength ranges, with opacity spectral index as a free parameter. Temperature and column density are the average value over the clump footprint. To ensure consistency between SPIRE, PACS, and SCUBA-2 data, the background has been subtracted from all fields. The errors are the standard deviation of the distribution. This table includes only those fields for which distances (to calculate mass) are available.    }
\end{table*}

In the main paper we relied mainly on SED fits to 250-500\,$\mu$m observations because we do not have 160\,$\mu$m data for all fields and the inclusion of the shorter wavelength might bias the column density estimates. In Sect. \ref{sec:results_Herschel_spectra_and_850um} we also studied the hypothesis that the flattening of the dust emission spectrum at long wavelengths could be visible already in the 850\,$\mu$m observations. Also in this case we used fits to the 250-500\,$\mu$m range, comparing the 850\,$\mu$m flux densities to the MBB SEDs fitted to the shorter wavelength observations. The flux calculation in this analysis differs slightly from that used in the main paper, as here we subtract the background value in an annulus around the clump from the flux of the clump. As a result of this background subtraction, MBB fits are also performed on a clump-by-clump basis instead of on the field as a whole. The background is removed from the clump in an annulus of width 0.5\arcmin, 20\arcsec\, from the edge of the clump.

\begin{figure*}
        \sidecaption
        \includegraphics[width=12cm]{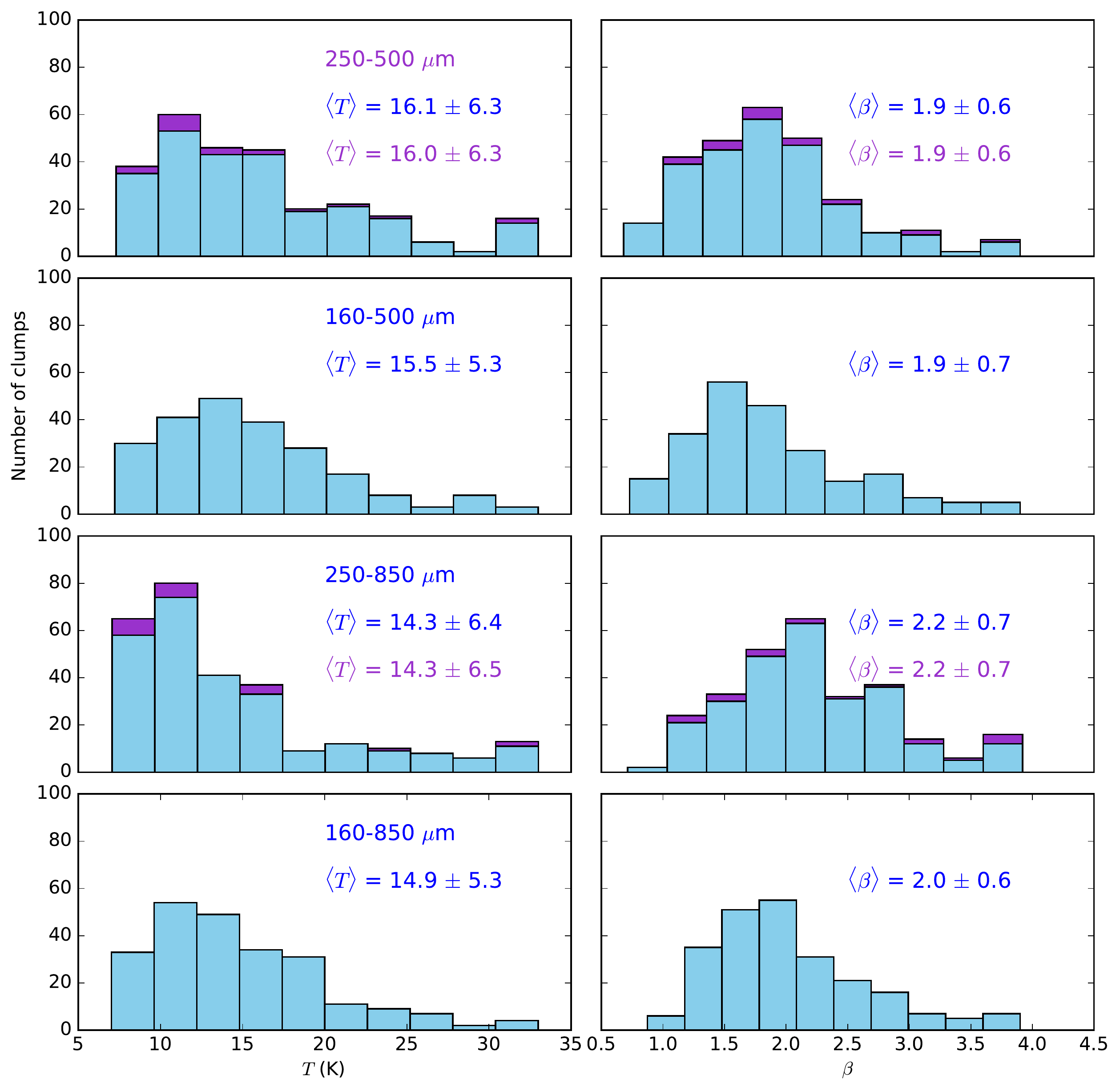}
        \caption{ Comparison of MBB fits of clump spectra performed using different band combinations. The purple histograms correspond to all the clumps and the blue histograms to the subsets of clumps for which 160\,$\mu$m measurements also exist. The mean values and standard deviations of the temperature and the dust opacity spectral index distributions are given in the frames, written with the same colors.   \label{fig:all_clumps} }
\end{figure*}

For completeness, we show in Fig.~\ref{fig:all_clumps} SED fits for all wavelength combinations 250-500\,$\mu$m, 160-500\,$\mu$m, 250-850\,$\mu$m, and 160-850\,$\mu$m. We also derive temperatures, column densities, and masses for clumps with 250-850\,\micro\, data using the MBB SED fit described in Section \ref{sec:methods_general_ISM} with a variable opacity spectral index (Table \ref{tbl:SED_SCUBA}). The mass is calculated as in Eq. (\ref{eq:mass}). Owing to SCUBA-2 filtering, we remove the background on all data analyzed. This results in slightly different predicted quantities in the 160--500\,\micro\, and 250--50\,\micro\, bands than we derive in the main paper.

Compared to the SPIRE-only fits, the inclusion of the 160\,$\mu$m point increases the temperature estimates by an average of 0.5\,K, whether the 850\,$\mu$m data point is included in the fit or not. For dust temperatures close to 10\,K, this corresponds to more than a factor of 2 decrease in the mass estimates, thus indicating a potentially significant bias. In contrast, the inclusion of the 850\,$\mu$m point decreases temperatures by close to 1\,K, thereby leading to significantly higher derived column densities. Mass derivation is not strongly affected by the inclusion of 850\,$\mu$m data. We find an increase in opacity spectral index with the inclusion of the 850\,\micro\  data point, both with and without 160\,\micro\, data.

\section{Data coverage \label{sec:appendix_data_coverage}  }
In Table \ref{tbl:data_coverage} we present what data we have for each field. In the case of \textit{Herschel}, this also includes fields that are only partially covered by observations. The table also lists the YSO catalogs used.

\begin{table*}
        \caption{Coverage of each SCUBA-2 field.   \label{tbl:data_coverage}}
        \begin{tabular}{l|lll|llllllll}
                \hline
                \hline
                & \multicolumn{3}{c|}{ Data }& \multicolumn{6}{c}{YSO catalog} \\
                SCUBA-2 & \multicolumn{2}{c}{\textit{Herschel}}  & TRAO & \multicolumn{2}{c}{Spitzer} & \multicolumn{2}{c}{Marton} & \multicolumn{2}{c}{AKARI} & \multicolumn{2}{c}{\textit{Herschel}} \\
                Field & 160\micro & SPIRE & & type I & type II & type I/II & type III & FIS & PSC & 70\,\micro & 100\,\micro\\
                \hline
                G001.36+20.96 & X & X &   &   &   & X &   & X & X & X &   \\ 
                G005.93-01.00 & X & X & X &   &   & X &   & X & X &   &   \\
                G006.01+36.74 & X & X &   &   &   &   &   & X & X &   &   \\
                G010.21+02.40 & X & X & X &   &   & X & X & X & X &   & X \\ 
                G014.14-00.55 & X & X\tablefootmark{a} & X & X & X & X &   & X & X & X & \\
                G014.15-00.55 & X & X\tablefootmark{a} & X & X & X &   X &   & X & X & X &   \\ 
                G014.22-00.19 & X & X &   &   & X & X &   & X & X & X &   \\ 
                G014.72-00.19 & X & X &   &   & X & X &   & X & X & X &   \\
                G014.73-00.19 & X & X &   &   & X & X &   & X & X & X &   \\ 
                G016.28-00.45 & X & X &   & X & X & X & X & X & X & X &   \\     
                G016.37-00.61 & X & X & X &   &   & X &   & X & X & X &   \\ 
                G016.43-00.62 & X & X & X & X & X & X &   & X & X & X &   \\ 
                G016.97+00.28 & X & X & X &   & X & X &   & X & X & X &   \\ 
                G017.23-01.47 &   & X & X &   &   & X &   & X & X &   &   \\ 
                G017.38+02.26 &   &   &   &   &   & X &   & X & X &   &   \\     
                G023.35-00.26 & X & X &   &   & X & X &   & X & X & X &   \\
                G023.68+00.58 & X & X & X &   & X & X &   & X & X & X &   \\ 
                G023.69+00.58 & X & X & X &   & X & X &   & X & X & X &   \\  
                G024.04+00.27 & X & X & X &   & X & X &   & X & X & X &   \\ 
                G026.54+00.72 & X & X & X &   &   & X &   & X & X & X &   \\     
                G033.73-00.01 & X & X & X & X & X & X & X & X & X & X &   \\ 
                G034.74-01.39 & X & X & X &   &   & X &   & X & X & X &   \\ 
                G035.38-01.77 & X & X\tablefootmark{a} & X &   &   & X &   & X & X & X &   \\
                G035.50-00.31 & X & X &   &   & X & X &   & X & X & X &   \\
                G037.93+02.18 & X & X &   &   &   & X & X & X & X &   & X \\ 
                G039.74+01.98 & X & X & X &   &   & X & X & X & X &   &   \\     
                G057.12+03.63 &   &   &   &   &   & X &   & X & X &   &   \\   
                G069.81-01.67 & X & X &   &   &   & X &   & X & X &   & X \\  
                G070.40-01.40 & X & X &   &   &   & X &   & X & X &   & X \\ 
                G074.13+00.11 & X & X &   &   &   & X &   & X & X & X &   \\
                G082.40-01.84 & X & X & X &   &   &   & X & X & X &   & X \\ 
                G082.42-01.84 & X & X & X &   &   &   & X & X & X &   & X \\
                G087.07-04.20 & X & X &   &   &   & X &   & X & X &   &   \\ 
                G089.66-06.62 & X & X &   &   &   & X &   & X & X &   & X \\ 
                G091.88+04.17 &   & X &   &   &   & X &   & X & X &   &   \\
                G092.04+03.92 & X & X &   &   &   & X &   & X & X &   & X \\ 
                G092.28+03.79 & X & X &   &   &   & X &   & X & X &   & X \\ 
                G093.51-04.31 & X & X &   &   &   & X &   & X & X & X & X \\ 
                G105.44+09.88 &   & X & X &   &   & X &   & X & X &   &   \\ 
                G107.18+05.44 & X & X & X &   &   & X &   & X & X &   & X \\ 
                G107.26+05.71 & X & X &   &   &   & X &   & X & X &   &   \\
                G109.70+02.52 & X & X &   &   &   & X &   & X & X & X &   \\  
                G109.81+02.72 & X & X & X &   &   & X &   & X & X &   &   \\     
                G139.14-03.23 & X & X & X &   &   & X &   & X & X &   & X \\
                G159.23-20.09 & X & X & X &   &   & X &   & X & X & X & X \\  
                G162.46-08.69 & X & X &   &   &   & X &   & X & X & X &   \\ 
                G171.53-14.91 & X & X &   &   &   & X &   & X & X & X & X \\
                G172.89+02.28 &   &   &   &   &   & X &   & X & X &   &   \\ 
                G173.15+02.40 & X &   & X &   &   & X &   & X & X & X & X \\
                G178.28-00.60 & X & X &   &   &   & X &   & X & X & X &   \\
                G195.73-02.28 & X & X & X &   &   & X &   & X & X &   & X \\      
                G202.31+02.53 & X & X & X &   &   &   &   & X & X &   & X \\ 
                G215.87-17.50 &   &   & X &   &   & X &   & X & X &   &   \\ 
                \hline
        \end{tabular}
        \tablefoot{'X' denotes fields covered by \textit{Herschel} or TRAO data, or the YSO catalog. Spitzer and Marton types I--III refer to YSO classifications.\\ \tablefoottext{a}{This field is located at the edge of the \textit{Herschel} map; the available data are not sufficient to calculate column densities and temperatures. }   }
\end{table*}

\section{Distances to sources}
\label{sec:appendix_distance}
The following tables list the distance estimates used in this study. Table \ref{tbl:kinematic_distance} shows new kinematic distances derived for sources with \element[][12]{CO} TRAO data, as well as the galactocentric and both near- and far heliocentric distances. Field G105.44+09.88 uses \element[][12]{CO} line data from \cite{ke_wang_sampling_data_release}. The distances are derived as explained in Sect. \ref{sec:methods_distance_estimate}. Table \ref{tbl:distances_GCCIV_2} summarizes all distance estimates, including the values from \citet{GCC_IV_cold_submm} and \citet{PGCC_catalog}.

\begin{table*}[h]
        \centering
        \caption{Input values and results of kinematic distance calculations.  \label{tbl:kinematic_distance} }
        \begin{tabular}{l|llll|l}
                \hline
                \hline
                & \multicolumn{4}{|l|}{$^{12}$CO} & Adopted \\
                & $v_{LSR}$ & \textit{R} & \textit{d} & \textit{d} & distance\\
                Field & (km\,s$^{-1}$) &  (kpc)&  (kpc)&  (kpc) &  (kpc) \\
                \hline
                
                G005.91-01.00  &  14.341  &  5.227  &  3.301  &  13.608  & 3.3 \\ 
                G010.20+2.39  &  5.582  &  7.53  &  0.988  &  15.744  & 1.0  \\ 
                G014.14-0.55  &  20.403  &  6.214  &  2.386  &  14.099  &  2.4 \\ 
                G016.36-0.62  &  22.111  &  6.321  &  2.306  &  14.006  & 2.3 \\ 
                G016.42-0.63  &  22.483  &  6.299  &  2.331  &  13.976  &  2.3\\ 
                G016.96+0.27  &  23.609  &  6.269  &  2.373  &  13.888  & 2.4 \\ 
                G017.21-01.46  &  42.355  &  5.17  &  3.602  &  12.637  &  3.6 \\ 
                G023.68+0.57  &  9.068  &  7.813  &  0.756  &  14.812  & 0.8 \\ 
                G024.04+0.26  &  7.988  &  7.912  &  0.649  &  14.877  & 0.6 \\ 
                G026.53+0.71  &  8.174  &  7.956  &  0.613  &  14.597  &  0.6 \\ 
                G033.72-0.027  &  10.887  &  7.912  &  0.719  &  13.421  & 0.7 \\ 
                G034.73-1.39  &  14.149  &  7.74  &  0.948  &  13.023  & 0.9 \\ 
                G035.37-01.76  &  42.86  &  6.419  &  2.809  &  11.054  & 2.8 \\ 
                G035.49-0.31  &  13.238  &  7.806  &  0.873  &  12.969 &  0.9 \\ 
                G039.38-1.98  &  31.112  &  7.03  &  2.061  &  11.079  & 2.1 \\ 
                G082.39-1.84  &  -3.231  &  8.763  &  -1.284  &  3.535  & 3.5 \\  
                G105.44+09.88\tablefootmark{a} & -10.48 & 9.087 & -6.188 & 1.669 & 1.7 \\
                G107.17+5.44  &  -9.502  &  9.047  &  -6.495  &  1.477 &  1.5 \\ 
                G109.79+2.71  &  -9.805  &  9.067  &  -7.15  &  1.394 &  1.4 \\ 
                G139.19-3.29  &  -31.792  &  11.121  &  -16.068  &  3.201  & 3.2 \\
                G159.23-34.51  &  -4.927  &  9.223  &  -16.664  &  0.769  & 0.8 \\ 
                G173.15+02.38  &  -17.722  &  26.556  &  -34.976  &  18.098  &  18.1 \\ 
                G195.74-2.29  &  4.548  &  9.355  &  -17.248  &  0.885  & 0.9 \\ 
                G202.32+02.51  &  6.825  &  9.412  &  -16.704  &  0.978  & 1.0 \\ 
                G215.8-17.51  &  10.082  &  9.377  &  -14.845  &  1.056  & 1.1  \\
                \hline

        \end{tabular}
        \tablefoot{Velocities have been found from TRAO CO line data.  \\ \tablefoottext{a}{Velocity estimate derived using the \element[][12]{CO} line measured by \cite{ke_wang_sampling_data_release}}    }
\end{table*}

\begin{table}[h]
        \centering
        \caption{Distances of the fields \label{tbl:distances_GCCIV_2}}
        \begin{tabular}{l|lll}
         \hline
         \hline
         SCUBA-2 & \multicolumn{3}{c}{Distance}\\
         field & \multicolumn{3}{c}{(kpc)}\\
         \hline
         (1) & (2) & (3) & (4) \\
         & (GCCIV) & PGCC& CO-line data\\
        \hline
        G005.93-01.00 & - & -  & 3.3$^{+6.6}_{-1.7}$\\ 
        G006.01+36.74 & 0.110 $\pm$ 0.010 & -  & -\\
        G010.21+02.40 & 0.830 $\pm$ 0.400 & - & 1.0$^{+1.0}_{-0.5}$ \\
        G014.14-00.55 & -& - &  2.4$^{+2.4}_{-1.2}$ \\ 
        G014.15-00.55 & -& - &   2.4$^{+2.4}_{-1.2}$ \\ 
        G016.37-00.61  & - & -  & 2.3$^{+2.3}_{-1.2}$ \\
        G016.43-00.62  &  - & -  &  2.3$^{+2.3}_{-1.2}$  \\ 
        G016.97+00.28  & - & -  &  2.4$^{+2.4}_{-1.2}$ \\ 
        G017.21-01.46  &  - & -  &  3.6$^{+3.6}_{-1.8}$ \\ 
        G023.68+00.58  &  - & -  & 0.8$^{+0.8}_{-0.4}$ \\ 
        G023.69+00.58  &  - & -  &  0.8$^{+0.8}_{-0.4}$ \\ 
        G024.04+00.27 & - &     4.460 & 0.6$^{+0.6}_{-0.3}$ \\  
        G026.54+00.72 & - &     4.220 & 0.6$^{+0.6}_{-0.3}$ \\
        G033.73-00.01& - &      4.220 & 0.7$^{+0.7}_{-0.4}$ \\
        G034.74-01.39& - &      4.220 & 0.9$^{+0.9}_{-0.5}$ \\
        G035.38-01.77& - &      4.220 & 2.8$^{+2.8}_{-1.4}$ \\
        G035.50-00.31& - &      4.220 & -\\
        G037.93+02.18 & 1.060 $\pm$ 0.790 & 1.060 & -\\ 
        G039.74+01.98 & 0.990 $\pm$ 0.480 & - & 2.1$^{+2.1}_{-1.1}$ \\ 
        G069.81-01.67 & 1.780 $\pm$ 0.810 & - & -\\
        G070.40-01.40 & 2.090 $\pm$ 0.830 & - & -\\ 
        G082.40-01.84 & $1.000^{+1.000}_{-0.600}$ & - & 3.5$^{+3.5}_{-1.8}$ \\ 
        G082.42-01.84 & $1.000^{+1.000}_{-0.600}$ & - & 3.5$^{+3.5}_{-1.8}$ \\
        G087.07-04.20 & 0.700 $\pm$ 0.100 & - & -\\ 
        G089.66-06.62 & 1.210 $\pm$ 1.210 & - & -\\ 
        G092.04+03.92 & 0.800 $\pm$ 0.100 & - & -\\ 
        G092.28+03.79 & 0.800 $\pm$ 0.100 & - & -\\ 
        G105.44+09.88   & - & - & 1.7 $\pm$ 0.3 \\ 
        G107.18+05.44 & 0.800 $\pm$ 0.100 & 0.900 & 1.5$^{+1.5}_{-0.8}$ \\
        G107.26+05.71 & 0.800 $\pm$ 0.100 & 0.900 & -\\ 
        G109.70+02.52 & 0.800 $\pm$ 0.100 & - & -\\ 
        G109.81+02.72 & 0.800 $\pm$ 0.100 & 0.700 & 1.4$^{+1.4}_{-0.7}$ \\ 
        G139.14-03.23 & 2.500 $\pm$ 0.500 & - & -\\
        G139.19-3.29  &  - & -  & 3.2 \\
        G159.23-20.09  &  - & -  &  0.8$^{+0.8}_{-0.4}$ \\
        G162.46-08.69 & 0.450 $\pm$ 0.023& - & -\\ 
        G173.15+02.40 & 2.000 $\pm$ 0.400& - & 18.1$^{+18.1}_{-9.1}$ \\
        G195.73-02.28 & 1.000 $\pm$ 0.500& - & 0.9$^{+0.9}_{-0.5}$ \\ 
        G202.31+02.53 & 0.760 $\pm$ 0.100 & - & 1.0$^{+1.0}_{-0.5}$ \\
        G215.87-17.50 & 0.425 $\pm$ 0.100 & -  & 1.1$^{+1.1}_{-0.6}$ \\
        \hline
        \end{tabular}
        \tablefoot{Distances estimated in \cite{GCC_IV_cold_submm} (column 2), from the PGCC catalog \citep{PGCC_catalog} (column 3), and estimated from CO line data in this paper (column 4). In case of discrepancy between estimates from the literature and kinematic distances, the former is adopted owing to lower uncertainty. }
\end{table}

\section{FellWalker input values \label{sec:appendix_FellWalker_input}}
The input values used for the FellWalker algorithm are listed in Table \ref{tbl:FellWalker_arguments}. See Sect. \ref{sec:methods_FellWalker} for details of the RMS values. 

\begin{table}
        \caption{Input values used for FellWalker algorithm. \label{tbl:FellWalker_arguments}}
                \begin{tabular}{ll}
                        \hline
                        \hline
                        Variable & Value \\
                        \hline
                        RMS & Varies  \\
                        MinHeight & 3$\times$RMS  \\
                        AllowEdge & 0 \\
                        MinPix & 7 \\
                        CleanIter & 5 \\
                        Noise &  1$\times$RMS \\
                        FWHMBeam &   1 \\
                        MaxJump &  3 \\
                        deconv &  True \\
                        MinDip  & 1.5$\times$RMS\\
                        \hline          
                \end{tabular}
                \tablefoot{ See section \ref{sec:methods_FellWalker} for details on RMS values. FlatSlope, VeloRes, and MaxBad values were left to their default values. }
\end{table}

\section{Centers and reference regions of SCUBA-2 fields \label{sec:appendix_scuba_centers}}

Table \ref{tbl:SCUBA_centers_1} lists the central coordinates of each SCUBA-2 field and the coordinates of the reference regions used for background subtraction. Sources without \textit{Herschel} data (or with insufficient data) do not need background subtraction.

\begin{table}
        \caption{Centers of the SCUBA-2 fields and reference regions \label{tbl:SCUBA_centers_1} }
        \begin{tabular}{l|ll|ll}
                \hline
                \hline
                SCUBA-2 & \multicolumn{2}{l|}{Center of}&\multicolumn{2}{l}{Center of}\\
                field & \multicolumn{2}{l|}{SCUBA-2 field}&\multicolumn{2}{l}{reference region}\\
                \hline
                & RA & DEC & RA & DEC\\
                %& (h:m:s) & (h:m:s) & (h:m:s) & (h:m:s)\\
                & (\degr) & (\degr) & (\degr) & (\degr)\\
                \hline
                G001.36+20.96 & 248.625 & -15.787 & 248.630 &  -15.814\\
                G005.93-01.00 & 270.732 & -24.33 & 270.707 & -24.277 \\
                G006.01+36.74 & 238.534 & -2.877 & 238.575  & -2.894\\
                G010.21+02.40 & 269.831 & -18.926 & 269.858 & -18.99\\
                G014.14-00.55 & 274.55 & -16.921 & 274.477 & -16.907 \\
                G014.15-00.55 & 274.552 & -16.911 & 274.467  & -16.913\\
                G014.22-00.19 & 274.254 & -16.675  &274.307 & -16.712   \\              
                G014.72-00.19 & 274.499 & -16.236 & 274.464 &   -16.180 \\
                G014.73-00.19 & 274.503 & -16.231 & 274.458 &  -16.189   \\
                G016.28-00.45 & 275.502 & -14.99 & 275.586 & -15.073\\
                G016.37-00.61 & 275.694 & -14.986 &  275.675 &  -15.036 \\
                G016.43-00.62 & 275.735 & -14.937 & 275.814 & -15.015\\
                G016.97+00.28 & 275.178 & -14.034 &  275.120 &  -14.016 \\
                G017.23-01.47 & 276.89 & -14.625  &  276.923 &  -14.661 \\
                G017.38+02.26 & 273.58 & -12.731 & &  \\
                G023.35-00.26 & 278.69 & -8.639  & 278.625 &  -8.601 \\
                G023.68+00.58 & 278.094 & -7.958  &  278.041 &  -7.955 \\
                G023.69+00.58 & 278.101 & -7.947 & 278.046 &  -7.944  \\
                G024.04+00.27 & 278.541 & -7.782 & &  \\
                G026.54+00.72 & 279.292 & -5.356  &  279.245  & -5.341 \\
                G033.73-00.01 & 283.236 & 0.706  &  283.280 &  0.719 \\
                G034.74-01.39 & 284.922 & 0.976 & 284.867 &  0.991  \\
                G035.38-01.77 & 285.547 & 1.372 & 285.519 &  1.418\\
                G035.50-00.31 & 284.311 & 2.146 & 284.320 &  2.187 \\
                G037.93+02.18 & 283.195 & 5.438 &  283.188 &  5.438 \\
                G039.74+01.98 & 284.2 & 6.962 & 284.214 &  6.9292 \\
                G057.12+03.63 & 290.992 & 23.131 & &  \\
                G069.81-01.67 & 303.393 & 31.365 &  303.413 &  31.337\\
                G070.40-01.40 & 303.513 & 32.015 &  303.492 &  32.067\\
                G074.13+00.11 & 304.504 & 35.955 &  304.472 &  36.008  \\
                G082.40-01.84 & 312.858 & 41.406 &  312.902 &  41.379\\
                G082.42-01.84 & 312.874 & 41.425 &  312.884 &  41.394\\
                G087.07-04.20 & 319.447 & 43.31 &  319.444 &  43.2741\\
                G089.66-06.62 & 324.313 & 43.358 &  324.352 &  43.350 \\
                G091.88+04.17 & 315.126 & 52.522  & 315.200  &  52.512 \\
                G092.04+03.92 & 315.616 & 52.475 &  315.669  &  52.472 \\
                G092.28+03.79 & 316.033 & 52.569 &  315.951 &  52.573 \\
                G093.51-04.31 & 326.236 & 47.624 &  326.131  &  47.631 \\
                G105.44+09.88 & 325.816 & 66.131 &  325.935  &  66.152  \\
                G107.18+05.44 & 335.376 & 63.626  &  335.266 &  63.605  \\
                G107.26+05.71 & 335.206 & 63.903 &  335.116 &  63.884  \\
                G109.70+02.52 & 343.374 & 62.326 &  343.240 &  62.329 \\
                G109.81+02.72 & 343.391 & 62.546 &  343.267  &  62.570 \\
                G139.14-03.23 & 42.617 & 55.849 & 42.487  &  55.859 \\
                G159.23-20.09 & 53.363 & 31.127 &  53.404 &  31.134 \\
                G162.46-08.69 & 65.388 & 37.567 &  65.434  &  37.620 \\
                G171.53-14.91 & 67.179 & 26.832 &  67.221 &  26.843  \\
                G172.89+02.28 & 84.235 & 36.177  & & \\
                G173.15+02.40 & 84.533 & 36.019  & & \\
                G178.28-00.60 & 84.784 & 30.081 &  84.720  &  30.108 \\
                G195.73-02.28 & 92.751 & 14.171 &  92.761 & 14.198 \\
                G202.31+02.53 & 100.255 & 10.602 & 100.189 &  10.599 \\
                G215.87-17.50 & 88.416 & -10.401 \\

                \hline
        \end{tabular}
        \tablefoot{ (Columns 2-3) Right ascension and declination (J2000) of the center of the SCUBA-2 maps. (Columns 4-5) The right ascension and declination (J2000) for the centers of empty reference regions used for background subtraction. The background subtraction uses a radius of 0.02\degr.  }
\end{table}

\section{Comparison of clump-finding algorithms\label{sec:appendix_clumpFind}}

\begin{table}
        \caption{Input values used for ClumpFind and GaussClumps algorithms. \label{tbl:Clumpfind_arguments}}
                \begin{tabular}{ll|ll}
                        \hline
                        \hline
                        \multicolumn{2}{l|}{ClumpFind} & \multicolumn{2}{l}{GaussClumps} \\
                        Variable & Value & Variable & Value  \\
                        \hline
                        RMS & Varies  & RMS & Varies   \\
                        Allowedge & 0 & Extracols & 0 \\
                        Minpix & 16 & Minpix & 3\\
                        Maxbad & 0.05 & Maxbad  & 0.05\\
                        FWHMbeam & 2 & FWHMbeam  & 2 \\
                        Velores & 2 & Velores  & 2\\
                        IDLalg & 1 & Wmin & 0.05\\
                        DeltaT & 2$\times$RMS & Npeak & 9 \\
                        Naxis & 3 & Maxskip & 10 \\
                        Noise & 2$\times$RMS & Nsigma & 3\\
                        Tlow & 2.5$\times$RMS & Thresh & 5\\
                        & & Width & 2\\
                        \hline          
                \end{tabular}
                \tablefoot{ See section \ref{sec:methods_FellWalker} for details on RMS values, which are the same as used for FellWalker analysis. The values used are the same as those used in \citep{2019arXiv191007692L}. }
\end{table}

There are many methods for finding dense clumps, for example, ClumpFind \citep{clumpFind}, GaussClumps
\citep{1990ApJ...356..513S,2014ascl.soft06018S}, and AstroDendro\footnote{\url{http://www.dendrograms.org/}} \citep{2019ascl.soft07016R}. Compared to the others, AstroDendro is more recent but has been used in several studies in the past year alone \citep[i.e.][]{2019ApJ...873...31K,2019AAS...23325308H}. ClumpFind is an older, more established algorithm that has been used in many of the GBS papers \citep{2017PASJ...69...18T,2019A&A...623A.141F}.

\begin{figure}
        \resizebox{\hsize}{!}{\includegraphics{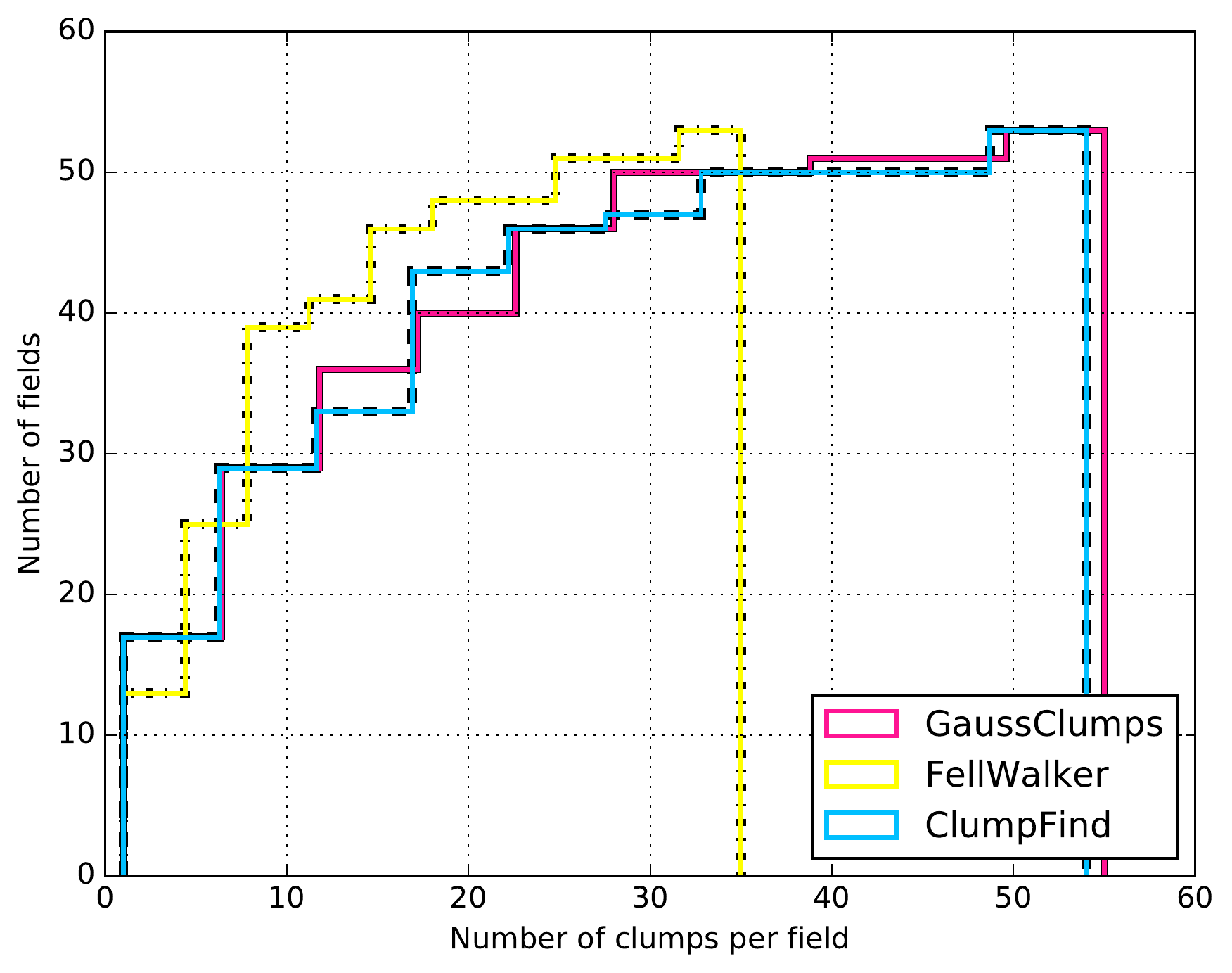}}
        \caption{Cumulative histograms of number of clumps found in each field using the GaussClumps (pink), FellWalker (yellow), and ClumpFind (blue) algorithms.  \label{fig:FW_CF_nClump} }
\end{figure}

\begin{figure}
        \resizebox{\hsize}{!}{\includegraphics{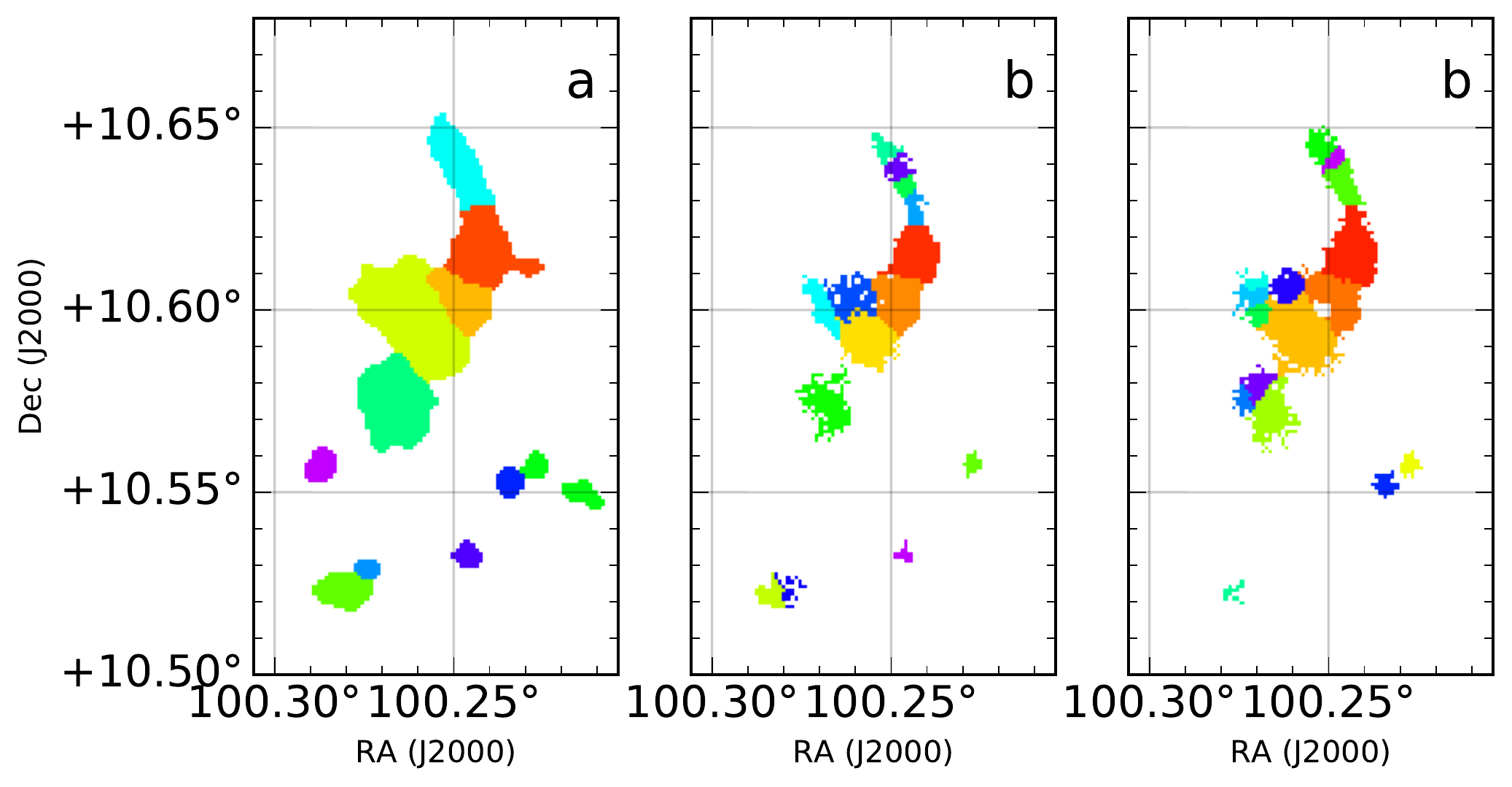}}
        \caption{Clumps produced by FellWalker (a), ClumpFind (b), and GaussClumps (c) for field G202.31+02.53 \label{fig:G202_fw_cf_clumps} }
\end{figure}

We examined the importance of the clump-finding algorithm used by comparing ClumpFind and GaussClumps with FellWalker. The parameters used in the ClumpFind and GaussClumps algorithms are listed in Table \ref{tbl:Clumpfind_arguments}. Figure \ref{fig:FW_CF_nClump} shows cumulative histograms of the number of clumps per field found using all three algorithms. The results for GaussClumps and ClumpFind are very similar. 
FellWalker found 529 clumps, but ClumpFind found 779 and GaussClumps 785. Likewise, with ClumpFind each field has between 1 and 54 clumps, and both ClumpFind and GaussClumps have a mean of just under 15 clumps per field, compared to a mean of 10 clumps per field with FellWalker. This is also visible in Fig. \ref{fig:G202_fw_cf_clumps}, which shows the 11 FellWalker, 14 ClumpFind, and 16 GaussClumps clumps found in the field G202.31+02.53. Compared to FellWalker, the other two algorithms divided the central structure into smaller pieces but detected fewer clumps in other parts of the map.

\citet{2018ApJ...853..160C} compared the results of AstroDendro with those of ClumpFind and also find a difference in number of clumps; however, this difference is slightly smaller than that which we find for field G202.31+02.53. ClumpFind is also found by the authors to overestimate the number of clumps when compared with other clump-finding algorithms. Fellwalker, Dendrograms, and GaussClumps have been ranked as the best clump-finding algorithms by \citet{2019arXiv191007692L}. These three had better detection completeness on a large sample of simulated clumps. Furthermore the authors found that FellWalker was best at extracting the total flux of the clumps.

The differences in the size and the number of the detected clumps would have a significant consequences for the analysis. The derived masses and radii of ClumpFind- or GaussClumps-derived clumps would be smaller, thus probably also decreasing the number of clumps that would be classified as prestellar. In addition to this, the parameters used and the determination of the RMS value also affects the number and type of clumps detected.

\clearpage
\section{Extended emission maps \label{sec:appendix_extended_emission}}

\begin{figure}
        \centering
        \begin{minipage}{\linewidth}
                \includegraphics[width=\linewidth]{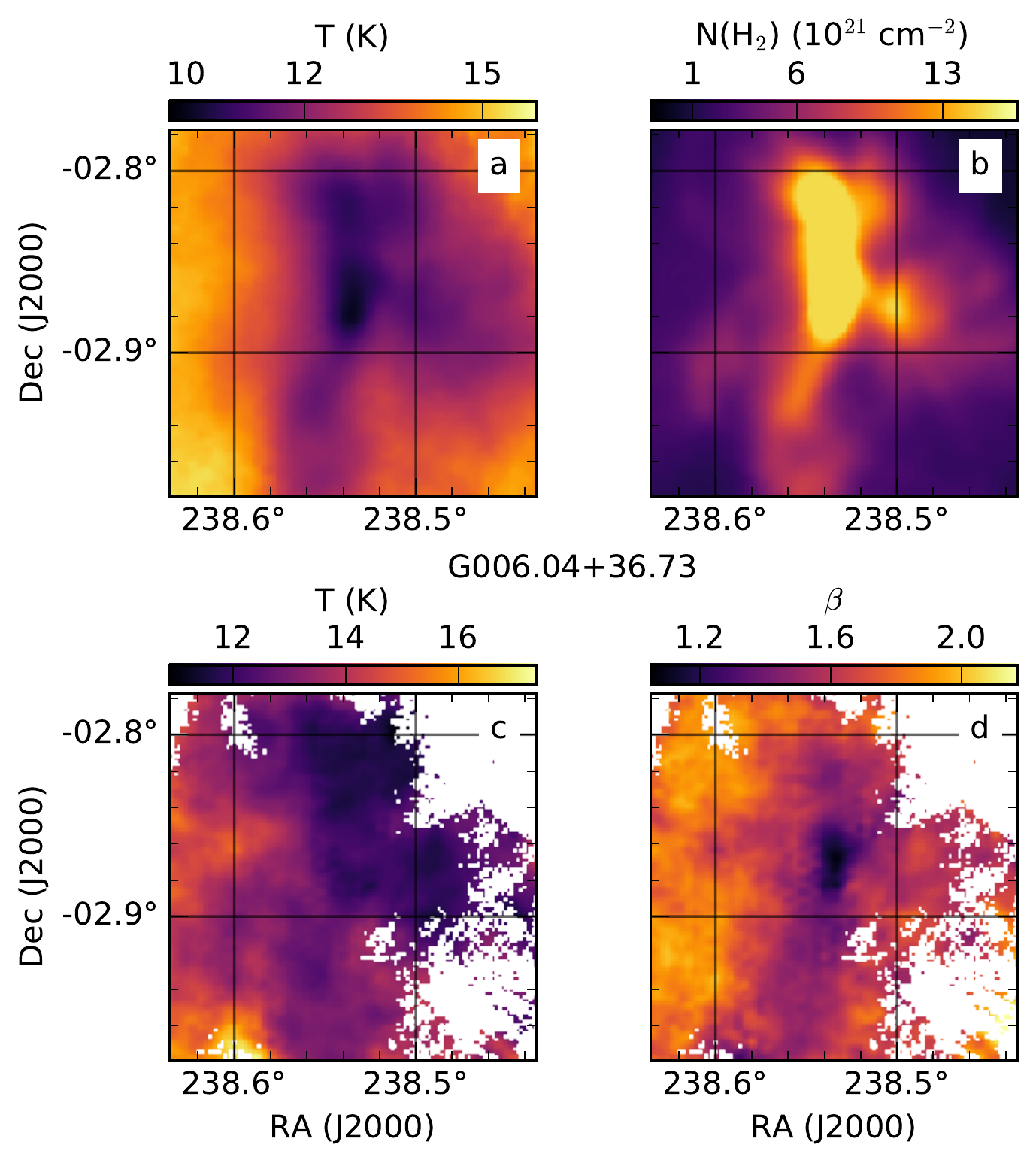}
        \end{minipage}  
        \begin{minipage}{\linewidth}
                \line(1,0){250}
        \end{minipage}
        \begin{minipage}{\linewidth}
                \centering
                \includegraphics[width=\linewidth]{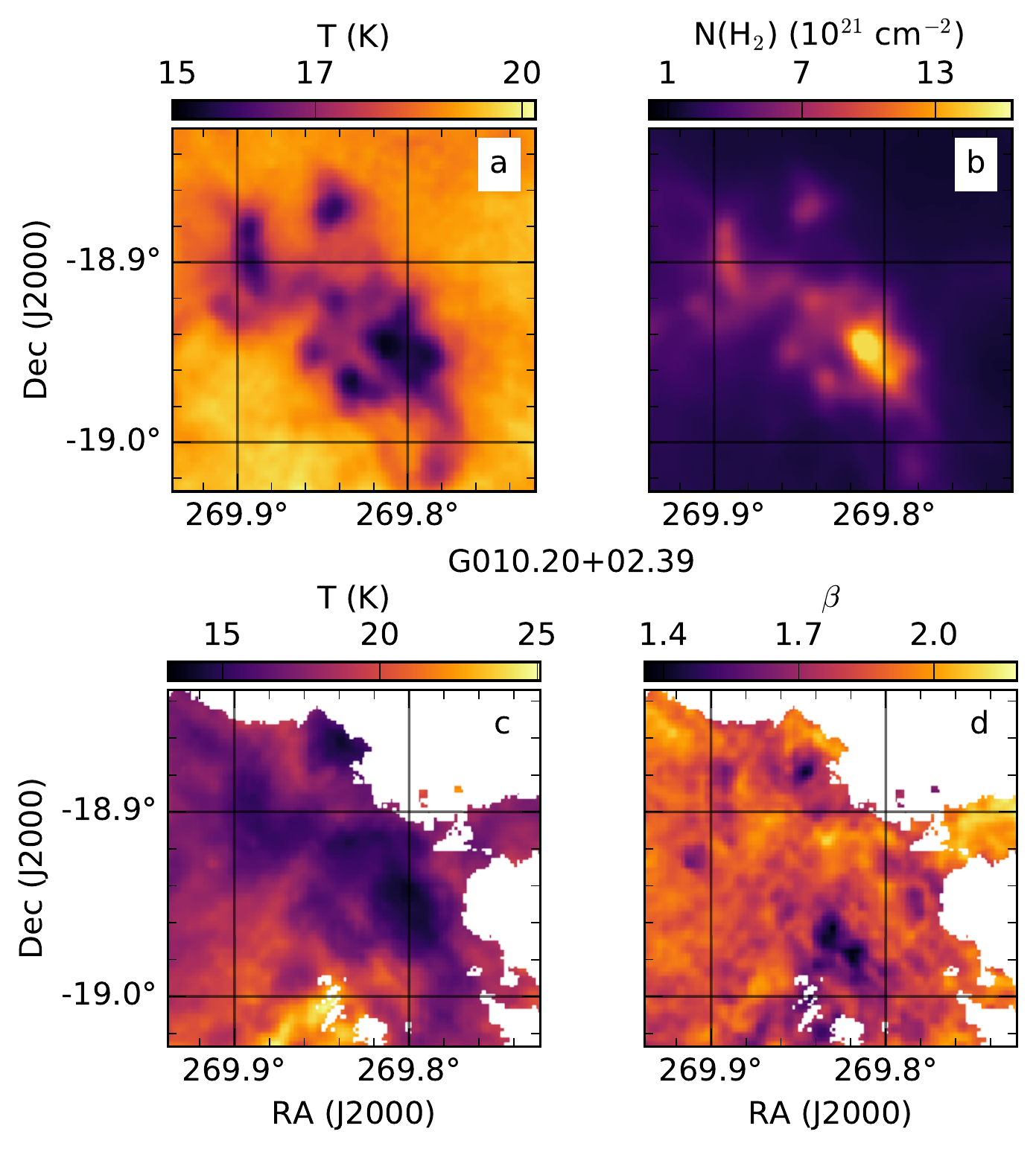}
        \end{minipage}
        \caption{Extended emission maps of the full \textit{Herschel} fields. The name of the field is written below frames a and b. The two fields are divided by the horizontal line. (a) 250--500\,\micro\spc temperature with constant $\beta$ = 1.8. (b) 250--500\,\micro\spc column density \textit{N}(H$_{2}$) with constant $\beta$ = 1.8. (c) 160--500\,\micro\spc temperature with background subtraction and varied spectral index. (d) 160--500\,\micro\spc spectral index with background subtraction. \label{fig:appendix_extended_1}} 
\end{figure}

We show temperature, column density, and spectral index maps for Herschel fields with SPIRE and PACS 160 \micro\spc data in Figs. \ref{fig:appendix_extended_1} -- \ref{fig:appendix_extended_last}. The temperatures and column densities of the upper frames are calculated using the SPIRE bands and a fixed value of $\beta$=1.8. The lower frames correspond to fits that include the PACS 160 \micro\spc band and have spectral index as a free parameter. All \textit{Herschel} fields are cropped to the size of the SCUBA-2 field. Fields with insufficient PACS 160\,\micro\spc data only show frames a and b. Pixels with negative PACS 160\,\micro\spc intensities are masked in frames c and d.

\begin{figure}
        \centering
        \begin{minipage}{\linewidth}
                \includegraphics[width=\linewidth]{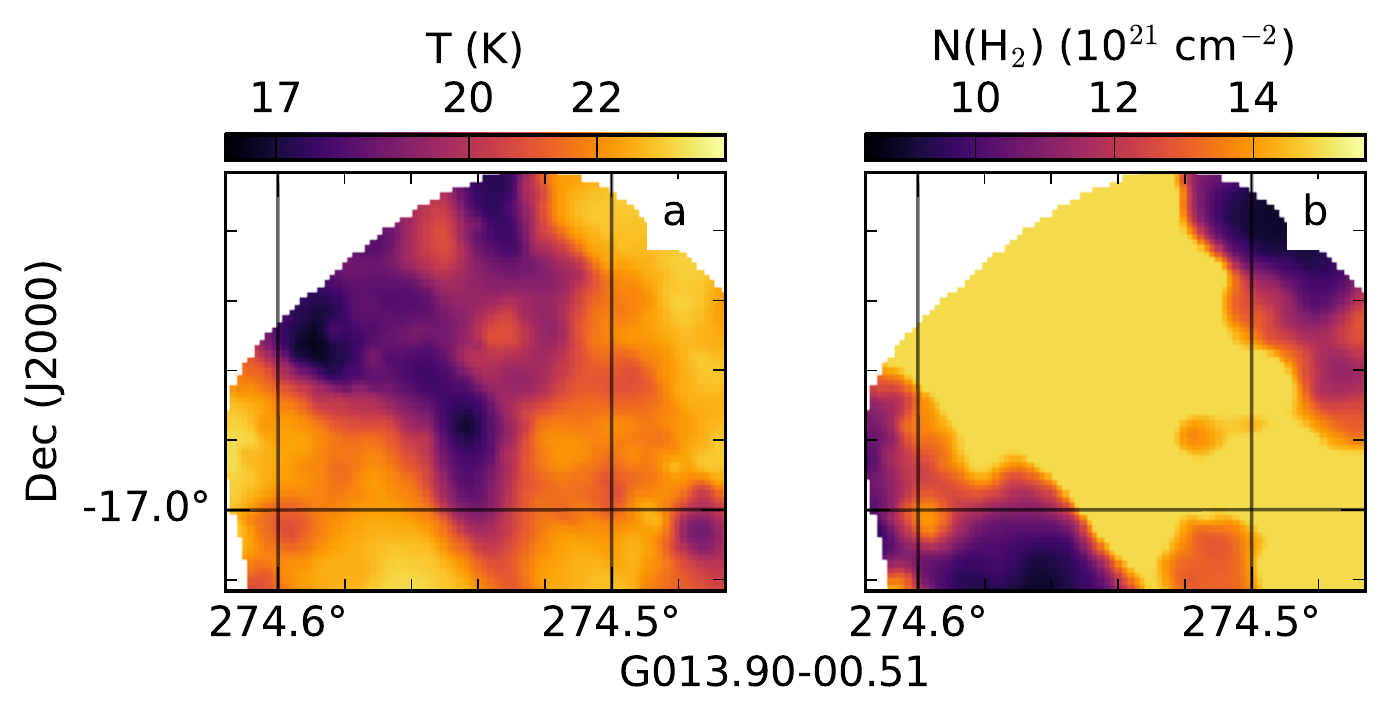}
        \end{minipage}
        \begin{minipage}{\linewidth}
                \line(1,0){250}
        \end{minipage}
        \begin{minipage}{\linewidth}
                \centering
                \includegraphics[width=\linewidth]{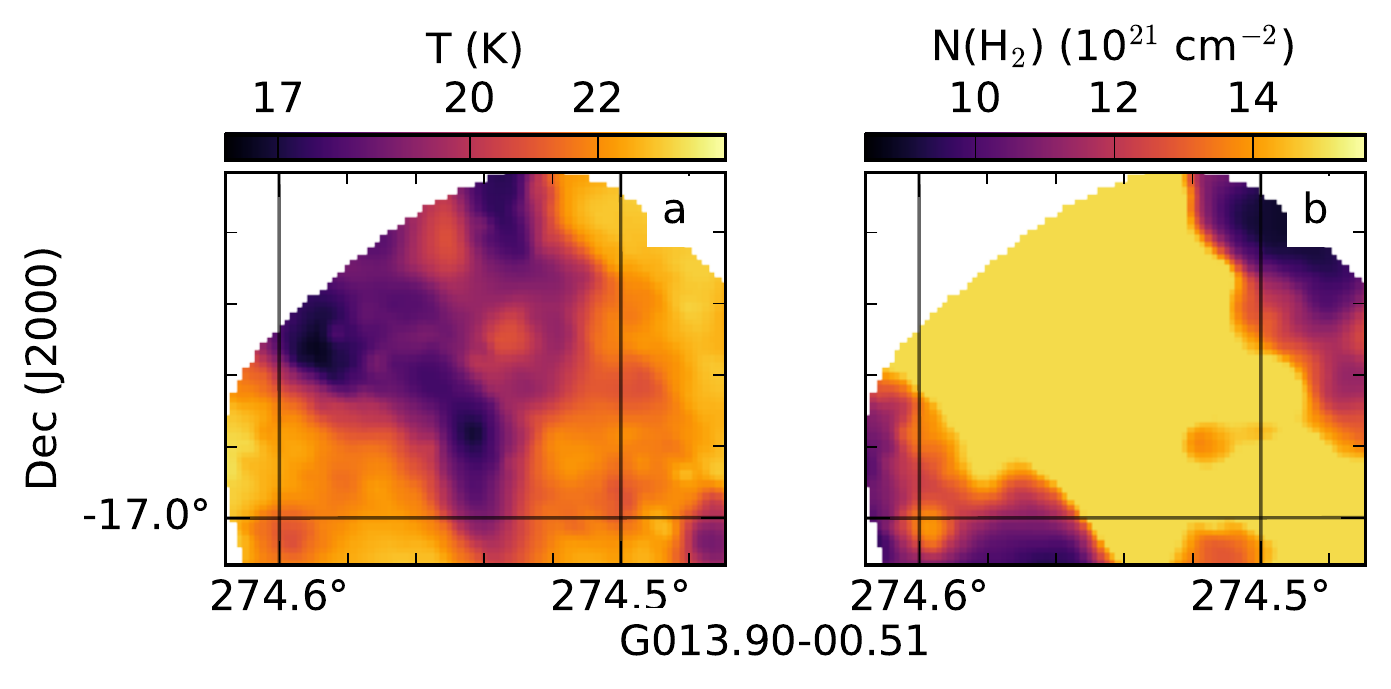}
        \end{minipage}
        \begin{minipage}{\linewidth}
                \line(1,0){250}
        \end{minipage}
        \begin{minipage}{\linewidth}
                \includegraphics[width=\linewidth]{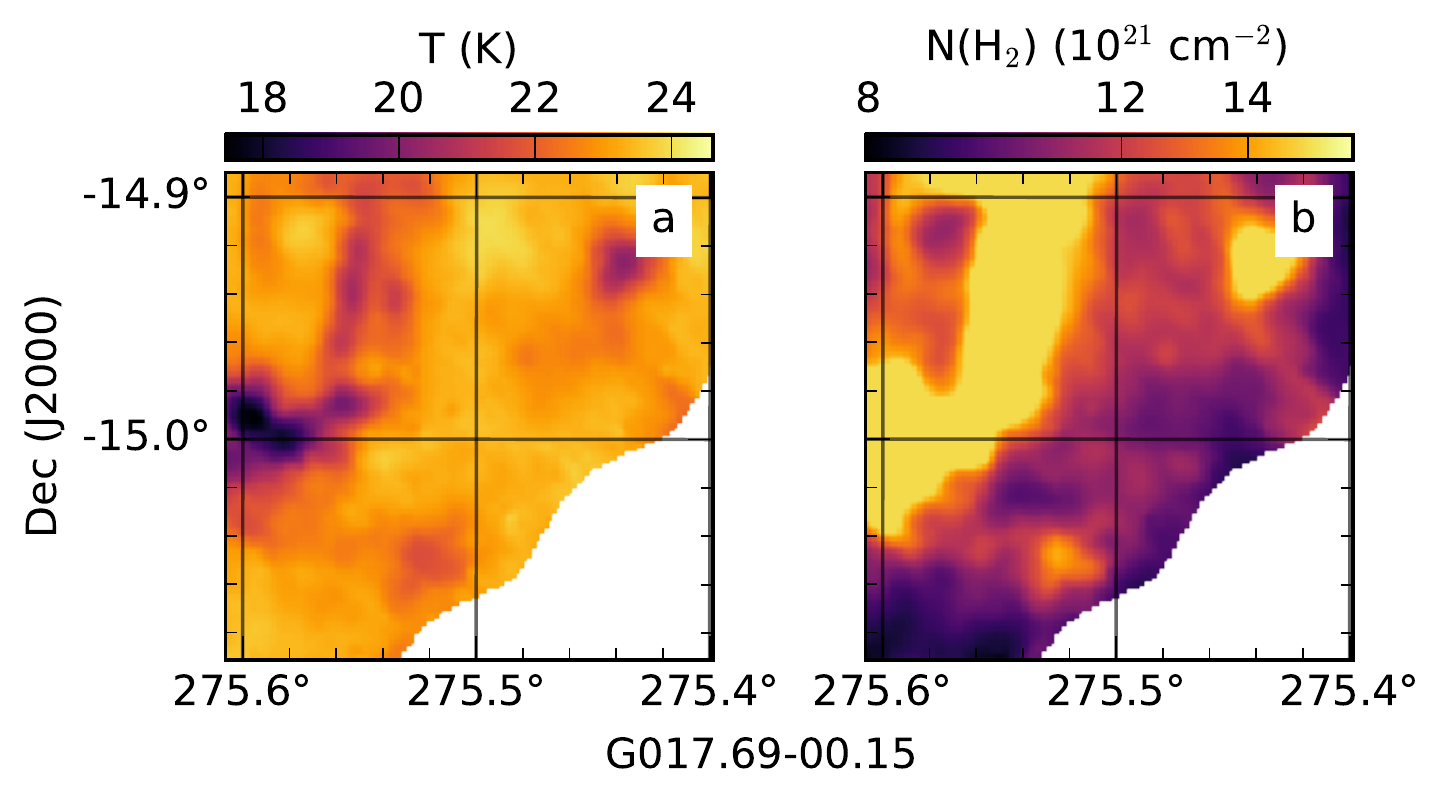}
        \end{minipage}
        \begin{minipage}{\linewidth}
                \line(1,0){250}
        \end{minipage}
        \begin{minipage}{\linewidth}
                \centering
                \includegraphics[width=\linewidth]{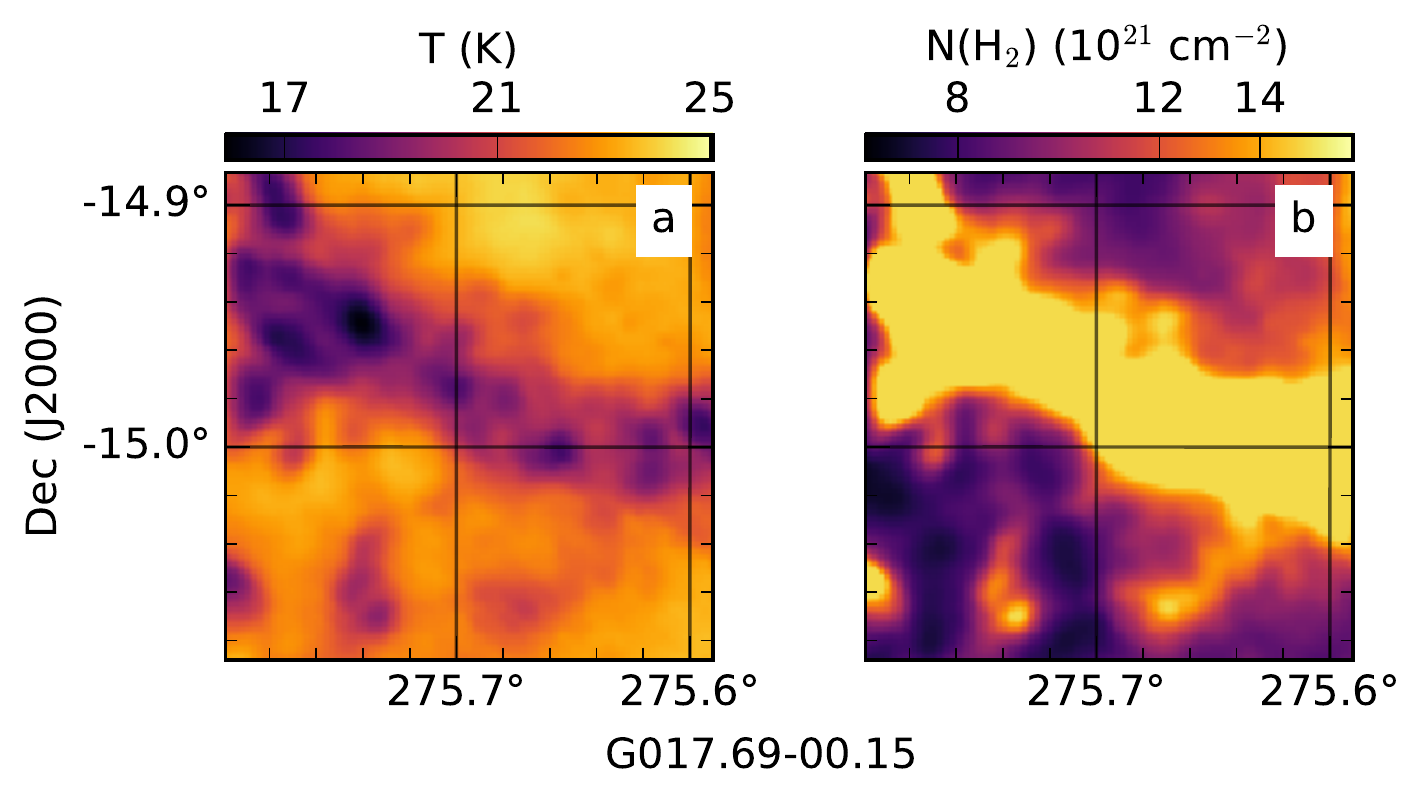}
        \end{minipage}
        \caption{Same as Fig. \ref{fig:appendix_extended_1} for fields G014.14-00.55 and G014.15-00.55, both located on \textit{Herschel} field G013.90, and  G016.28-00.45 and G016.37-00.61, both located on \textit{Herschel} field G017.69.}
\end{figure}

\begin{figure}
        \centering
        \begin{minipage}{\linewidth}
                \includegraphics[width=\linewidth]{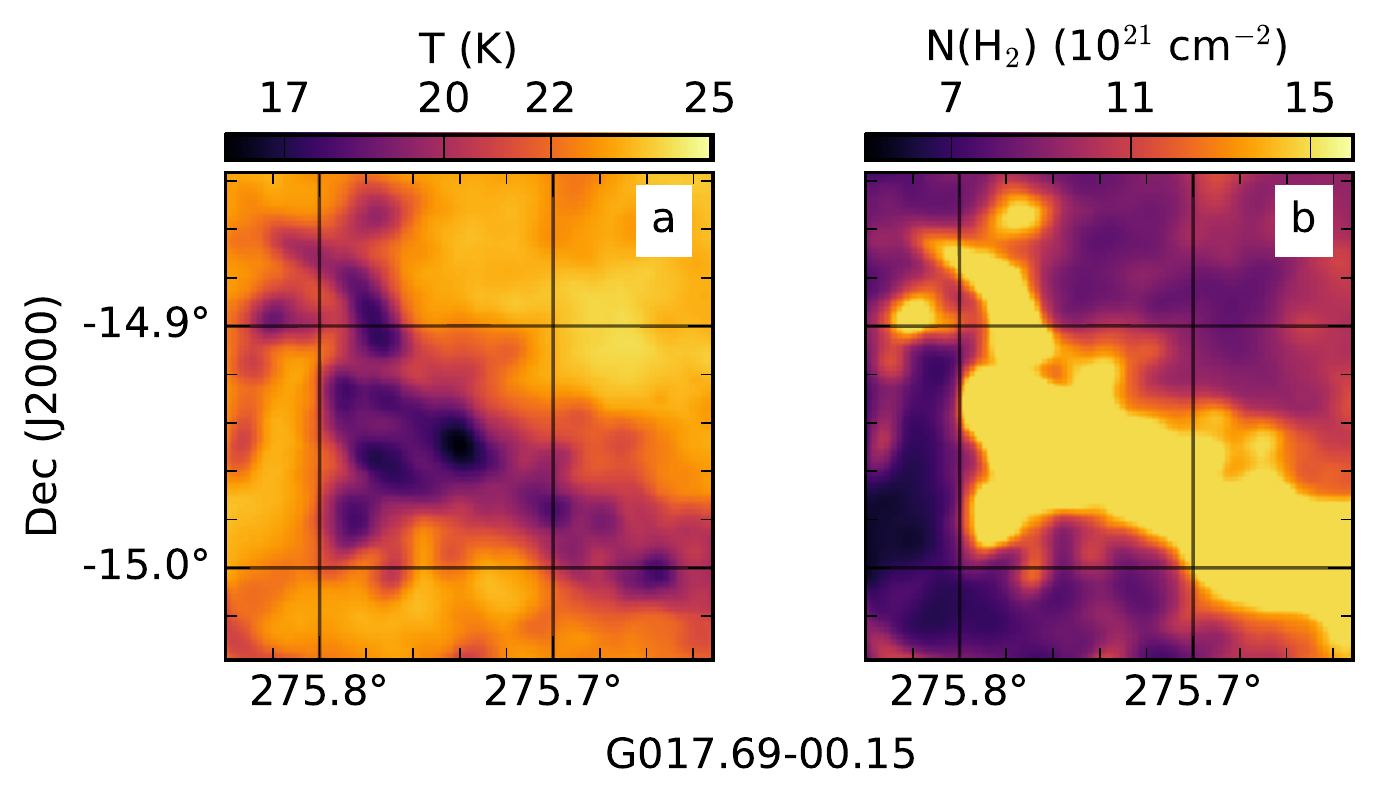}
        \end{minipage}
        \begin{minipage}{\linewidth}
                \line(1,0){250}
        \end{minipage}
        \begin{minipage}{\linewidth}
                \centering
                \includegraphics[width=\linewidth]{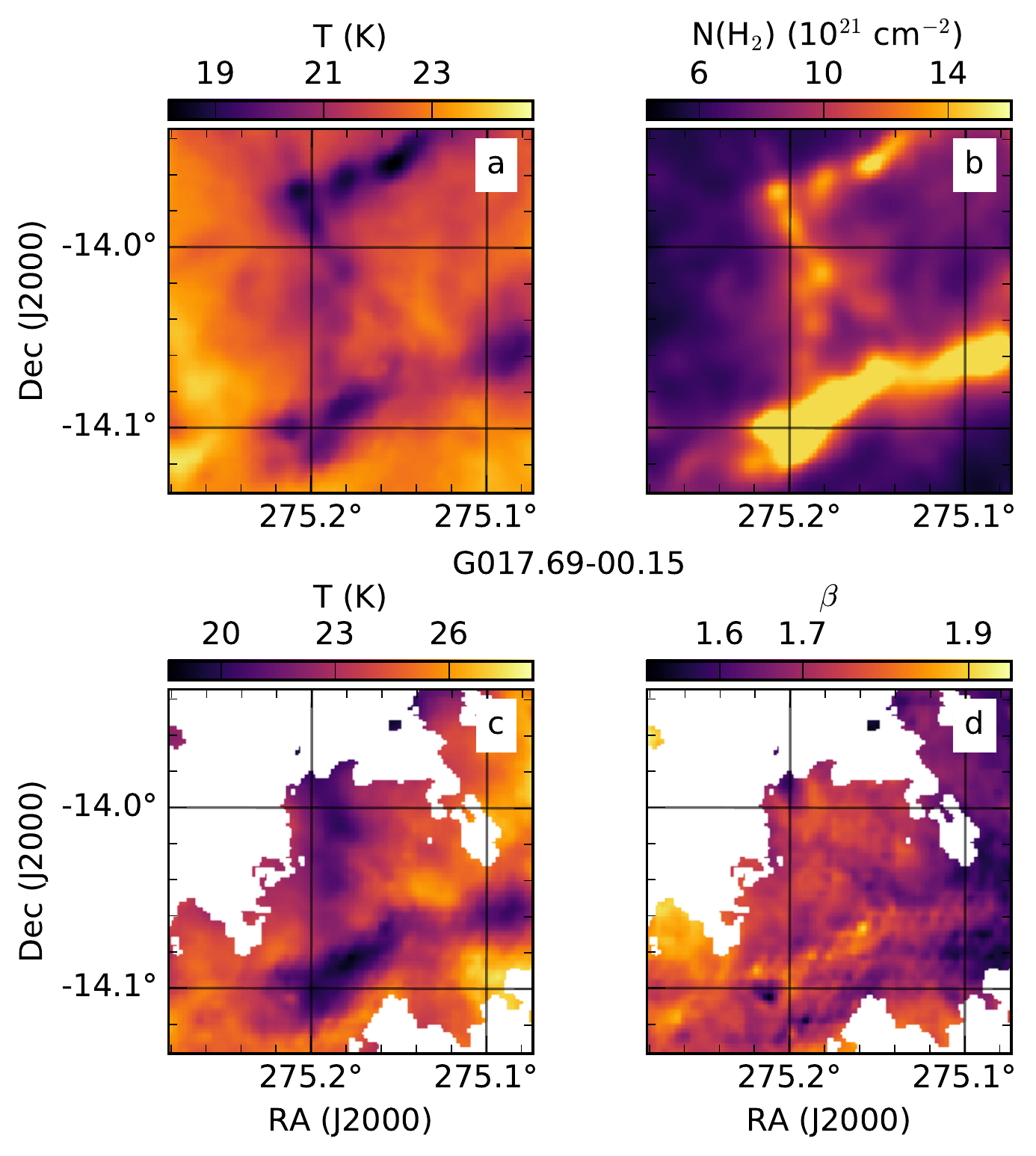}
        \end{minipage}
        \begin{minipage}{\linewidth}
                \line(1,0){250}
        \end{minipage}
        \begin{minipage}{\linewidth}
                \includegraphics[width=\linewidth]{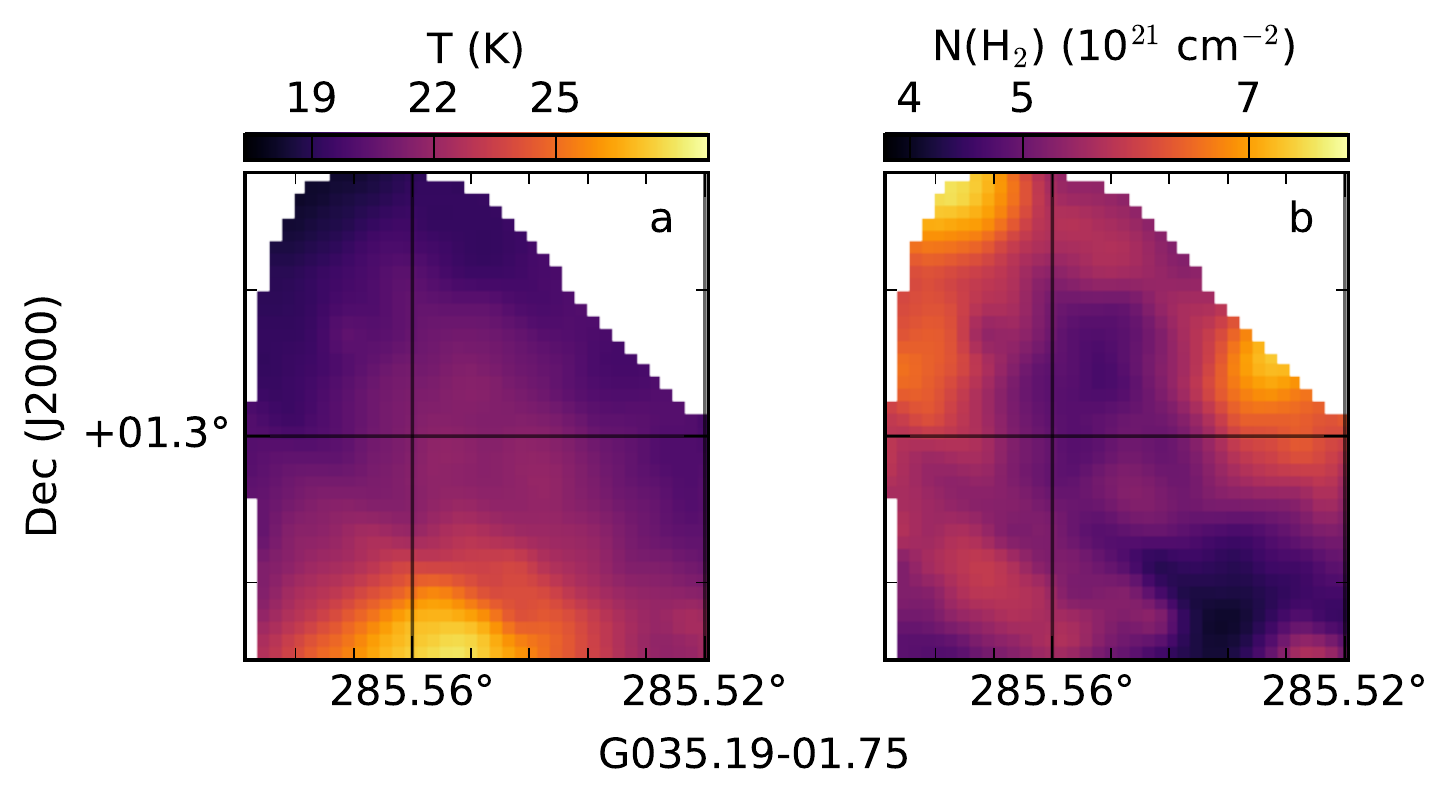}
        \end{minipage}

        \caption{Same as Fig. \ref{fig:appendix_extended_1} for fields G016.43-00.62, G016.97+00.28 (both located on \textit{Herschel} field G017.69), and G035.38-01.77.}
\end{figure}

\begin{figure}
        \centering
        \begin{minipage}{\linewidth}
                \centering
                \includegraphics[width=\linewidth]{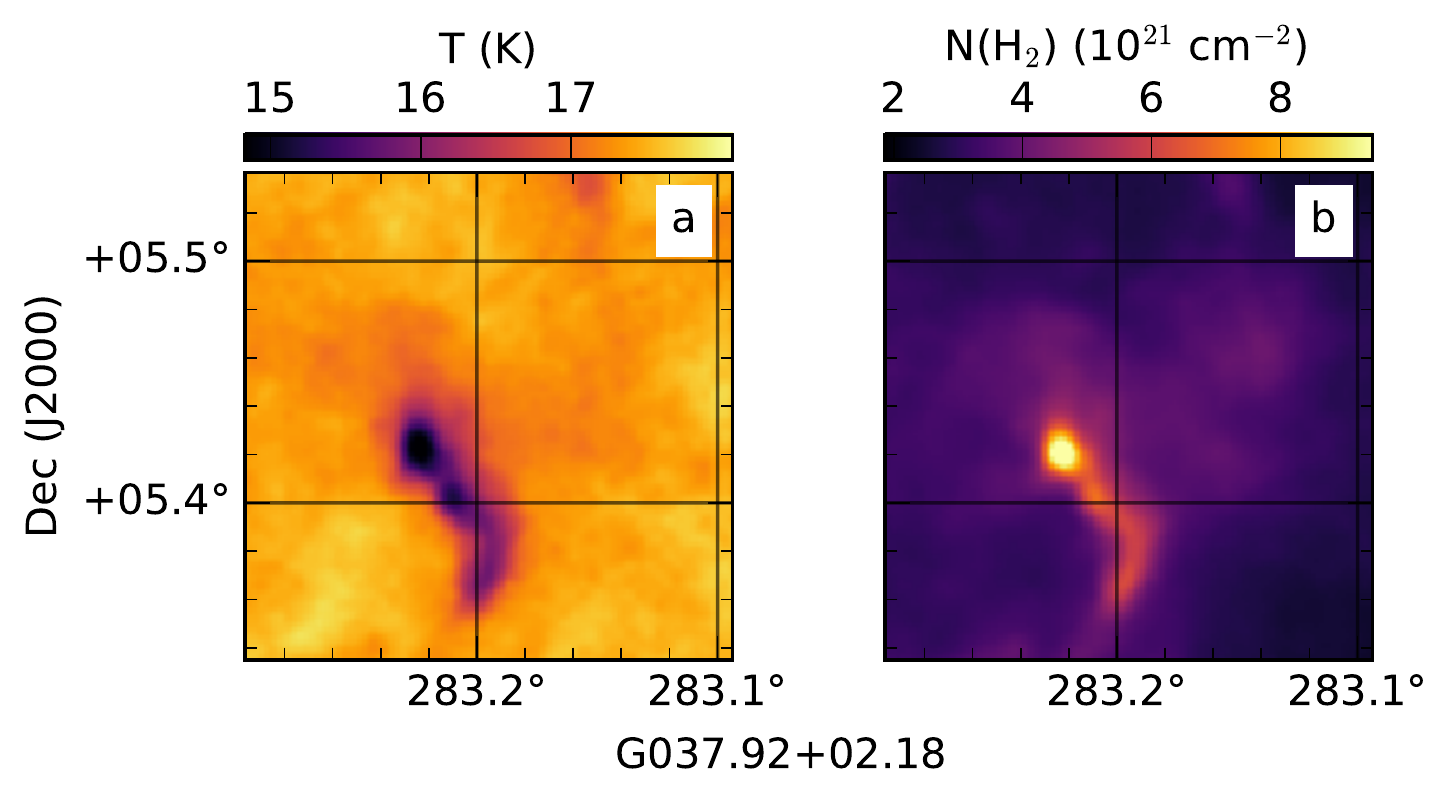}
        \end{minipage}
        \begin{minipage}{\linewidth}
                \line(1,0){250}
        \end{minipage}
        \begin{minipage}{\linewidth}
                \includegraphics[width=\linewidth]{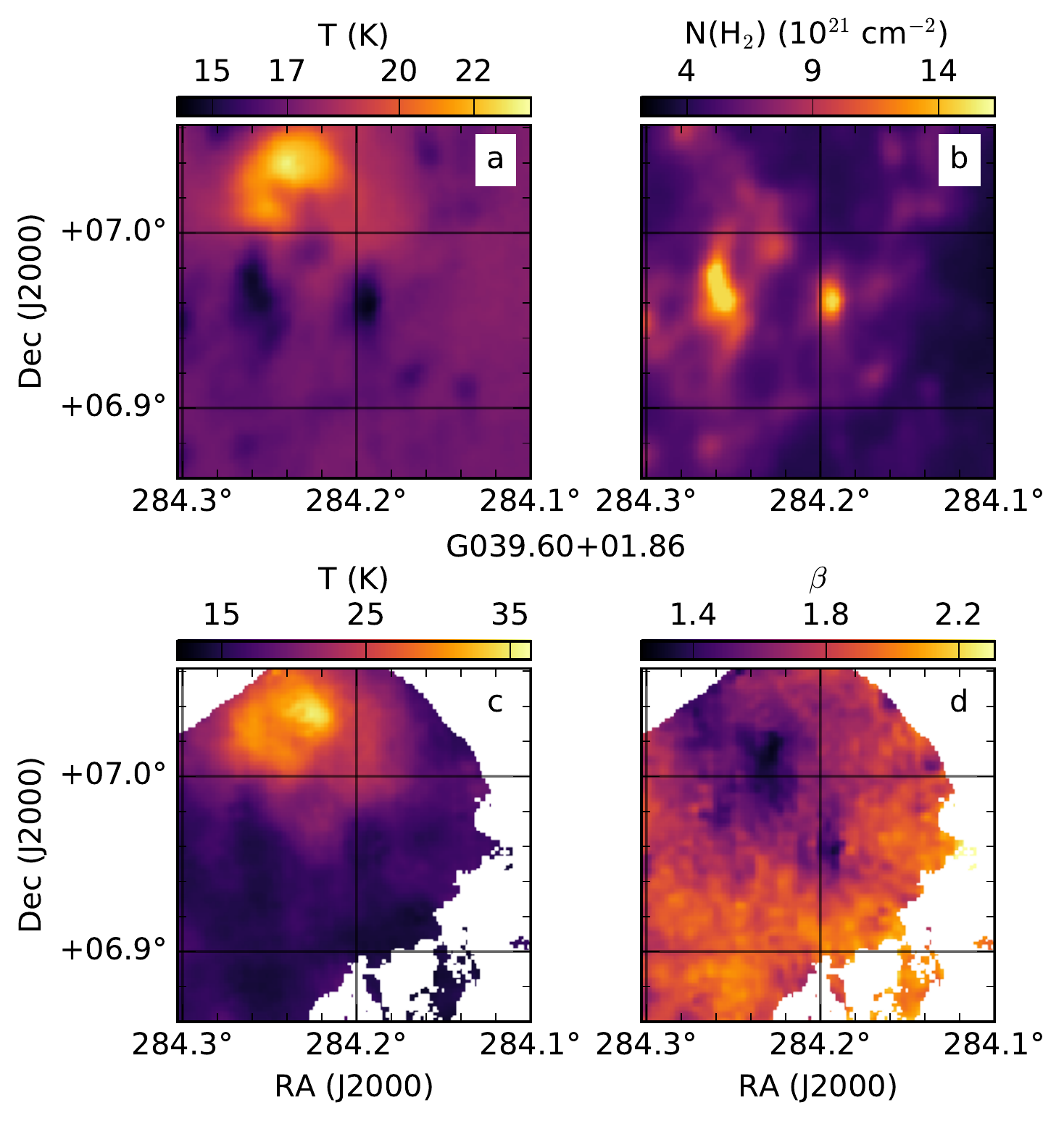}
        \end{minipage}
        \begin{minipage}{\linewidth}
                \line(1,0){250}
        \end{minipage}
        \begin{minipage}{\linewidth}
                \centering
                \includegraphics[width=\linewidth]{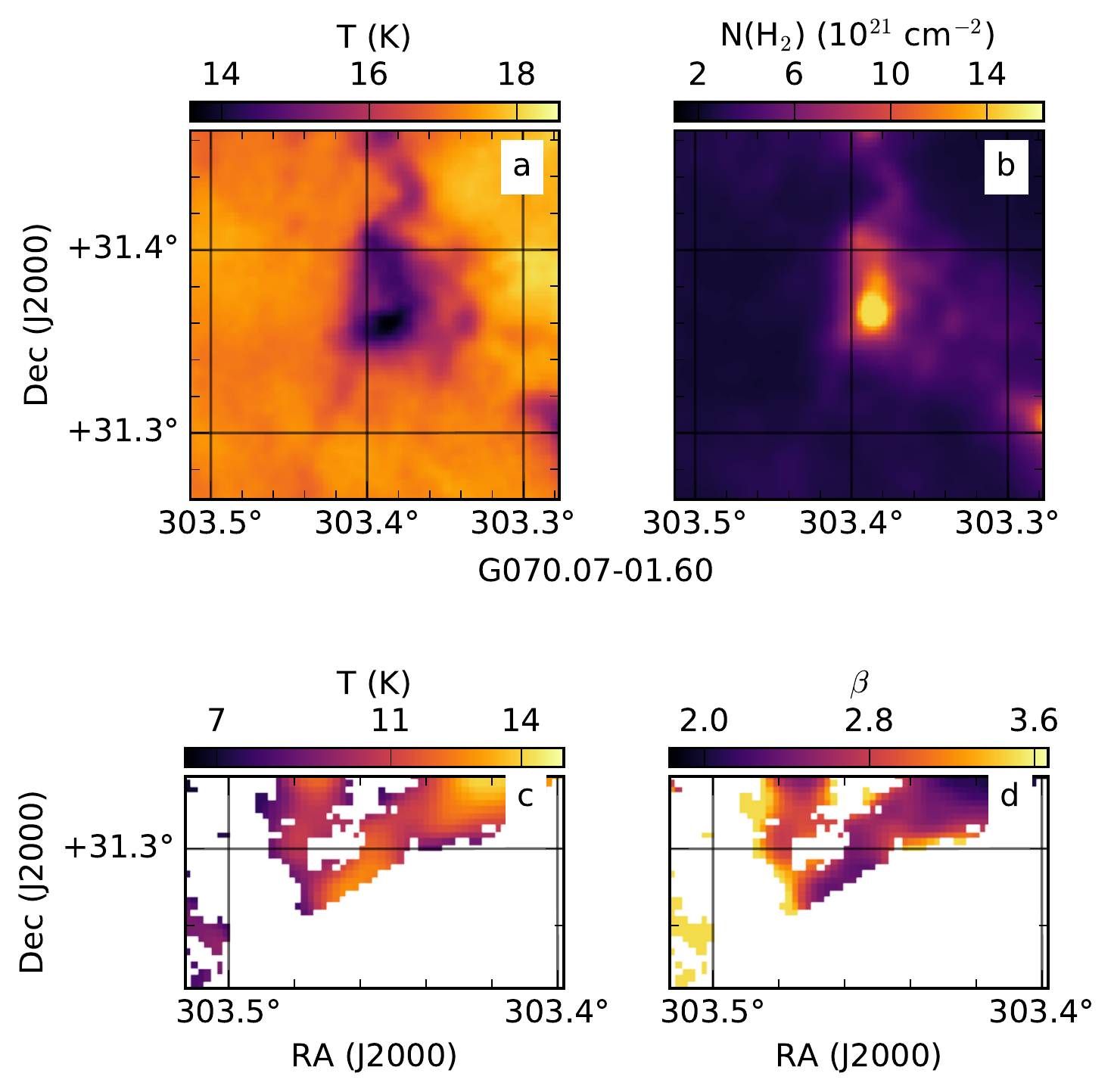}
        \end{minipage}

        \caption{Same as Fig. \ref{fig:appendix_extended_1} for fields G037.93+02.18, G039.74+01.98, and G069.81-01.67.}
\end{figure}

\begin{figure}
        \centering
        \begin{minipage}{\linewidth}
                \includegraphics[width=\linewidth]{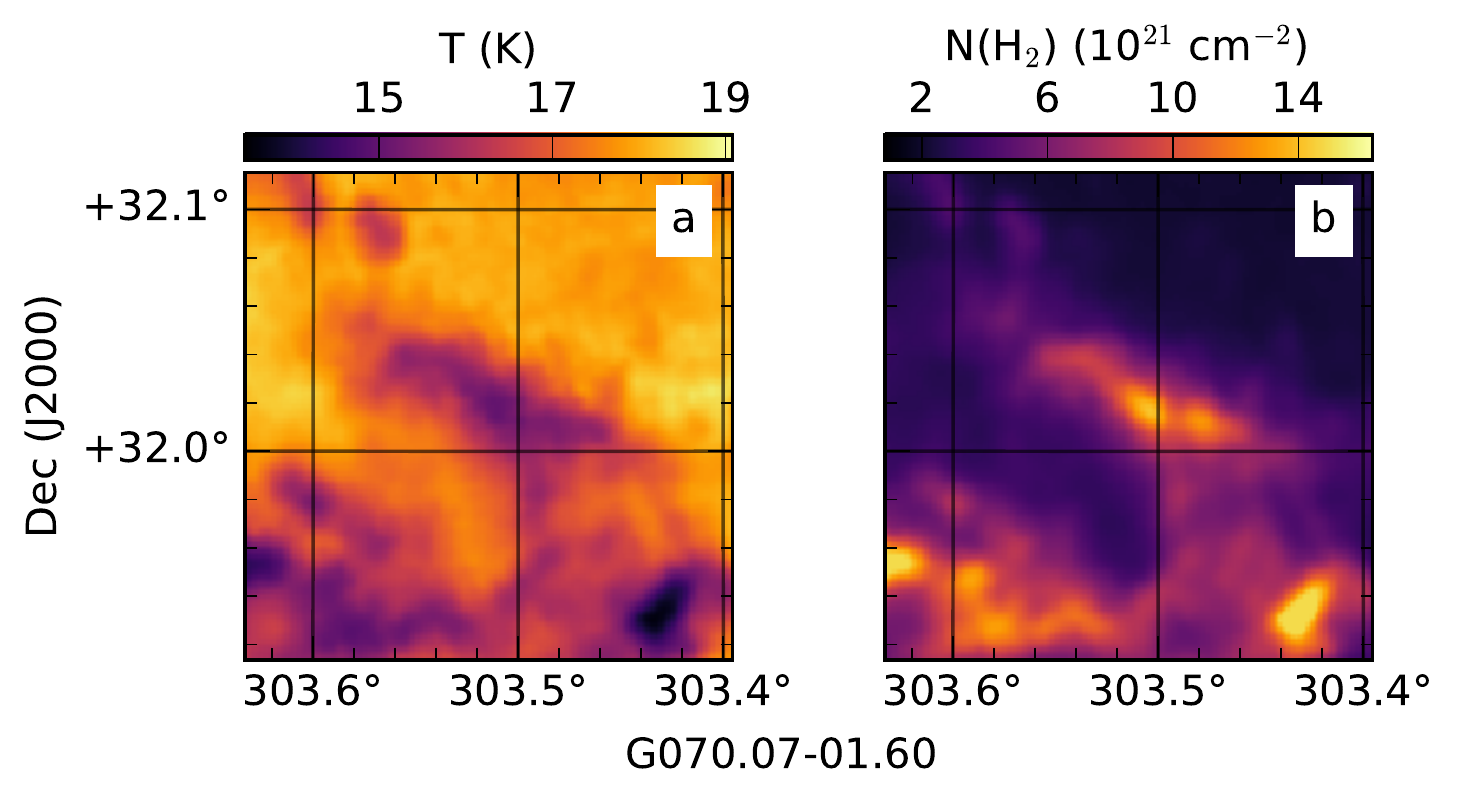}
        \end{minipage}
        \begin{minipage}{\linewidth}
                \line(1,0){250}
        \end{minipage}
        \begin{minipage}{\linewidth}
                \centering
                \includegraphics[width=\linewidth]{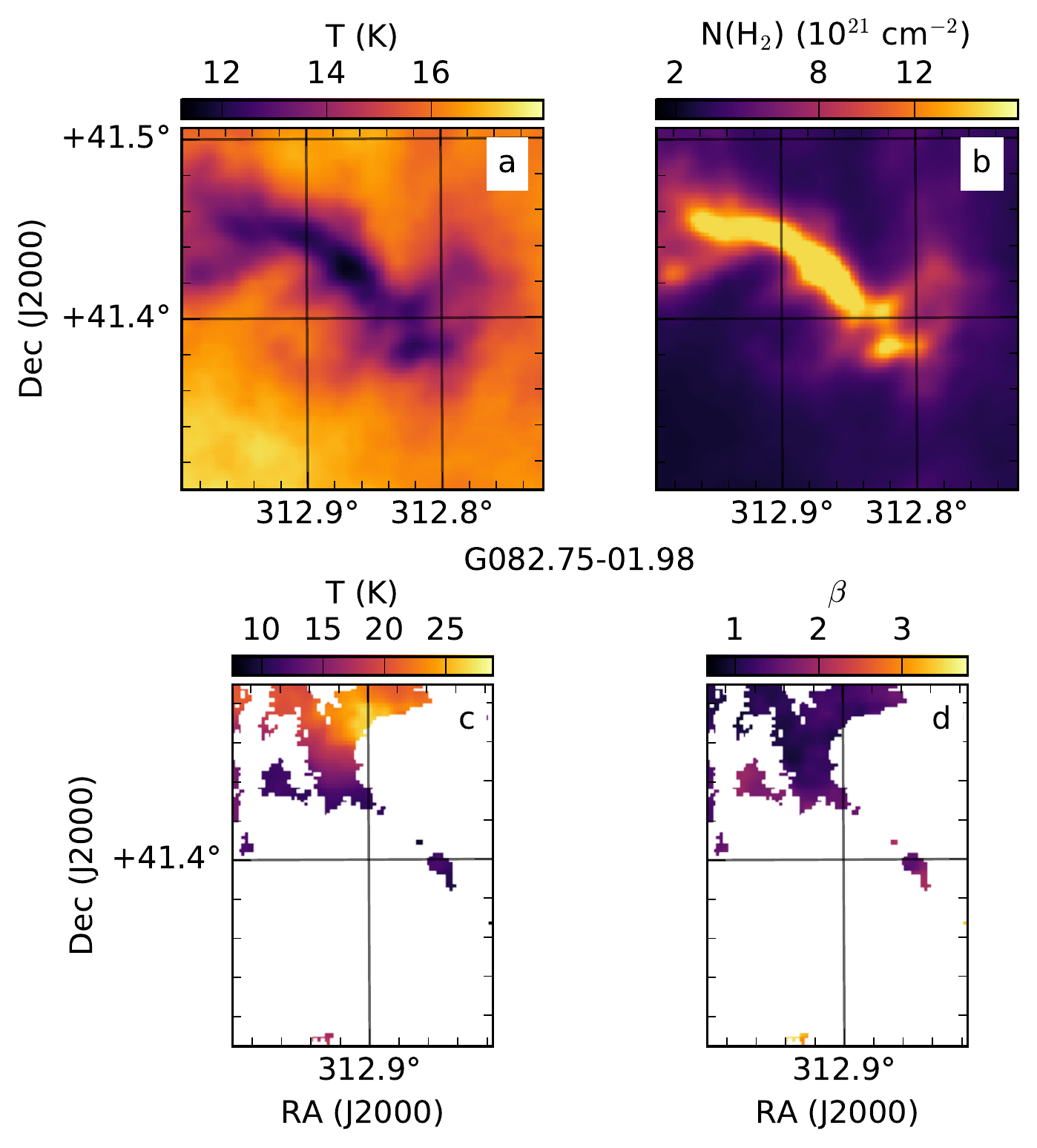}
        \end{minipage}  
        \caption{Same as Fig. \ref{fig:appendix_extended_1} for fields G070.40-01.40 and G082.40-01.84.}
\end{figure}

\begin{figure}
        \centering
        \begin{minipage}{\linewidth}
                \includegraphics[width=\linewidth]{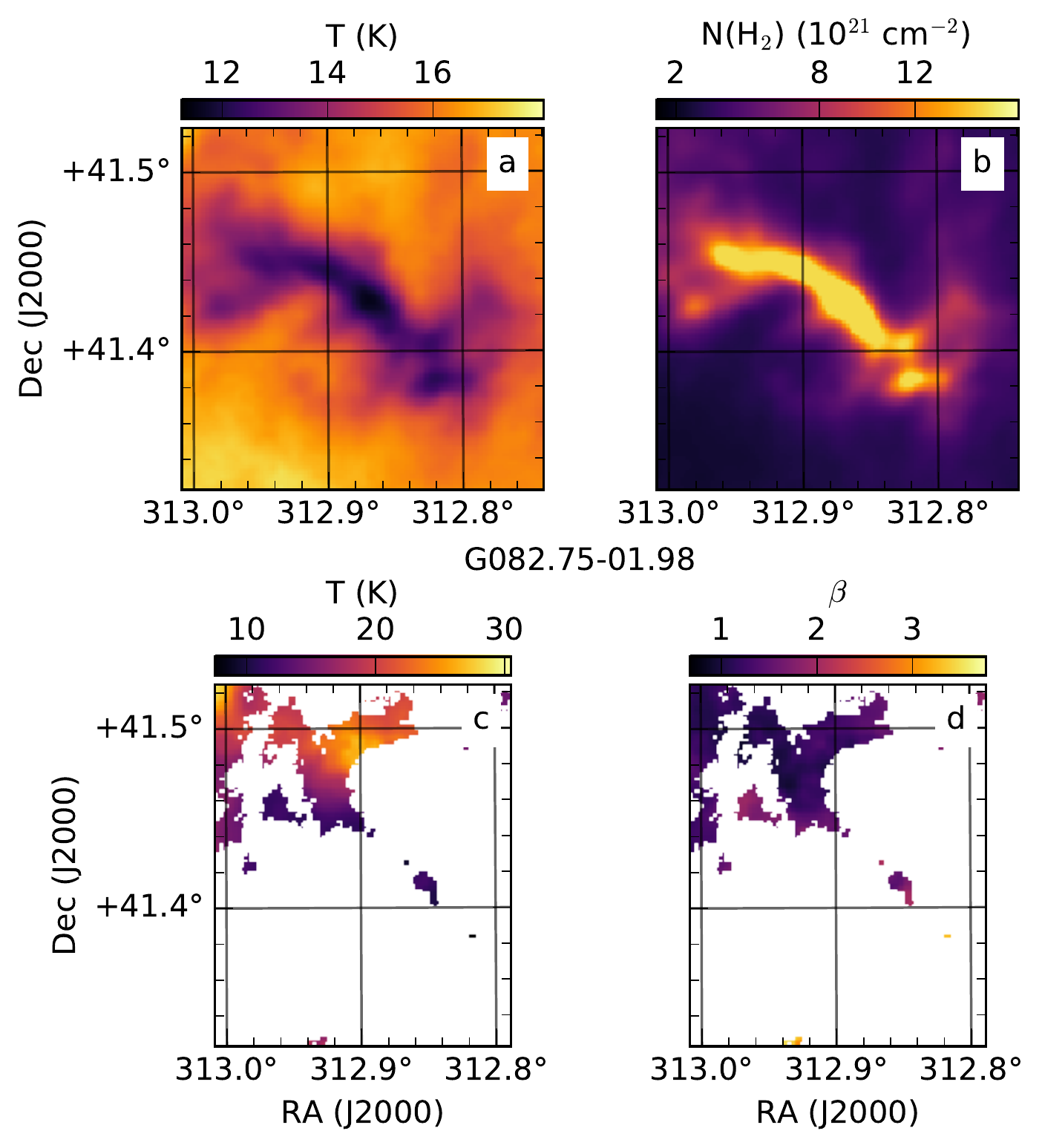}
        \end{minipage}
        \begin{minipage}{\linewidth}
                \line(1,0){250}
        \end{minipage}
        \begin{minipage}{\linewidth}
                \centering
                \includegraphics[width=\linewidth]{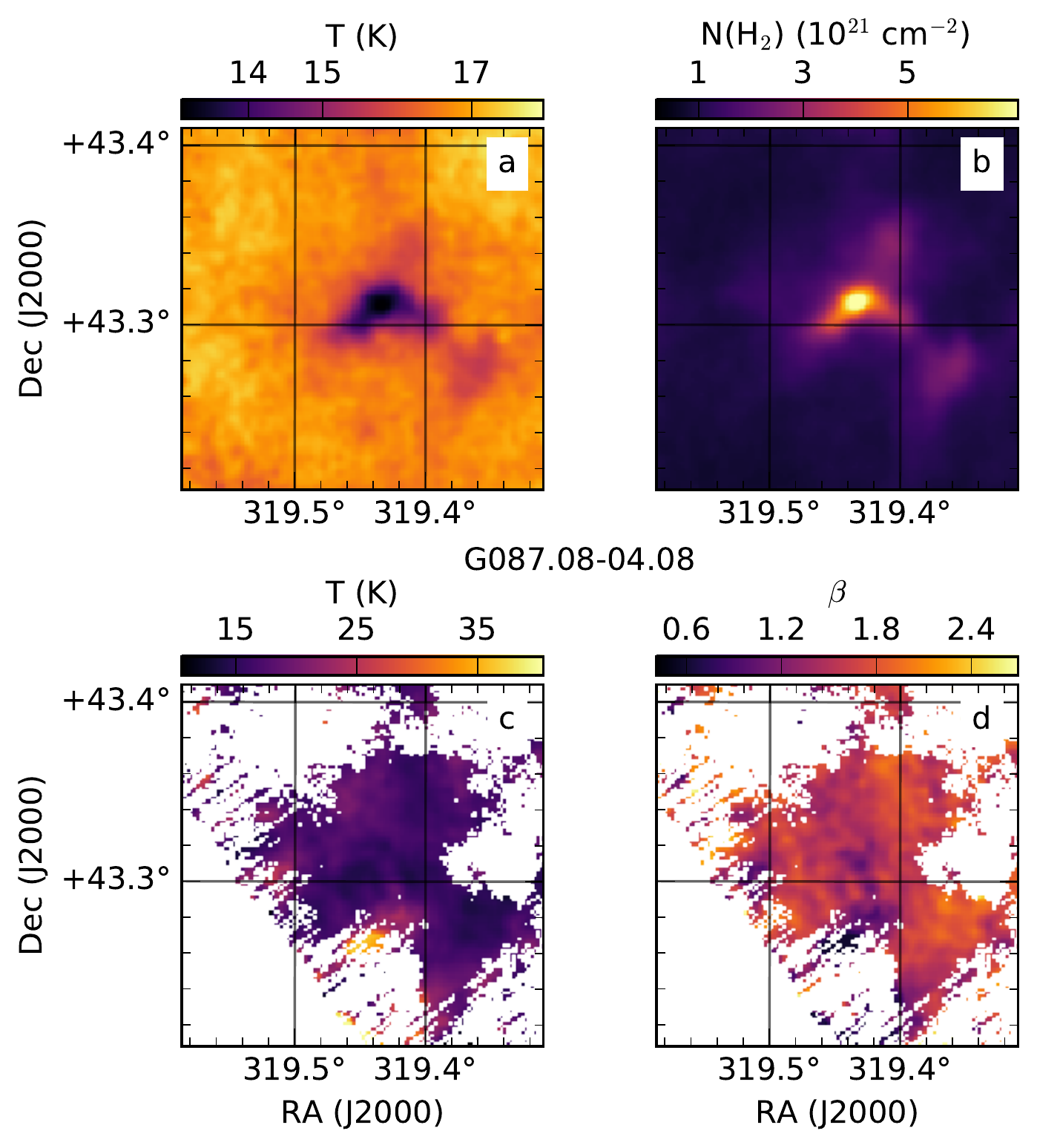}
        \end{minipage}
        \caption{Same as Fig. \ref{fig:appendix_extended_1} for fields G082.42-01.84 and G087.07-04.20.}
\end{figure}

\begin{figure}
        \centering
        \begin{minipage}{\linewidth}
                \includegraphics[width=\linewidth]{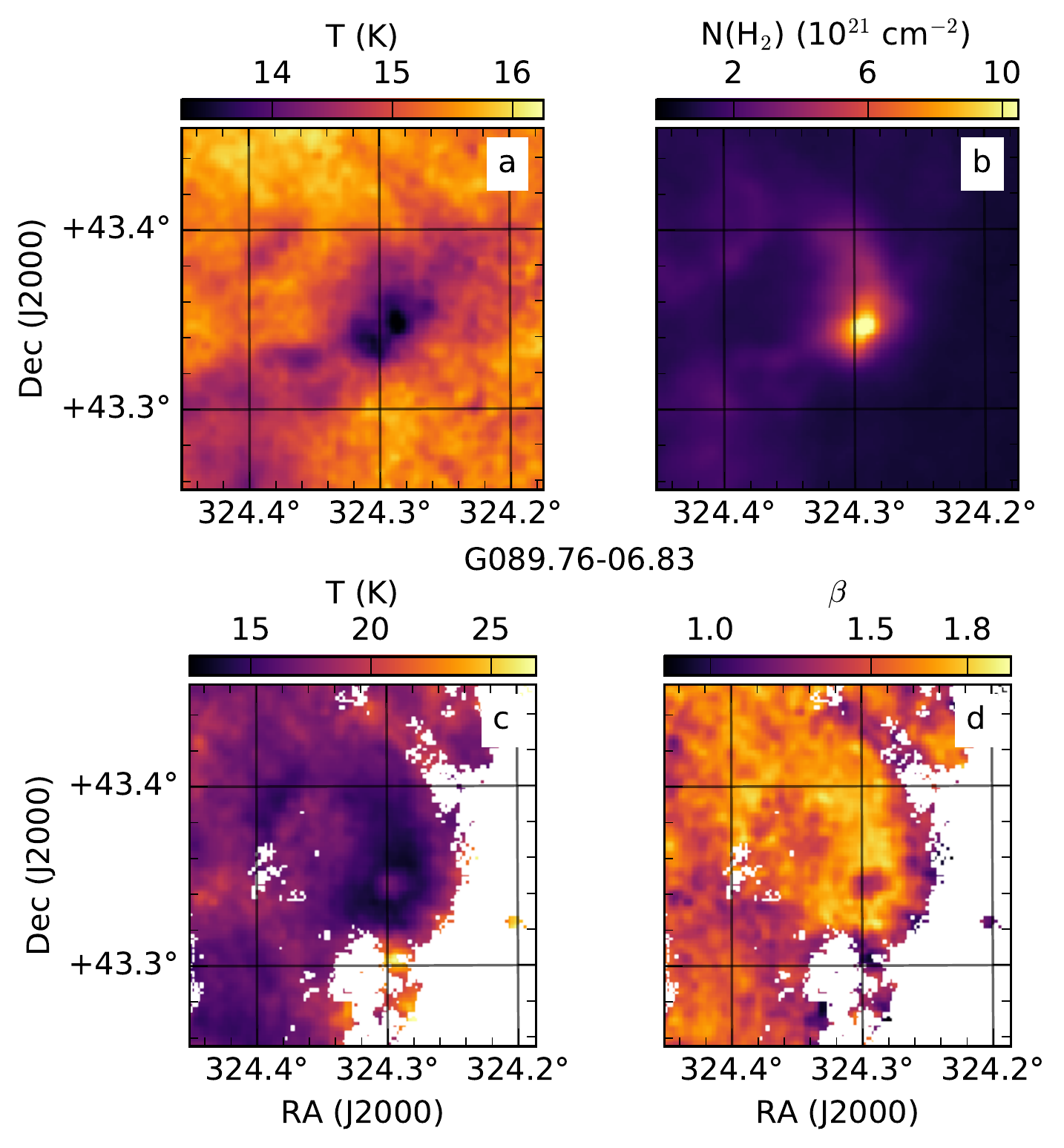}
        \end{minipage}
        \begin{minipage}{\linewidth}
                \line(1,0){250}
        \end{minipage}
        \begin{minipage}{\linewidth}
                \centering
                \includegraphics[width=\linewidth]{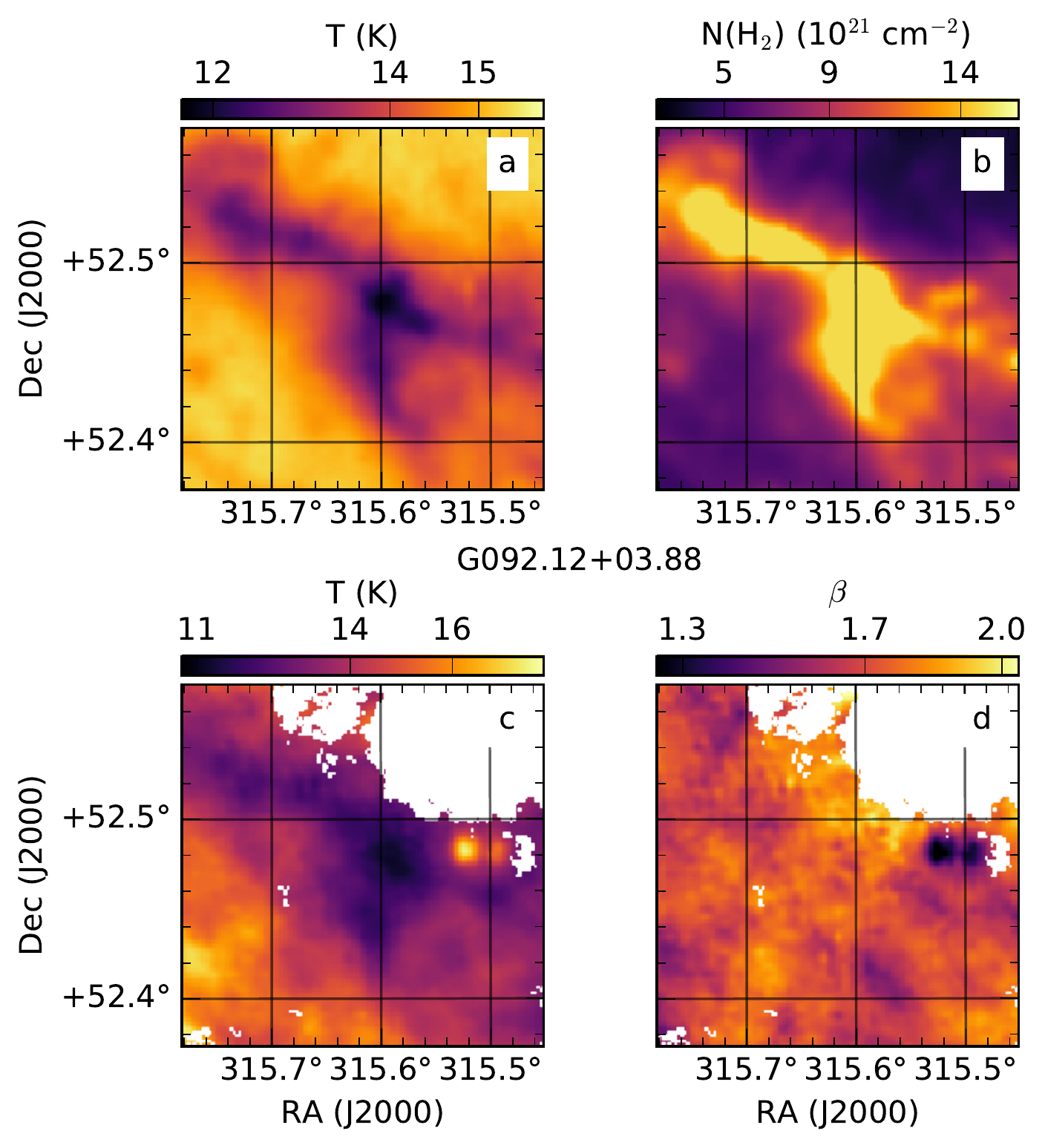}
        \end{minipage}
        \caption{Same as Fig. \ref{fig:appendix_extended_1} for fields G089.66-06.62 and G092.04+03.92.}
\end{figure}

\begin{figure}
        \centering
        \begin{minipage}{\linewidth}
                \includegraphics[width=\linewidth]{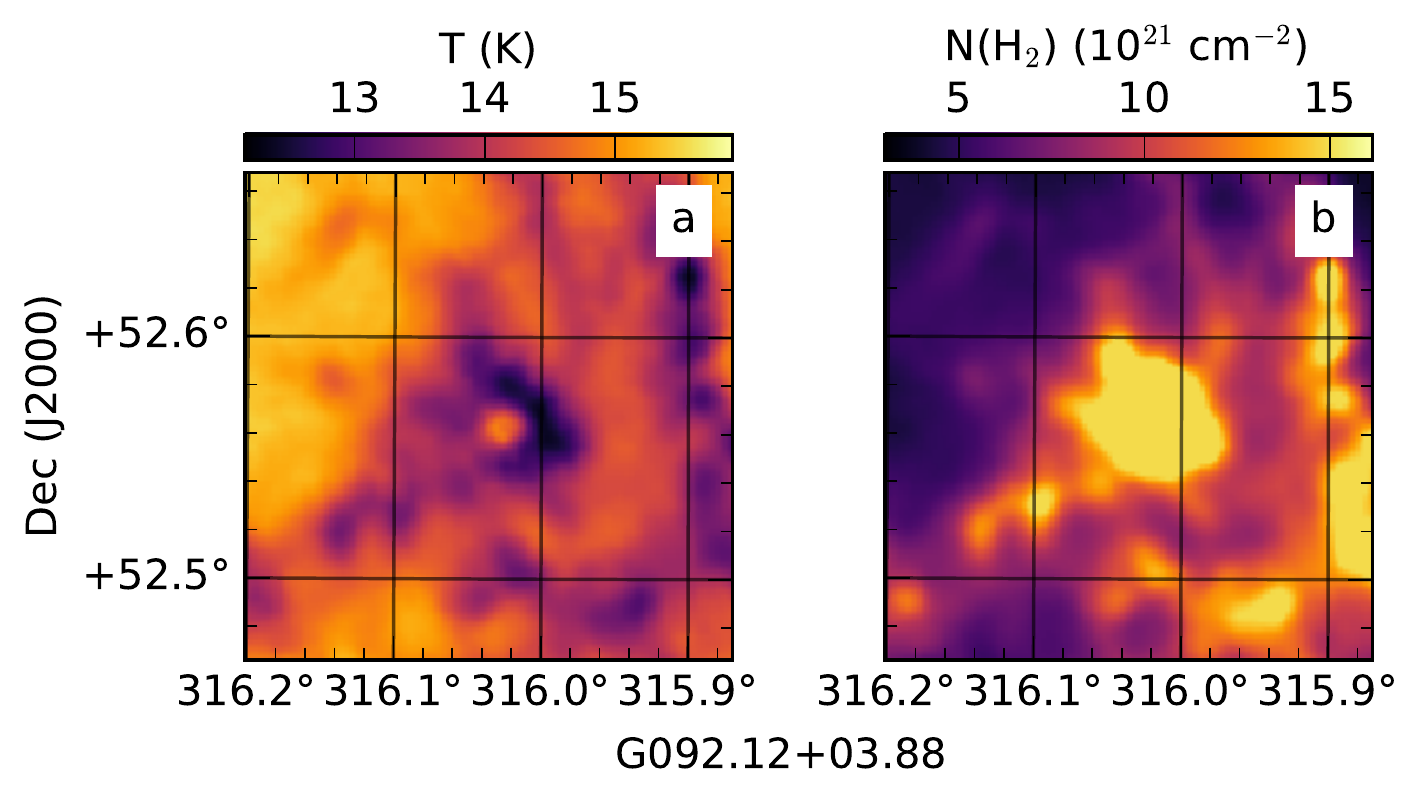}
        \end{minipage}
        \begin{minipage}{\linewidth}
                \line(1,0){250}
        \end{minipage}
        \begin{minipage}{\linewidth}
                \centering
                \includegraphics[width=\linewidth]{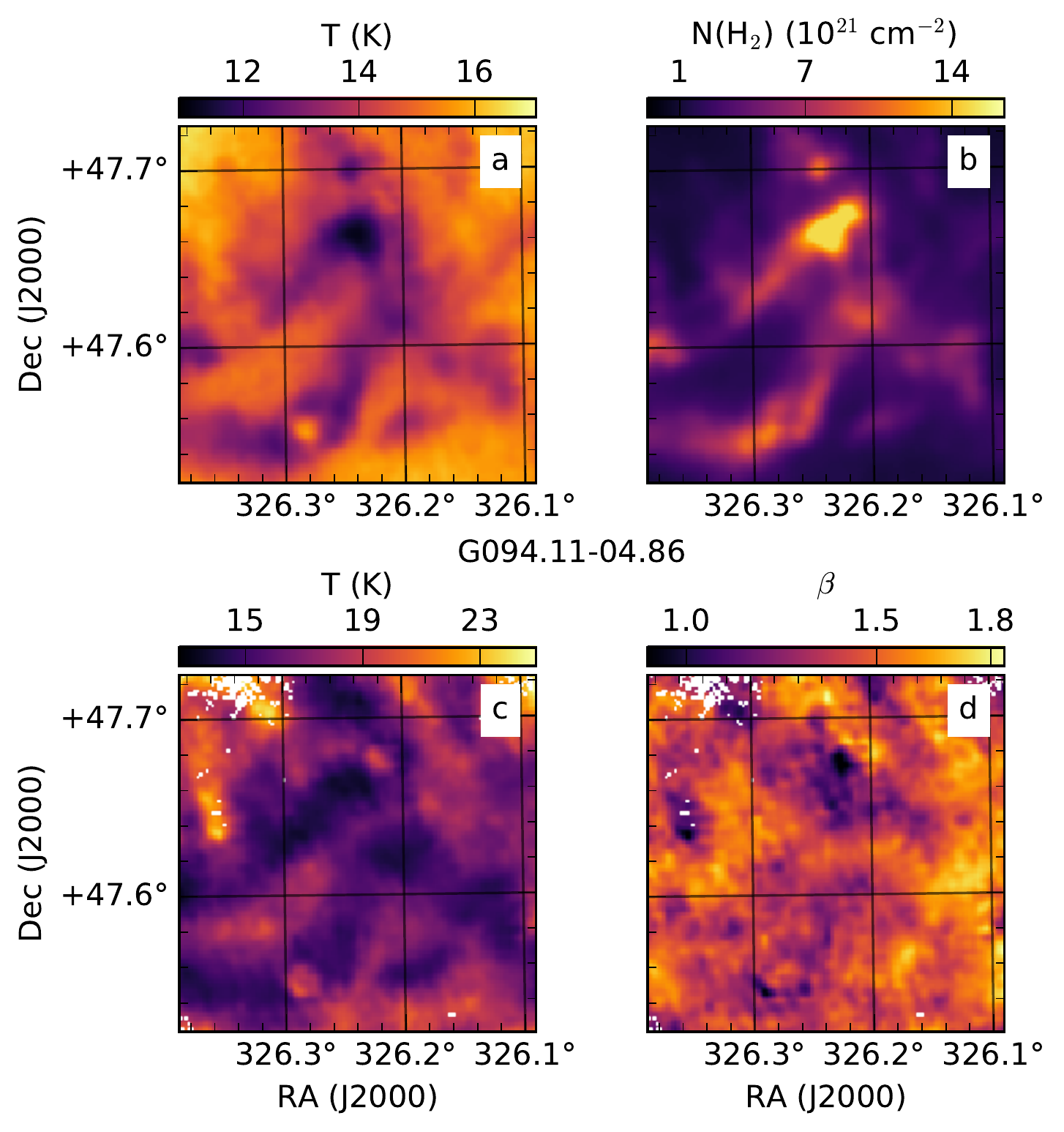}
        \end{minipage}
        \caption{Same as Fig. \ref{fig:appendix_extended_1} for fields G092.28+03.79 and G093.51-04.31.}
\end{figure}

\begin{figure}
        \centering
        \begin{minipage}{\linewidth}
                \includegraphics[width=\linewidth]{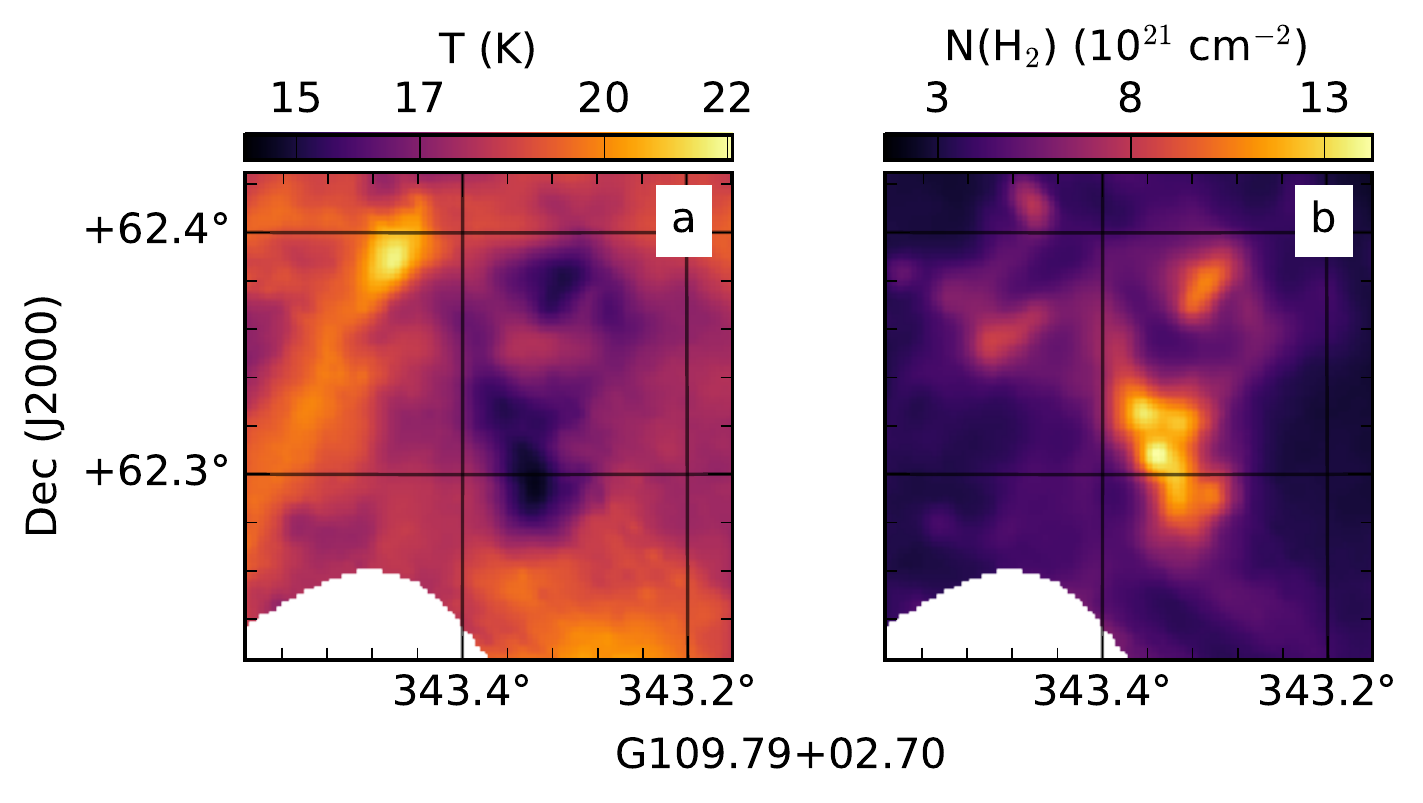}
        \end{minipage}
        \begin{minipage}{\linewidth}
                \line(1,0){250}
        \end{minipage}
        \begin{minipage}{\linewidth}
                \centering
                \includegraphics[width=\linewidth]{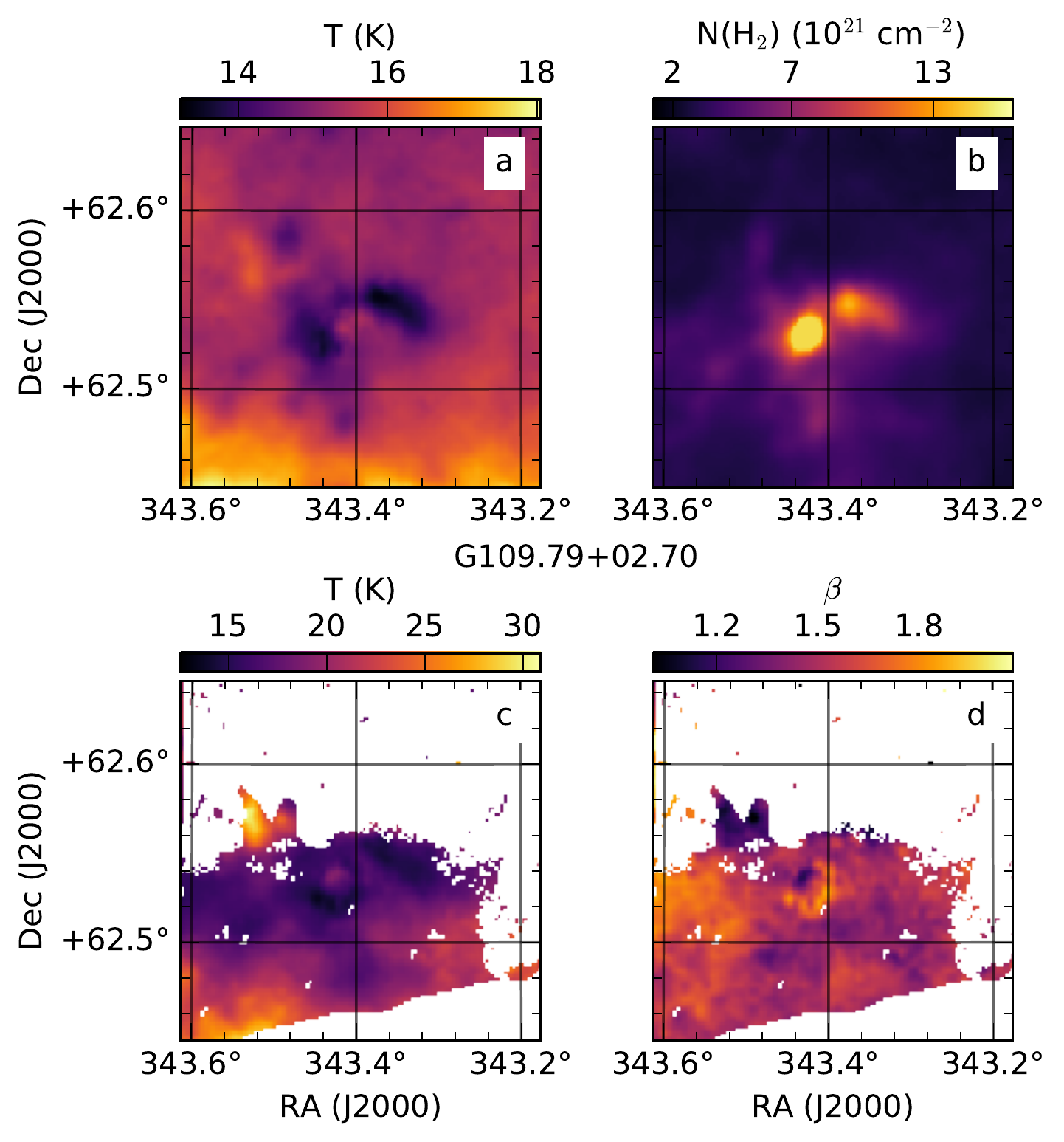}
        \end{minipage}
        \caption{Same as Fig. \ref{fig:appendix_extended_1} for fields G109.70+02.52 and G109.81+02.72.}
\end{figure}

\begin{figure}
        \centering
        \begin{minipage}{\linewidth}
                \includegraphics[width=\linewidth]{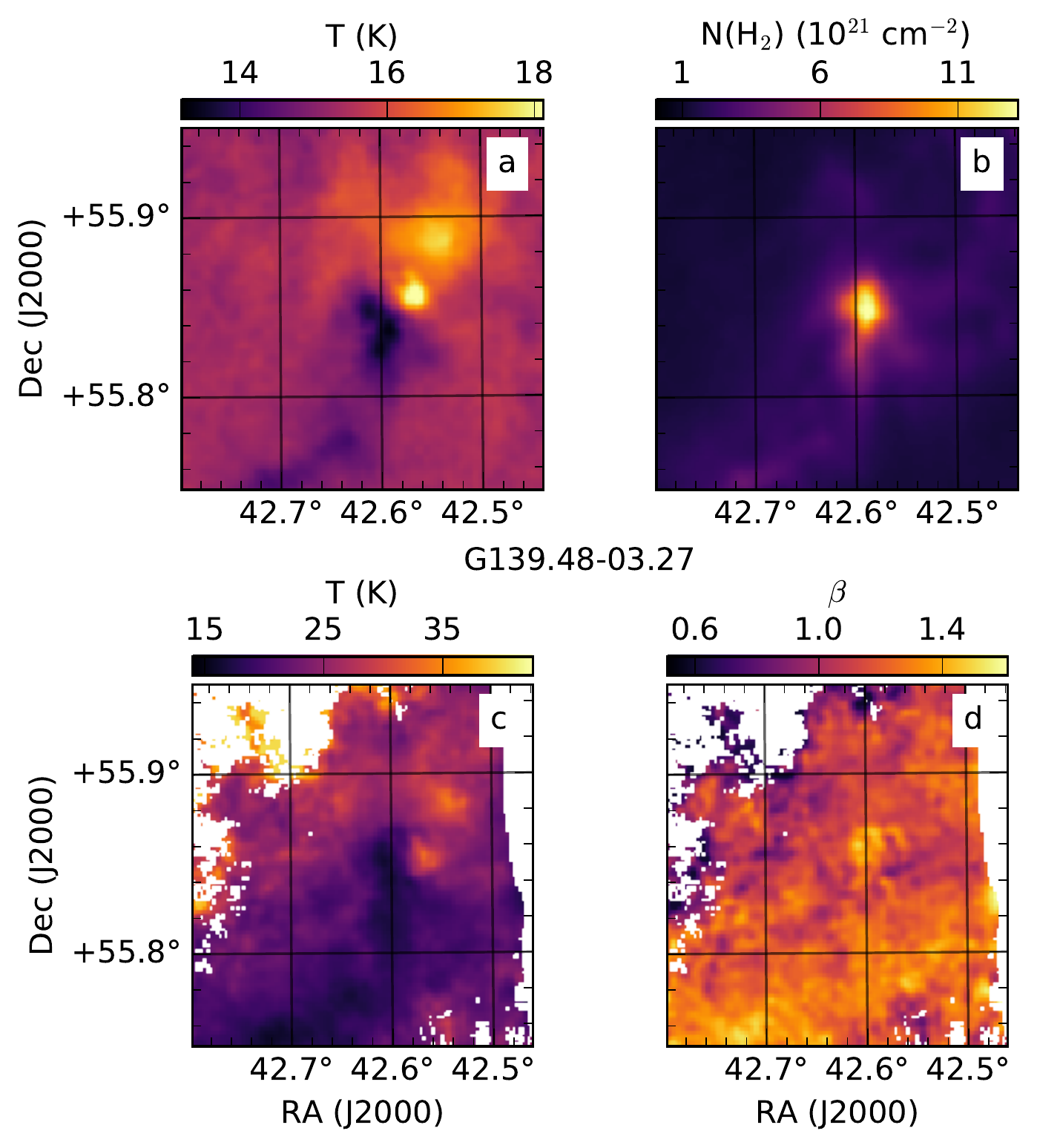}
        \end{minipage}
        \begin{minipage}{\linewidth}
                \line(1,0){250}
        \end{minipage}
        \begin{minipage}{\linewidth}
                \centering
                \includegraphics[width=\linewidth]{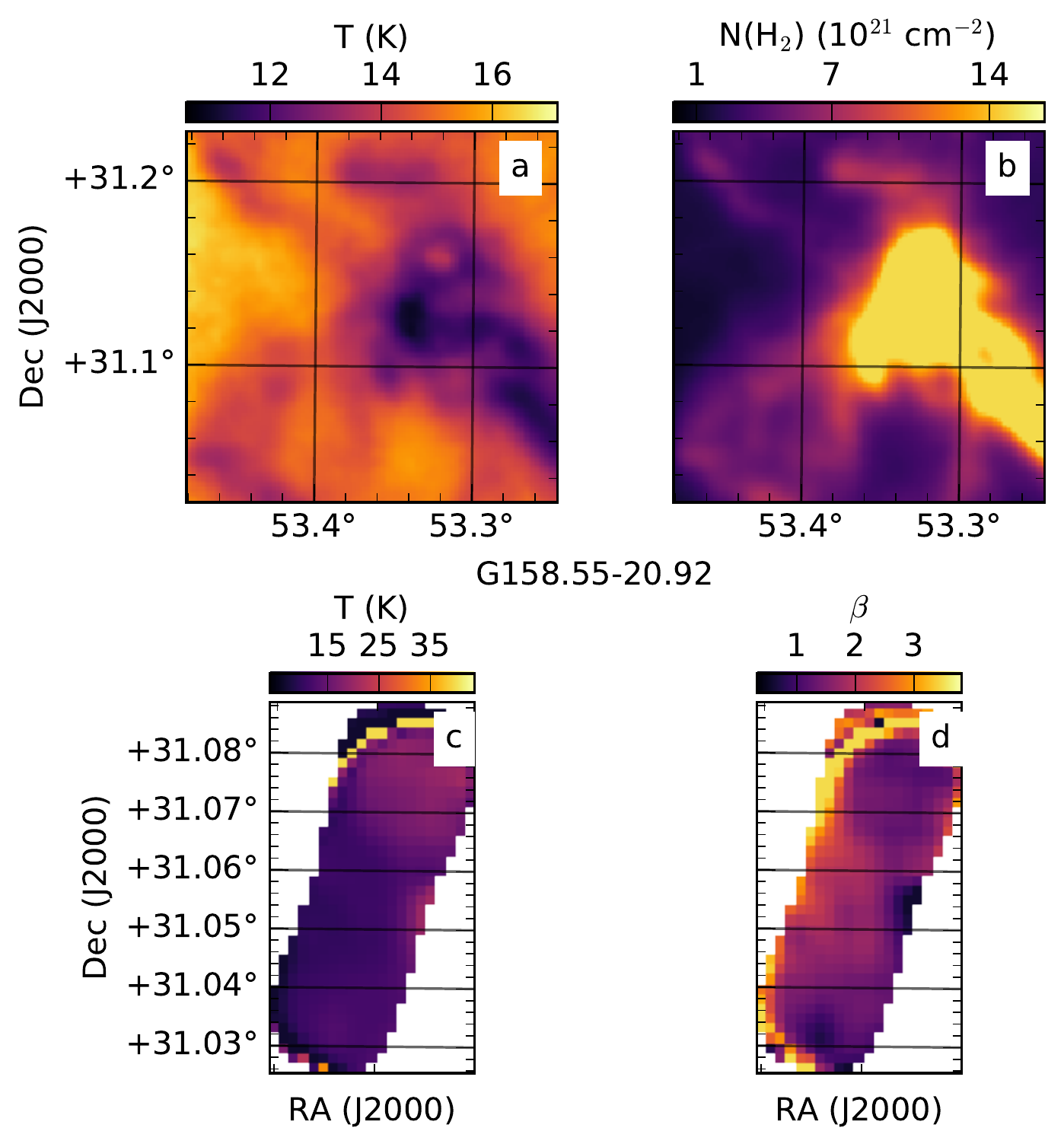}
        \end{minipage}
        \caption{Same as Fig. \ref{fig:appendix_extended_1} for fields G139.14-03.23 and G159.23-20.09.}
\end{figure}

\begin{figure}
        \centering
        \begin{minipage}{\linewidth}
                \includegraphics[width=\linewidth]{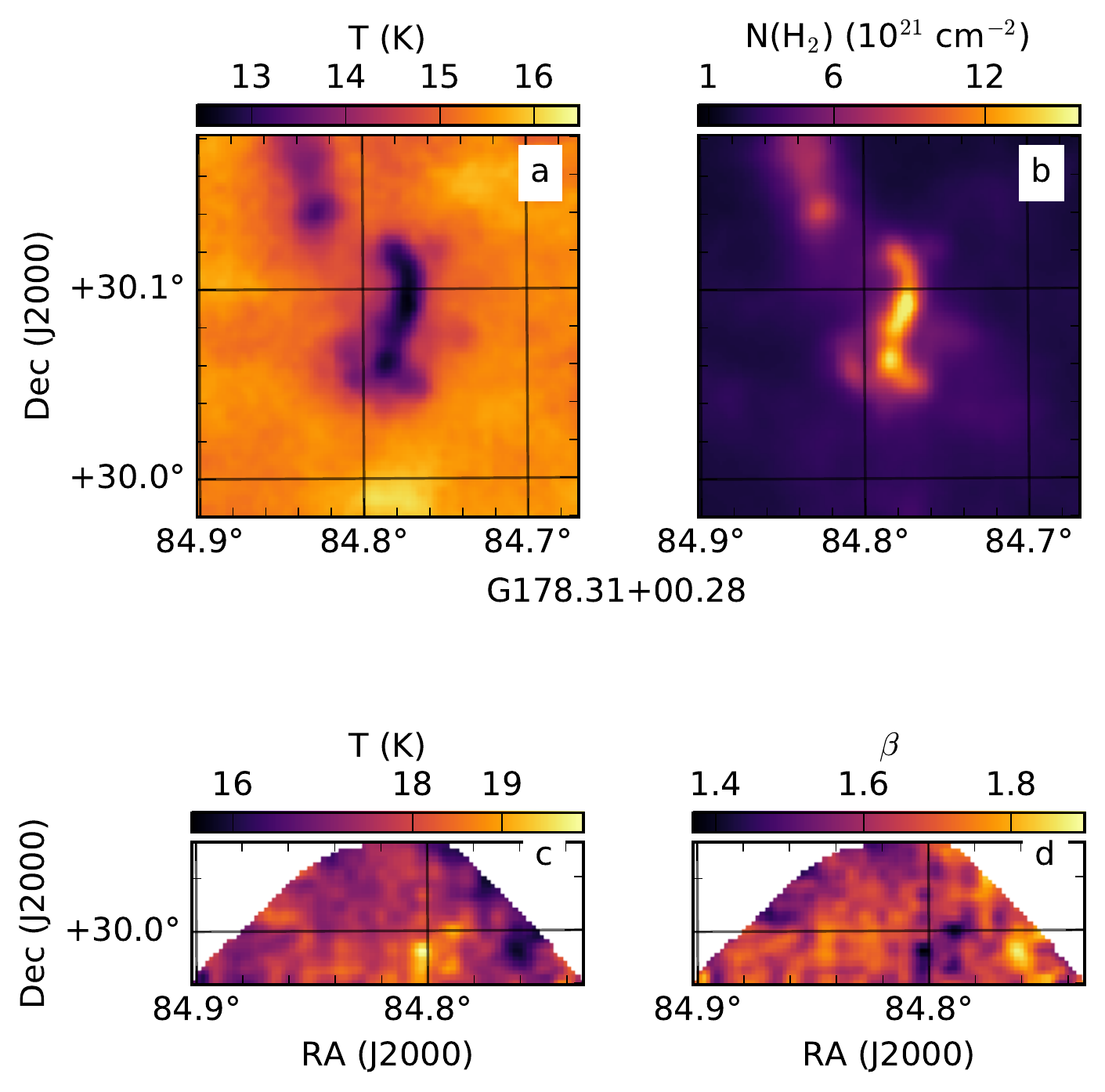}
        \end{minipage}
        \begin{minipage}{\linewidth}
                \line(1,0){250}
        \end{minipage}
        \begin{minipage}{\linewidth}
                \centering
                \includegraphics[width=\linewidth]{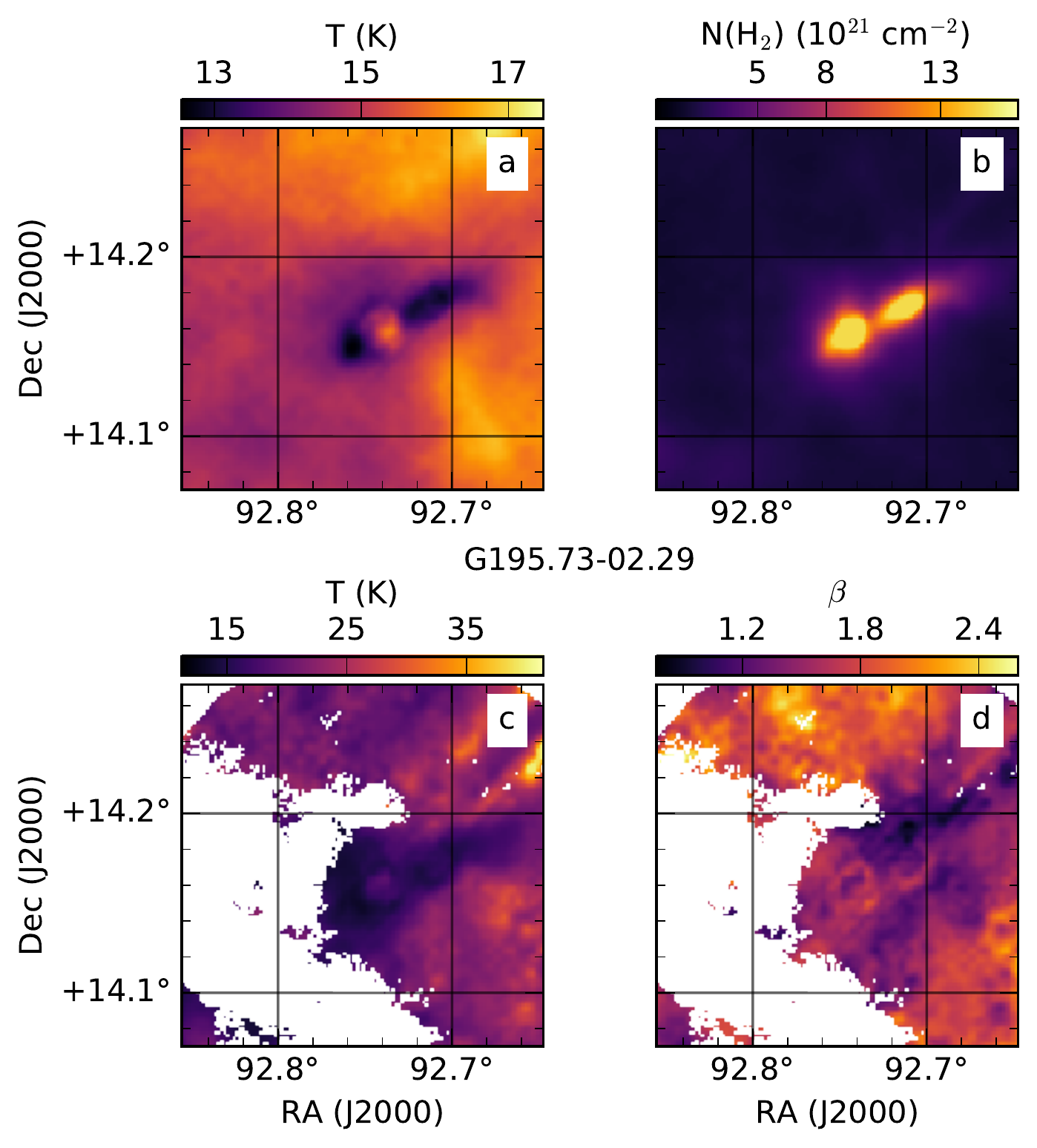}
        \end{minipage}
        \caption{Same as Fig. \ref{fig:appendix_extended_1} for fields G178.28-00.60 and G195.73-02.28.}
\end{figure}

\begin{figure}[h]
        \resizebox{\hsize}{!}{\includegraphics{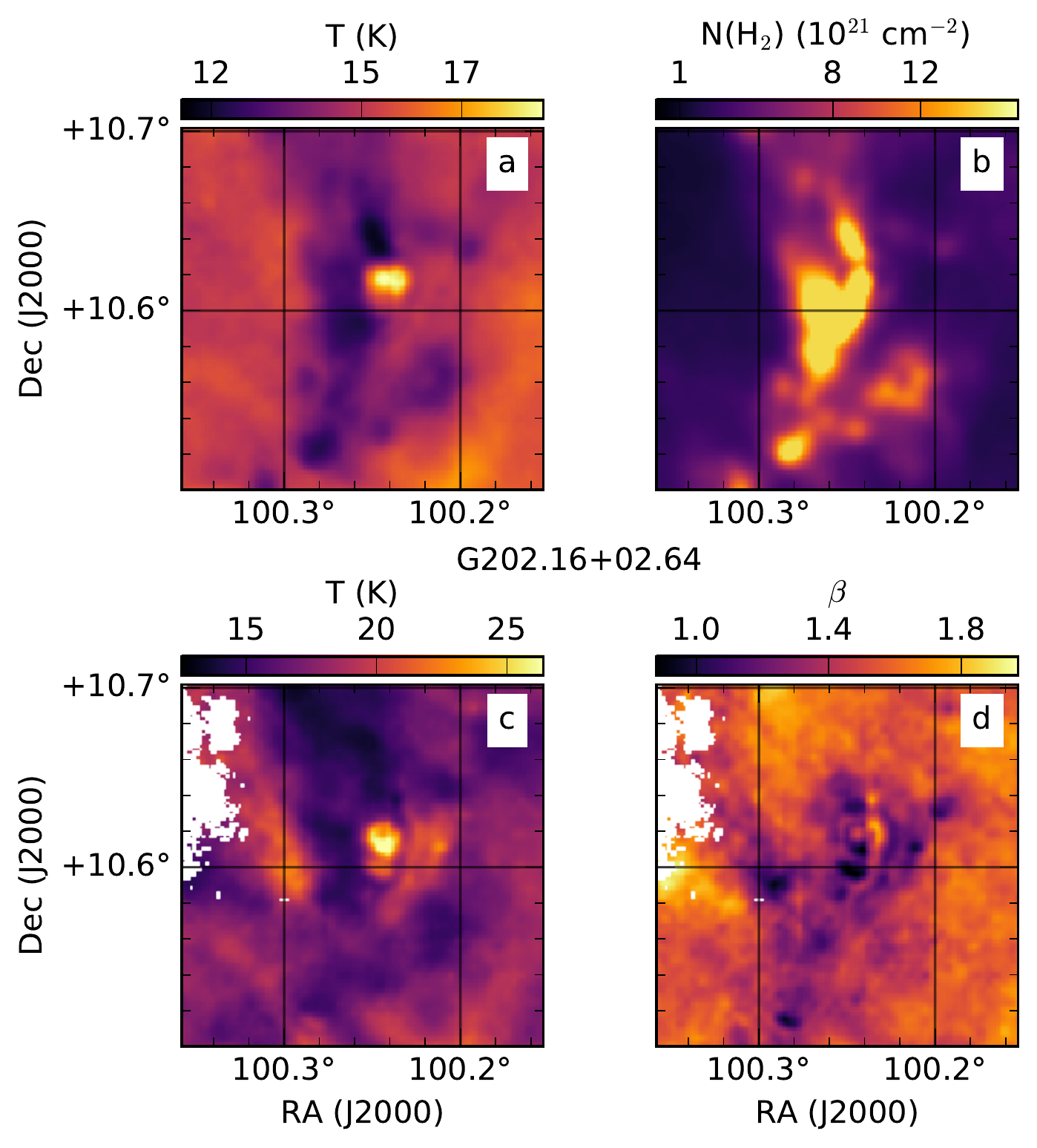}}
        \caption{Same as Fig. \ref{fig:appendix_extended_1} for field G202.31+02.53. \label{fig:appendix_extended_last} }
\end{figure}

%-------------------------------------------------------------

\clearpage

\section{\textit{Herschel} spectra \label{sec:appendix_herschel_spectra}}
\begin{figure*}
        \resizebox{\hsize}{!}{\includegraphics{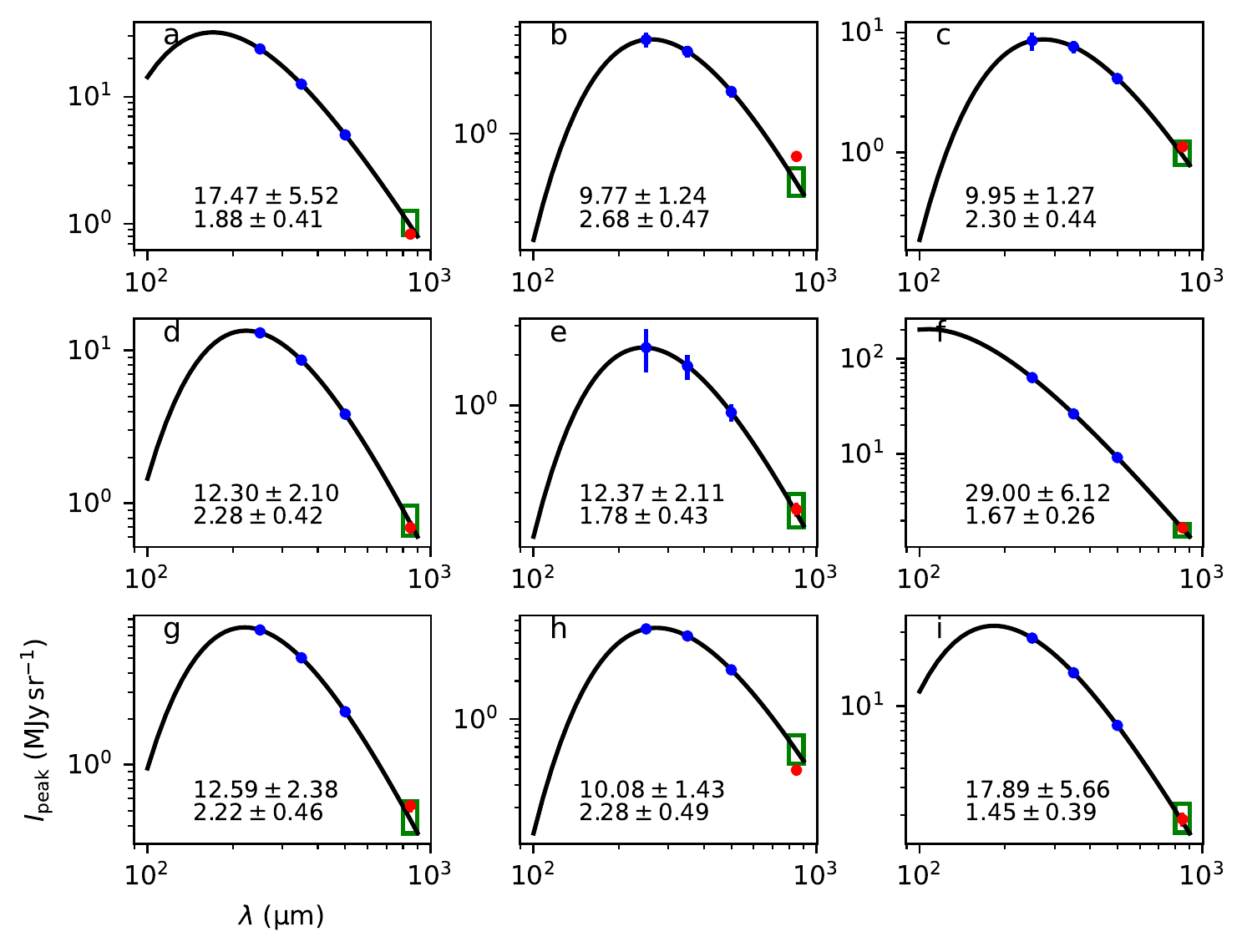}}
        \caption{SEDs of the peak pixels of nine randomly-selected clumps with S/N $>3$ in all bands. The blue dots are at 250, 350, and 500\,\micro, respectively. The green box indicates the 5\%-95\% confidence interval of the extrapolated 850\,\micro\spc value, and the red dot the observed value at 850\,\micro. \label{fig:SED} }
\end{figure*}

In Fig. \ref{fig:SED} we show as an example the SEDs of nine randomly selected clumps used for studying 850\,\micro\, excess. See Sect.  \ref{sec:results_Herschel_spectra_and_850um} for further details.

\section{YSOs and clumps on 850 \micro\spc intensity maps}
\label{sec:appendix_YSO_clump}
\begin{figure}
        \centering
        \begin{minipage}{\linewidth}
                \centering
                \includegraphics[width=\linewidth]{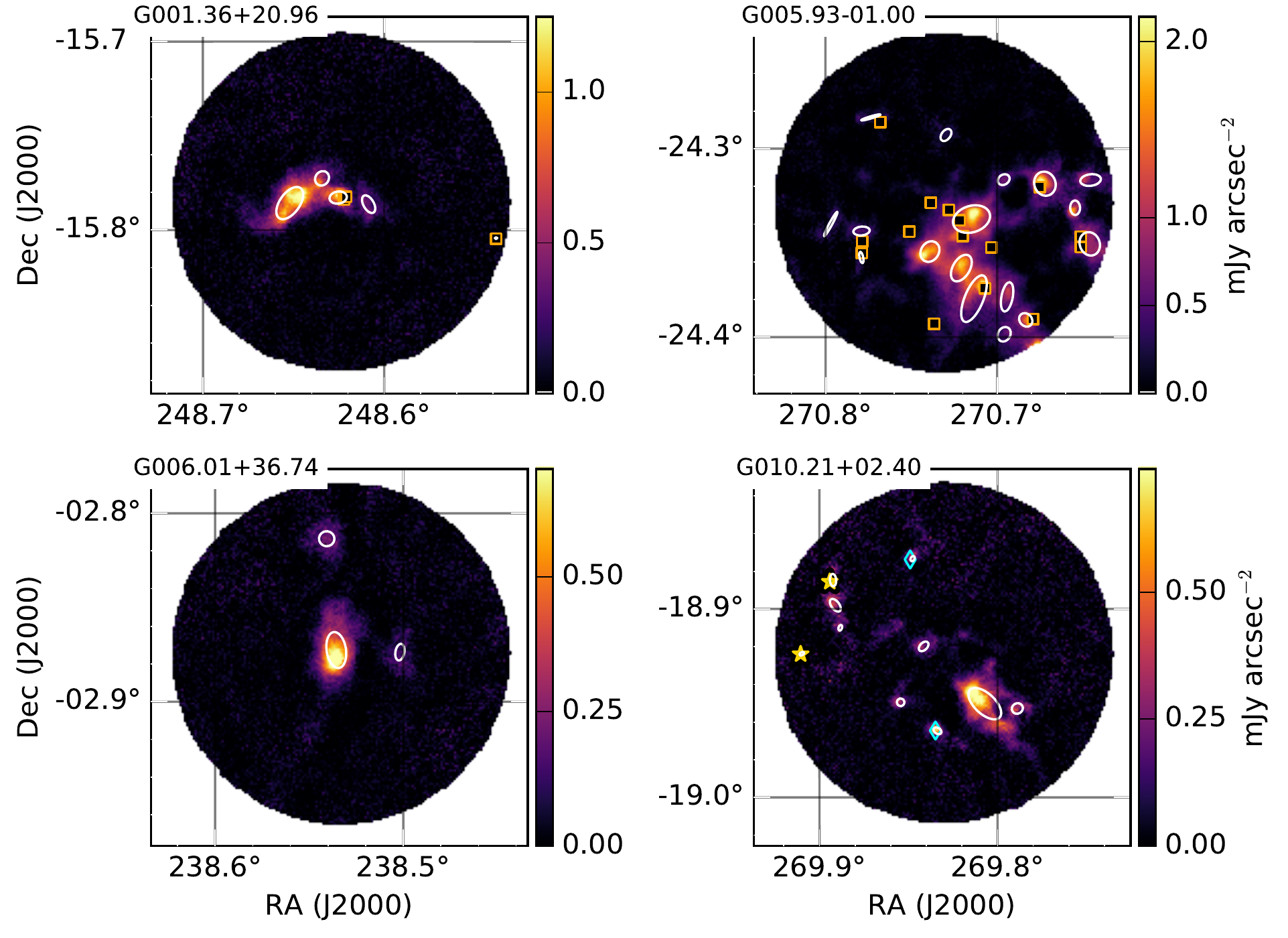}
        \end{minipage}
        \begin{minipage}{\linewidth}
                \centering
                \includegraphics[width=\linewidth]{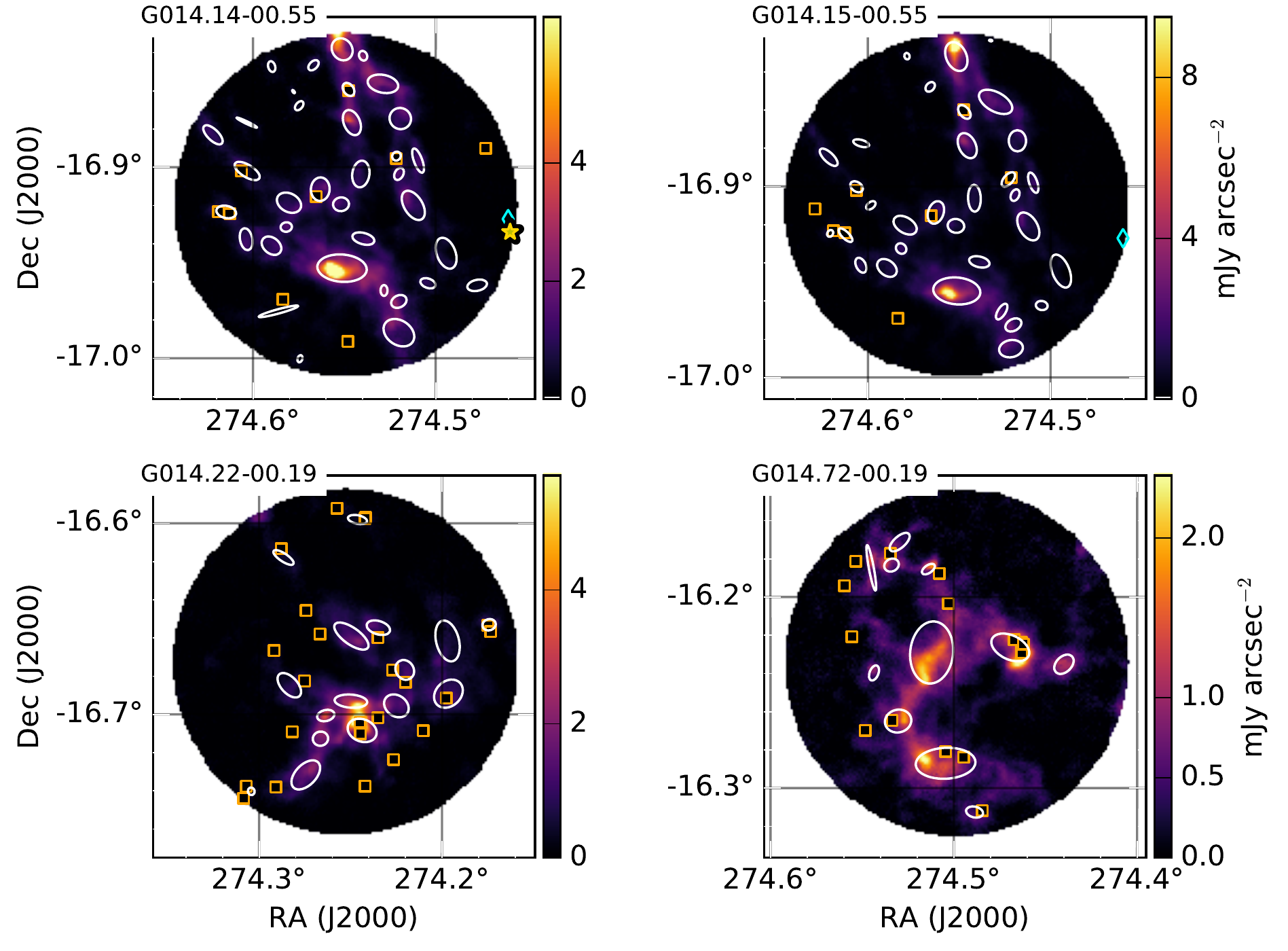}
        \end{minipage}
        \begin{minipage}{\linewidth}
                \centering
                \includegraphics[width=\linewidth]{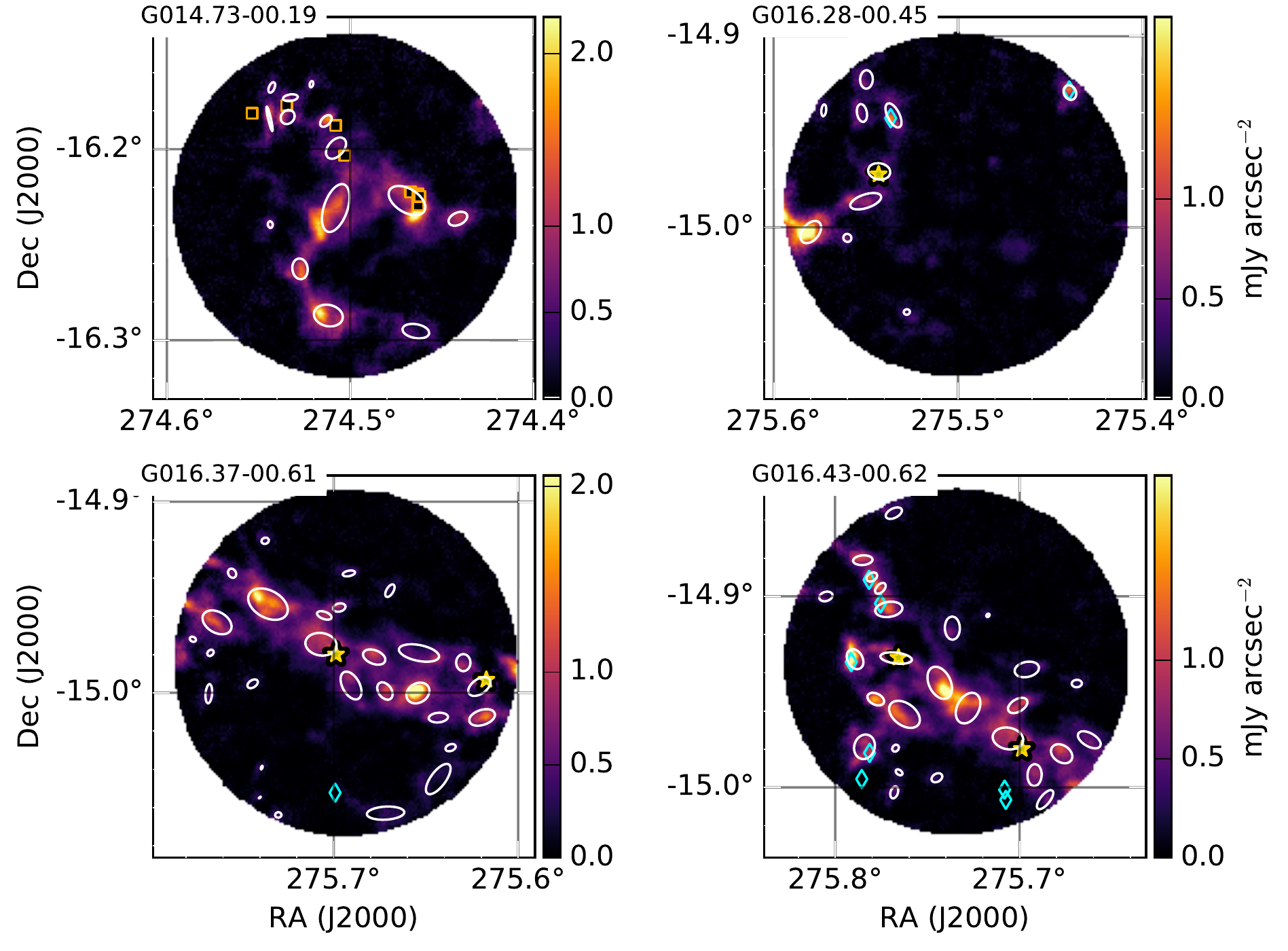}
        \end{minipage}
        \caption{Surface brightness maps of the SCUBA-2 850 \micro\spc intensity. Clumps found by FellWalker are plotted with white ellipses. The orange squares with black centers show the locations of unconfirmed YSOs for which \textit{Herschel} intensity ratio was not possible to calculate. The cyan diamonds show those YSOs that do not show significant increase in $\Delta C$, and yellow stars those YSOs that have been confirmed with $\Delta C$. Field YSOs are not plotted and no distinction is made between YSOs from different catalogs. The field names are in the top left. \label{fig:YSO_contours_first}}
\end{figure}

All candidate YSOs associated with the clumps are shown on SCUBA-2 intensity maps in Figs. \ref{fig:YSO_contours_first} -- \ref{fig:YSO_contours_last}. The clumps are plotted with white ellipses.

\begin{figure}
        \centering
        \begin{minipage}{\linewidth}
                \centering
                \includegraphics[width=\linewidth]{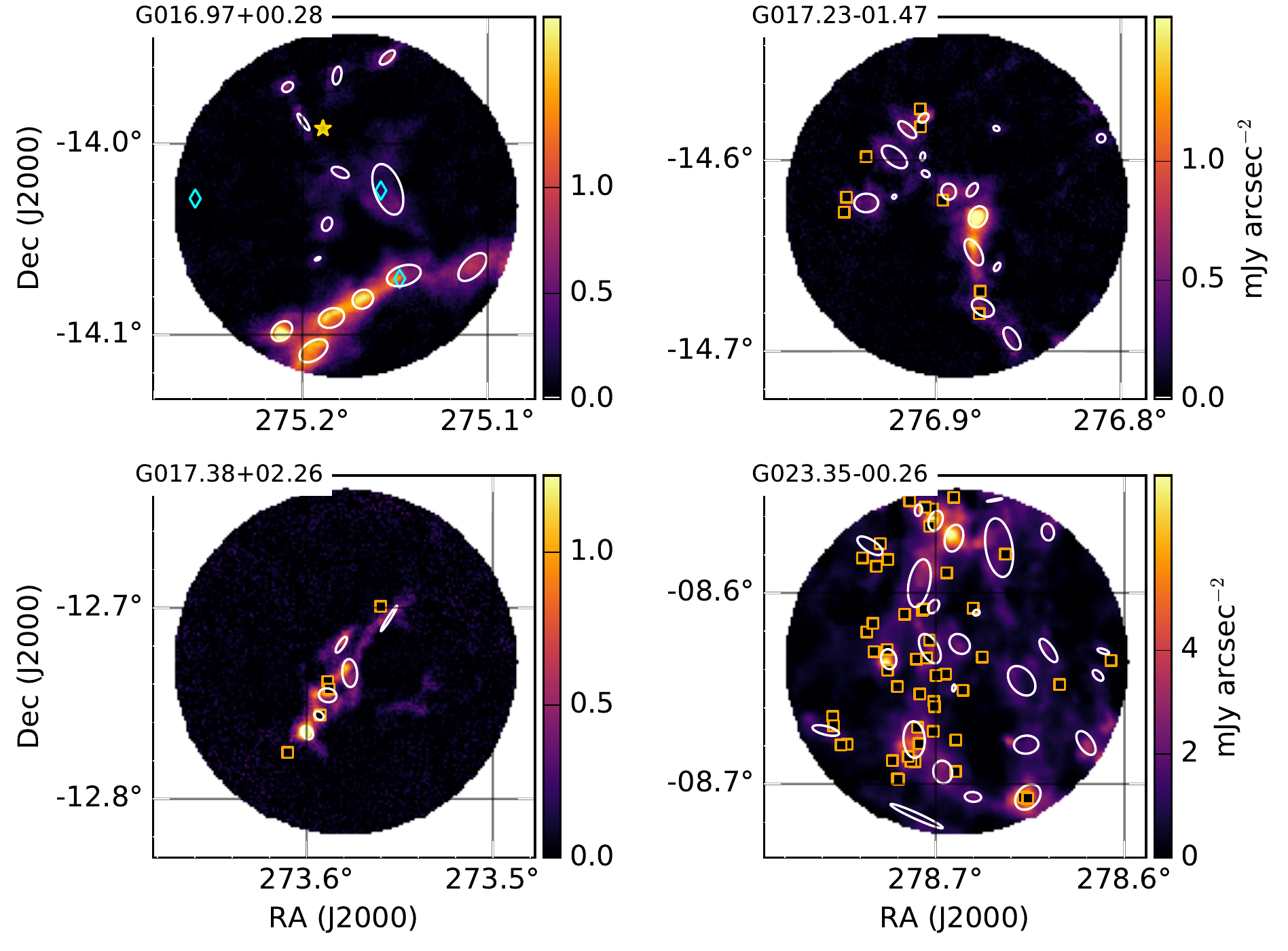}
        \end{minipage}
        \begin{minipage}{\linewidth}
                \centering
                \includegraphics[width=\linewidth]{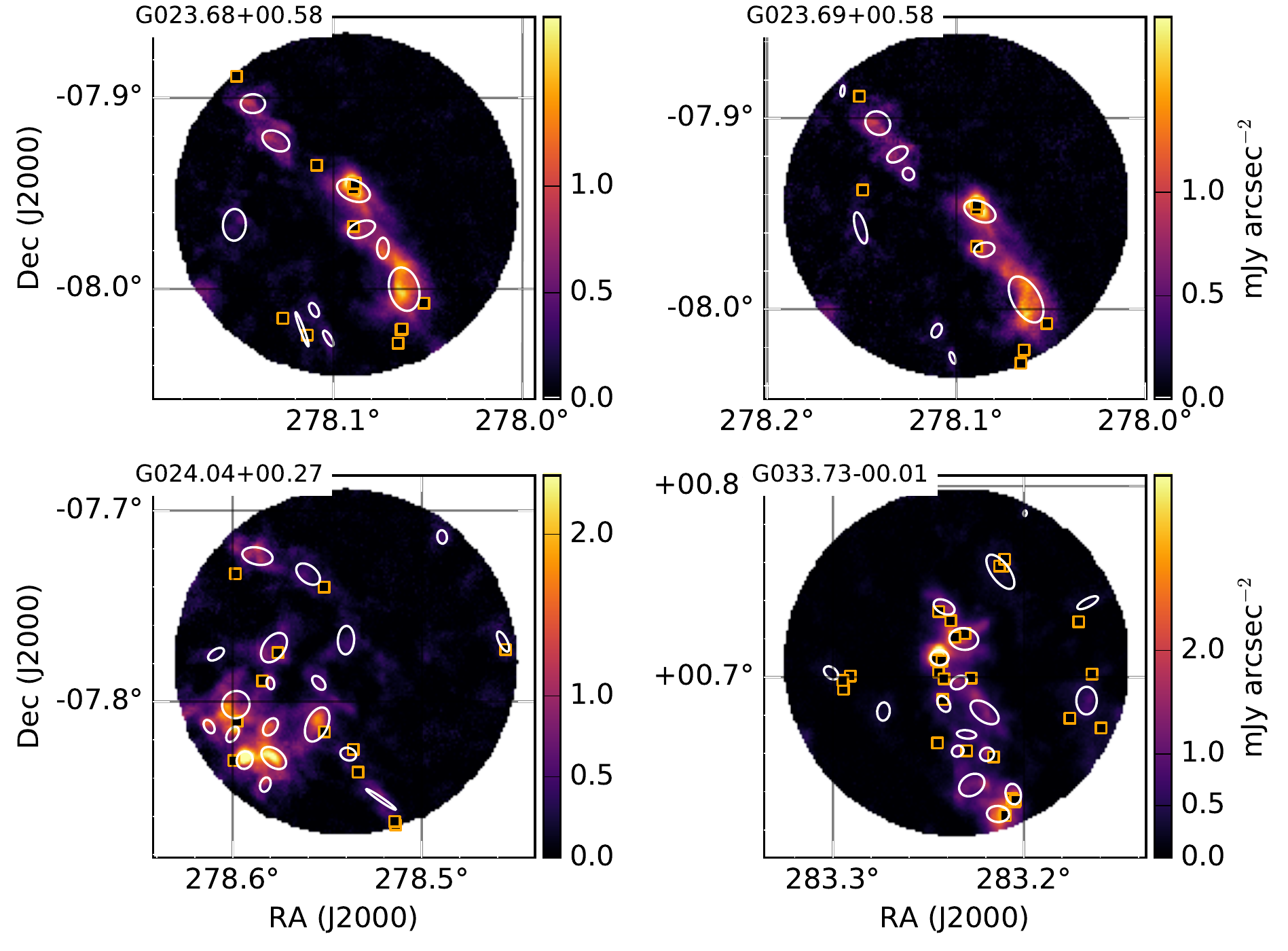}
        \end{minipage}
        \begin{minipage}{\linewidth}
                \centering
                \includegraphics[width=\linewidth]{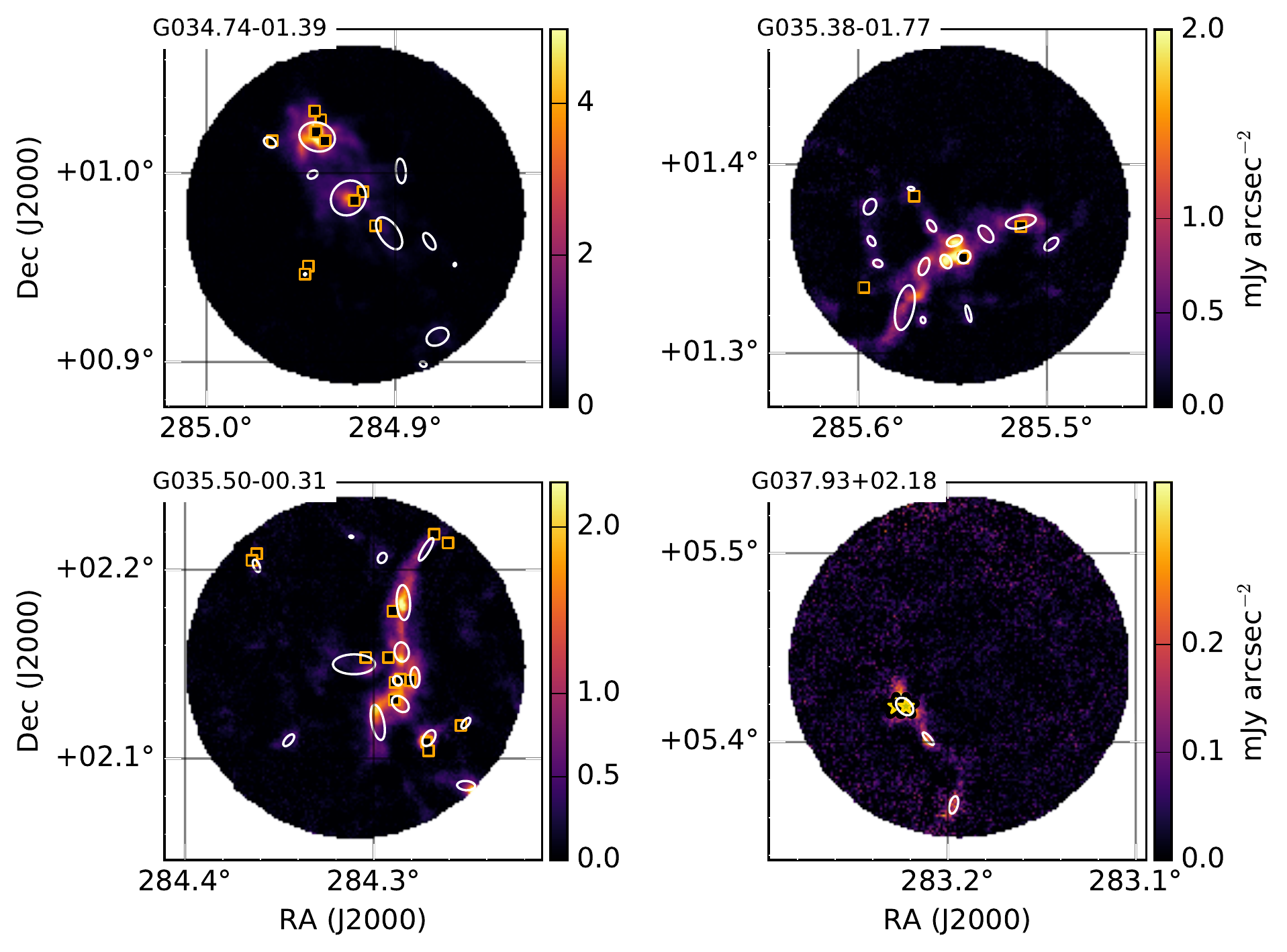}
        \end{minipage}
        \caption{Same as Fig. \ref{fig:YSO_contours_first} for fields G016.97+00.28 -- G037.93+02.18.}
\end{figure}

\begin{figure}
        \centering
        \begin{minipage}{\linewidth}
                \centering
                \includegraphics[width=\linewidth]{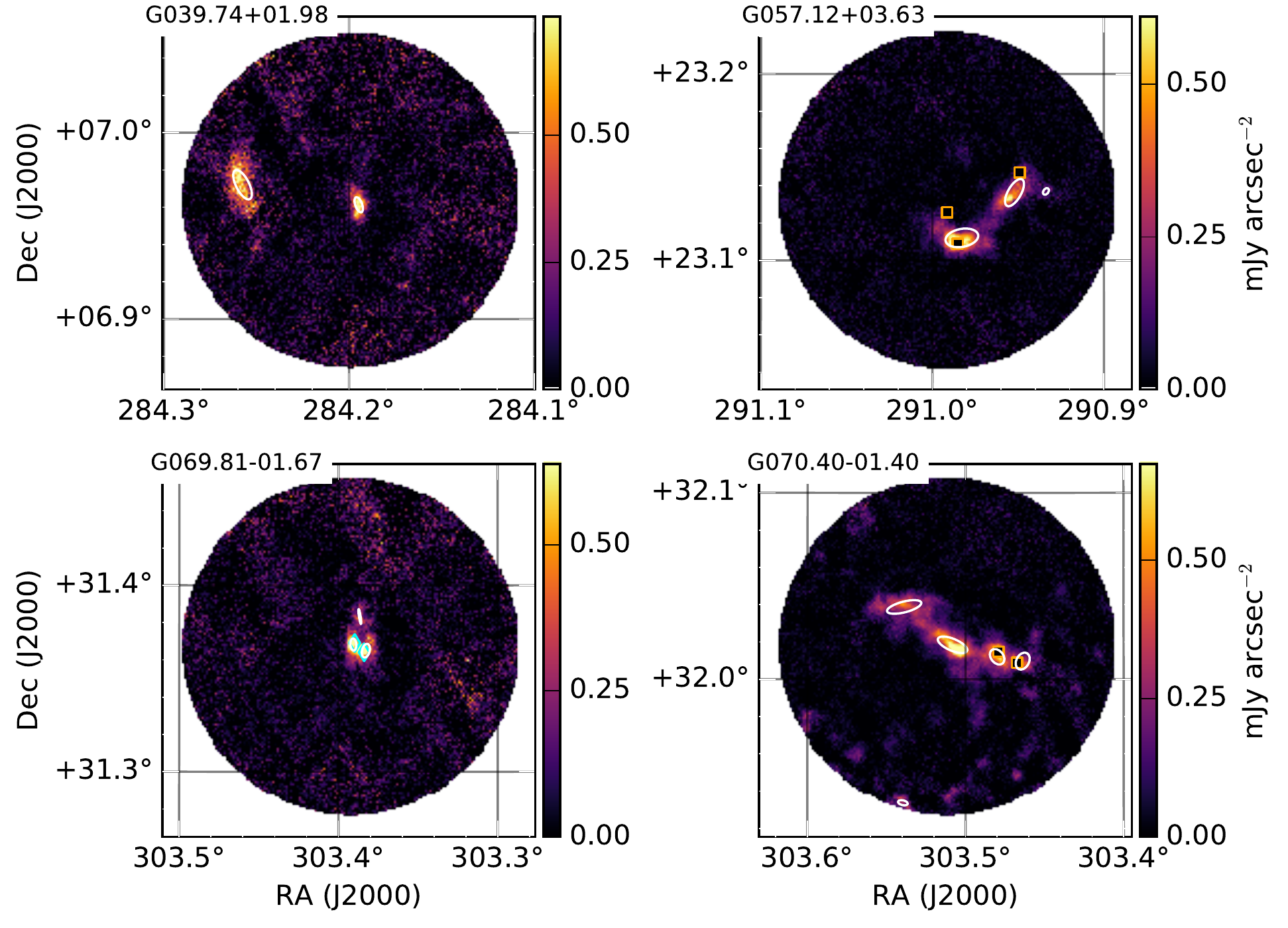}
        \end{minipage}
        \begin{minipage}{\linewidth}
                \centering
                \includegraphics[width=\linewidth]{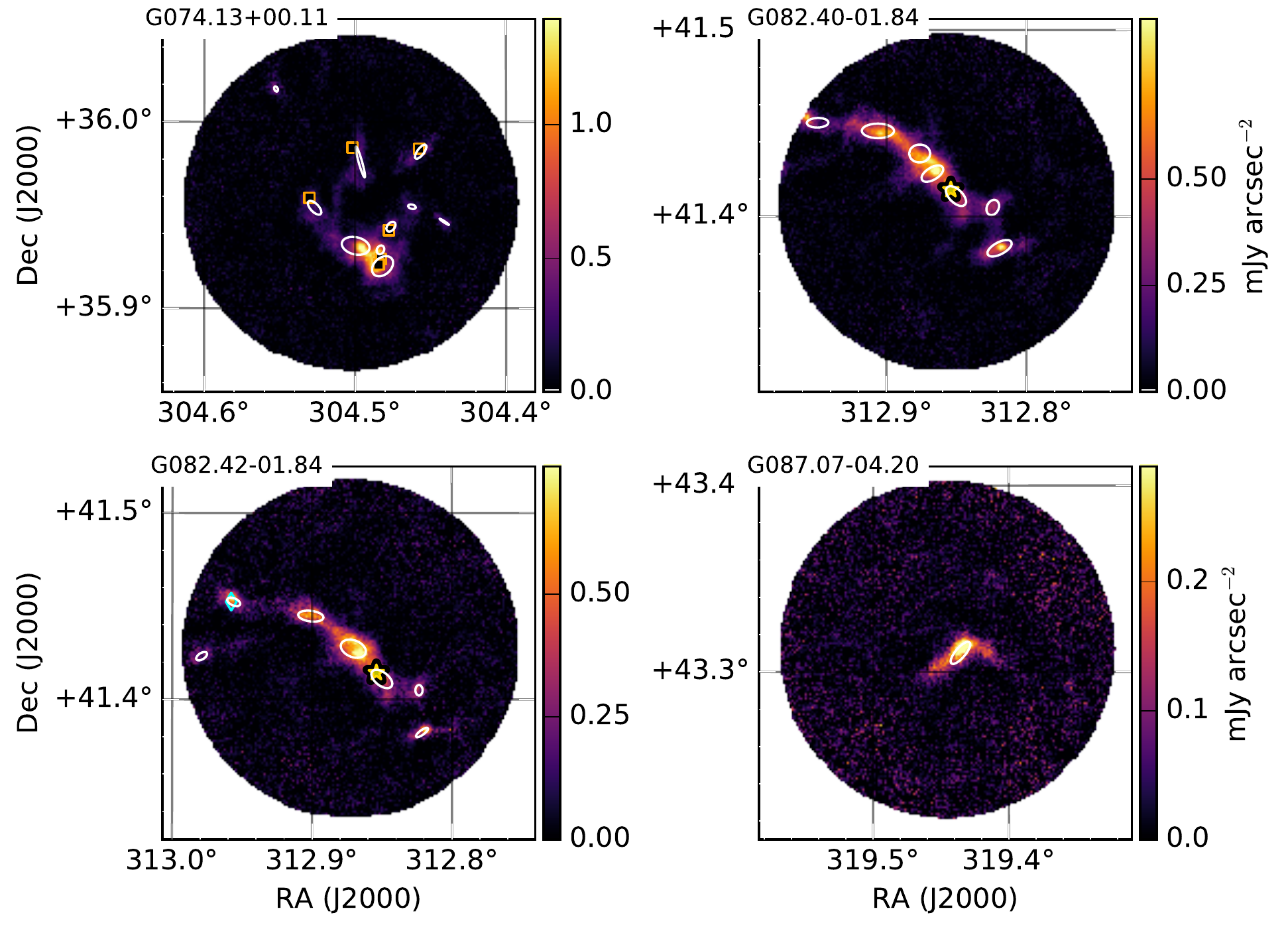}
        \end{minipage}
        \begin{minipage}{\linewidth}
                \centering
                \includegraphics[width=\linewidth]{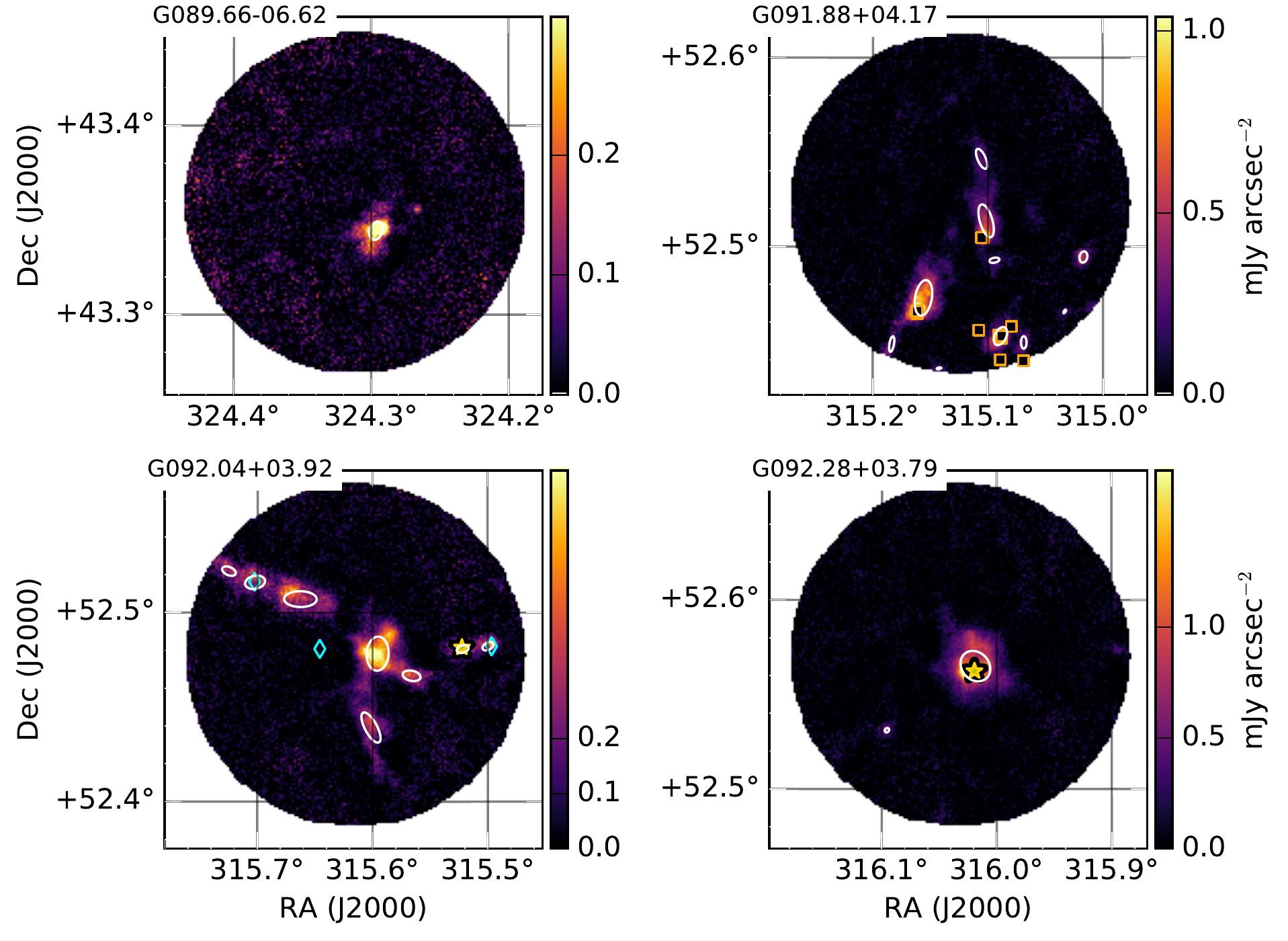}
        \end{minipage}
        \caption{Same as Fig. \ref{fig:YSO_contours_first} for fields G039.74+01.98 -- G092.28+03.79.}
\end{figure}

\begin{figure}
        \centering
        \begin{minipage}{\linewidth}
                \centering
                \includegraphics[width=\linewidth]{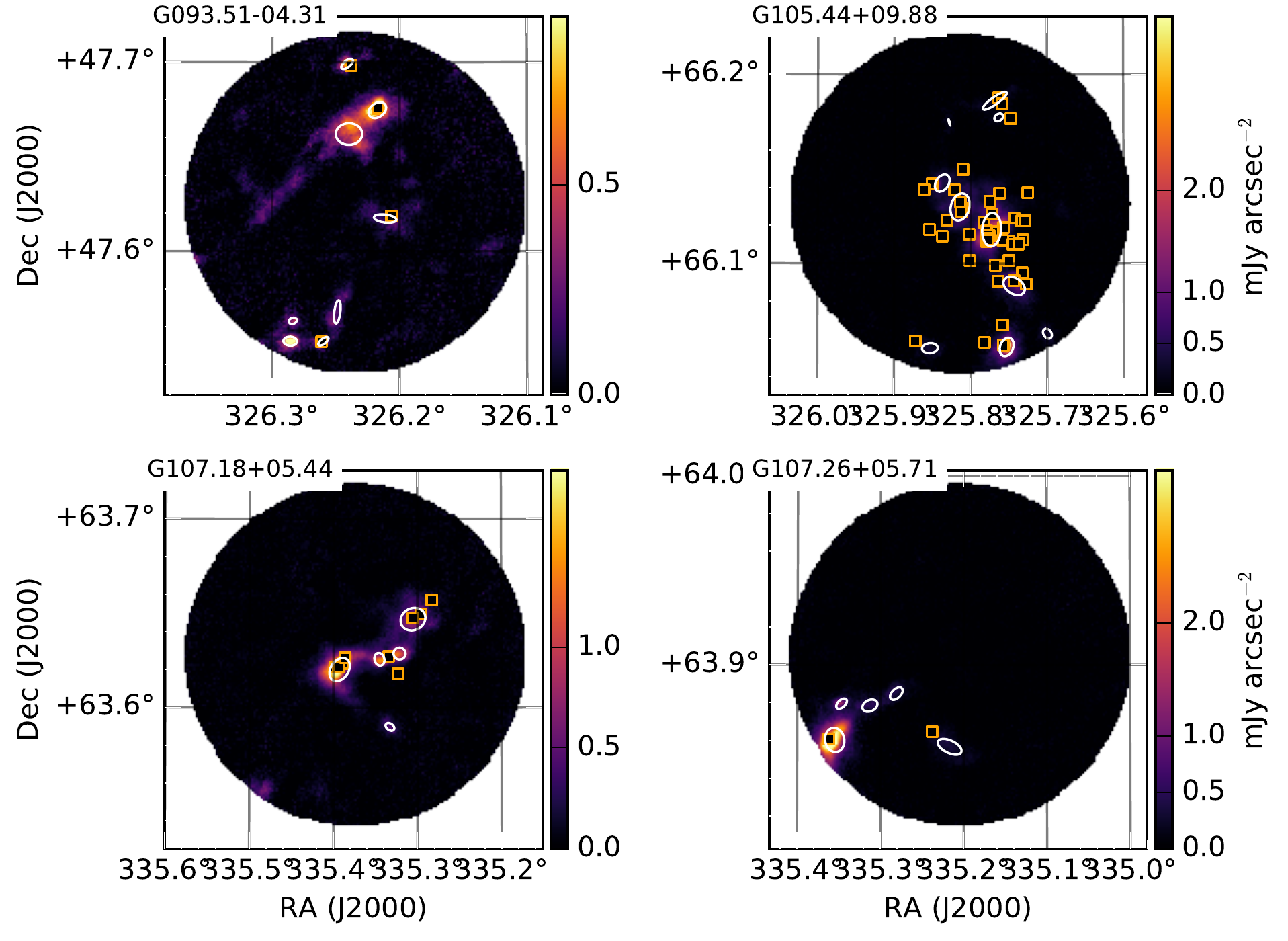}
        \end{minipage}
        \begin{minipage}{\linewidth}
                \centering
                \includegraphics[width=\linewidth]{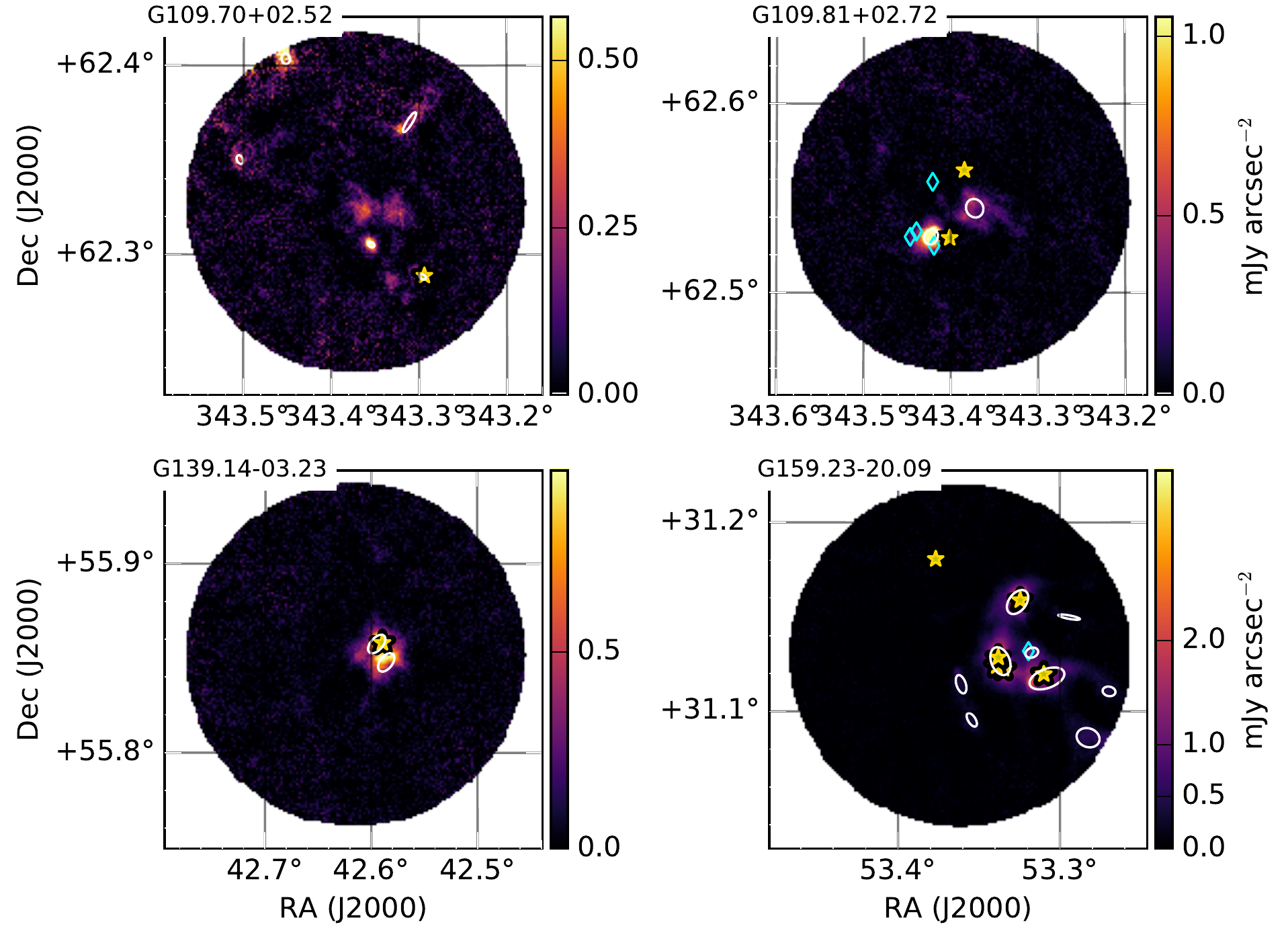}
        \end{minipage}
        \begin{minipage}{\linewidth}
                \centering
                \includegraphics[width=\linewidth]{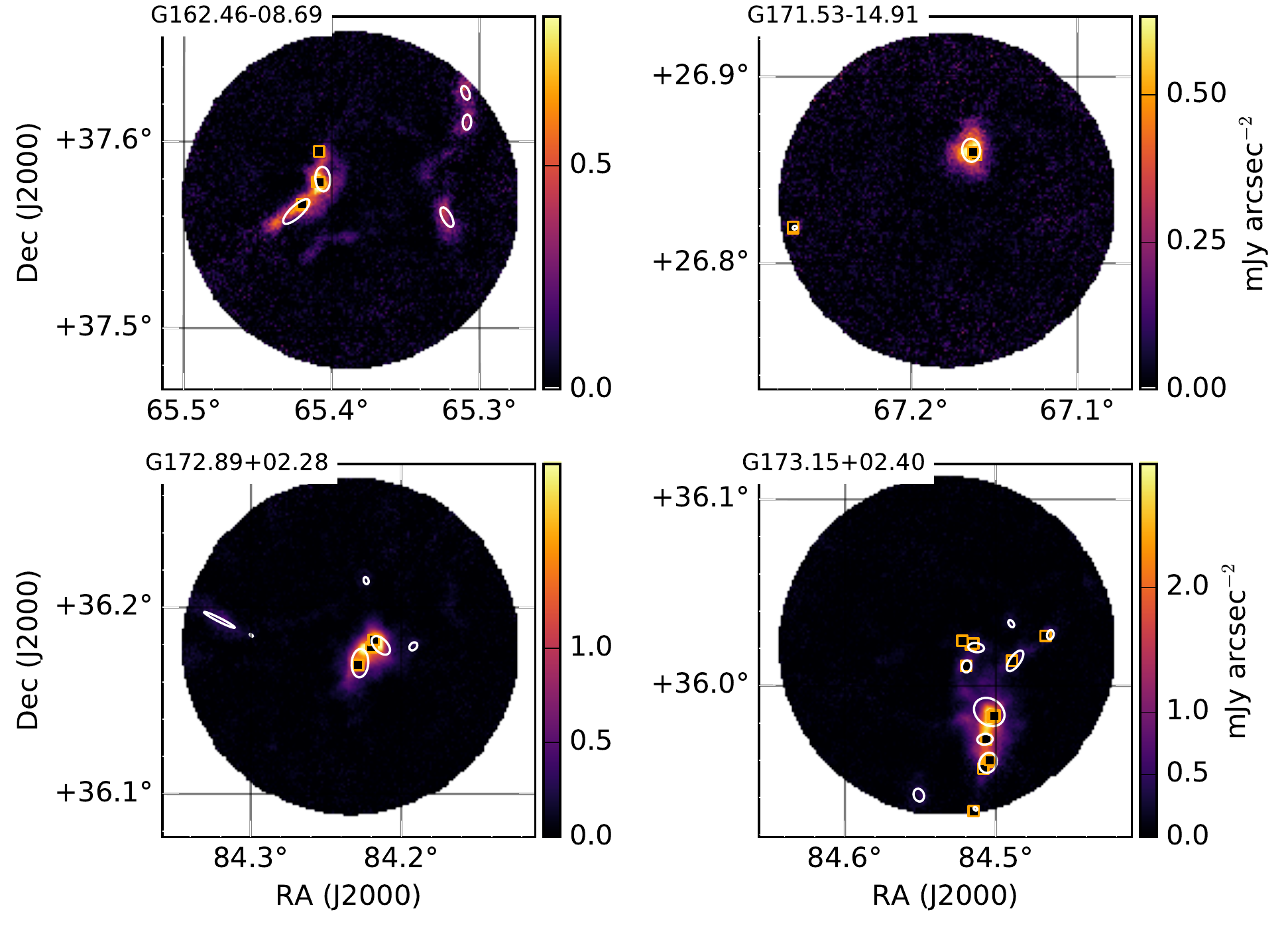}
        \end{minipage}
        \caption{Same as Fig. \ref{fig:YSO_contours_first} for fields G093-04.31 -- G173.15+02.40.}
\end{figure}

\begin{figure}
        \centering
        \begin{minipage}{\linewidth}
                \centering
                \includegraphics[width=\linewidth]{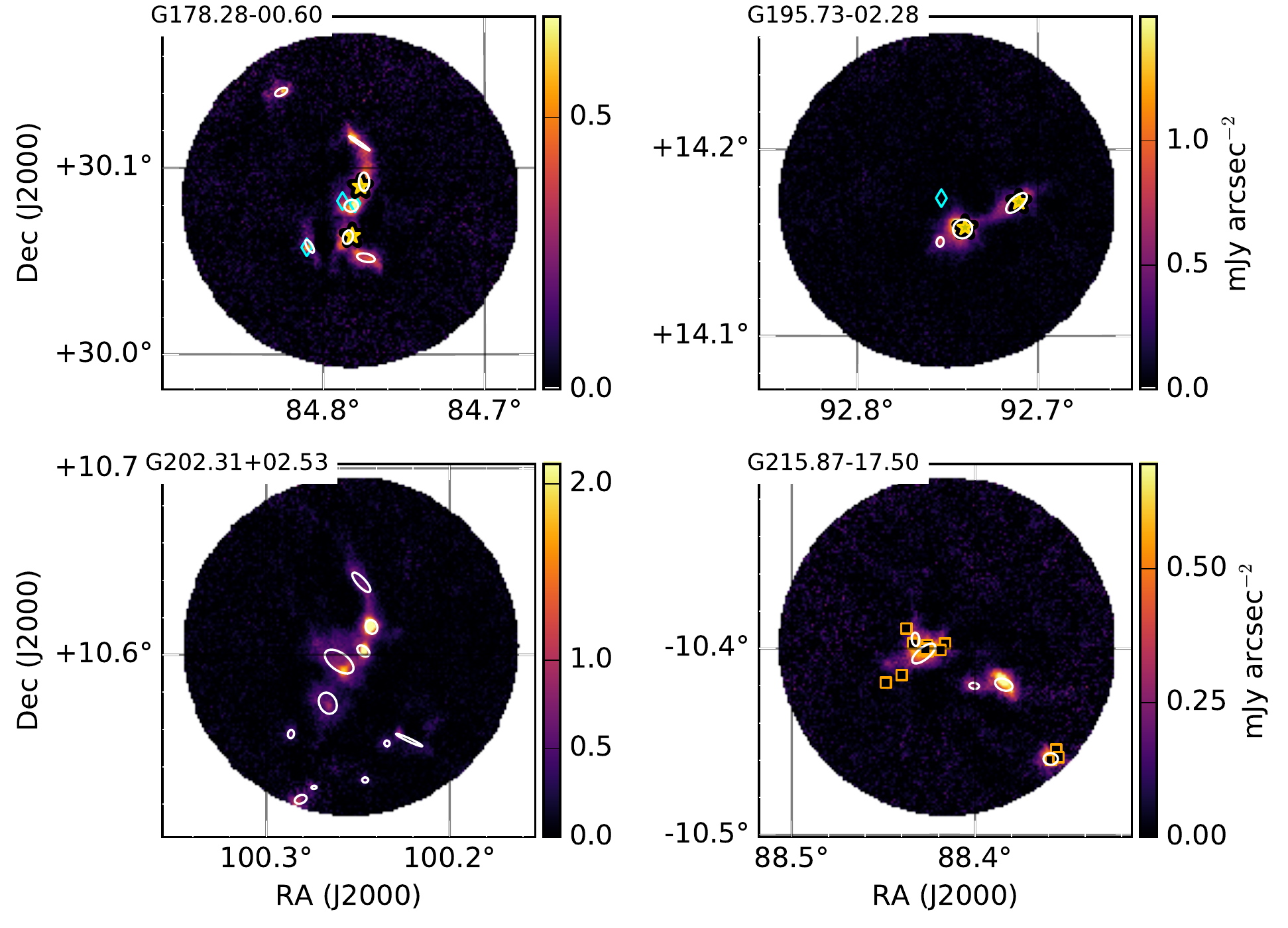}
        \end{minipage}
        \caption{Same as Fig. \ref{fig:YSO_contours_first} for fields G178.28-00.60 -- G215.87-17.50. \label{fig:YSO_contours_last}}
\end{figure}

\clearpage
\section{Catalog of derived clump properties\label{sec:appendix_clump_properties}}
We derive up to 14 properties for each clump. The temperature, flux density, $I_{\rm 250}$,  \textit{N}(\MH), and $N_{\rm YSO, \Delta C}$ were not possible to calculate for clumps without \textit{Herschel} data. The effective radius, mass, and virial parameters were not possible to calculate for clumps without distance estimates. These are denoted with a "-" when a value was not available.

\begin{sidewaystable*}
        \caption{Properties of field G001.36+20.96}
        \label{tbl:properties_G001.36+20.96}
        \centering
\begin{tabular}{l|l l l l l l l l l l l l l l l l l}
                \hline\hline
                (1) & (2) & (3) & (4) & (5) & (6) & (7) & (8) & (9)  & (10)  & (11) & (12) & (13) & (14) & (15) & (16) & (17) & (18) \\ 
                Clump & RA & DEC & Position & Angular & $R_{\rm eff}$ & $\langle \beta\rangle$ & $F_{\rm 250}$ & $\langle T\rangle$  & $I_{\rm 250}$ & $\langle$\textit{N}(\MH)$\rangle$  & \textit{N}(\MH)$_{\rm peak}$ & \textit{M} & $M_{\rm vir}$ & $\alpha_{\rm vir}$ & $N_{\rm *}$ & $N_{\rm *, \Delta C}$ & type\\ 
                &   &  & angle & size &  & &  &   &  &   &  &  & &  &  &  & \\ 
                ID & (J2000) & (J2000) & (\degr) & (ra\arcsec, dec\arcsec) & (pc) &  & (Jy) & (K) & (MJy\,sr$^{-1}$)  & (10$^{21}$\,cm$^{-2}$)  & (10$^{21}$\,cm$^{-2}$) & (\msun) & (\msun) & \\ 
                \hline
                \multicolumn{17}{l}{Field G001.36+20.96}\\ 
                \hline
                1 & -111.375 & -15.783 & 264.4 & (16.8,11.6) & - & 1.7 & 1.9 & 15.861 & 616.7 & 13.1 & 36.8 & - & - & - & 1 & 1 & PS-C\\ 
                2 & -111.348 & -15.786 & 310.3 & (35.3,19.1) & - & 1.9 & 5.8 & 12.303 & 304.9 & 19.4 & 46.7 & - & - & - & 0 & 0 & SL\\ 
                3 & -111.366 & -15.773 & 308.4 & (14.5,12.8) & - & 2.0 & 1.2 & 12.7434 & 290.9 & 15.5 & 34.5 & - & - & - & 0 & 0 & SL\\
\hline 
        \end{tabular}
                \tablefoot{$N_{\rm YSO, \Delta C}$ refers to number of YSOs confirmed by \textit{Herschel} color. Clumps are starless (SL), prestellar (PRE), candidate protostellar (PS), or confirmed protostellar (PS-C). }
\end{sidewaystable*}

\begin{sidewaystable*}
        \caption{Properties of field G014.14-00.55}
        \label{tbl:properties_G014.14-00.55}
        \centering
                \begin{tabular}{l|l l l l l l l l l l l l l l l l l}
                \hline\hline
                (1) & (2) & (3) & (4) & (5) & (6) & (7) & (8) & (9)  & (10)  & (11) & (12) & (13) & (14) & (15) & (16) & (17) & (18) \\ 
                Clump & RA & DEC & Position & Angular & $R_{\rm eff}$ & $\langle \beta\rangle$ & $F_{\rm 250}$ & $\langle T\rangle$  & $I_{\rm 250}$ & $\langle$\textit{N}(\MH)$\rangle$  & \textit{N}(\MH)$_{\rm peak}$ & \textit{M} & $M_{\rm vir}$ & $\alpha_{\rm vir}$ & $N_{\rm *}$ & $N_{\rm *, \Delta C}$ & type\\ 
                &   &  & angle & size &  & &  &   &  &   &  &  & &  &  &  & \\ 
                ID & (J2000) & (J2000) & (\degr) & (ra\arcsec, dec\arcsec) & (pc) &  & (Jy) & (K) & (MJy\,sr$^{-1}$)  & (10$^{21}$\,cm$^{-2}$)  & (10$^{21}$\,cm$^{-2}$) & (\msun) & (\msun) & \\ 
                \hline
                \multicolumn{17}{l}{Field G014.14-00.55}\\ 
                \hline
                1 & -85.449 & -16.953 & 251.4 & (46.8,25.7) & 0.42 & 1.7 & 36.5 & - & - & - & 223.6 & 1582.95 & 23.6 & 0.0 & 0 & 0 & PRE\\ 
                2 & -85.449 & -16.838 & 201.2 & (21.8,18.8) & 0.24 & nan & 11.2 & - & - & - & 123.9 & 487.88 & 13.3 & 0.0 & 2 & - & PS\\ 
                3 & -85.454 & -16.877 & 188.6 & (24.9,15.3) & 0.23 & nan & 6.6 & - & - & - & 73.1 & 286.47 & 12.9 & 0.0 & 0 & 0 & PRE\\
\hline 
        \end{tabular}
                \tablefoot{$N_{\rm YSO, \Delta C}$ refers to number of YSOs confirmed by \textit{Herschel} color. Clumps are SL, PRE, candidate PS, or PS-C. }
\end{sidewaystable*}

\begin{sidewaystable*}
        \caption{Properties of field G014.15-00.55}
        \label{tbl:properties_G014.15-00.55}
        \centering
                \begin{tabular}{l|l l l l l l l l l l l l l l l l l}
                \hline\hline
                (1) & (2) & (3) & (4) & (5) & (6) & (7) & (8) & (9)  & (10)  & (11) & (12) & (13) & (14) & (15) & (16) & (17) & (18) \\ 
                Clump & RA & DEC & Position & Angular & $R_{\rm eff}$ & $\langle \beta\rangle$ & $F_{\rm 250}$ & $\langle T\rangle$  & $I_{\rm 250}$ & $\langle$\textit{N}(\MH)$\rangle$  & \textit{N}(\MH)$_{\rm peak}$ & \textit{M} & $M_{\rm vir}$ & $\alpha_{\rm vir}$ & $N_{\rm *}$ & $N_{\rm *, \Delta C}$ & type\\ 
                &   &  & angle & size &  & &  &   &  &   &  &  & &  &  &  & \\ 
                ID & (J2000) & (J2000) & (\degr) & (ra\arcsec, dec\arcsec) & (pc) &  & (Jy) & (K) & (MJy\,sr$^{-1}$)  & (10$^{21}$\,cm$^{-2}$)  & (10$^{21}$\,cm$^{-2}$) & (\msun) & (\msun) & \\ 
                \hline
                \multicolumn{17}{l}{Field G014.15-00.55}\\ 
                \hline
                1 & -85.448 & -16.832 & 190.1 & (28.6,19.2) & 0.27 & nan & 26.8 & - & - & - & 313.0 & 1164.17 & 15.4 & 0.0 & 1 & - & PS\\ 
                2 & -85.449 & -16.955 & 250.7 & (44.6,25.3) & 0.4 & 1.7 & 36.1 & - & - & - & 230.6 & 1568.01 & 22.8 & 0.0 & 1 & - & PS\\ 
                3 & -85.454 & -16.879 & 191.5 & (25.2,15.9) & 0.23 & nan & 6.0 & - & - & - & 74.3 & 259.27 & 13.2 & 0.1 & 0 & 0 & PRE\\
\hline 
        \end{tabular}
                \tablefoot{$N_{\rm YSO, \Delta C}$ refers to number of YSOs confirmed by \textit{Herschel} color. Clumps are SL, PRE, candidate PS, or PS-C. }
\end{sidewaystable*}

\begin{sidewaystable*}
        \caption{Properties of field G014.22-00.19}
        \label{tbl:properties_G014.22-00.19}
        \centering
                \begin{tabular}{l|l l l l l l l l l l l l l l l l l}
                \hline\hline
                (1) & (2) & (3) & (4) & (5) & (6) & (7) & (8) & (9)  & (10)  & (11) & (12) & (13) & (14) & (15) & (16) & (17) & (18) \\ 
                Clump & RA & DEC & Position & Angular & $R_{\rm eff}$ & $\langle \beta\rangle$ & $F_{\rm 250}$ & $\langle T\rangle$  & $I_{\rm 250}$ & $\langle$\textit{N}(\MH)$\rangle$  & \textit{N}(\MH)$_{\rm peak}$ & \textit{M} & $M_{\rm vir}$ & $\alpha_{\rm vir}$ & $N_{\rm *}$ & $N_{\rm *, \Delta C}$ & type\\ 
                &   &  & angle & size &  & &  &   &  &   &  &  & &  &  &  & \\ 
                ID & (J2000) & (J2000) & (\degr) & (ra\arcsec, dec\arcsec) & (pc) &  & (Jy) & (K) & (MJy\,sr$^{-1}$)  & (10$^{21}$\,cm$^{-2}$)  & (10$^{21}$\,cm$^{-2}$) & (\msun) & (\msun) & \\ 
                \hline
                \multicolumn{17}{l}{Field G014.22-00.19}\\ 
                \hline
                1 & -85.756 & -16.708 & 234.7 & (28.2,21.0) & - & 1.7 & 17.8 & 19.6177 & 3036.5 & 34.1 & 210.7 & - & - & - & 1 & - & PS\\ 
                2 & -85.75 & -16.693 & 251.7 & (30.9,12.4) & - & 1.6 & 10.6 & 18.4574 & 2621.0 & 35.2 & 121.0 & - & - & - & 0 & 0 & SL\\ 
                3 & -85.736 & -16.701 & 268.9 & (16.7,10.5) & - & 1.6 & 3.9 & 18.8398 & 2265.1 & 27.6 & 63.1 & - & - & - & 0 & 0 & SL\\
\hline 
        \end{tabular}
                \tablefoot{$N_{\rm YSO, \Delta C}$ refers to number of YSOs confirmed by \textit{Herschel} color. Clumps are SL, PRE, candidate PS, or PS-C. }
\end{sidewaystable*}

\begin{sidewaystable*}
        \caption{Properties of field G016.28-00.45}
        \label{tbl:properties_G016.28-00.45}
        \centering
                \begin{tabular}{l|l l l l l l l l l l l l l l l l l}
                \hline\hline
                (1) & (2) & (3) & (4) & (5) & (6) & (7) & (8) & (9)  & (10)  & (11) & (12) & (13) & (14) & (15) & (16) & (17) & (18) \\ 
                Clump & RA & DEC & Position & Angular & $R_{\rm eff}$ & $\langle \beta\rangle$ & $F_{\rm 250}$ & $\langle T\rangle$  & $I_{\rm 250}$ & $\langle$\textit{N}(\MH)$\rangle$  & \textit{N}(\MH)$_{\rm peak}$ & \textit{M} & $M_{\rm vir}$ & $\alpha_{\rm vir}$ & $N_{\rm *}$ & $N_{\rm *, \Delta C}$ & type\\ 
                &   &  & angle & size &  & &  &   &  &   &  &  & &  &  &  & \\ 
                ID & (J2000) & (J2000) & (\degr) & (ra\arcsec, dec\arcsec) & (pc) &  & (Jy) & (K) & (MJy\,sr$^{-1}$)  & (10$^{21}$\,cm$^{-2}$)  & (10$^{21}$\,cm$^{-2}$) & (\msun) & (\msun) & \\ 
                \hline
                \multicolumn{17}{l}{Field G016.28-00.45}\\ 
                \hline
                1 & -84.42 & -15.003 & 305.9 & (24.1,15.4) & - & 1.9 & 6.2 & - & - & - & 47.4 & - & - & - & 0 & 0 & SL\\ 
                2 & -84.457 & -14.971 & 246.2 & (21.2,15.2) & - & 1.8 & 2.1 & - & - & - & 26.7 & - & - & - & 1 & - & PS\\ 
                3 & -84.465 & -14.941 & 192.7 & (25.2,11.2) & - & 1.8 & 2.0 & - & - & - & 26.8 & - & - & - & 1 & - & PS\\
\hline 
        \end{tabular}
                \tablefoot{$N_{\rm YSO, \Delta C}$ refers to number of YSOs confirmed by \textit{Herschel} color. Clumps are starless (SL), prestellar (PRE), candidate protostellar (PS), or confirmed protostellar (PS-C). }
\end{sidewaystable*}

\begin{sidewaystable*}
        \caption{Properties of field G016.43-00.62}
        \label{tbl:properties_G016.43-00.62}
        \centering
                \begin{tabular}{l|l l l l l l l l l l l l l l l l l}
                \hline\hline
                (1) & (2) & (3) & (4) & (5) & (6) & (7) & (8) & (9)  & (10)  & (11) & (12) & (13) & (14) & (15) & (16) & (17) & (18) \\ 
                Clump & RA & DEC & Position & Angular & $R_{\rm eff}$ & $\langle \beta\rangle$ & $F_{\rm 250}$ & $\langle T\rangle$  & $I_{\rm 250}$ & $\langle$\textit{N}(\MH)$\rangle$  & \textit{N}(\MH)$_{\rm peak}$ & \textit{M} & $M_{\rm vir}$ & $\alpha_{\rm vir}$ & $N_{\rm *}$ & $N_{\rm *, \Delta C}$ & type\\ 
                &   &  & angle & size &  & &  &   &  &   &  &  & &  &  &  & \\ 
                ID & (J2000) & (J2000) & (\degr) & (ra\arcsec, dec\arcsec) & (pc) &  & (Jy) & (K) & (MJy\,sr$^{-1}$)  & (10$^{21}$\,cm$^{-2}$)  & (10$^{21}$\,cm$^{-2}$) & (\msun) & (\msun) & \\ 
                \hline
                \multicolumn{17}{l}{Field G016.43-00.62}\\ 
                \hline
                1 & -84.211 & -14.933 & 194.1 & (20.1,14.1) & 0.19 & 1.8 & 3.8 & - & - & - & 50.5 & 152.1 & 10.7 & 0.1 & 1 & - & PS\\ 
                2 & -84.257 & -14.945 & 191.4 & (32.2,20.5) & 0.29 & 1.8 & 5.6 & - & - & - & 42.8 & 224.38 & 16.3 & 0.1 & 0 & 0 & PRE\\ 
                3 & -84.272 & -14.959 & 317.8 & (30.8,20.6) & 0.29 & 1.8 & 6.3 & - & - & - & 31.3 & 249.23 & 16.5 & 0.1 & 2 & - & PS\\
\hline 
        \end{tabular}
                \tablefoot{$N_{\rm YSO, \Delta C}$ refers to number of YSOs confirmed by \textit{Herschel} color. Clumps are SL, PRE, candidate PS, or PS-C. }
\end{sidewaystable*}

\begin{sidewaystable*}
        \caption{Properties of field G017.23-01.47}
        \label{tbl:properties_G017.23-01.47}
        \centering
                \begin{tabular}{l|l l l l l l l l l l l l l l l l l}
                \hline\hline
                (1) & (2) & (3) & (4) & (5) & (6) & (7) & (8) & (9)  & (10)  & (11) & (12) & (13) & (14) & (15) & (16) & (17) & (18) \\ 
                Clump & RA & DEC & Position & Angular & $R_{\rm eff}$ & $\langle \beta\rangle$ & $F_{\rm 250}$ & $\langle T\rangle$  & $I_{\rm 250}$ & $\langle$\textit{N}(\MH)$\rangle$  & \textit{N}(\MH)$_{\rm peak}$ & \textit{M} & $M_{\rm vir}$ & $\alpha_{\rm vir}$ & $N_{\rm *}$ & $N_{\rm *, \Delta C}$ & type\\ 
                &   &  & angle & size &  & &  &   &  &   &  &  & &  &  &  & \\ 
                ID & (J2000) & (J2000) & (\degr) & (ra\arcsec, dec\arcsec) & (pc) &  & (Jy) & (K) & (MJy\,sr$^{-1}$)  & (10$^{21}$\,cm$^{-2}$)  & (10$^{21}$\,cm$^{-2}$) & (\msun) & (\msun) & \\ 
                \hline
                \multicolumn{17}{l}{Field G017.23-01.47}\\ 
                \hline
                1 & -83.123 & -14.63 & 317.6 & (20.8,16.5) & 0.33 & 2.4 & 4.3 & 16.6599 & 438.5 & 8.8 & 53.2 & 473.76 & 18.5 & 0.0 & 0 & 0 & PRE\\ 
                2\tablefootmark{a}\tablefootmark{b} & -83.12 & -14.648 & 194.7 & (27.2,13.4) & 0.34 & nan & 3.4 & - & - & - & 29.9 & 332.32 & 19.4 & 0.1 & 0 & 0 & PRE\\ 
                3 & -83.093 & -14.578 & 295.9 & (11.0,7.5) & 0.16 & nan & 0.8 & 17.4609 & 415.0 & 6.3 & 23.4 & 80.45 & 9.2 & 0.1 & 0 & 0 & PRE\\
\hline 
        \end{tabular}
                \tablefoot{$N_{\rm YSO, \Delta C}$ refers to number of YSOs confirmed by \textit{Herschel} color. Clumps are SL, PRE, candidate (PS), or PS-C. \\ \tablefoottext{a}{Denotes clumps that are located off \textit{Herschel} fields; the values derived with \textit{Herschel} data are erroneous. (\textit{T}, $I_{\rm 250}$, \textit{N}(\MH)) } \\ \tablefoottext{b}{The estimate for \textit{N}(\MH)$_{\rm peak}$ uses clump average temperature of 14\,K. } }
\end{sidewaystable*}

\begin{sidewaystable*}
        \caption{Properties of field G023.35-00.26}
        \label{tbl:properties_G023.35-00.26}
        \centering
\begin{tabular}{l|l l l l l l l l l l l l l l l l l}
                \hline\hline
                (1) & (2) & (3) & (4) & (5) & (6) & (7) & (8) & (9)  & (10)  & (11) & (12) & (13) & (14) & (15) & (16) & (17) & (18) \\ 
                Clump & RA & DEC & Position & Angular & $R_{\rm eff}$ & $\langle \beta\rangle$ & $F_{\rm 250}$ & $\langle T\rangle$  & $I_{\rm 250}$ & $\langle$\textit{N}(\MH)$\rangle$  & \textit{N}(\MH)$_{\rm peak}$ & \textit{M} & $M_{\rm vir}$ & $\alpha_{\rm vir}$ & $N_{\rm *}$ & $N_{\rm *, \Delta C}$ & type\\ 
                &   &  & angle & size &  & &  &   &  &   &  &  & &  &  &  & \\ 
                ID & (J2000) & (J2000) & (\degr) & (ra\arcsec, dec\arcsec) & (pc) &  & (Jy) & (K) & (MJy\,sr$^{-1}$)  & (10$^{21}$\,cm$^{-2}$)  & (10$^{21}$\,cm$^{-2}$) & (\msun) & (\msun) & \\ 
                \hline
                \multicolumn{17}{l}{Field G023.35-00.26}\\ 
                \hline
                1 & -81.3 & -8.562 & 326.3 & (19.4,12.7) & - & 1.8 & 14.4 & 26.2099 & 9101.8 & 42.0 & 98.7 & - & - & - & 0 & 0 & SL\\ 
                2 & -81.275 & -8.635 & 357.6 & (19.2,14.2) & - & 1.7 & 11.3 & 21.6586 & 4097.5 & 34.4 & 117.6 & - & - & - & 0 & 0 & SL\\ 
                3 & -81.31 & -8.571 & 329.4 & (25.9,16.8) & - & 1.8 & 21.0 & 24.7782 & 9747.6 & 51.9 & 94.8 & - & - & - & 1 & - & PS\\
\hline 
        \end{tabular}
                \tablefoot{$N_{\rm YSO, \Delta C}$ refers to number of YSOs confirmed by \textit{Herschel} color. Clumps are SL, PRE, candidate PS, or PS-C. }
\end{sidewaystable*}

\begin{sidewaystable*}
        \caption{Properties of field G023.69+00.58}
        \label{tbl:properties_G023.69+00.58}
        \centering
                \begin{tabular}{l|l l l l l l l l l l l l l l l l l}
                \hline\hline
                (1) & (2) & (3) & (4) & (5) & (6) & (7) & (8) & (9)  & (10)  & (11) & (12) & (13) & (14) & (15) & (16) & (17) & (18) \\ 
                Clump & RA & DEC & Position & Angular & $R_{\rm eff}$ & $\langle \beta\rangle$ & $F_{\rm 250}$ & $\langle T\rangle$  & $I_{\rm 250}$ & $\langle$\textit{N}(\MH)$\rangle$  & \textit{N}(\MH)$_{\rm peak}$ & \textit{M} & $M_{\rm vir}$ & $\alpha_{\rm vir}$ & $N_{\rm *}$ & $N_{\rm *, \Delta C}$ & type\\ 
                &   &  & angle & size &  & &  &   &  &   &  &  & &  &  &  & \\ 
                ID & (J2000) & (J2000) & (\degr) & (ra\arcsec, dec\arcsec) & (pc) &  & (Jy) & (K) & (MJy\,sr$^{-1}$)  & (10$^{21}$\,cm$^{-2}$)  & (10$^{21}$\,cm$^{-2}$) & (\msun) & (\msun) & \\ 
                \hline
                \multicolumn{17}{l}{Field G023.69+00.58}\\ 
                \hline
                1 & -81.912 & -7.949 & 232.1 & (31.1,18.4) & 0.1 & 1.9 & 7.7 & 20.6192 & 1730.7 & 16.0 & 68.2 & 30.09 & 5.6 & 0.2 & 2 & - & PS\\ 
                2 & -81.937 & -7.995 & 195.7 & (47.8,25.9) & 0.14 & 1.8 & 12.1 & 20.7561 & 1600.4 & 15.1 & 25.2 & 47.28 & 7.9 & 0.2 & 1 & - & PS\\ 
                3 & -81.915 & -7.969 & 268.9 & (19.7,13.2) & 0.06 & 1.6 & 2.6 & 21.5104 & 1564.5 & 12.4 & 20.2 & 9.57 & 3.5 & 0.4 & 1 & - & PS\\
\hline 
        \end{tabular}
                \tablefoot{$N_{\rm YSO, \Delta C}$ refers to number of YSOs confirmed by \textit{Herschel} color. Clumps are SL, PRE, candidate PS, or PS-C. }
\end{sidewaystable*}

\begin{sidewaystable*}
        \caption{Properties of field G033.73-00.01}
        \label{tbl:properties_G033.73-00.01}
        \centering
                \begin{tabular}{l|l l l l l l l l l l l l l l l l l}
                \hline\hline
                (1) & (2) & (3) & (4) & (5) & (6) & (7) & (8) & (9)  & (10)  & (11) & (12) & (13) & (14) & (15) & (16) & (17) & (18) \\ 
                Clump & RA & DEC & Position & Angular & $R_{\rm eff}$ & $\langle \beta\rangle$ & $F_{\rm 250}$ & $\langle T\rangle$  & $I_{\rm 250}$ & $\langle$\textit{N}(\MH)$\rangle$  & \textit{N}(\MH)$_{\rm peak}$ & \textit{M} & $M_{\rm vir}$ & $\alpha_{\rm vir}$ & $N_{\rm *}$ & $N_{\rm *, \Delta C}$ & type\\ 
                &   &  & angle & size &  & &  &   &  &   &  &  & &  &  &  & \\ 
                ID & (J2000) & (J2000) & (\degr) & (ra\arcsec, dec\arcsec) & (pc) &  & (Jy) & (K) & (MJy\,sr$^{-1}$)  & (10$^{21}$\,cm$^{-2}$)  & (10$^{21}$\,cm$^{-2}$) & (\msun) & (\msun) & \\ 
                \hline
                \multicolumn{17}{l}{Field G033.73-00.01}\\ 
                \hline
                1 & -76.756 & 0.71 & 253.7 & (17.2,14.4) & 0.32 & 1.6 & 8.1 & 19.1302 & 1952.9 & 25.1 & 110.4 & 991.66 & 18.2 & 0.0 & 2 & 1 & PS-C\\ 
                2 & -76.794 & 0.639 & 359.5 & (19.8,14.2) & 0.34 & 1.8 & 3.8 & 19.7245 & 2206.5 & 22.1 & 66.7 & 446.77 & 19.4 & 0.0 & 2 & 2 & PS-C\\ 
                3 & -76.786 & 0.628 & 251.8 & (21.2,15.1) & 0.37 & 1.8 & 7.6 & 18.0211 & 2388.1 & 32.7 & 71.4 & 1014.71 & 20.9 & 0.0 & 1 & 1 & PS-C\\
\hline 
        \end{tabular}
                \tablefoot{$N_{\rm YSO, \Delta C}$ refers to number of YSOs confirmed by \textit{Herschel} color. Clumps are SL, PRE, candidate PS, or PS-C. }
\end{sidewaystable*}

\begin{sidewaystable*}
        \caption{Properties of field G035.38-01.77}
        \label{tbl:properties_G035.38-01.77}
        \centering
                \begin{tabular}{l|l l l l l l l l l l l l l l l l l}
                \hline\hline
                (1) & (2) & (3) & (4) & (5) & (6) & (7) & (8) & (9)  & (10)  & (11) & (12) & (13) & (14) & (15) & (16) & (17) & (18) \\ 
                Clump & RA & DEC & Position & Angular & $R_{\rm eff}$ & $\langle \beta\rangle$ & $F_{\rm 250}$ & $\langle T\rangle$  & $I_{\rm 250}$ & $\langle$\textit{N}(\MH)$\rangle$  & \textit{N}(\MH)$_{\rm peak}$ & \textit{M} & $M_{\rm vir}$ & $\alpha_{\rm vir}$ & $N_{\rm *}$ & $N_{\rm *, \Delta C}$ & type\\ 
                &   &  & angle & size &  & &  &   &  &   &  &  & &  &  &  & \\ 
                ID & (J2000) & (J2000) & (\degr) & (ra\arcsec, dec\arcsec) & (pc) &  & (Jy) & (K) & (MJy\,sr$^{-1}$)  & (10$^{21}$\,cm$^{-2}$)  & (10$^{21}$\,cm$^{-2}$) & (\msun) & (\msun) & \\ 
                \hline
                \multicolumn{17}{l}{Field G035.38-01.77}\\ 
                \hline
                1 & -74.456 & 1.351 & 302.0 & (13.1,11.2) & 0.25 & nan & 2.7 & - & - & - & 66.7 & 365.97 & 14.1 & 0.0 & 2 & - & PS\\ 
                2 & -74.447 & 1.349 & 196.0 & (13.9,9.7) & 0.24 & nan & 2.1 & - & - & - & 46.4 & 283.75 & 13.5 & 0.1 & 0 & 0 & PRE\\ 
                3 & -74.451 & 1.359 & 279.4 & (15.6,9.3) & 0.25 & nan & 2.0 & - & - & - & 41.5 & 272.77 & 14.0 & 0.1 & 0 & 0 & PRE\\
\hline 
        \end{tabular}
                \tablefoot{$N_{\rm YSO, \Delta C}$ refers to number of YSOs confirmed by \textit{Herschel} color. Clumps are SL, PRE, candidate PS, or PS-C. }
\end{sidewaystable*}

\begin{sidewaystable*}
        \caption{Properties of field G037.93+02.18}
        \label{tbl:properties_G037.93+02.18}
        \centering
                \begin{tabular}{l|l l l l l l l l l l l l l l l l l}
                \hline\hline
                (1) & (2) & (3) & (4) & (5) & (6) & (7) & (8) & (9)  & (10)  & (11) & (12) & (13) & (14) & (15) & (16) & (17) & (18) \\ 
                Clump & RA & DEC & Position & Angular & $R_{\rm eff}$ & $\langle \beta\rangle$ & $F_{\rm 250}$ & $\langle T\rangle$  & $I_{\rm 250}$ & $\langle$\textit{N}(\MH)$\rangle$  & \textit{N}(\MH)$_{\rm peak}$ & \textit{M} & $M_{\rm vir}$ & $\alpha_{\rm vir}$ & $N_{\rm *}$ & $N_{\rm *, \Delta C}$ & type\\ 
                &   &  & angle & size &  & &  &   &  &   &  &  & &  &  &  & \\ 
                ID & (J2000) & (J2000) & (\degr) & (ra\arcsec, dec\arcsec) & (pc) &  & (Jy) & (K) & (MJy\,sr$^{-1}$)  & (10$^{21}$\,cm$^{-2}$)  & (10$^{21}$\,cm$^{-2}$) & (\msun) & (\msun) & \\ 
                \hline
                \multicolumn{17}{l}{Field G037.93+02.18}\\ 
                \hline
                1 & -76.777 & 5.419 & 209.8 & (19.6,13.1) & 0.09 & 1.7 & 1.0 & 15.394 & 321.9 & 7.8 & 23.2 & 10.75 & 4.8 & 0.5 & 1 & 1 & PS-C\\ 
                2 & -76.79 & 5.402 & 207.0 & (15.4,4.9) & 0.06 & 1.7 & 0.2 & 15.4302 & 286.8 & 6.8 & 8.7 & 2.24 & 3.1 & 1.4 & 1 & 1 & PS-C\\ 
                3 & -76.803 & 5.367 & 330.5 & (17.2,7.9) & 0.07 & 1.8 & 0.2 & 16.0973 & 296.4 & 6.0 & 6.8 & 2.32 & 3.8 & 1.6 & 0 & 0 & SL\\
\hline 
        \end{tabular}
                \tablefoot{$N_{\rm YSO, \Delta C}$ refers to number of YSOs confirmed by \textit{Herschel} color. Clumps are SL, PRE, PS, or PS-C. }
\end{sidewaystable*}

\begin{sidewaystable*}
        \caption{Properties of field G082.40-01.84}
        \label{tbl:properties_G082.40-01.84}
        \centering
                \begin{tabular}{l|l l l l l l l l l l l l l l l l l}
                \hline\hline
                (1) & (2) & (3) & (4) & (5) & (6) & (7) & (8) & (9)  & (10)  & (11) & (12) & (13) & (14) & (15) & (16) & (17) & (18) \\ 
                Clump & RA & DEC & Position & Angular & $R_{\rm eff}$ & $\langle \beta\rangle$ & $F_{\rm 250}$ & $\langle T\rangle$  & $I_{\rm 250}$ & $\langle$\textit{N}(\MH)$\rangle$  & \textit{N}(\MH)$_{\rm peak}$ & \textit{M} & $M_{\rm vir}$ & $\alpha_{\rm vir}$ & $N_{\rm *}$ & $N_{\rm *, \Delta C}$ & type\\ 
                &   &  & angle & size &  & &  &   &  &   &  &  & &  &  &  & \\ 
                ID & (J2000) & (J2000) & (\degr) & (ra\arcsec, dec\arcsec) & (pc) &  & (Jy) & (K) & (MJy\,sr$^{-1}$)  & (10$^{21}$\,cm$^{-2}$)  & (10$^{21}$\,cm$^{-2}$) & (\msun) & (\msun) & \\ 
                \hline
                \multicolumn{17}{l}{Field G082.40-01.84}\\ 
                \hline
                1 & -47.15 & 41.411 & 211.5 & (24.3,12.8) & 0.09 & 1.8 & 1.9 & 12.9277 & 318.7 & 16.1 & 35.9 & 24.81 & 5.3 & 0.2 & 1 & - & PS\\ 
                2 & -47.133 & 41.423 & 286.7 & (23.5,11.8) & 0.08 & 1.7 & 1.7 & 12.8341 & 284.8 & 16.0 & 36.2 & 22.77 & 4.7 & 0.2 & 0 & 0 & PRE\\ 
                3 & -47.181 & 41.383 & 283.8 & (26.0,11.6) & 0.09 & 1.8 & 1.1 & 12.8866 & 265.7 & 13.5 & 31.5 & 13.81 & 4.9 & 0.4 & 0 & 0 & PRE\\
\hline 
        \end{tabular}
                \tablefoot{$N_{\rm YSO, \Delta C}$ refers to number of YSOs confirmed by \textit{Herschel} color. Clumps are SL, PRE, candidate PS, or PS-C. }
\end{sidewaystable*}

\begin{sidewaystable*}
        \caption{Properties of field G092.04+03.92}
        \label{tbl:properties_G092.04+03.92}
        \centering
                \begin{tabular}{l|l l l l l l l l l l l l l l l l l}
                \hline\hline
                (1) & (2) & (3) & (4) & (5) & (6) & (7) & (8) & (9)  & (10)  & (11) & (12) & (13) & (14) & (15) & (16) & (17) & (18) \\ 
                Clump & RA & DEC & Position & Angular & $R_{\rm eff}$ & $\langle \beta\rangle$ & $F_{\rm 250}$ & $\langle T\rangle$  & $I_{\rm 250}$ & $\langle$\textit{N}(\MH)$\rangle$  & \textit{N}(\MH)$_{\rm peak}$ & \textit{M} & $M_{\rm vir}$ & $\alpha_{\rm vir}$ & $N_{\rm *}$ & $N_{\rm *, \Delta C}$ & type\\ 
                &   &  & angle & size &  & &  &   &  &   &  &  & &  &  &  & \\ 
                ID & (J2000) & (J2000) & (\degr) & (ra\arcsec, dec\arcsec) & (pc) &  & (Jy) & (K) & (MJy\,sr$^{-1}$)  & (10$^{21}$\,cm$^{-2}$)  & (10$^{21}$\,cm$^{-2}$) & (\msun) & (\msun) & \\ 
                \hline
                \multicolumn{17}{l}{Field G092.04+03.92}\\ 
                \hline
                1 & -44.478 & 52.481 & 273.5 & (11.7,7.1) & 0.04 & 1.7 & 0.5 & 14.1199 & 396.3 & 13.5 & 26.4 & 3.49 & 2.0 & 0.6 & 1 & - & PS\\ 
                2 & -44.404 & 52.478 & 342.3 & (32.5,21.0) & 0.1 & 1.8 & 3.4 & 12.4435 & 420.5 & 25.6 & 28.1 & 30.34 & 5.8 & 0.2 & 1 & - & PS\\ 
                3 & -44.5 & 52.482 & 284.0 & (11.6,6.9) & 0.04 & 1.8 & 0.4 & 13.5341 & 311.7 & 12.6 & 20.5 & 2.74 & 2.0 & 0.7 & 1 & - & PS\\
\hline 
        \end{tabular}
                \tablefoot{$N_{\rm YSO, \Delta C}$ refers to number of YSOs confirmed by \textit{Herschel} color. Clumps are SL, PRE, candidate PS, or PS-C. \\ \tablefoottext{a}{Denotes clumps that are located off \textit{Herschel} fields; the values derived with \textit{Herschel} data are erroneous. (\textit{T}, $I_{\rm 250}$, \textit{N}(\MH)) } \\ \tablefoottext{b}{The estimate for \textit{N}(\MH)$_{\rm peak}$ uses clump average temperature of 14\,K. } }
\end{sidewaystable*}

\begin{sidewaystable*}
        \caption{Properties of field G107.18+05.44}
        \label{tbl:properties_G107.18+05.44}
        \centering
                \begin{tabular}{l|l l l l l l l l l l l l l l l l l}
                \hline\hline
                (1) & (2) & (3) & (4) & (5) & (6) & (7) & (8) & (9)  & (10)  & (11) & (12) & (13) & (14) & (15) & (16) & (17) & (18) \\ 
                Clump & RA & DEC & Position & Angular & $R_{\rm eff}$ & $\langle \beta\rangle$ & $F_{\rm 250}$ & $\langle T\rangle$  & $I_{\rm 250}$ & $\langle$\textit{N}(\MH)$\rangle$  & \textit{N}(\MH)$_{\rm peak}$ & \textit{M} & $M_{\rm vir}$ & $\alpha_{\rm vir}$ & $N_{\rm *}$ & $N_{\rm *, \Delta C}$ & type\\ 
                &   &  & angle & size &  & &  &   &  &   &  &  & &  &  &  & \\ 
                ID & (J2000) & (J2000) & (\degr) & (ra\arcsec, dec\arcsec) & (pc) &  & (Jy) & (K) & (MJy\,sr$^{-1}$)  & (10$^{21}$\,cm$^{-2}$)  & (10$^{21}$\,cm$^{-2}$) & (\msun) & (\msun) & \\ 
                \hline
                \multicolumn{17}{l}{Field G107.18+05.44}\\ 
                \hline
                1 & -24.607 & 63.62 & 311.1 & (24.0,17.0) & 0.08 & 1.6 & 6.0 & 15.4136 & 550.7 & 12.7 & 83.0 & 37.35 & 4.5 & 0.1 & 2 & - & PS\\ 
                2 & -24.655 & 63.626 & 181.6 & (12.4,9.5) & 0.04 & 1.7 & 1.1 & 15.3403 & 483.7 & 11.7 & 37.5 & 6.84 & 2.4 & 0.3 & 0 & 0 & PRE\\ 
                3 & -24.679 & 63.629 & 231.0 & (11.6,11.4) & 0.04 & 1.4 & 1.2 & 14.7195 & 365.0 & 10.6 & 35.0 & 8.33 & 2.5 & 0.3 & 0 & 0 & PRE\\ 
                4 & -24.695 & 63.647 & 283.6 & (24.3,20.8) & 0.09 & 1.8 & 2.7 & 15.564 & 454.8 & 10.6 & 18.2 & 16.23 & 4.9 & 0.3 & 0 & 0 & PRE\\ 
                5 & -24.667 & 63.59 & 219.8 & (9.5,6.2) & 0.03 & 2.0 & 0.3 & 15.4366 & 316.6 & 7.5 & 13.2 & 1.73 & 1.8 & 1.0 & 1 & - & PS\\ 
                \hline
                \multicolumn{17}{l}{Field G107.26+05.71}\\ 
                \hline
                1 & -24.645 & 63.86 & 355.3 & (23.3,18.3) & 0.08 & 1.6 & 12.3 & 17.7039 & 1193.3 & 14.2 & 234.4 & 60.91 & 4.5 & 0.1 & 0 & 0 & PRE\\ 
                2 & -24.653 & 63.879 & 299.0 & (12.3,7.0) & 0.04 & 1.7 & 0.8 & 15.7338 & 577.4 & 12.6 & 36.4 & 4.91 & 2.1 & 0.4 & 0 & 0 & PRE\\ 
                3 & -24.687 & 63.878 & 282.2 & (15.4,10.7) & 0.05 & 1.7 & 0.8 & 15.0939 & 437.0 & 11.3 & 16.2 & 5.06 & 2.8 & 0.6 & 0 & 0 & PRE\\ 
                4 & -24.719 & 63.885 & 300.7 & (14.9,8.6) & 0.05 & 1.8 & 0.5 & 15.9557 & 327.4 & 6.9 & 12.1 & 3.18 & 2.6 & 0.8 & 0 & 0 & PRE\\ 
                5 & -24.783 & 63.857 & 230.2 & (24.7,11.9) & 0.07 & 1.8 & 0.9 & 15.9058 & 201.1 & 4.3 & 10.8 & 5.47 & 4.2 & 0.8 & 0 & 0 & PRE\\ 
                \hline
                \multicolumn{17}{l}{Field G109.70+02.52}\\ 
                \hline
                1 & -16.548 & 62.404 & 352.9 & (10.2,8.1) & 0.04 & 1.4 & 0.6 & 20.9461 & 628.6 & 5.3 & 16.8 & 2.2 & 2.0 & 0.9 & 0 & 0 & PRE\\ 
                2 & -16.645 & 62.305 & 224.5 & (7.0,4.7) & 0.02 & 1.6 & 0.3 & 15.7442 & 547.7 & 12.2 & 26.2 & 1.94 & 1.3 & 0.7 & 0 & 0 & PRE\\ 
                3 & -16.704 & 62.288 & 225.7 & (7.3,4.7) & 0.02 & 1.7 & 0.3 & 17.1763 & 498.7 & 8.1 & 18.1 & 1.46 & 1.4 & 0.9 & 1 & - & PS\\ 
                4 & -16.496 & 62.35 & 191.3 & (8.9,5.8) & 0.03 & 1.4 & 0.2 & 20.109 & 703.5 & 6.8 & 9.3 & 0.86 & 1.6 & 1.8 & 0 & 0 & SL\\ 
                5 & -16.689 & 62.37 & 313.9 & (22.4,5.2) & 0.06 & 1.6 & 0.4 & 16.078 & 415.3 & 8.5 & 10.9 & 2.31 & 3.6 & 1.5 & 0 & 0 & SL\\ 
                \hline
                \multicolumn{17}{l}{Field G109.81+02.72}\\ 
                \hline
                1 & -16.577 & 62.53 & 312.8 & (15.1,12.8) & 0.05 & 1.9 & 2.4 & 14.4664 & 541.8 & 15.9 & 48.5 & 16.21 & 3.0 & 0.2 & 4 & 4 & PS-C\\ 
                2 & -16.627 & 62.545 & 194.9 & (18.3,16.2) & 0.07 & 1.8 & 1.2 & 14.3503 & 323.4 & 10.3 & 15.5 & 8.39 & 3.8 & 0.5 & 2 & 2 & PS-C\\ 
                \hline
                \multicolumn{17}{l}{Field G139.14-03.23}\\ 
                \hline
                1 & 42.595 & 55.858 & 304.3 & (21.4,11.6) & 0.21 & 1.4 & 1.9 & 15.0279 & 300.1 & 7.7 & 40.0 & 122.39 & 11.6 & 0.1 & 1 & 1 & PS-C\\ 
                2 & 42.586 & 55.848 & 307.8 & (20.3,11.7) & 0.2 & 1.3 & 1.9 & 15.1504 & 392.0 & 10.4 & 25.3 & 119.29 & 11.3 & 0.1 & 0 & 0 & PRE\\ 
                \hline
                \multicolumn{17}{l}{Field G159.23-20.09}\\ 
                \hline
                1 & 53.326 & 31.158 & 308.7 & (25.8,16.5) & 0.08 & 1.6 & 7.3 & 12.4203 & 556.4 & 31.0 & 250.7 & 65.31 & 4.7 & 0.1 & 1 & 1 & PS-C\\ 
                2 & 53.337 & 31.127 & 187.7 & (27.7,17.5) & 0.09 & 1.7 & 10.6 & 11.4716 & 510.9 & 47.9 & 254.4 & 110.17 & 4.8 & 0.0 & 3 & - & PS\\ 
                3 & 53.308 & 31.117 & 276.2 & (34.9,18.2) & 0.1 & 1.7 & 7.1 & 11.9293 & 371.9 & 28.0 & 80.3 & 68.57 & 5.6 & 0.1 & 2 & 2 & PS-C\\ 
                4 & 53.317 & 31.131 & 276.2 & (12.6,9.3) & 0.04 & 1.6 & 1.2 & 12.4728 & 551.0 & 31.8 & 43.1 & 11.1 & 2.4 & 0.2 & 1 & 1 & PS-C\\ 
                5 & 53.361 & 31.114 & 184.3 & (18.4,8.9) & 0.05 & 1.7 & 0.7 & 13.2261 & 427.2 & 19.3 & 23.8 & 5.68 & 2.8 & 0.5 & 0 & 0 & PRE\\ 
                6 & 53.283 & 31.086 & 235.2 & (22.4,17.8) & 0.08 & 1.7 & 1.7 & 12.1367 & 311.4 & 22.0 & 25.2 & 16.15 & 4.4 & 0.3 & 0 & 0 & PRE\\ 
                7 & 53.27 & 31.111 & 245.0 & (12.5,9.2) & 0.04 & 1.6 & 0.5 & 11.8015 & 244.0 & 18.8 & 23.6 & 5.02 & 2.4 & 0.5 & 0 & 0 & PRE\\ 
                8 & 53.354 & 31.096 & 198.0 & (14.7,7.1) & 0.04 & 1.6 & 0.4 & 12.5835 & 308.4 & 17.4 & 17.2 & 3.72 & 2.4 & 0.6 & 0 & 0 & PRE\\ 
                9 & 53.294 & 31.15 & 245.2 & (21.4,3.3) & 0.05 & 1.6 & 0.2 & 12.5056 & 273.0 & 15.8 & 12.1 & 1.85 & 2.8 & 1.5 & 0 & 0 & SL\\ 
                \hline 
        \end{tabular}
                \tablefoot{$N_{\rm YSO, \Delta C}$ refers to number of YSOs confirmed by \textit{Herschel} color. Clumps are SL, PRE, candidate PS, or PS-C. }
\end{sidewaystable*}

\begin{sidewaystable*}
        \caption{Properties of field G162.46-08.69}
        \label{tbl:properties_G162.46-08.69}
        \centering
                \begin{tabular}{l|l l l l l l l l l l l l l l l l l}
                \hline\hline
                (1) & (2) & (3) & (4) & (5) & (6) & (7) & (8) & (9)  & (10)  & (11) & (12) & (13) & (14) & (15) & (16) & (17) & (18) \\ 
                Clump & RA & DEC & Position & Angular & $R_{\rm eff}$ & $\langle \beta\rangle$ & $F_{\rm 250}$ & $\langle T\rangle$  & $I_{\rm 250}$ & $\langle$\textit{N}(\MH)$\rangle$  & \textit{N}(\MH)$_{\rm peak}$ & \textit{M} & $M_{\rm vir}$ & $\alpha_{\rm vir}$ & $N_{\rm *}$ & $N_{\rm *, \Delta C}$ & type\\ 
                &   &  & angle & size &  & &  &   &  &   &  &  & &  &  &  & \\ 
                ID & (J2000) & (J2000) & (\degr) & (ra\arcsec, dec\arcsec) & (pc) &  & (Jy) & (K) & (MJy\,sr$^{-1}$)  & (10$^{21}$\,cm$^{-2}$)  & (10$^{21}$\,cm$^{-2}$) & (\msun) & (\msun) & \\ 
                \hline
                \multicolumn{17}{l}{Field G162.46-08.69}\\ 
                \hline
                1 & 65.406 & 37.58 & 349.6 & (23.9,14.2) & 0.04 & 1.8 & 2.3 & 13.8594 & 314.2 & 10.8 & 40.9 & 5.34 & 2.3 & 0.4 & 0 & 0 & PRE\\ 
                2 & 65.424 & 37.563 & 298.3 & (32.9,11.4) & 0.05 & 1.8 & 1.9 & 12.0374 & 207.8 & 14.3 & 37.0 & 5.82 & 2.9 & 0.5 & 0 & 0 & PRE\\ 
                3 & 65.309 & 37.626 & 186.2 & (13.9,7.4) & 0.02 & 1.8 & 0.4 & 12.459 & 129.6 & 7.7 & 21.3 & 1.18 & 1.2 & 1.1 & 0 & 0 & SL\\ 
                4 & 65.322 & 37.559 & 194.2 & (21.1,8.9) & 0.03 & 1.9 & 0.7 & 11.9219 & 162.2 & 11.9 & 19.5 & 2.16 & 1.8 & 0.8 & 0 & 0 & PRE\\ 
                5 & 65.308 & 37.61 & 342.7 & (14.7,8.3) & 0.03 & 1.6 & 0.4 & 12.2831 & 117.7 & 7.5 & 15.4 & 1.22 & 1.4 & 1.2 & 0 & 0 & SL\\ 
                \hline
                \multicolumn{17}{l}{Field G171.53-14.91}\\ 
                \hline
                1 & 67.164 & 26.861 & 351.0 & (22.3,17.4) & - & 1.9 & 2.3 & 11.6226 & 242.2 & 20.3 & 39.9 & - & - & - & 1 & - & PS\\ 
                2 & 67.27 & 26.819 & 278.1 & (4.4,3.7) & - & 1.7 & 0.2 & 14.1608 & 94.4 & 3.1 & 33.4 & - & - & - & 0 & 0 & SL\\ 
                \hline
                \multicolumn{17}{l}{Field G172.89+02.28}\\ 
                \hline
                1 & 84.227 & 36.17 & 341.7 & (27.4,16.0) & - & - & 5.0 & - & - & - & 45.0 & - & - & - & 0 & - & SL\\ 
                2 & 84.214 & 36.18 & 210.2 & (22.1,12.5) & - & - & 4.2 & - & - & - & 45.3 & - & - & - & 0 & - & SL\\ 
                3 & 84.321 & 36.193 & 228.9 & (501.0,3.5) & - & - & 0.9 & - & - & - & 9.5 & - & - & - & 0 & - & SL\\ 
                4 & 84.192 & 36.179 & 302.2 & (8.9,6.6) & - & - & 0.2 & - & - & - & 7.1 & - & - & - & 0 & - & SL\\ 
                5 & 84.3 & 36.185 & 218.5 & (4.0,1.6) & - & - & 0.0 & - & - & - & 12.2 & - & - & - & 0 & - & SL\\ 
                6 & 84.223 & 36.214 & 182.4 & (7.3,5.0) & - & - & 0.1 & - & - & - & 7.1 & - & - & - & 0 & - & SL\\ 
                \hline
                \multicolumn{17}{l}{Field G173.15+02.40}\\ 
                \hline
                1 & 84.507 & 35.971 & 261.2 & (15.2,10.3) & 0.12 & - & 4.2 & - & - & - & 84.2 & 125.5 & 6.8 & 0.1 & 1 & - & PS\\ 
                2 & 84.504 & 35.986 & 220.5 & (31.5,25.1) & 0.28 & - & 10.4 & - & - & - & 73.6 & 313.55 & 15.6 & 0.1 & 1 & - & PS\\ 
                3 & 84.505 & 35.959 & 321.3 & (20.2,16.4) & 0.18 & - & 3.7 & - & - & - & 42.6 & 111.61 & 10.1 & 0.1 & 2 & - & PS\\ 
                4 & 84.519 & 36.01 & 337.5 & (11.8,8.9) & 0.1 & - & 1.0 & - & - & - & 22.9 & 29.38 & 5.7 & 0.2 & 1 & - & PS\\ 
                5 & 84.513 & 36.021 & 248.5 & (15.5,8.8) & 0.12 & - & 0.9 & - & - & - & 19.0 & 27.08 & 6.6 & 0.2 & 1 & - & PS\\ 
                6 & 84.487 & 36.013 & 309.9 & (23.7,10.0) & 0.18 & - & 1.1 & - & - & - & 11.4 & 34.64 & 9.9 & 0.3 & 1 & - & PS\\ 
                7 & 84.551 & 35.941 & 185.2 & (12.8,10.0) & 0.11 & - & 0.5 & - & - & - & 10.3 & 13.81 & 6.2 & 0.5 & 0 & - & PRE\\ 
                8 & 84.464 & 36.027 & 333.5 & (9.4,6.4) & 0.08 & - & 0.3 & - & - & - & 10.7 & 9.2 & 4.3 & 0.5 & 0 & - & PRE\\ 
                9 & 84.514 & 35.934 & 214.8 & (4.2,3.1) & 0.04 & - & 0.1 & - & - & - & 15.5 & 2.98 & 2.0 & 0.7 & 0 & - & PRE\\ 
                10 & 84.49 & 36.033 & 199.1 & (8.0,4.7) & 0.06 & - & 0.2 & - & - & - & 10.5 & 5.65 & 3.4 & 0.6 & 0 & - & PRE\\ 
                \hline
                \multicolumn{17}{l}{Field G178.28-00.60}\\ 
                \hline
                1 & 84.775 & 30.093 & 346.3 & (18.0,10.0) & - & 1.5 & 1.2 & 12.9794 & 251.3 & 12.4 & 34.6 & - & - & - & 1 & - & PS\\ 
                2 & 84.783 & 30.08 & 267.4 & (13.3,11.4) & - & 1.5 & 0.9 & 13.2726 & 277.2 & 12.2 & 27.0 & - & - & - & 2 & 1 & PS-C\\ 
                3 & 84.785 & 30.063 & 328.4 & (13.4,9.4) & - & 1.6 & 0.8 & 13.0667 & 252.7 & 12.0 & 26.0 & - & - & - & 1 & - & PS\\ 
                4 & 84.778 & 30.113 & 221.2 & (363.3,2.6) & - & 1.4 & 0.7 & 13.1613 & 228.5 & 10.5 & 24.8 & - & - & - & 0 & 0 & SL\\ 
                5 & 84.826 & 30.141 & 281.8 & (13.0,6.6) & - & 1.7 & 0.4 & 13.7319 & 226.5 & 8.6 & 21.1 & - & - & - & 0 & 0 & SL\\ 
                6 & 84.808 & 30.058 & 195.7 & (14.0,5.5) & - & 1.5 & 0.3 & 14.6594 & 251.3 & 7.3 & 17.9 & - & - & - & 1 & - & PS\\ 
                7 & 84.774 & 30.052 & 243.6 & (17.7,7.9) & - & 1.6 & 0.6 & 14.0983 & 233.3 & 8.1 & 14.3 & - & - & - & 0 & 0 & SL\\ 
                \hline 
        \end{tabular}
                \tablefoot{$N_{\rm YSO, \Delta C}$ refers to number of YSOs confirmed by \textit{Herschel} color. Clumps are starless (SL), prestellar (PRE), candidate protostellar (PS), or confirmed protostellar (PS-C). }
\end{sidewaystable*}

\begin{sidewaystable*}
        \caption{Properties of field G195.73-02.28}
        \label{tbl:properties_G195.73-02.28}
        \centering
                \begin{tabular}{l|l l l l l l l l l l l l l l l l l}
                \hline\hline
                (1) & (2) & (3) & (4) & (5) & (6) & (7) & (8) & (9)  & (10)  & (11) & (12) & (13) & (14) & (15) & (16) & (17) & (18) \\ 
                Clump & RA & DEC & Position & Angular & $R_{\rm eff}$ & $\langle \beta\rangle$ & $F_{\rm 250}$ & $\langle T\rangle$  & $I_{\rm 250}$ & $\langle$\textit{N}(\MH)$\rangle$  & \textit{N}(\MH)$_{\rm peak}$ & \textit{M} & $M_{\rm vir}$ & $\alpha_{\rm vir}$ & $N_{\rm *}$ & $N_{\rm *, \Delta C}$ & type\\ 
                &   &  & angle & size &  & &  &   &  &   &  &  & &  &  &  & \\ 
                ID & (J2000) & (J2000) & (\degr) & (ra\arcsec, dec\arcsec) & (pc) &  & (Jy) & (K) & (MJy\,sr$^{-1}$)  & (10$^{21}$\,cm$^{-2}$)  & (10$^{21}$\,cm$^{-2}$) & (\msun) & (\msun) & \\ 
                \hline
                \multicolumn{17}{l}{Field G195.73-02.28}\\ 
                \hline
                1 & 92.742 & 14.157 & 290.5 & (18.7,17.8) & 0.09 & 1.8 & 4.0 & 14.1334 & 505.7 & 15.6 & 87.2 & 44.59 & 5.0 & 0.1 & 2 & 1 & PS-C\\ 
                2 & 92.712 & 14.171 & 298.7 & (25.4,11.5) & 0.09 & 1.7 & 2.5 & 13.4369 & 321.1 & 13.5 & 60.4 & 30.71 & 5.2 & 0.2 & 1 & 1 & PS-C\\ 
                3 & 92.754 & 14.15 & 336.7 & (9.4,6.9) & 0.04 & 1.7 & 0.5 & 12.9199 & 226.2 & 11.3 & 29.8 & 6.48 & 2.2 & 0.3 & 0 & 0 & PRE\\ 
                \hline
                \multicolumn{17}{l}{Field G202.31+02.53}\\ 
                \hline
                1 & 100.243 & 10.615 & 186.9 & (13.7,11.3) & 0.05 & 1.4 & 4.3 & 15.9279 & 929.1 & 17.3 & 178.5 & 22.86 & 2.6 & 0.1 & 2 & 2 & PS-C\\ 
                2 & 100.247 & 10.602 & 216.9 & (13.6,8.7) & 0.04 & 1.4 & 2.2 & 15.2933 & 807.2 & 19.0 & 94.0 & 12.16 & 2.3 & 0.2 & 1 & 1 & PS-C\\ 
                3 & 100.26 & 10.596 & 220.0 & (31.9,16.8) & 0.09 & 1.7 & 4.3 & 13.0406 & 344.3 & 16.9 & 57.4 & 31.91 & 5.3 & 0.2 & 1 & 1 & PS-C\\ 
                4 & 100.281 & 10.523 & 277.4 & (12.0,8.0) & 0.04 & 1.6 & 0.9 & 12.7943 & 319.9 & 16.4 & 48.1 & 6.77 & 2.0 & 0.3 & 1 & 1 & PS-C\\ 
                5 & 100.222 & 10.554 & 230.7 & (424.7,3.0) & 0.07 & 1.7 & 0.5 & 13.6768 & 290.9 & 11.3 & 31.0 & 3.39 & 3.9 & 1.1 & 0 & 0 & SL\\ 
                6 & 100.266 & 10.574 & 192.1 & (21.4,16.3) & 0.07 & 1.6 & 1.9 & 13.2184 & 329.8 & 14.9 & 28.9 & 14.12 & 3.9 & 0.3 & 1 & - & PS\\ 
                7 & 100.248 & 10.639 & 208.3 & (24.5,8.2) & 0.06 & 1.6 & 1.3 & 12.3008 & 211.6 & 13.4 & 31.1 & 10.61 & 3.6 & 0.3 & 1 & 1 & PS-C\\ 
                8 & 100.274 & 10.529 & 260.7 & (5.3,3.3) & 0.02 & 1.6 & 0.1 & 12.8667 & 252.0 & 12.8 & 38.1 & 1.14 & 0.9 & 0.8 & 0 & 0 & PRE\\ 
                9 & 100.234 & 10.553 & 357.3 & (6.0,5.3) & 0.02 & 1.7 & 0.2 & 14.0961 & 321.3 & 11.0 & 23.1 & 1.46 & 1.2 & 0.8 & 1 & 1 & PS-C\\ 
                10 & 100.246 & 10.533 & 259.3 & (6.1,5.3) & 0.02 & 1.7 & 0.2 & 13.6556 & 250.7 & 9.8 & 23.6 & 1.48 & 1.2 & 0.8 & 0 & 0 & PRE\\ 
                11 & 100.287 & 10.558 & 336.2 & (8.1,6.0) & 0.03 & 1.4 & 0.3 & 13.5932 & 229.8 & 9.1 & 20.1 & 1.85 & 1.5 & 0.8 & 0 & 0 & PRE\\ 
                \hline
                \multicolumn{17}{l}{Field G215.87-17.50}\\ 
                \hline
                1 & 88.384 & -10.419 & 235.0 & (17.7,10.9) & 0.03 & - & 1.3 & - & - & - & 21.7 & 1.77 & 1.7 & 1.0 & 1 & - & PS\\ 
                2 & 88.428 & -10.403 & 294.7 & (27.3,11.3) & 0.04 & - & 1.6 & - & - & - & 15.8 & 2.21 & 2.3 & 1.1 & 2 & - & PS\\ 
                3 & 88.432 & -10.395 & 350.8 & (13.1,7.5) & 0.02 & - & 0.5 & - & - & - & 14.5 & 0.73 & 1.2 & 1.6 & 0 & - & SL\\ 
                4 & 88.359 & -10.459 & 256.0 & (14.0,11.3) & 0.03 & - & 0.7 & - & - & - & 12.2 & 0.94 & 1.5 & 1.6 & 0 & - & SL\\ 
                5 & 88.401 & -10.42 & 247.6 & (9.6,5.8) & 0.02 & - & 0.2 & - & - & - & 8.3 & 0.28 & 0.9 & 3.3 & 1 & - & PS\\ 
                \hline 
        \end{tabular}
                \tablefoot{$N_{\rm YSO, \Delta C}$ refers to number of YSOs confirmed by \textit{Herschel} color. Clumps are SL, PRE, candidate PS, or PS-C. }
\end{sidewaystable*}

\section{\co\spc line graphs for individual clumps \label{sec:appendix_CO_line_graphs}}

\begin{figure}[h]
        \resizebox{\hsize}{!}{\includegraphics{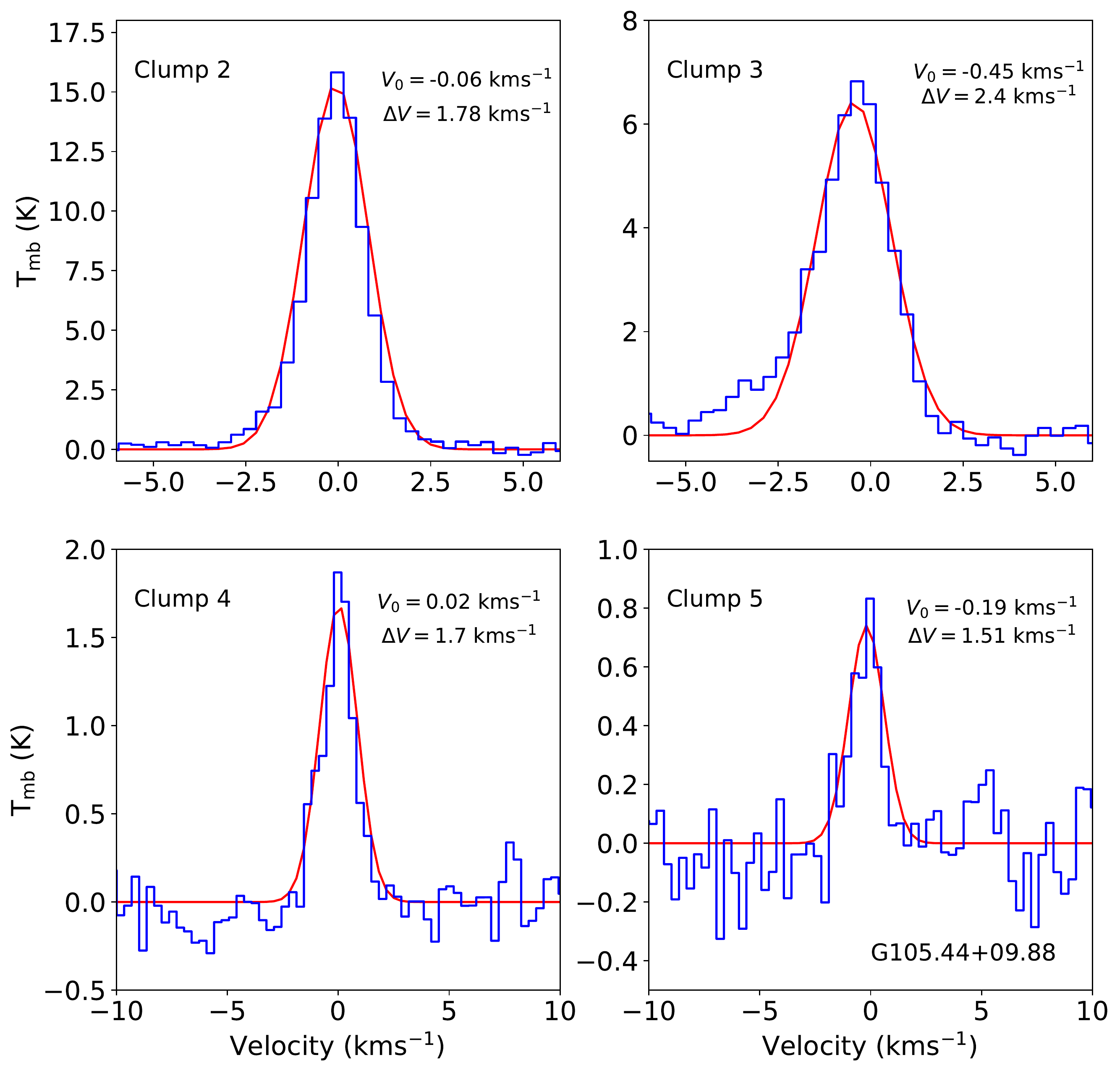}}
        \caption{\co\spc spectra for clumps 2, 3, 4, and 5 in field G105.44+09.88.  \label{fig:13CO_G105} }
\end{figure}

\begin{figure}[h]
        \resizebox{\hsize}{!}{\includegraphics{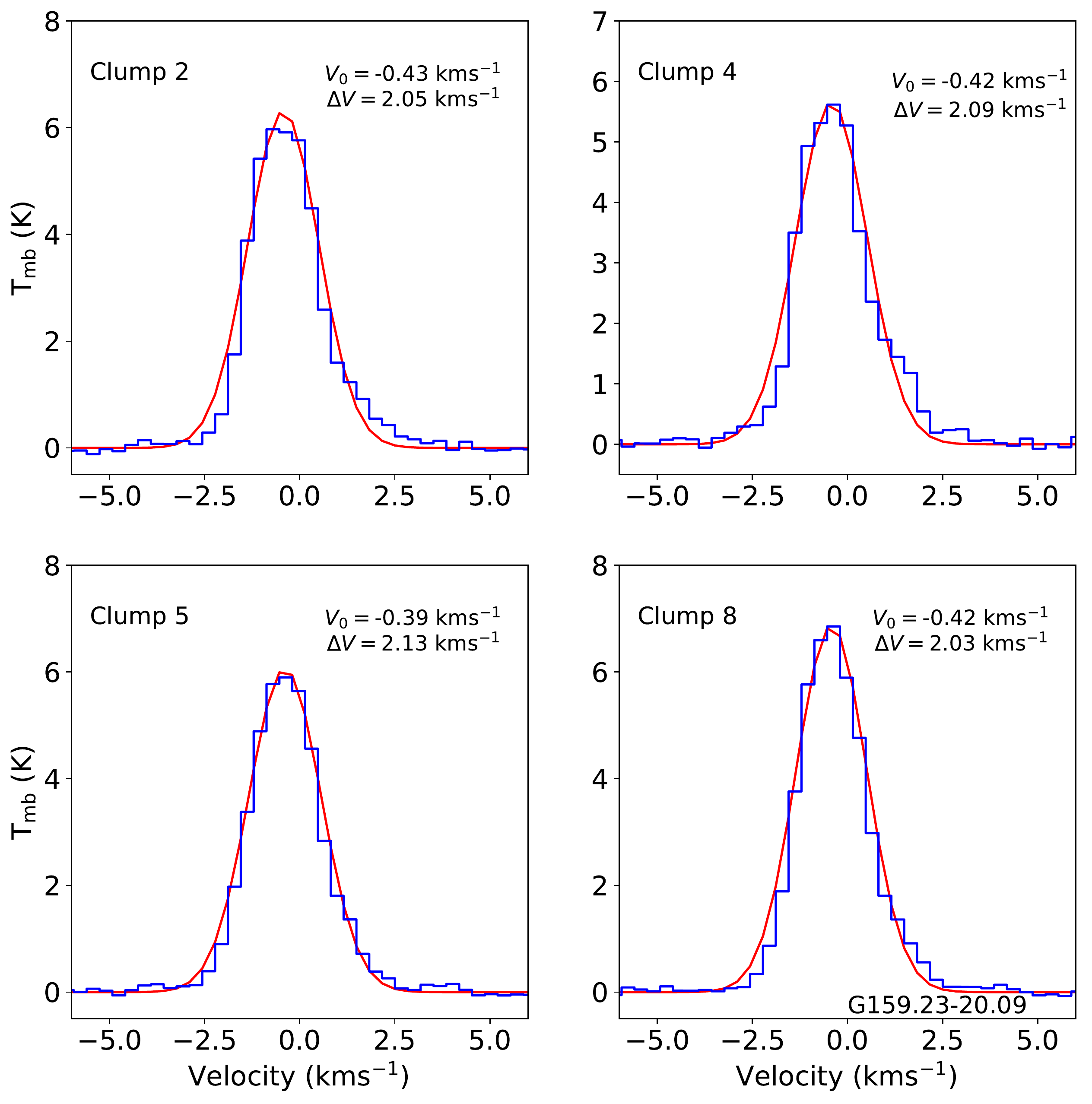}}
        \caption{\co\spc spectra for clumps 2, 4, 5, and 8 in field G159.23-20.09.  \label{fig:13CO_G159} }
\end{figure}

\begin{figure}[h]
        \resizebox{\hsize}{!}{\includegraphics{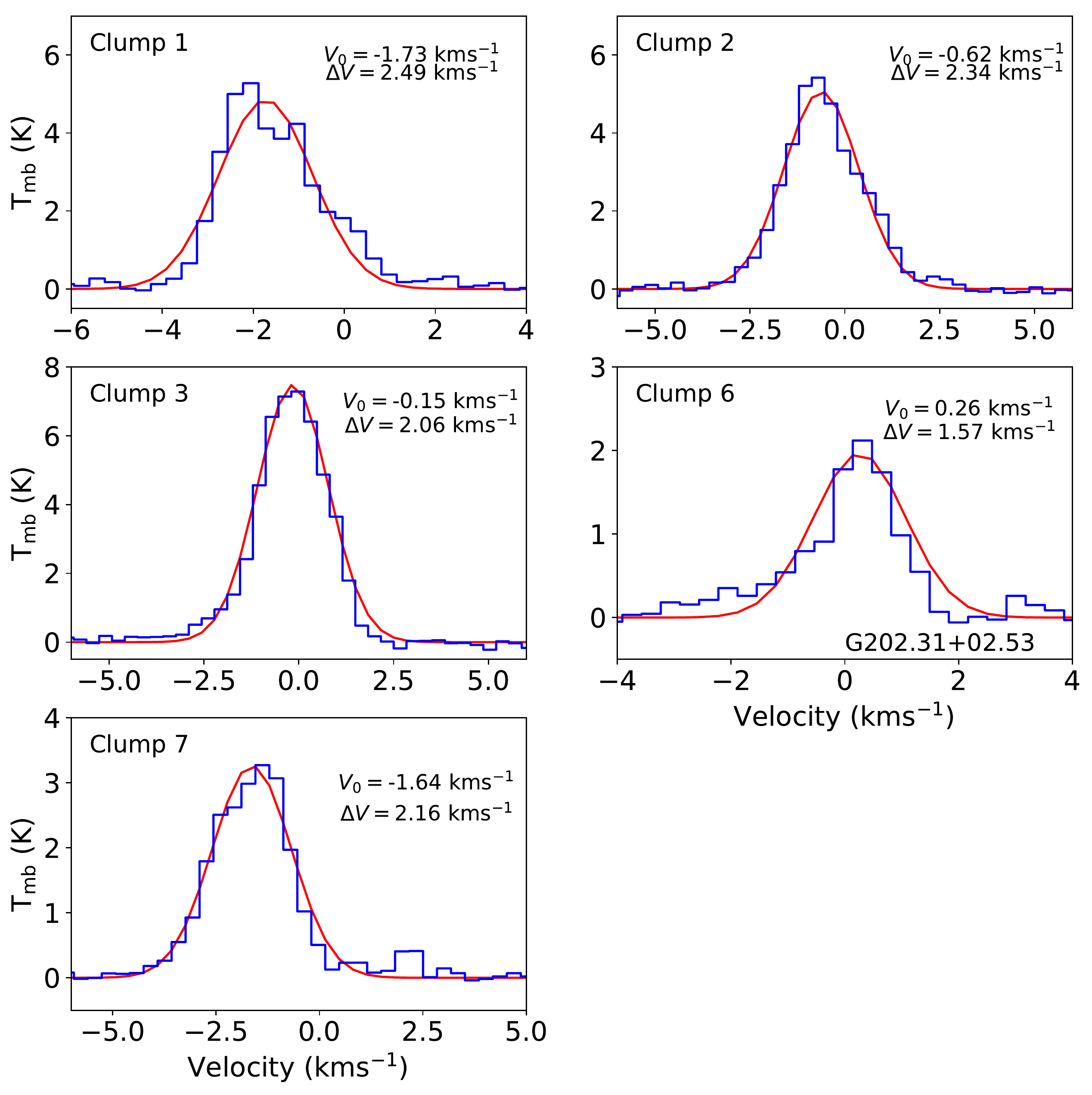}}
        \caption{\co\spc spectra for clumps 1, 2, 3, 6, and 7 in field G202.31+02.53.  \label{fig:13CO_G202} }
        \label{fig:13_CO_clumps}
\end{figure}

The \co\spc spectra of the 13 clumps in the fields G105.44+09.88, G159.23-20.09, and G202.31+02.53 are shown in Figs. \ref{fig:13CO_G105}-\ref{fig:13CO_G202}.  The radial velocity of the peak emission $v_{0}$ and the line widths $\Delta V$ are listed in the frames. 


\clearpage
\section{Clump data \label{sec:appendix_clump_data}}
\begin{figure*}
        \centering
        \includegraphics[width=17cm]{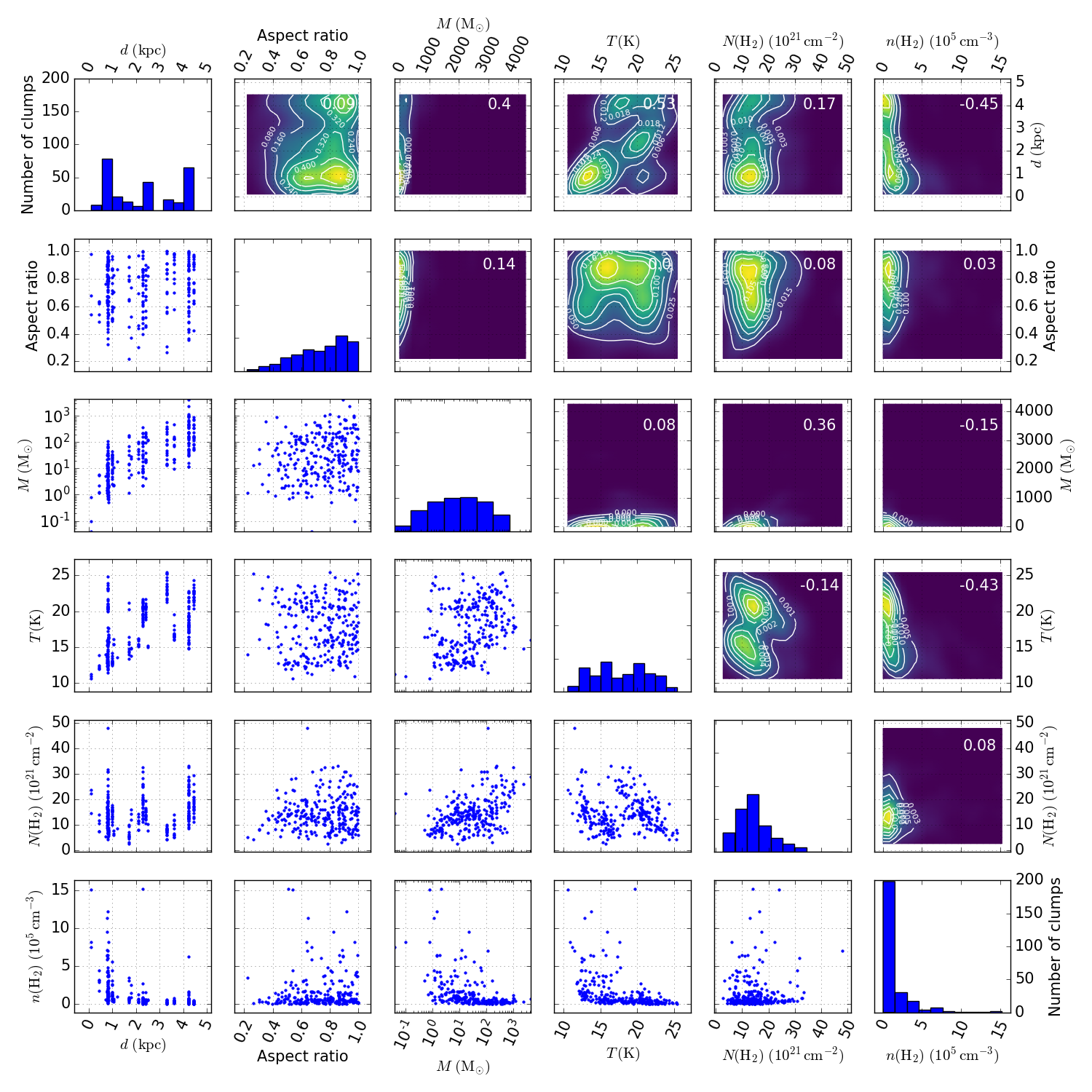}
        \caption{Histograms of clump distance, aspect ratio (minor/major axis), mass, temperature, column density \textit{N}(\MH), and number density \textit{n} are shown on the diagonal, with scatter plots relating each parameter to each other on the lower frames. The upper frames show a contour plot of the same relation; the correlation coefficient is quoted in the frame. These values are derived using SPIRE data with constant $\beta=1.8$. In this figure, 242 clumps with distance estimates and SPIRE data are included. Mass has a logarithmic scale for only the histogram and scatter plots.  \label{fig:clump_chars}}
\end{figure*}

\begin{figure}
        \resizebox{\hsize}{!}{\includegraphics{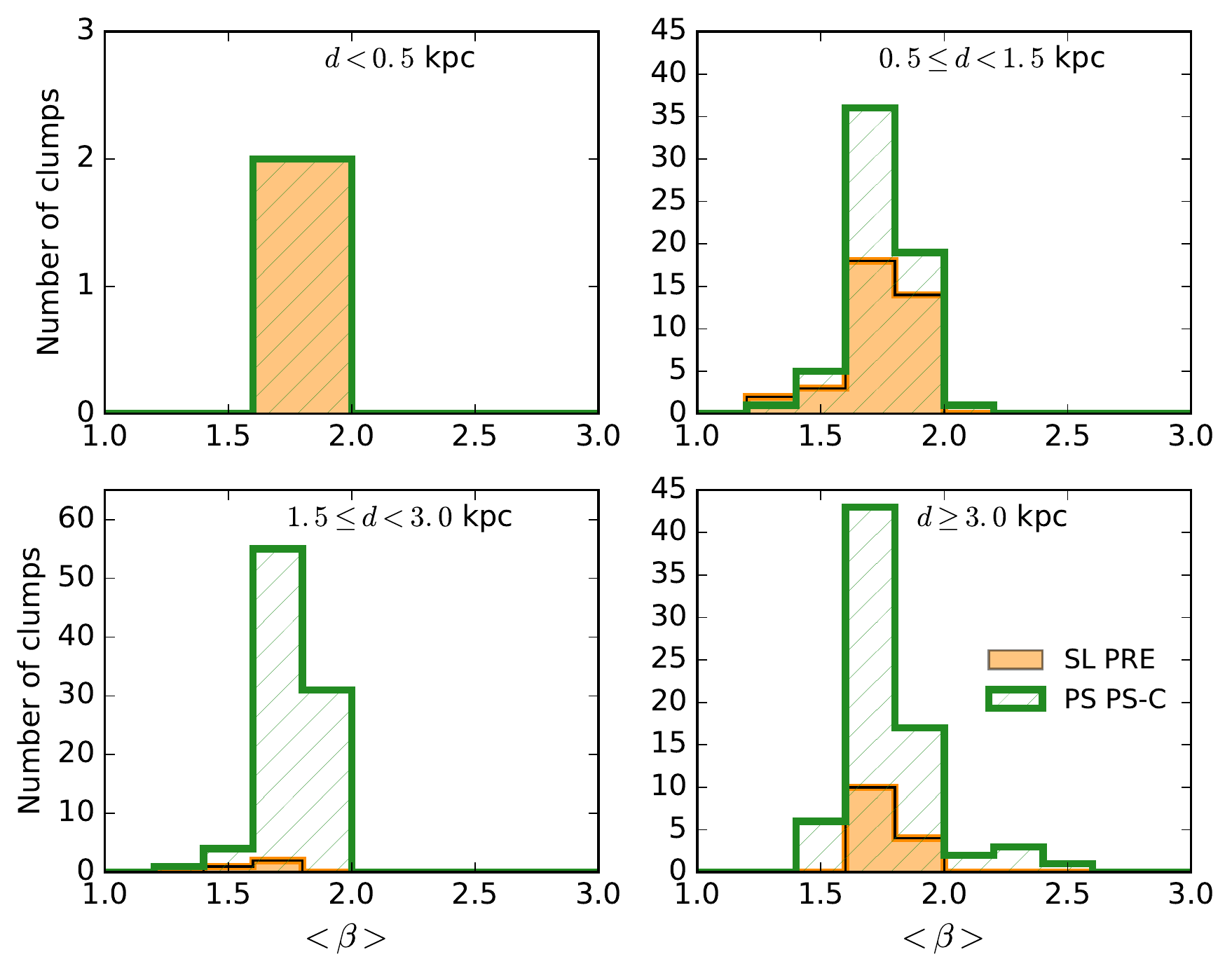}}
        \caption{Histograms for mean spectral index of all starless (SL PRE), and all protostellar (PS PS-C) clumps in distance bins of $d < 0.5$\,kpc, $0.5 \leq d < 1.5$\,kpc, $1.5 \leq d < 3.0$\,kpc, and $d \geq 3.0$\,kpc, written on the images.  Includes all 242 clumps with distance estimates and SPIRE data. The number of clumps in each bin is written in column 11 of Table \ref{tbl:starless_vs_prestellar_values}.   \label{fig:comparisons_beta}  }
\end{figure}

\begin{figure}
        \resizebox{\hsize}{!}{\includegraphics{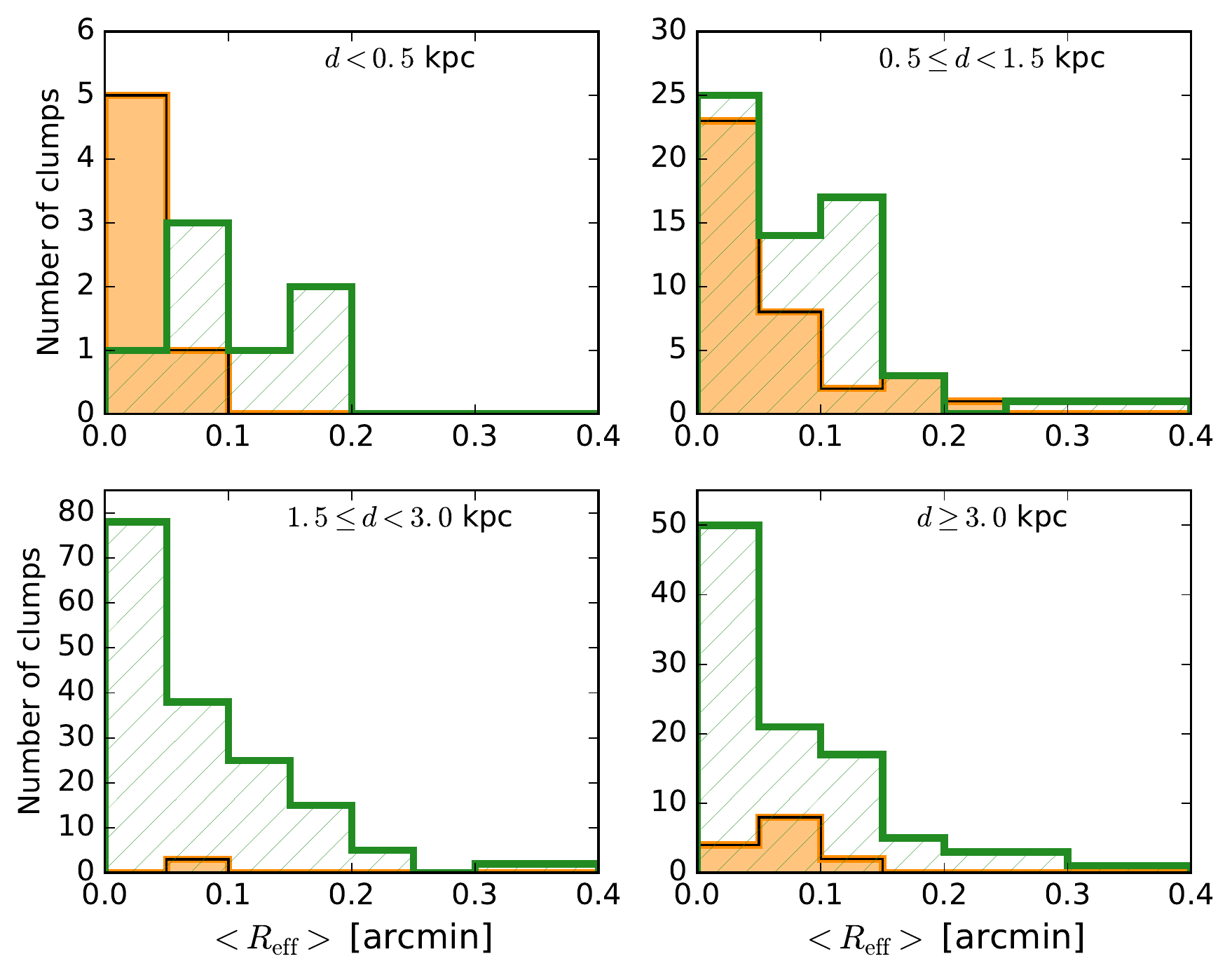}}
        \caption{Same as Fig. \ref{fig:comparisons_beta} for mean clump angular effective radius (Eq. (\ref{eq:R_eff})). Includes all 336 clumps with distance estimates. The number of clumps in each bin is written in column 11 of Table \ref{tbl:starless_vs_prestellar_values}.   \label{fig:comparisons_Reff}  }
\end{figure}

\begin{figure}
        \resizebox{\hsize}{!}{\includegraphics{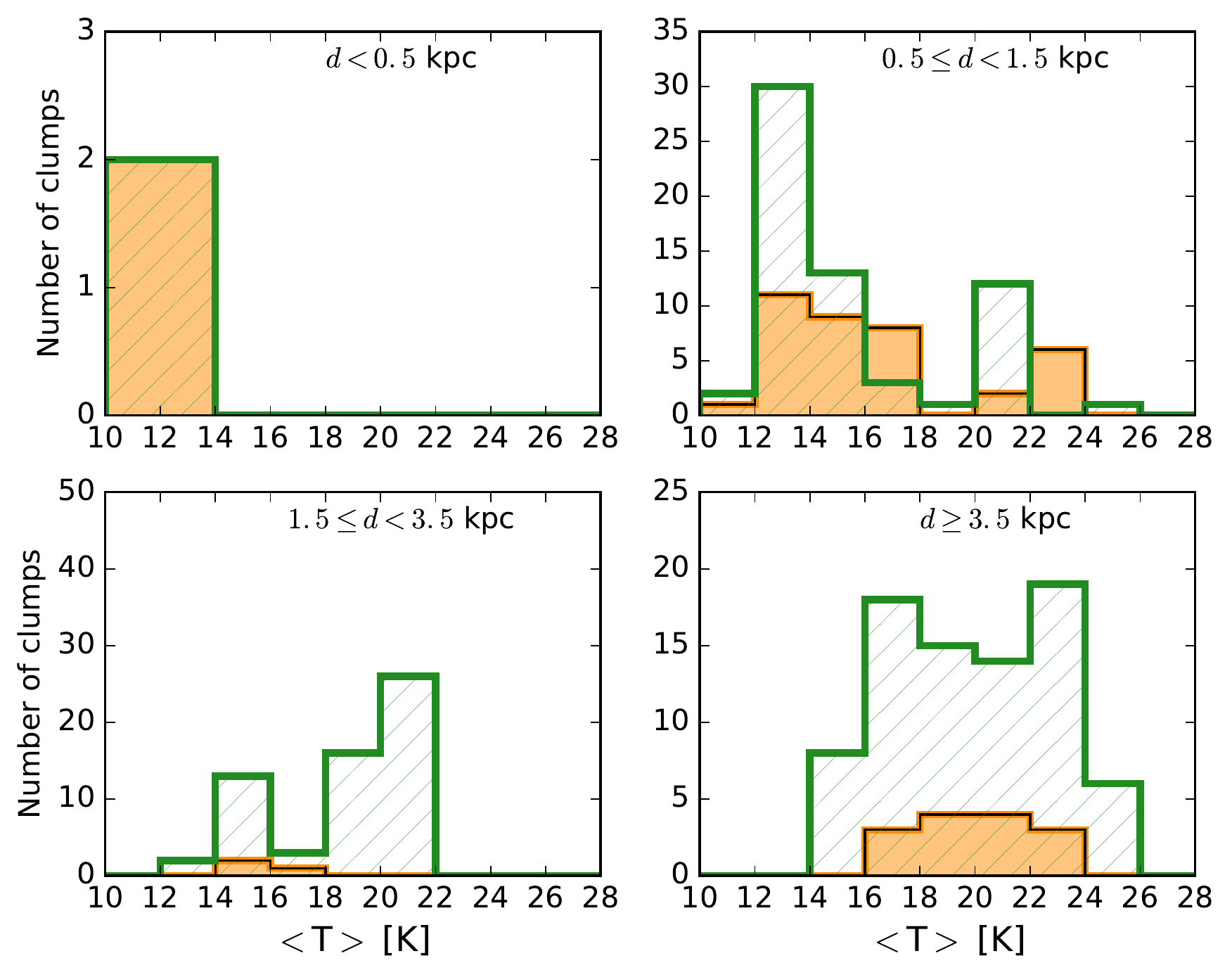}}
        \caption{Same as Fig. \ref{fig:comparisons_beta} for temperature. The number of clumps in each bin is written in column 10 of Table \ref{tbl:starless_vs_prestellar_values}.   \label{fig:comparisons_T}  }
\end{figure}

\begin{figure}
        \resizebox{\hsize}{!}{\includegraphics{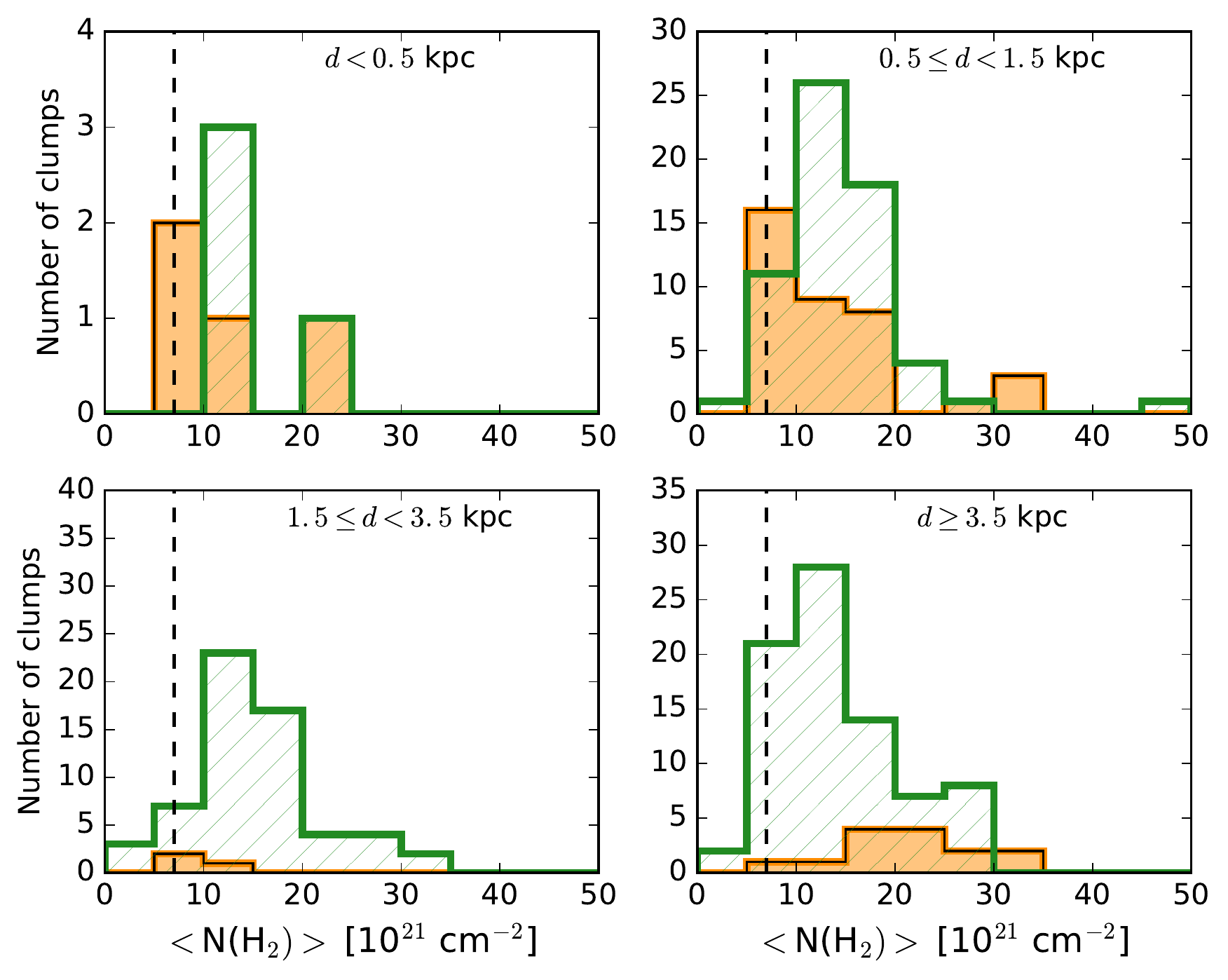}}
        \caption{Same as Fig. \ref{fig:comparisons_beta} for mean clump column density. The dotted line corresponds to $A_{\rm v}$ = 7. Includes all 242 clumps with distance estimates and SPIRE data. The number of clumps in each bin is written in column 10 of Table \ref{tbl:starless_vs_prestellar_values}.  \label{fig:comparisons_NH2}  }
\end{figure}

\begin{figure}
        \resizebox{\hsize}{!}{\includegraphics{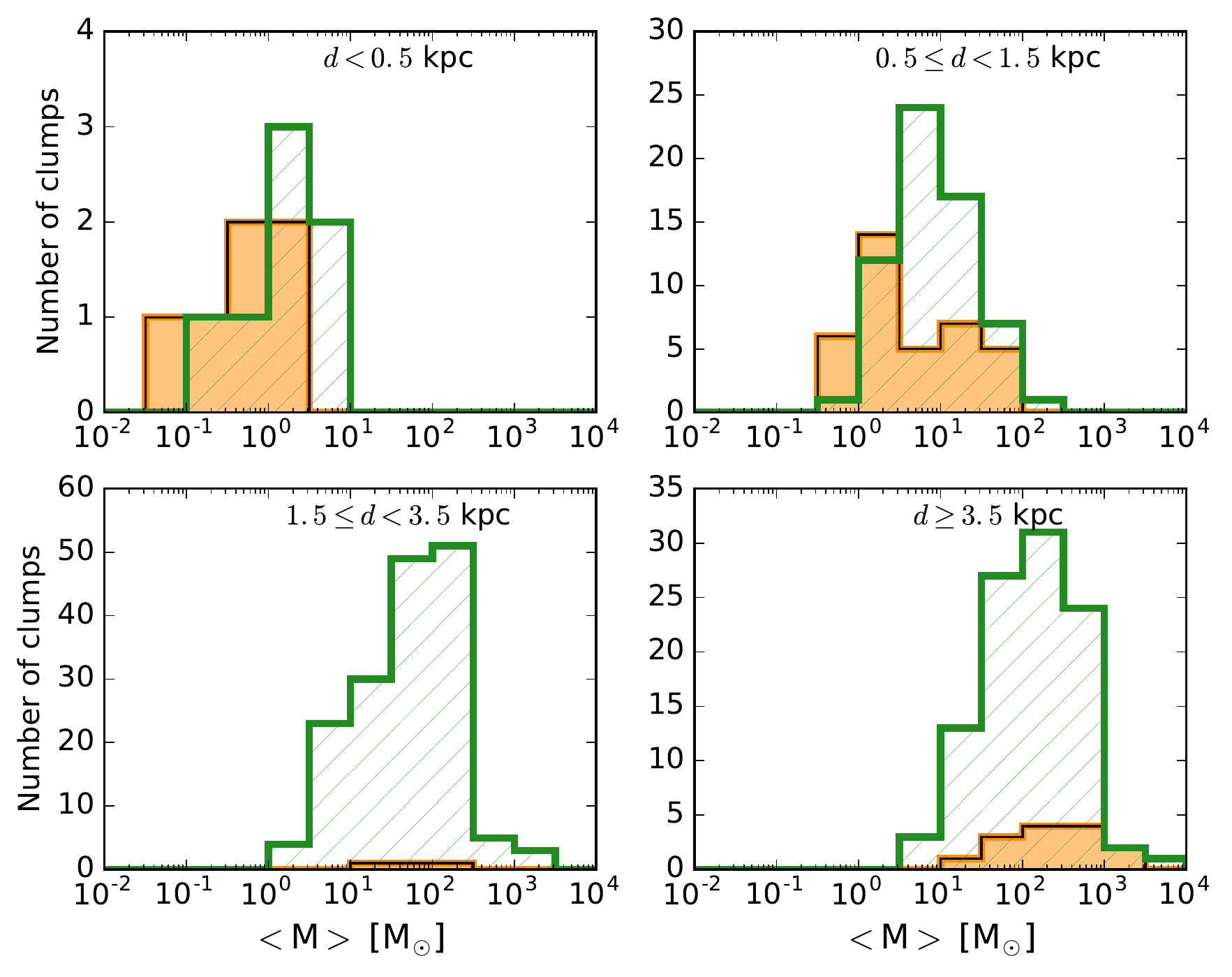}}
        \caption{Same as Fig. \ref{fig:comparisons_beta} for clump mass.  Includes all 336 clumps with distance estimates. The number of clumps in each bin is written in column 11 of Table \ref{tbl:starless_vs_prestellar_values}.   \label{fig:comparisons_M}  }
\end{figure}

\begin{figure}
        \resizebox{\hsize}{!}{\includegraphics{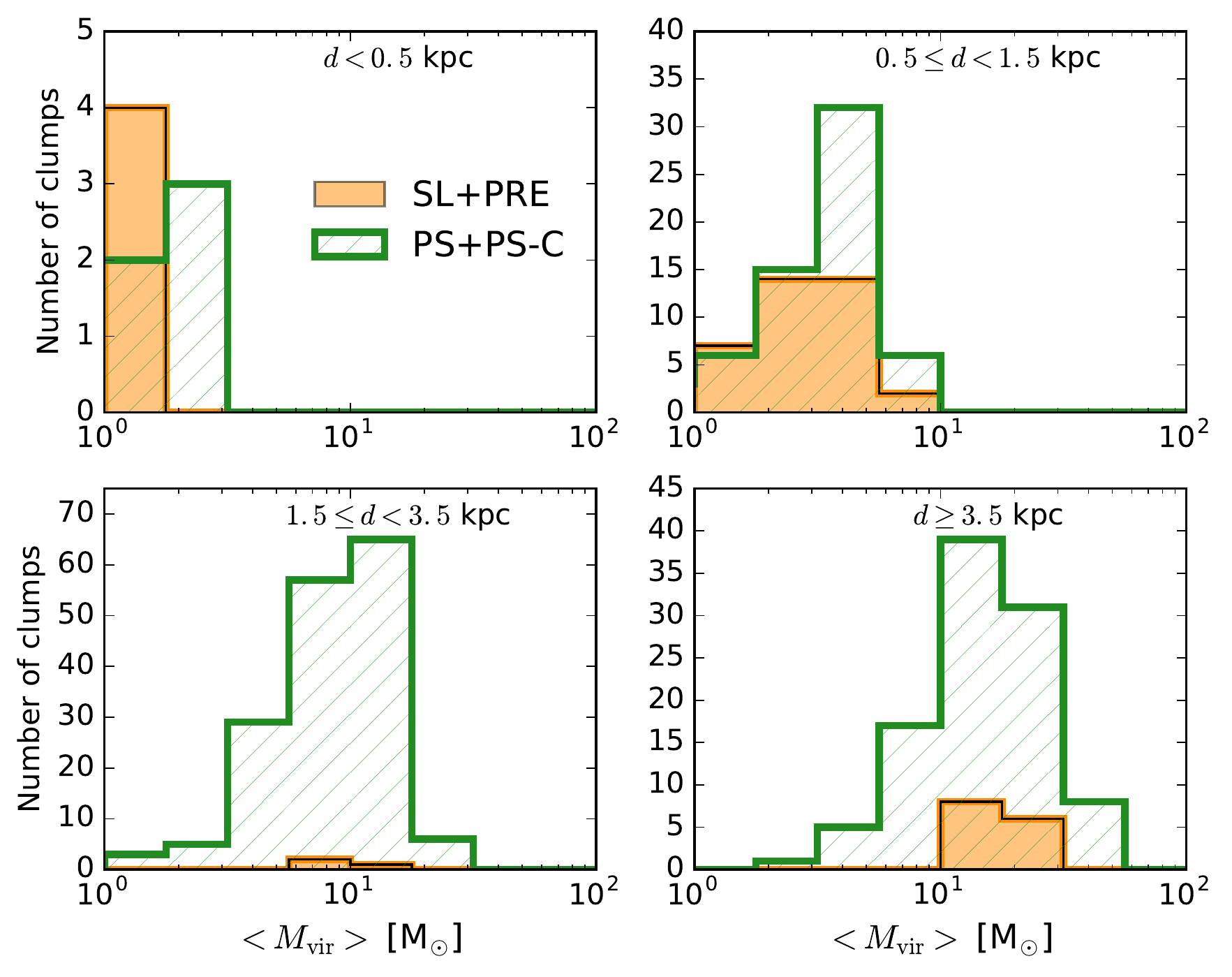}}
        \caption{Same as Fig. \ref{fig:comparisons_beta} for clump virial mass. Includes all 336 clumps with distance estimates. The number of clumps in each bin is written in column 11 of Table \ref{tbl:starless_vs_prestellar_values}.   \label{fig:comparisons_Mvir}  }
\end{figure}

In this appendix,we show further plots for the clump properties. Figure \ref{fig:clump_chars} shows a correlation plot including the clump distances, aspect ratios, masses, temperatures, column densities, and volume densities. Figures \ref{fig:comparisons_beta} -- \ref{fig:comparisons_Mvir}   show histograms for the individual clump properties, categorized according to the estimated source distances.


%
%
%\bibliographystyle{aa} % style aa.bst
%\bibliography{paper_bib} % your references Yourfile.bib
%

%\end{figure*}
%
% Online Material
%_____________________________________________________________
%        Online appendices have to be placed at the end, after
%                                        \end{thebibliography}
%-------------------------------------------------------------
%\end{thebibliography}

%\Online

%\begin{appendix} %First online appendix

%\begin{appendix} %Second online appendix
%These studies, however, have faced...
%\end{appendix}

\end{document}